\documentclass[12pt]{article}
\pdfoutput=1
\usepackage{color}
\usepackage{epsfig, palatino}
\usepackage{epic}
\usepackage{dsfont}
\usepackage{colonequals}
\usepackage{mathrsfs}
\usepackage{ae} 
\usepackage[T1]{fontenc}
\usepackage[utf8]{inputenc}
\usepackage{amsmath}
\usepackage{amssymb}
\usepackage{graphicx}
\usepackage[normalem]{ulem}
\usepackage{xcolor}
\definecolor{darkblue}{cmyk}{0.9,0.9,0,0}
\usepackage[colorlinks=true,linkcolor=darkblue,citecolor=darkblue,urlcolor=darkblue]{hyperref}
\usepackage{cite}
\usepackage{hyperref}
\usepackage{varioref}
\usepackage{makeidx}
\usepackage[english]{babel}
\usepackage{simplewick}
\usepackage{array}
\usepackage{multirow}
\usepackage[utf8]{inputenc}

\usepackage{tikzit}
\tikzstyle{point}=[fill=black, draw=black, shape=circle, inner sep=0pt, minimum size=3pt]
\tikzstyle{lr_point}=[fill=red, draw=red, shape=circle, inner sep=0pt, minimum size=2pt]
\tikzstyle{sh_point}=[fill={rgb,255: red,128; green,128; blue,128}, draw={rgb,255: red,128; green,128; blue,128}, shape=circle, inner sep=0pt, minimum size=3pt]
\tikzstyle{sh_lr_point}=[fill={red!40}, draw={red!40}, shape=circle, inner sep=0pt, minimum size=2pt, tikzit fill={rgb,255: red,255; green,128; blue,0}, tikzit draw={rgb,255: red,255; green,128; blue,0}]

\tikzstyle{dashed_st}=[-, dashed]
\tikzstyle{arrow}=[->]
\tikzstyle{dashed_grey}=[-, draw=black, dotted]
\tikzstyle{blue_line}=[draw=blue, ->]
\tikzstyle{energy}=[draw={rgb,255: red,255; green,128; blue,0}, decoration={{snake,amplitude=1pt,segment length=6pt,post length=1pt}}, decorate, ->]
\tikzstyle{blue_line_0}=[-, draw=blue]
\tikzstyle{redline}=[-, draw=red]
\tikzstyle{blue dot}=[scale=0.3, draw=blue, shape=circle, fill=blue]
\tikzstyle{textdot}=[shape=circle, scale=0.7]
\tikzstyle{point}=[fill=black, draw=black, shape=circle, inner sep=0pt, minimum size=3pt]
\tikzstyle{lr_point}=[fill=red, draw=red, shape=circle, inner sep=0pt, minimum size=2pt]
\tikzstyle{sh_point}=[fill={rgb,255: red,128; green,128; blue,128}, draw={rgb,255: red,128; green,128; blue,128}, shape=circle, inner sep=0pt, minimum size=3pt]
\tikzstyle{sh_lr_point}=[fill={red!40}, draw={red!40}, shape=circle, inner sep=0pt, minimum size=2pt, tikzit fill={rgb,255: red,255; green,128; blue,0}, tikzit draw={rgb,255: red,255; green,128; blue,0}]
\tikzstyle{blue_point}=[fill=blue, draw=blue, shape=circle, inner sep=0pt, minimum size=3pt]
\tikzstyle{circle dot}=[fill=white, draw=black, shape=circle, dashed, inner sep=0pt, minimum size=120pt]

\tikzstyle{new edge style 0}=[-, dashed]
\tikzstyle{arrow end}=[->]
\tikzstyle{dashed_st}=[-, dashed]
\tikzstyle{arrow}=[->]
\tikzstyle{dashed_grey}=[-, draw=black, dotted]
\tikzstyle{blue_line}=[draw=blue, ->]
\tikzstyle{energy}=[draw={rgb,255: red,255; green,128; blue,0}, decoration={{snake,amplitude=1pt,segment length=6pt,post length=1pt}}, decorate, ->]
\tikzstyle{blue_line_0}=[-, draw=blue]
\tikzstyle{redline}=[-, draw=red]
 
\usepackage{subcaption}

\newcommand{\cancel}{\sout}

\usepackage[font={small}]{caption}

\newcommand{\comment}[1]{}

\newcommand{\begBvR}[1]{\begin{#1}} 
\newcommand{\beq}{\begBvR{equation}}
\newcommand{\eeq}{\end{equation}}

\newcommand{\eeqq}{\end{equation*}}

\newcommand\eeqaa{\end{eqnarray*}}

\newcommand\eeqa{\end{array}}

\newcommand{\eea}{\end{eqnarray}}

\renewcommand{\Re}{\operatorname{Re}}

\newcommand{\nn}{\nonumber}

\newcommand{\diag}[1]{{\rm diag}(#1)} 
\newcommand{\neqa}{\nonumber\end{eqnarray}} 
\newcommand{\la}[1]{\label{#1}}

\renewcommand{\d}{\partial}

\newcommand{\<}{{\langle}}
\renewcommand{\>}{{\rangle}}

\newcommand{\re}{\relax{\rm I\kern-.18em R}}

\renewcommand{\sp}{p\hspace{-.40em}/}

\definecolor{darkgreen}{rgb}{0.0, 0.45, 0.0}
\definecolor{mathematicablue}{RGB}{94,130,182}

\def\XXint#1#2#3{{\setbox0=\hbox{$#1{#2#3}{\int}$}
\vcenter{\hbox{$#2#3$}}\kern-.5\wd0}}

\def\tr{{\rm tr~}}

\def\su2{{SU(2)}}

\def\eps{{\epsilon}}

\def\[{\left[}
\def\]{\right]}

\def\({\left(}
\def\){\right)}
\def\[{\left[}
\def\]{\right]}

\def\<{\langle}
\def\>{\rangle}

\def\i2{\frac{i}{2}}

\def\O{{\mathcal O}}

\def\spi{\relax{\rm \pi\kern-0.5em /}}
\def\sA{\relax{\rm A\kern-0.5em /}}
\def\sp{\relax{\rm p\kern-0.5em /}}
\def\sd{\relax{\rm \d\kern-0.5em /}}
\def\sk{\relax{\rm k\kern-0.5em /}}
\def\sn{\relax{\rm n\kern-0.5em /}}
\def\sl{\relax{\rm l\kern-0.5em /}}
\def\sP{\relax{\rm P\kern-0.7em /}}
\def\sBethe{\relax{\rm \Bethe\kern-0.5em /}}

\def\cO{{\cal O}}

\def\2F1{\,_2{\rm F}_1}

\def\I{{\rm I}}
\def\II{{\rm II}}
\def\III{{\rm III}}

\def\Re{{\rm Re}}

\makeindex

\def \la {\langle}
\def \ra {\rangle}
\def \pa {\partial}

\def \eps {\epsilon}

\begin{document}

\thispagestyle{empty}

\renewcommand{\thefootnote}{\fnsymbol{footnote}}
\setcounter{page}{1}
\setcounter{footnote}{0}
\setcounter{figure}{0}

\begin{center}
$$$$

{\Large\textbf{\mathversion{bold}
Nonperturbative Mellin Amplitudes: \\
Existence, Properties, Applications
}\par}
\vspace{1.0cm}

\vspace{1.0cm}

\textrm{ Joao Penedones$^\text{\tiny 1}$, Joao A. Silva$^\text{\tiny 1}$, Alexander Zhiboedov$^\text{\tiny 2}$}
\\ \vspace{1.2cm}
\footnotesize{\textit{ 
$^\text{\tiny 1}$
Fields and Strings Laboratory, Institute of Physics, \'Ecole Polytechnique F\'ed\'erale de Lausanne (EPFL)\\
 CH-1015 Lausanne,
Switzerland\\
$^\text{\tiny 2}$ CERN, Theoretical Physics Department, 1211 Geneva 23, Switzerland
}  
\vspace{4mm}
}

\par\vspace{1.5cm}

\textbf{Abstract}\vspace{2mm}
\end{center}

We argue that nonperturbative CFT correlation functions admit a Mellin amplitude representation. Perturbative Mellin representation readily follows. We discuss the main properties of nonperturbative CFT Mellin amplitudes:  subtractions, analyticity, unitarity, Polyakov conditions and polynomial boundedness at infinity.
Mellin amplitudes are particularly simple for large $N$ CFTs and 2D rational CFTs. We discuss these examples to illustrate our general discussion.
We consider subtracted dispersion relations for Mellin amplitudes and use them to derive bootstrap bounds on CFTs. We combine crossing, dispersion relations and Polyakov conditions to write down a set of extremal functionals that act on the OPE data. We check these functionals using the known 3d Ising model OPE data and other known bootstrap constraints. We then apply them to holographic theories. 
%

%

%
%
%
%

%

\noindent

\setcounter{page}{1}
\renewcommand{\thefootnote}{\arabic{footnote}}
\setcounter{footnote}{0}

\setcounter{tocdepth}{2}

 \def\nref#1{{(\ref{#1})}}

\newpage

\tableofcontents

\parskip 5pt plus 1pt   \jot = 1.5ex

\section{Introduction}

The basic observables in conformal field theory (CFT) are correlation functions of local operators
\beq
\label{eq:corrfunctions}
\langle \cO_1 (x_1) \dots \cO_n(x_n) \rangle \ .
\eeq
These correlation functions are constrained by conformal invariance and the associativity of the operator product expansion (OPE).  Recent studies of such correlation functions using conformal bootstrap methods and the AdS/CFT correspondence have experienced tremendous progress, see e.g  \cite{Poland:2018epd,Aharony:1999ti}. In particular, by studying nonperturbative consistency conditions of the correlation functions  (\ref{eq:corrfunctions}) we hope to get new insights into nonperturbative aspects of quantum gravity \cite{Heemskerk:2009pn,ElShowk:2011ag}. 

The main properties of conformal correlation functions are nicely encoded in the Mellin representation  \cite{Mack:2009mi} of the connected part of the correlation function
\begin{eqnarray}\label{def Mellin general}
\langle \cO_1 (x_1) \dots \cO_n(x_n) \rangle_{conn} = \int \[ d \gamma_{ij} \]  M(\gamma_{ij})
\prod_{i<j}^n \frac{\Gamma(\gamma_{ij})}{|x_i-x_j|^{2 \gamma_{ij}}}\,,
\end{eqnarray}
where $M(\gamma_{ij})$ is the {\it Mellin amplitude}, the integration variables obey the constraints
\beq
\sum_{j=1}^n \gamma_{ij}=0\,,\qquad
\gamma_{ij}=\gamma_{ji}\,,\qquad
\gamma_{ii}=-\Delta_i\, ,
\eeq
and the integration contours typically (but not necessarily) run parallel to the imaginary axis.

Mellin amplitudes have been particularly useful for holographic CFTs \cite{Penedones:2010ue,Fitzpatrick:2011ia,Paulos:2011ie,Yuan:2018qva,Rastelli:2017udc}, as well as in perturbative computations \cite{Symanzik:1972wj,Usyukina:1992jd,Belitsky:2013xxa,Belitsky:2013ofa}. One might wonder if Mellin space is particularly fit for bootstrapping holographic CFTs? We will present some evidence that it is indeed the case in the present paper. However, to start addressing this question we need first to ask a more basic question:

{\it When does the Mellin representation of the correlator (\ref{def Mellin general}) exist and what is the physical input that goes into it?}

It is a purpose of this paper to study this question in the context of a nonperturbative four-point function of scalar identical operators in a generic CFT. We will argue that the four-point function of light operators\footnote{We will be more specific about the precise meaning of that below.} admits the following representation
\begin{eqnarray}
\label{eq:correlatorIntro}
&&\la {\O}(x_1) {\O}(x_2)  {\O}(x_3) {\O}(x_4) \ra  ={ 1 \over \left( x_{12}^{2} x_{34}^{2} \right)^\Delta } + {1 \over  \left( x_{13}^{2} x_{24}^{2} \right)^\Delta}  + { 1 \over  \left( x_{14}^{2} x_{23}^{2} \right)^\Delta } + {F_{conn}(u,v) \over \left( x_{13}^{2} x_{24}^{2} \right)^\Delta } \nonumber , \\
&&F_{conn}(u, v) =  \int_{{\cal C}} {d \gamma_{12} d \gamma_{14} \over (2 \pi i)^2} \ u^{- \gamma_{12}} v^{- \gamma_{14} }\Gamma(\gamma_{12})^2 \Gamma(\gamma_{14})^2 \Gamma(\Delta - \gamma_{12} - \gamma_{14})^2 M (\gamma_{12}, \gamma_{14})\nonumber , \\
&&u = {x_{12}^2 x_{34}^2 \over x_{13}^2 x_{24}^2 }  \  , ~~~v  = {x_{14}^2 x_{23}^2 \over x_{13}^2 x_{24}^2 } \ ,
\end{eqnarray}
where the integration contour ${\cal C}$ will be specified below. In the case of the four-point function the Mellin amplitude (\ref{def Mellin general}) is a function of two independent variables $M(\gamma_{12}, \gamma_{14})$. Crossing symmetry simply becomes
\beq
\label{eq:crossingMellin}
M(\gamma_{12}, \gamma_{14}) = M(\gamma_{14}, \gamma_{12})  = M(\gamma_{12}, \gamma_{13}) , ~~~ \gamma_{12} + \gamma_{13} + \gamma_{14} = \Delta \ .
\eeq
As usual we expect the cases of non-identical scalar and spinning operators to be very similar, though we do not establish this rigorously in the present paper.

In fact, a certain version of the Mellin bootstrap approach \cite{Gopakumar:2016wkt,Gopakumar:2016cpb,Gopakumar:2018xqi} has already been used to reformulate and improve perturbative calculations. The ansatz for the correlation function that goes into this approach does not immediately follow from CFT axioms and it is not clear to us what is the range of applicability of these methods beyond certain perturbative examples. For that reason we follow a different path, where we develop bootstrap in Mellin space starting from the first principles.

As we will see the existence of nonperturbative Mellin amplitudes is not a given. In section \ref{sec:MelGen} we review the conditions for the existence of the Mellin amplitude which follow from its definition as an integral transform of the four-point function as a function of cross ratios $u$ and $v$. There are two basic properties that are needed for the existence of the Mellin amplitude, see section \ref{sec:MelGen} for details and definitions: 
\begin{itemize}
\item analyticity in a sectorial domain $({\rm arg}[u],{\rm arg}[v]) \in \Theta_{CFT}$, see section \ref{sec: analyticity};
\item polynomial boundedness, see (\ref{eq:polybound}) and appendix \ref{sec:polboundapp}.
\end{itemize}
As we will argue physical correlation functions have required analyticity properties but are not polynomial bounded in a required sense. Therefore, defining Mellin amplitudes always requires subtractions. The simplest possible subtraction involves considering the connected part of the correlation function as in the formula above. We systematically derive subtractions needed to define the nonperturbative Mellin amplitudes based on the general properties of the correlator. Essentially, it reduces to bounding the correlation function on the first and the second sheet in different limits which we do using the standard OPE techniques as well as some plausible but not fully rigorous assumptions. The main result of this analysis is that we establish (\ref{def Mellin general}) for a nonperturbative four-point function of light scalar operators in a generic CFT. 

Once existence of the nonperturbative Mellin amplitude is established we proceed to the study of its properties. This is the subject of sections \ref{sec: dif cont}, \ref{sec: unitarity and regge} and \ref{sec: polyakov}.  

In section \ref{sec: dif cont} we study analytic properties of Mellin amplitudes using a different approach. We define the Mellin amplitude as
\beq
M(\gamma_{12}, \gamma_{14}) = K(\gamma_{12}, \gamma_{14})+K(\gamma_{12}, \gamma_{13})+K(\gamma_{13}, \gamma_{14})\,,
\eeq
where the function $K$ is defined by the integral  \eqref{defK} of the four-point function. 
The integral defining $K(\gamma_{12}, \gamma_{14})$ converges for  $\Re\, \gamma_{12}>\Delta$ and $\Re\, \gamma_{14}>\Delta$. Therefore, there is no overlapping region where the 3 terms above are given by convergent integrals. 
In section \ref{sec: dif cont} we study the analytic continuation of the function $K$ and motivate the following conjecture:

\vspace{0.2cm}
\noindent {\bf Maximal Mellin Analyticity:} the only singularities of the nonperturbative Mellin amplitudes are the ones required by the OPE, namely (\ref{OPEpoles}). 
\vspace{0.2cm}

Notice that the  twist accumulation points (see section \ref{sec:twist} for a review) give rise to essential singularities of the Mellin amplitude where an infinite sequence of poles accumulates. This singularity is however of the type familiar from tree-level string theory amplitudes where it is located at infinity in the complex plane of the Mandelstam invariant $s$. We shall see in section \ref{sec: polyakov} that the analogy extends to the usual Reggeizetion of tree-level string amplitudes when $s\to \infty$.

In sections \ref{sec: unitarity and regge}  we review how unitarity, or the OPE, and Regge boundedness of the correlator constrain Mellin amplitudes. The OPE implies that the Mellin amplitude $M(\gamma_{ij})=M(\gamma_{12},\gamma_{13},\dots)$ has poles at 
\begin{eqnarray}
\label{OPEpoles}
\gamma_{ij}&=&\frac{\Delta_i+\Delta_j-\tau_k}{2}-m\,,
\qquad\qquad m=0,1,2,\dots , \nn \\
\tau_k &\equiv& \Delta_k-J_k \ ,
\end{eqnarray}
for each operator $\cO_k$, of dimension  $\Delta_k$ and spin $J_k$, that appears in the OPE $\cO_i \times \cO_j$.
The position of the poles is determined by the twist $\tau_k = \Delta_k-J_k$. Therefore, the analytic structure of the Mellin amplitude follows from the twist spectrum of the operators appearing in the OPE $\cO_i \times \cO_j$.
As we will review below, in a generic CFT the twist spectrum is very complicated (in $d\ge 3$ it contains infinite number of accumulation points \cite{Fitzpatrick:2012yx,Komargodski:2012ek} and for irrational $d=2$ CFTs it is continuous \cite{Collier:2016cls,Kusuki:2018wpa,Collier:2018exn,Benjamin:2019stq}).
Fortunately, in $d=2$ rational CFTs  and in the planar limit of large $N$ CFTs the twist spectrum is simpler and gives rise to meromorphic Mellin amplitudes.

The residues of the poles factorize into lower point Mellin amplitudes \cite{Fitzpatrick:2011ia}. In particular, the residues of the $\gamma_{12}$ poles of the four-point Mellin amplitude are given by
\begin{align}\ 
\label{eq:OPEintro}
M(\gamma_{ij}) &\approx {C_{12k}C_{34k} {\cal Q}_{J_k,m}^{\Delta_k}(\gamma_{13})\over \Delta_1+\Delta_2-\tau_k -   2 m -2\gamma_{12} } + ... \ , ~~~ m = 0,1,2,... \,,
\end{align}
where $C_{ijk}$ are OPE coefficients and  ${\cal Q}_{J,m}^{\Delta}(\gamma_{13})$ is a kinematical polynomial of degree $J$ in the variable $\gamma_{13}$. We review some basic properties of the Mack polynomials and point out some of their previously unnoticed positivity properties   which are very useful in  applications. For a more detailed and pedagogical introduction to Mellin amplitudes, see \cite{Penedones:2016voo} and references therein.

 We also translate the familiar Regge bounds of the four-point correlator to  statements about Mellin amplitudes. They become  statements about polynomial boundedness of the Mellin amplitude in the Regge limit \cite{Caron-Huot:2017vep, Maldacena:2015waa}
 \begin{eqnarray}
\label{eq:ReggeBoundIntro}
\lim_{|\gamma_{12}| \to \infty} | M(\gamma_{12}, \gamma_{14}) | \leq C(\gamma_{14}) |\gamma_{12}|, ~~~ {\rm Re}[\gamma_{14}] >  \Delta - {d-2 \over 2} \ . 
\end{eqnarray}
We establish this claim along the imaginary axis and assume that they hold in the whole complex plane.\footnote{Along the real axis they should be understood in a tauberian sense, as is common in the scattering amplitude literature.}  

In section \ref{sec: polyakov} we analyze twist accumulation points in Mellin space. We first explain how, despite the fact  we explicitly subtracted the disconnected piece in (\ref{eq:correlatorIntro}) and the statement that Mellin amplitude has singularities at the positions of all exchanged operators (except the identity operator), there is no problem of double counting upon closing the integration contour. We also examine how the statement about the absence of isolated operators with twist precisely $2 \Delta + 2n$ in the full nonperturbative CFT is encoded in the %
Mellin amplitude, see formula (\ref{eq:finalpolyakov}). 

To summarize: the nonperturbative Mellin amplitude $M(\gamma_{12}, \gamma_{14})$ (\ref{eq:correlatorIntro}) has simple analytic properties (\ref{eq:OPEintro}), it is bounded in the Regge limit (\ref{eq:ReggeBoundIntro}) and it satisfies nonperturbative Polyakov conditions (\ref{eq:finalpolyakov}).  

After the basic properties of the Mellin amplitudes are established we proceed to applications. In section   \ref{sec:application} we study dispersion relations with subtractions in Mellin space. We then impose crossing symmetry and the Polyakov condition. In this way we arrive at a set of analytic functionals that act on the OPE data and produce zero (\ref{eq:magicfunctionals}). These functionals have remarkable properties that we lay out, e.g. they are extremal functionals \cite{Poland:2010wg,ElShowk:2012hu}, and we use them to derive some bootstrap bounds. 

Finally, in section \ref{sec:MM}, we study Mellin amplitudes in 2D minimal models using the Coulomb gas formalism. We obtain expressions for any $n$-point Mellin amplitude and study their analytic structure, Regge limit and bulk point limit. We also check explicitly that correlators satisfy all the bounds discussed in section \ref{sec:MelGen}.

We end the paper with the summary of our results and assumptions, as well as open questions and many interesting directions to explore further.

\section{Nonperturbative Mellin Amplitudes}\label{sec:MelGen}

In this section we argue that a four-point correlation function of light scalar operators in a generic CFT admits a Mellin representation in the sense of (\ref{eq:correlatorIntro}). We start by reviewing basic facts about the multi-dimensional Mellin transform and a natural space of functions associated with it. These are functions analytic and polynomially bounded in a sectorial domain (see below for the precise definition). 

We then consider the Mellin transform of the four-point function of identical scalar primary operators in a generic CFT. We show that the Mellin transform has a physical interpretation of an integral over the principal Euclidean sheet, a connected space of conformally non-equivalent configurations for which all $x_{ij}^2$ are space-like.\footnote{Considered on a Lorentzian cylinder there are multiple loci of this type labeled by an integer which corresponds to the number of light-cone crossed starting from the Euclidean correlator. The principal Euclidean sheet corresponds to the one which contains the ordinary Euclidean correlator and does not involve any light-cone crossing.} Equipped with this understanding, we use OPE to identify the sectorial domain of analyticity of the physical correlator. We then argue (not fully rigorously) that upon appropriate subtractions the physical correlator also satisfies the required polynomial boundedness and therefore admits the Mellin representation (\ref{eq:correlatorIntro}).  

For simplicity, we consider the four-point function $\langle {\O} {\O} {\O} {\O} \rangle$ of equal scalar primary operators. Conformal symmetry restricts it as follows
\begin{eqnarray}
\label{eq:correlator}
&&\la {\O}(x_1) {\O}(x_2)  {\O}(x_3) {\O}(x_4) \ra  = {F(u,v) \over \left( x_{13}^{2} x_{24}^{2} \right)^\Delta } ,\nonumber \\
&&u = {x_{12}^2 x_{34}^2 \over x_{13}^2 x_{24}^2 } = z \bar z  \  , ~~~v  = {x_{14}^2 x_{23}^2 \over x_{13}^2 x_{24}^2 } =(1-z)(1-\bar z)\ ,
\end{eqnarray}
where $\Delta$ is the scaling dimension of $\O$ and the arbitrary function of cross ratios $F(u,v)$ satisfies crossing relations
\beq
F(u,v) =  F(v,u) = v^{-\Delta} F\left({u \over v}, {1 \over v}\right) \ .
\label{crossing}
\eeq
Unitarity implies that $\Delta\geq {d-2 \over 2}$.

\subsection{Two-dimensional Mellin Transform}
\label{sec:inversionMellin}

Here we review inversion theorems of the two-dimensional Mellin transform relevant for the four-point function. The generalization to an arbitrary number of dimensions is straightforward and the corresponding theorems and proofs can be found for example in \cite{Antipova2}. 

Consider a two-variable function $g(u,v)$. Its Mellin transform is defined as 
\beq
\mathcal{M}[g](\gamma_{12}, \gamma_{14}) \equiv \int_0^{\infty} {d u d v \over u v} \ u^{\gamma_{12}} v^{\gamma_{14}} g(u,v).
\label{mellindef}
\eeq 
Similarly, consider a two-variable function $\hat M(\gamma_{12}, \gamma_{14})$. We define the inverse Mellin transform as
\beq
\label{eq:inversemellin}
\mathcal{M}^{-1}[\hat M](u,v) = \int_{c - i \infty}^{c + i \infty} {d \gamma_{12} d \gamma_{14} \over (2 \pi i)^2} \ u^{- \gamma_{12}} v^{- \gamma_{14} } \hat M (\gamma_{12}, \gamma_{14}) .
\eeq   
There are two natural vector spaces of functions associated with these transforms: $M_{\Theta}^U$ and $W^\Theta_U$. Let us define them.

We say that function $g(u,v)$ belongs to the vector space of functions $M_{\Theta}^U$ if two conditions are satisfied. First, it is holomorphic in a sectorial domain $({\rm arg}[u], {\rm arg}[v]) \in \Theta \subset {\mathbb R}^2$ which we assume to be open and bounded, as well as to include the origin $(0,0) \in \Theta$. Second, in the region of holomorphy $g(u,v)$ obeys
\begin{eqnarray}
\label{polbound}
| g(u,v) | \leq C(c_u, c_v) {1 \over |u|^{c_u}} {1 \over |v|^{c_v}} , ~~~ ({\rm arg}[u], {\rm arg}[v]) \in \Theta, ~~~ (c_u , c_v) \in U \ ,
\end{eqnarray}
for some open region $U \in {\mathbb R}^2$. A typical example will be $a_u < c_u < b_u $ and $a_v < c_v < b_v$. 

Similarly, we say that $\hat M(\gamma_{12}, \gamma_{14})$ belongs to a vector space of functions $W^\Theta_U$ if two conditions are satisfied. First, $\hat M(\gamma_{12}, \gamma_{14})$ is holomorphic in a tube $U + i {\mathbb R}^2$ for some open region $U \in {\mathbb R}^2$. Second, in the holomorphic tube it decays exponentially fast in the imaginary directions
\begin{eqnarray}
\label{condition}
| \hat M(\gamma_{12}, \gamma_{14}) | \leq K({\rm Re}[\gamma_{12}], {\rm Re}[\gamma_{14}]) e^{- {\rm sup}_{\Theta} \left( {\rm arg}[u] {\rm Im}[\gamma_{12}] + {\rm arg}[v] {\rm Im}[\gamma_{14}]  \right)}, ~~~  (\gamma_{12}, \gamma_{14}) \in U + i {\mathbb R}^2.
\end{eqnarray}

Having introduced $M_{\Theta}^U$ and $W^\Theta_U$, we then have the following theorems \cite{Antipova,Antipova2}:

\vspace{0.3cm}

{\bf Theorem I:} Given $F(u,v) \in M_{\Theta}^U$, its Mellin transform $\mathcal{M}[F](\gamma_{12},\gamma_{14})$ exists and is in $W^\Theta_U$. Moreover, $\mathcal{M}^{-1}\mathcal{M}[F](u,v) = F(u,v)$ for any $ ({\rm arg}[u], {\rm arg}[v]) \in \Theta$.

\vspace{0.3cm}

{\bf Theorem II:} Given $\hat M(\gamma_{12},\gamma_{14}) \in W^{\Theta}_U$, its inverse Mellin transform $\mathcal{M}^{-1}[\hat M](u,v)$ exists and is in $M_\Theta^U$. Moreover, $\mathcal{M} \mathcal{M}^{-1}[\hat M](\gamma_{12},\gamma_{14}) = \hat M(\gamma_{12},\gamma_{14})$ for any $(\gamma_{12}, \gamma_{14}) \in U + i {\mathbb R}^n$.

\vspace{0.3cm}

Note that we are not saying that the function has to be in $M_{\Theta}^U$ to admit a Mellin representation,\footnote{As an example consider $u^{a} v^b \theta(0<u<1) \theta(0<v<1)$ which is obivously not holomorphic but it has a well-defined Mellin amplitude. We will use this loophole to define the intermediate objects that we call the $K$-functions in section \ref{sec: dif cont}.} but we will see that  $M_{\Theta}^U$ are $W^\Theta_U$ are indeed the relevant classes for the physical CFT correlators.

Since the discussion might look a little too abstract let us consider a couple of one-dimensional examples that illustrate the application of these theorems. The first example is $g(u) = e^{-u}$. This is an entire function which is polynomially bounded for $- {\pi \over 2} < {\rm arg}[u] < {\pi \over 2}$ for $c_u > 0$. According to the theorem I, the Mellin transform of $g(u)$ exists and decays as $e^{- {\pi \over 2} |{\rm Im}[\gamma_{12}]|}$, where we used (\ref{condition}). Indeed, the Cahen-Mellin integral takes the form
\beq
e^{-u} =  \int_{c_u - i \infty}^{c_u + i \infty} {d \gamma_{12}  \over 2 \pi i} \Gamma(\gamma_{12}) u^{-\gamma_{12}} , ~~~ c_u > 0,  ~ | {\rm arg}[u] | < {\pi \over 2}\ .
\eeq
It is a well-known fact that in the complex plane $\Gamma(\gamma_{12}) \sim e^{- {\pi \over 2} |{\rm Im}[\gamma_{12}]|}$ in agreement with the theorem above.

The second example is $g(u) = {1 \over (1 + \sqrt{u})^\alpha}$. Note that it is analytic for $|{\rm arg}[u] | < 2 \pi$ and satisfies 
(\ref{polbound}) for $0 < c_u < {\alpha \over 2}$. Using theorem I, we conclude that the Mellin transform of $g(u)$ has to decay as $e^{- 2 \pi |{\rm Im}[\gamma_{12}]|}$. The explicit formula takes the form
\beq
 {1 \over (1 + \sqrt{u})^\alpha} = \int_{c_u - i \infty}^{c_u + i \infty} {d \gamma_{12}  \over 2 \pi i} {2 \Gamma(2 \gamma_{12}) \Gamma(\alpha - 2 \gamma_{12}) \over \Gamma(\alpha) }u^{-\gamma_{12}} , ~~~ 0 < c_u < {\alpha \over 2},  ~ | {\rm arg}[u] | <  2 \pi .
\eeq
Again one can explicitly check that the Mellin amplitude decays with an expected exponential rate dictated by the analyticity region of the original function.

We show below that the physical CFT correlators are indeed analytic in a certain sectorial domain. However, they do not satisfy (\ref{polbound}) and therefore the Mellin transform of the full correlation function does not exist. We claim, however, that it does exist upon doing simple subtractions.\footnote{This is true in a generic, interacting CFT. We will discuss some special cases like $2d$ minimal models and higher $d$ free theories separately.} These subtractions are designed in such a way that they do not spoil the analytic structure of the correlator and at the same time they make it polynomially bounded in the sense described above.
The result of this analysis is (\ref{eq:correlatorIntro}). We will see that performing the subtractions beyond the disconnected part is equivalent\footnote{Up to some very special situations that we analyze separately.} to introducing the deformed Mellin contour ${\cal C}$.

\subsection{Principal Euclidean Sheet}\label{sec: integration}

Let us understand better the physical meaning of the integration region in \eqref{mellindef}. By conformal transformations, we can put the four points on a $2d$ Lorentzian plane with coordinates $(t^M, x^M)$. Furthermore, by performing conformal transformations, we put point $1$ at $(0, 0)$, point $3$ at $(0, 1)$ and point $4$ at $(0, \infty)$. The position of point $2$ is not fixed and we denote it by $(t^M_{2}, x^M_{2})$.

We map the plane to the Lorentzian cylinder. On the cylinder, we use coordinates $(t, \phi)$, which are related to coordinates on the plane by
\begin{eqnarray}
t^M= \frac{\sin t}{\cos t + \cos \phi}, ~~~ x^M= \frac{\sin \phi}{\cos t + \cos \phi }.
\end{eqnarray}
The metric on the cylinder is related to the metric on the plane by $- (dt^M)^2 + (dx^M)^2 = \frac{- dt^2 + d\phi^2}{(\cos t + \cos \phi)^2}$, so they are indeed conformal to one another. The cross ratios are given by
\begin{eqnarray}\label{crossratiosspecial}
u = - (t^M_{2})^2 + (x^M_{2})^2 = {\cos t_2 - \cos \phi_2 \over \cos t_2 + \cos \phi_2}, \\
v= - (t^M_{2})^2 + (x^M_{2}-1)^2 = 2 {\cos t_2 - \sin \phi_2 \over \cos t_2 + \cos \phi_2}.
\end{eqnarray}
Points $1$, $3$ and $4$ are mapped respectively to $(0, 0)$, $(0, \frac{\pi}{2})$ and $(0, \pi)$ on the cylinder. Note that cross ratios are invariant under $t_2 \to - t_2$.

\begin{figure}
  \centering
  \includegraphics[width=0.4\textwidth]{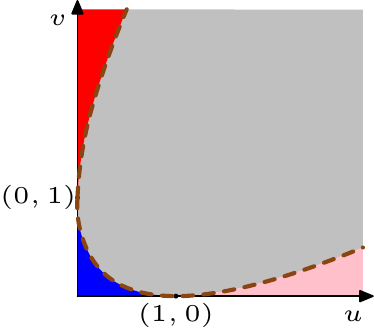}
  \caption{We divide the region $u, v >0$ into $4$ regions coloured in blue, pink, red and grey. The regions are separated by the curves $v=(1-\sqrt{u})^2$ and $v=(1+\sqrt{u})^2$ or, equiavelntly, $z = \bar z$. In the grey region $z$ and $\bar{z}$ are the complex conjugate of each other. 
  In the colored regions $z$ and $\bar{z}$ are real and independent variables. In the red region we have that $z, \bar{z} \in (-\infty, 0)$. In the blue and pink regions we have that $z, \bar{z} \in (0,1)$ and $z, \bar{z} \in (1,\infty)$ respectively.}
  \label{fig:u v plane}
\end{figure}

\begin{figure}[t!]
\centering
 \includegraphics[width=0.5\textwidth]{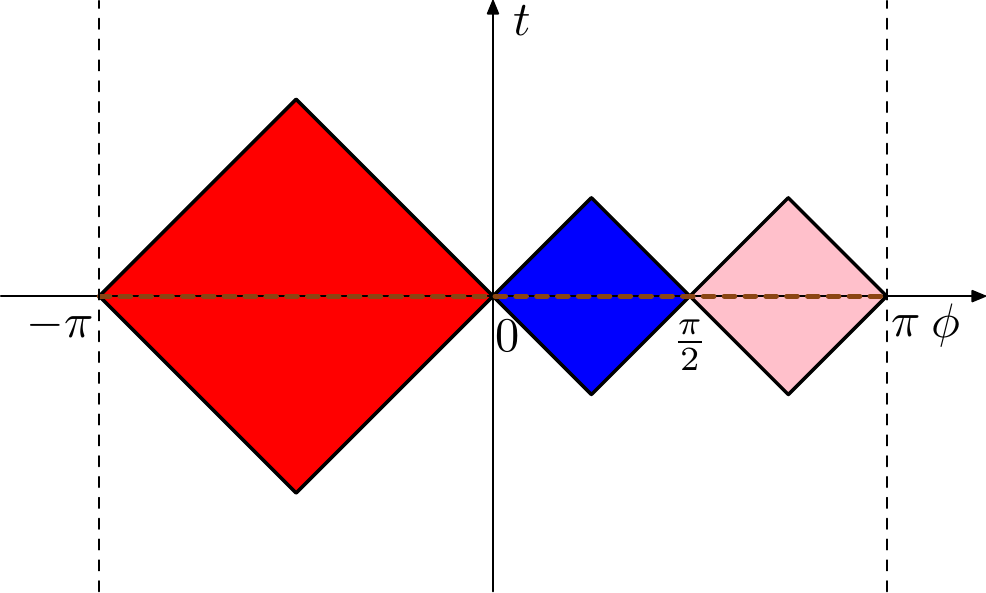}
  \caption{Cylinder picture of points $1$, $3$ and $4$. The point at $(t=0, \phi=\pi)$ should be identified with the point at $(t=0, \phi=-\pi)$. The colored area signifies regions where point $2$ is spacelike separated from three other points and no light-cones has been crossed. Note that the colored region is a double cover in the cross ratio space. Indeed changing $t_2 \to - t_2$ does not change the cross ratios (\ref{crossratiosspecial}).
  }
  \label{fig:cylinder}
\end{figure}

A simple observation is that the region $0 \leq u,v \leq \infty$ is  the region of Minkowski space (or of Lorentzian cylinder) of spacetime dimension $d\ge 3$,  for which the point $x_2$ is space-like separated from the three other points. Indeed, it is a well-known fact that for $u,v>0$ and four points in Euclidean space, cross ratios satisfy $(1-\sqrt{u})^2 \leq v \leq (1 + \sqrt{u})^2$, see figure \ref{fig:u v plane}. The rest of the quadrant $u,v>0$ is covered by the fully spacelike configurations on the $2d$ Lorentzian cylinder depicted in  figure \ref{fig:cylinder}. This fact allows us to use the OPE and analyze convergence of the integral in different regions. It will be natural for us to split the integral into three regions, each of which containing the region between the light cones emitted from points $x_1$,  $x_3$ and $x_4$  (see figure \ref{fig:cylinder}) and part of the Euclidean domain. The regions are defined as follows

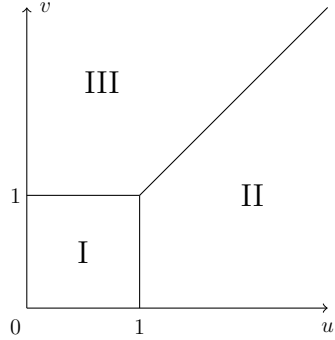
\begin{figure}[t!]
\centering
\begin{tikzpicture}
	\begin{pgfonlayer}{nodelayer}
		\node [style=none] (0) at (0, 0) {};
		\node [style=none] (1) at (0, 4) {};
		\node [style=none] (2) at (4, 0) {};
		\node [style=none] (3) at (1.5, 0) {};
		\node [style=none] (4) at (0, 1.5) {};
		\node [style=none] (5) at (1.5, 1.5) {};
		\node [style=none] (6) at (4, 4) {};
		\node [style=none] (7) at (0.75, 0.75) {${\rm I}$};
		\node [style=none] (8) at (3, 1.5) {${\rm II}$};
		\node [style=none] (9) at (1, 3) {${\rm III}$};
		\node [style=textdot] (10) at (4, -0.25) {$u$};
		\node [style=textdot] (11) at (0.25, 4) {$v$};
		\node [style=textdot] (12) at (-0.15, -0.25) {$0$};
		\node [style=textdot] (13) at (1.5, -0.25) {$1$};
		\node [style=textdot] (14) at (-0.15,1.5) {$1$};
	\end{pgfonlayer}
	\begin{pgfonlayer}{edgelayer}
		\draw [style=arrow end] (0.center) to (1.center);
		\draw [style=arrow end] (0.center) to (2.center);
		\draw (3.center) to (5.center);
		\draw (5.center) to (4.center);
		\draw (5.center) to (6.center);
	\end{pgfonlayer}
\end{tikzpicture}
  \caption{Different regions in the $(u,v)$ plane that we will find convenient to consider. Different regions are mapped into each other by crossing.}
  \label{fig:uvregions}
\end{figure}
\begin{eqnarray}
\label{regions}
&{\rm Region ~ I}: ~~~ &0 \leq u,v \leq 1 \ , \nonumber \\
&{\rm Region ~ II}: ~~~ &1 \leq u, ~ 0 \leq v \leq u \ , \nonumber \\
&{\rm Region ~ III}: ~~~ &1 \leq v, ~ 0 \leq u \leq v \ .
\end{eqnarray}
These three regions are mapped to each other via crossing transformations (\ref{crossing}).

\subsection{Analyticity in a Sectorial Domain}\label{sec: analyticity}

Let us study the analytic properties of $F(u,v)$ in the $(u,v)$ plane. Analyticity for real and positive $u,v >0$ is obvious from the discussion above. Indeed, for such cross ratios the correlation function describes a generic configuration of space-like separated operators, whereas non-analyticities of correlation functions can only occur when ``something happens'' \cite{Maldacena:2015iua}, say a pair of two points become light-like separated. A more rigorous argument relies on the exponential convergence of the OPE in CFTs which makes analyticity manifest \cite{Pappadopulo:2012jk}. 

To discuss Mellin amplitudes however we need to understand analytic properties of correlation functions in a sectorial domain, namely we would like to allow for ${\rm arg}[u], {\rm arg}[v] \neq 0$. 
In other words, consider the correlation function $F(|u| e^{i {\rm arg}[u]}, |v| e^{i  {\rm arg}[v]})$ with $|u|, |v| \in (0, \infty)$.

\begin{figure}[h]
  \centering
  \includegraphics[width=0.5\textwidth]{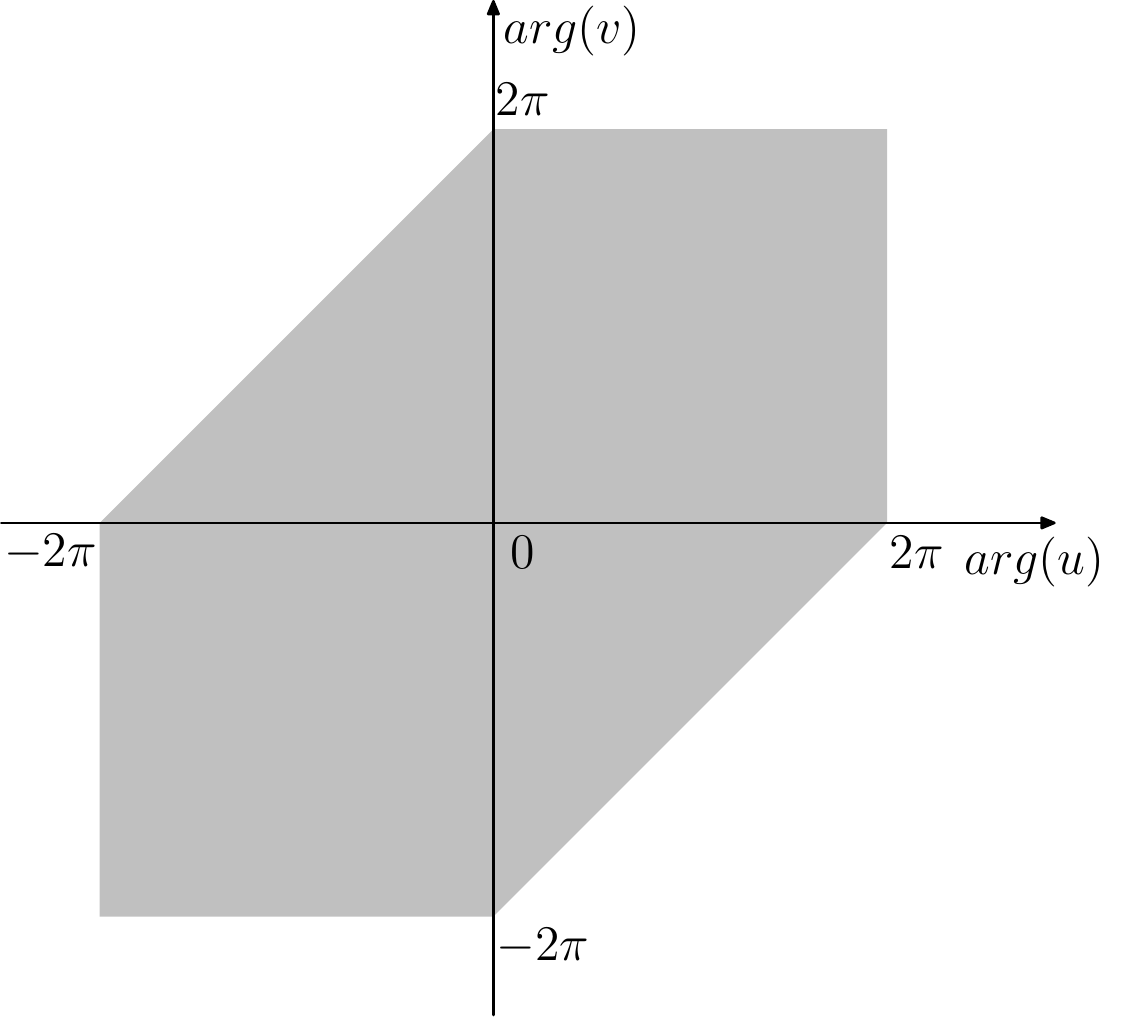} 
  \caption{Sectorial domain $\Theta_{CFT}$ of analyticity of a generic CFT correlation function $F(u,v)$.}
  \label{fig:thetaCFT}
\end{figure}

{\bf Claim:} Any physical correlator $F(u,v)$ is analytic in the convex sectorial domain $\Theta_{CFT} = ({\rm arg}[u] , {\rm arg}[v])$ defined as the union of the following 3 regions
\begin{eqnarray}
\label{eq:thetaCFT}
\Theta_{CFT}:  | {\rm arg}[u] | + | {\rm arg}[v]| &<& 2 \pi  , \nn \\
0 < {\rm arg}[u] , {\rm arg}[v] &<& 2 \pi , \nn \\
- 2 \pi < {\rm arg}[u] , {\rm arg}[v] &<& 0 , \nn 
\end{eqnarray}
where ${\rm arg}[u]={\rm arg}[v]=0$ corresponds to the principal Euclidean sheet.  $\Theta_{CFT}$ is depicted on \ref{fig:thetaCFT}. 

\begin{figure}[h]
  \centering
  \includegraphics[width=0.5\textwidth]{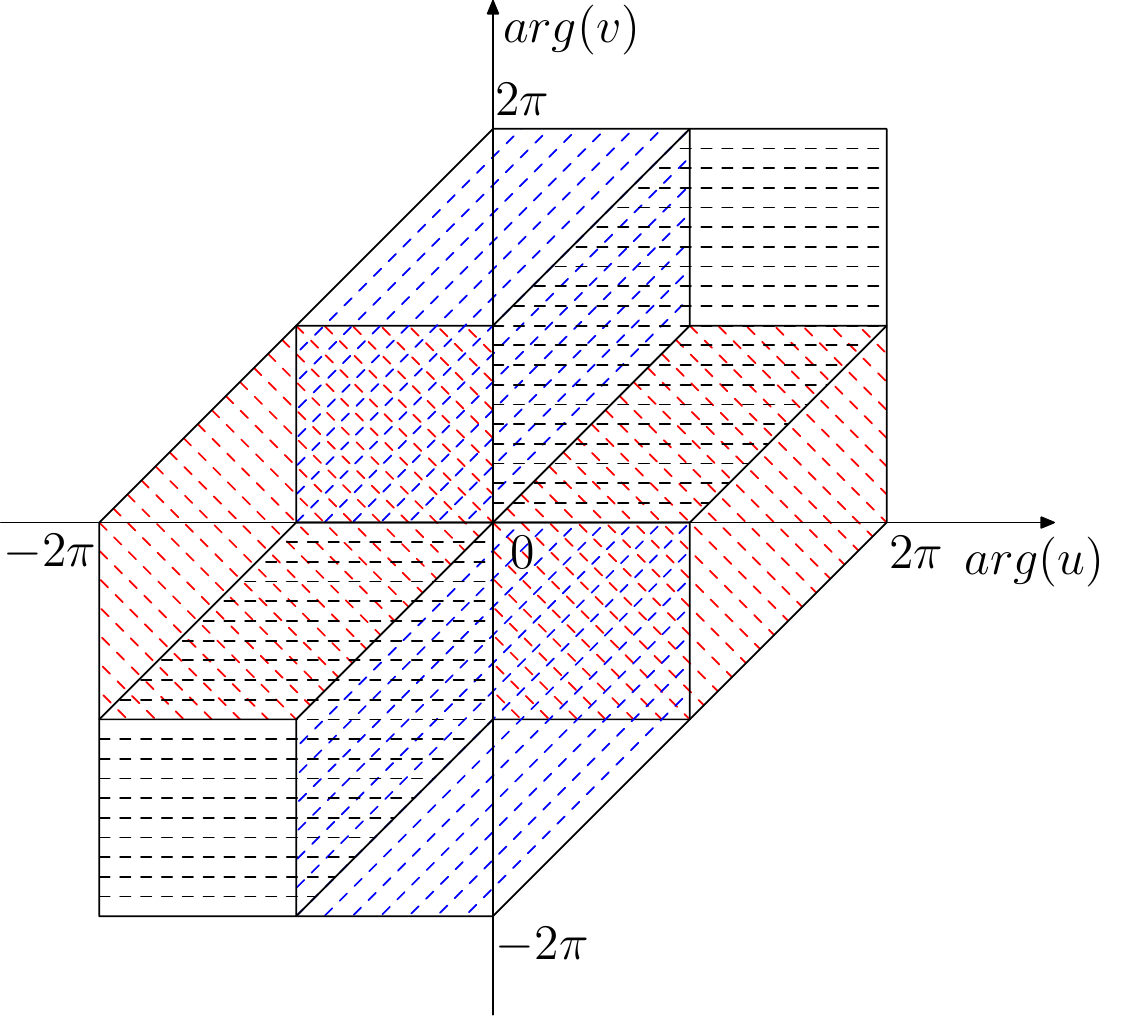} 
  \caption{We use red, blue and grey to colour the regions where the ${\cal O}(x_1) \times {\cal O}(x_2)$, ${\cal O}(x_1) \times {\cal O}(x_4)$ and ${\cal O}(x_1) \times {\cal O}(x_3)$ OPE channels converge respectively.}
  \label{fig:OPErhombus}
\end{figure}

Let us briefly outline the argument for our claim. We present all the details in appendix \ref{app: rhombus}. As we analytically continue $u \rightarrow |u| e^{i {\rm arg}[u]}$, $v \rightarrow |v| e^{i  {\rm arg}[v]}$ we are sure that we do not encounter any singularity provided there is an OPE channel that  converges. Indeed, we can then use the Cauchy-Schwarz inequality to bound the continued correlator by its value at ${\rm arg}[u] = {\rm arg}[v] = 0$ which then ensures analyticity.
The rhombus of figure \ref{fig:thetaCFT} is precisely the union of the regions where at least one OPE channel converges, see figure \ref{fig:OPErhombus}.

We expect that generic correlation functions will have a non-analyticty at the boundary of the $\Theta_{CFT}$ for some values of $|u|,|v|$. One example of such a non-analyticity is the bulk point singularity  \cite{Gary:2009ae, Maldacena:2015iua}.\footnote{Note that in terms of $(u,v)$ the bulk point locus $z = \bar z$ can lead to a non-analyticity of the correlator in terms of $(u,v)$ even in $2d$ CFTs. This does not contradict \cite{Maldacena:2015iua} which established analytic properties of the correlator as a function of $(z, \bar z)$ which are fully analytic at $z = \bar z$ and is related to the singular character of the Jacobian when going from $(z, \bar z)$ to $(u,v)$. See  appendix F for more on that. } In special cases this singularity may be absent. An example is  free field theory.

\subsection{Dangerous Limits}\label{sec: convergence}

To analyze the convergence of the integrals above let us understand what are the relevant regions when we integrate in $0 \leq u, v \leq 1$. There are several different regions involved:

a) {\bf Euclidean OPE} region, %
which corresponds to $u \to 0$ and $v \to 1$. The correlator in this limit behaves as
\beq \label{euclid}
\lim_{u \to 0, {1 - v \over \sqrt u} - \ {\rm fixed} }F(u, v) \sim u^{{\Delta_{min}\over 2} - \Delta} ,
\eeq
where $\Delta_{min}$ is the minimal scaling dimension of the operators that appear in the OPE  of ${\cal O}\times {\cal O}$. More generally, we can use the Euclidean OPE to bound the correlator in the vicinity of $u=0$ and $v=1$, not necessarily along the directions (\ref{euclid}).

b) {\bf Lorentzian OPE} region, %
which corresponds to $u \to 0$ and $v$ fixed and finite. The correlator in this limit behaves as
\beq \label{lorentz}
\lim_{u \to 0} F(u, v) \sim u^{{\tau_{min}\over 2} - \Delta} ,
\eeq
where $\tau_{min}$ is the minimal twist\footnote{We define twist as $\tau = \Delta - J$, where $J$ is the spin of the operator.} of the operator that appears in the OPE of ${\cal O} \times {\cal O}$. 

In unitary theories, the identity is the lowest dimension operator exchanged in ${\cal O}\times {\cal O}$, thus $\Delta_{min}=\tau_{min}=0$. 

c) {\bf Double light-cone limit}, %
which corresponds to $u,v \to 0$.\footnote{This limit was discussed for example in \cite{Alday:2010zy,Alday:2015ota}.} In general we do not know what is the behavior of the correlator in this limit. 
For the moment, we just bound  the correlator in that limit. 

As explained for example in \cite{Hartman:2015lfa}, the correlator can be expanded as
\begin{eqnarray}
\label{eq:schOPE}
F(u, v) = u^{- \Delta} \sum_{h, \bar{h}}a_{h, \bar{h}} z^h \bar{z}^{\bar{h}},
\end{eqnarray}
where $h=\frac{\Delta_{ex} \pm J}{2}$, $\bar{h}=\frac{\Delta_{ex} \mp J}{2}$ and we sum over all exchanged operator in the ${\cal O}\times {\cal O}$ OPE, both primaries and descendants. $\Delta_{ex}$ is the conformal dimension of the exchanged operator and $J$ is its spin. The coefficients $a_{h, \bar{h}}$ are non-negative in a unitary theory.  The usual relation between $z,\bar z$ and $u,v$ is reproduced in (\ref{eq:correlator}).

Suppose now that we are in the Lorentzian region $\sqrt{u}+\sqrt{v}<1$.\footnote{This region is given by a Lorentzian correlator.} In this region, $z$ and $\bar{z}$ are independent and real positive variables. Using unitarity and the OPE expansion (\ref{eq:schOPE}) we conclude that 
\begin{eqnarray} 
F(z_1, \bar{z}_1) (z_1 \bar{z}_1)^{\Delta}< F(z_2, \bar{z}_1 )(z_2 \bar{z}_1)^{\Delta},
\end{eqnarray}
provided that we pick $z_2$ such that $z_1<z_2<1$. 

The previous inequality can also be stated in the following manner. Pick $z_1$ and $\bar{z}_1$ as independent real variables between $0$ and $1$. Define $u_1$ and $v_1$ in the usual manner. Then, for any $z_2$ such that $z_1<z_2<1$,
\begin{eqnarray}\label{ineq useful}
F(u_1, v_1)< F\big(u_2, (1- \bar{z}_1)(1-z_2) \big) \frac{u_2^{\Delta}}{u_1^{\Delta}},
\end{eqnarray}
where $u_2=\bar{z}_1 z_2$. Now let us take the double light-cone limit $z_1 \to 0$, $\bar{z}_1 \rightarrow 1 $. 
Since $z_2$ is fixed we can use the lightcone limit in the RHS to get
(\ref{ineq useful}): $F \big( u_2, (1- \bar{z_1})(1-z_2) \big) \sim (1- \bar{z_1})^{-\Delta}$
\begin{eqnarray}\label{ineq useful 2}
F(u, v)< {c_0 \over u^\Delta v^\Delta }, ~~~ 0 < u, v < c,
\end{eqnarray}
where $c<1$ and $c_0$ are some constants. This bound is saturated in the 2d Ising model (see (\ref{f 4sigma})). So, we cannot improve it without making further assumptions.

\subsection{Subtractions and Polynomial Boundedness}

By combining the small $u,v$ analysis of the previous section with crossing (\ref{crossing}) we can find the power-like bounds and asymptotics of the full correlator $F(u,v)$ for any $0<u,v<\infty$. Plugging them in the definition of the Mellin transform (\ref{mellindef}) it is clear that the integral diverges for any $\gamma_{12}$ and $\gamma_{14}$.

To improve the convergence of the Mellin transform we consider the subtracted correlator
\begin{eqnarray}
\label{eq:subtractionF}
&&F_{sub}(u,v) \equiv F(u,v) - (1 + u^{- \Delta} + v^{- \Delta})  \\
&-& \sum_{\tau_{gap} \leq \tau \leq \tau_{sub}} \sum_{J=0}^{J_{max}} \sum_{m=0}^{ \[ \frac{\tau_{sub} - \tau}{2}\]} C_{\tau, J}^2  \Big( u^{- \Delta + \frac{\tau}{2}+m} g^{(m)}_{\tau, J}(v) + v^{- \Delta + \frac{\tau}{2}+m} g^{(m)}_{\tau, J}(u) + v^{- \frac{\tau}{2}-m} g^{(m)}_{\tau, J}(\frac{u}{v}) \Big) \nn \ ,
\end{eqnarray}
where $g^{(m)}_{\tau, J}(v)$ is defined as the $m$-th term in the small $u$ expansion of a single conformal block $g_{\tau, J}(u,v)$
\begin{eqnarray}
\label{eq:mfunctiondef}
F(u,v) =u^{-\Delta} \sum_{\tau, J - {\rm even}} C_{\tau, J}^2 g_{\tau, J}(u,v)  \ , ~~~g_{\tau, J}(u,v) = u^{{\tau \over 2}}\sum_{m=0}^{+\infty} u^{m} g^{(m)}_{\tau, J}(v) .
\end{eqnarray}
These functions satisfy the following useful identity $v^{- \frac{\tau}{2}-m} g_{\tau, J}^{(m)}(\frac{1}{v})= (-1)^J g_{\tau, J}^{(m)}(v) $, see e.g. \cite{Costa:2012cb}.   The subtraction (\ref{eq:subtractionF}) makes every OPE limit of the correlator less singular. More precisely, we explicitly subtract the contribution of operators with twists $\tau \leq \tau_{sub} < \tau_*$ in every channel. Here $\tau_*$ is the smallest twist accumulation point that is exchanged in the OPE of ${\cal O} \times {\cal O}$. On general grounds \cite{Fitzpatrick:2012yx,Komargodski:2012ek} $\tau_* \leq 2 \Delta$. The choice of our twist cut-off $\tau_{sub} < \tau_*$ guarantees that in (\ref{eq:subtractionF}) $J_{max} < \infty$.

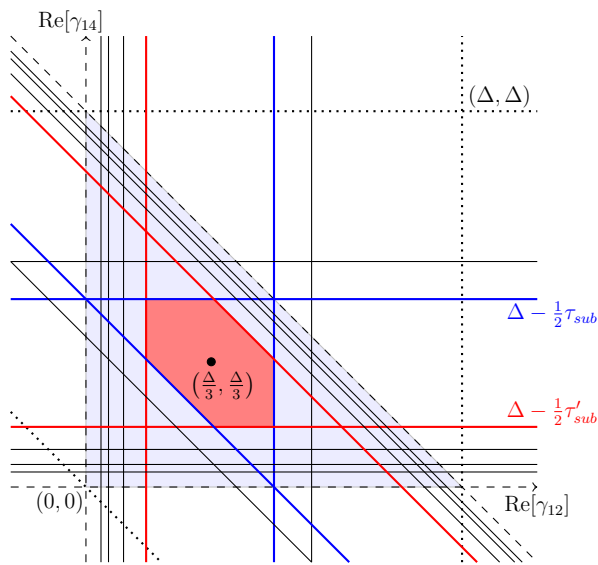
\begin{figure}[h]
  \centering
  \begin{tikzpicture}
	\begin{pgfonlayer}{nodelayer}
		\node [style=none] (0) at (0, -1) {};
		\node [style=none] (1) at (0, 6) {};
		\node [style=none] (2) at (-1, 0) {};
		\node [style=none] (3) at (6, 0) {};
		\node [style=textdot,red ] (4) at (6.2, 1) {$\Delta-\frac{1}{2}\tau'_{sub}$};
		\node [style=textdot,blue ] (4) at (6.2, 2.3) {$\Delta-\frac{1}{2}\tau_{sub}$};
		\node [style=textdot] (4) at (6, -0.25) {${\rm Re}[\gamma_{12}]$};
		\node [style=textdot] (5) at (-0.2, 6.2) {${\rm Re}[\gamma_{14}]$};
		\node [style=point] (10) at (1.666, 1.666) {};
		\node [style=none] (11) at (5, -1) {};
		\node [style=none] (12) at (5, 6) {};
		\node [style=none] (13) at (-1, 5) {};
		\node [style=none] (14) at (6, 5) {};
		\node [style=textdot]  at (-0.35, -0.2) {$(0,0)$};
		\node [style=textdot]  at (5.5, 5.2) {$(\Delta, \Delta)$};
		\node [style=textdot]  at (1.82, 1.35) {$\left(\frac{\Delta}{3}, \frac{\Delta}{3}\right)$};
	\end{pgfonlayer}
	\begin{pgfonlayer}{edgelayer}
	\draw[fill=blue!30, nearly transparent]  (0,0) -- (5,0) -- (0,5) -- cycle;
		\draw[fill=red!50]  (.8,2.5) -- (1.7,2.5)-- (2.5,1.7) -- (2.5,.8)-- (1.7,0.8) -- (0.8,1.7) -- cycle;
		\draw [style=dashed, arrow end] (2.center) to (3.center);
		\draw [style=dashed, arrow end] (0.center) to (1.center);
		\draw [style=dotted,thick] (13.center) to (14.center);
		\draw [style=dotted,thick] (11.center) to (12.center);
		\draw [style=dotted,thick] (-1,1) to (1,-1);
		\draw [style=dashed] (-1,6) to (6,-1);
		\draw [style=black] (3,-1) to (3,6);
		\draw [style=blue,thick] (2.5,-1) to (2.5,6);
		\draw [style=red,thick] (.8,-1) to (.8,6);
		\draw [style=black] (.5,-1) to (.5,6);
		\draw [style=black] (.3,-1) to (.3,6);
		\draw [style=black] (.2,-1) to (.2,6);
		\draw [style=black] (-1,3) to (6,3);
		\draw [style=blue,thick] (-1,2.5) to (6,2.5);
		\draw [style=red,thick] (-1,.8) to (6,.8);
		\draw [style=black] (-1,.5) to (6,.5);
		\draw [style=black] (-1,.3) to (6,.3);
		\draw [style=black] (-1,.2) to (6,.2);
		\draw [style=black] (-1,3) to (3,-1);
		\draw [style=blue,thick] (-1,3.5) to (3.5,-1);
		\draw [style=red,thick] (-1,5.2) to (5.2,-1);
		\draw [style=black] (-1,5.5) to (5.5,-1);
		\draw [style=black] (-1,5.7) to (5.7,-1);
		\draw [style=black] (-1,5.8) to (5.8,-1);
	\end{pgfonlayer}
\end{tikzpicture}
  \caption{ Domain $U_{CFT}$ is shown in red with the crossing symmetric point $\gamma_{12}=\gamma_{13}=\gamma_{14}=\frac{\Delta}{3}$ at the center. In this picture we assumed that the first twist accumulation point is $\tau_* = 2\Delta$ (marked with dashed lines). The {\it Mellin-Mandelstam triangle} $\Re (\gamma_{12},\gamma_{13},\gamma_{14})>0$ is depicted in light blue.
   The blue lines correspond to  $\Re (\gamma_{1j}) = \Delta -\tau_{sub}/2$ for $j=2,3,4$.
   Similarly, the red lines correspond to  $\Re (\gamma_{1j}) = \Delta -\tau_{sub}'/2$.
   The black lines correspond to $\Re (\gamma_{1j}) = \Delta -\tau/2$ for other twists $\tau < \tau_*$ in the spectrum. The dotted lines corresponds to the identity operator with $\tau=0$.  
  }
  \label{fig:UCFT}
\end{figure}

The subtracted correlator $F_{sub}(u,v)$ still satisfies crossing (\ref{crossing}) and it is still analytic in the sectorial domain $\Theta_{CFT}$ due to the analytic properties of the $g_{\tau, J}^{(m)}$ functions. However, on top of that we claim that $F_{sub}(u,v)$ is polynomially bounded as follows
\beq
\label{eq:polybound}
| F_{sub}(u,v) | \leq C(\gamma_{12}, \gamma_{14}) {1 \over |u|^{\gamma_{12}}} {1 \over |v|^{\gamma_{14}}}  , ~~~ ({\rm arg} [u],{\rm arg} [v]) \in \Theta_{CFT}, ~~~ (\Re [\gamma_{12}], \Re [\gamma_{14}]) \in U_{CFT} \ ,
\eeq
where $U_{CFT}$ is given by
\begin{eqnarray}
\label{eq:Ucft}
U_{CFT}:&&  \Delta - {\tau_{sub} \over 2} > {\rm Re}[\gamma_{12}, \gamma_{13}, \gamma_{14}] > \Delta - {\tau_{sub}' \over 2} \ ,
\end{eqnarray}
where  $\tau_{sub}'$ is the next twist after $\tau_{sub}$ appearing in the OPE $\mathcal{O} \times \mathcal{O}$ and recall that $\gamma_{13} = \Delta - \gamma_{12} - \gamma_{14}$. 
The condition $\Re [\gamma_{12}] > \Delta - {\tau_{sub}' \over 2}$ follows trivially  from the light-cone limit 
$F_{sub}(u,v) \sim u^{\Delta-\tau'_{sub}/2}$ when $u \to 0$ with fixed $v>0$.
The condition $ \gamma_{12}+\gamma_{14}=\Delta - \gamma_{13}  > \tau_{sub}/2$ follows from 
the behaviour of the last subtraction term in \eqref{eq:subtractionF} in double light-cone limit $u\sim v \to 0$. Finally, the remaining conditions in \eqref{eq:Ucft} are obtained from these two by crossing symmetry.
The domain $U_{CFT}$ is depicted in figure \ref{fig:UCFT}. One may also think of the domain $U_{CFT}$ as the region of analyticity surrounding the crossing symmetric point $\gamma_{12}=\gamma_{13}=\gamma_{14}=\frac{\Delta}{3}$.

We conclude that $F_{sub}(u,v) \in M_{\Theta_{CFT}}^{U_{CFT}}$ and we can write its Mellin representation with the straight contour
\beq
\label{eq:mellinsub}
F_{sub}(u,v) = \int_{U_{CFT} - i \infty}^{U_{CFT} + i \infty} {d \gamma_{12} d \gamma_{14} \over (2 \pi i)^2} \ u^{- \gamma_{12}} v^{- \gamma_{14} } \hat M (\gamma_{12}, \gamma_{14}) .
\eeq
In section \ref{sec: dif cont} we show (\ref{eq:mellinsub}) implies (\ref{eq:correlatorIntro}) with ${\cal C}$ being a deformed contour.

Few comments are in order. First of all a necessary condition for the non-emptiness of $U_{CFT}$ (\ref{eq:Ucft}) is $\frac{\Delta}{3}  > \Delta - \frac{ \tau_{*}}{2}$ or equivalently
\beq
\label{eq:twistboundlight}
 \Delta < \frac{3}{4} \tau_* \ .
\eeq
In an interacting CFT we expect that $\tau_{*} = 2 \tau_{lightest}$, where $\tau_{lightest}$ is scaling dimension of the lightest operator present in the theory. It is in this precise sense that our construction concerns only the correlation functions of the light operators in the theory. 
In section \ref{sec: bent}, we will present a different argument that allows us to generalize this construction beyond (\ref{eq:twistboundlight}). Unfortunately, we do not establish (\ref{eq:polybound}) rigorously. Nevertheless we believe that it is a true property of physical correlators. We present some arguments for why we believe that it has to be true in appendix \ref{sec:polboundapp}. Proving (\ref{eq:polybound}) rigorously is the an important missing step in our analysis.

At this point we can also introduce the notion of {\it  Mellin-Mandelstam triangle} (see figure \ref{fig:UCFT}). Consider a CFT where the lightest primary operator is a scalar with $\Delta_{lightest} < d-2$. In this case, if we consider the Mellin amplitude for the lightest scalar, the first twist accumulation point that appears in the OPE is $\tau_* = 2 \Delta_{lightest}$. %
Therefore, in the region 
\beq
\label{eq:Mandelstam}
{\rm Re}[\gamma_{12}, \gamma_{13}, \gamma_{14}]  >0 \,,
\eeq
there are no twist accumulation points. 
This is the analog of the Mandelstam triangle in the context of flat space scattering amplitudes. 
The twist accumulation point at $\gamma_{12}=\Delta -\tau_*/2=0$ is the analogue of the two-particle branch point (in the 12 channel) of the flat space scattering amplitude.

According to theorem I $\hat M(\gamma_{12}, \gamma_{14})$ decays exponentially fast in the complex plane
\beq
| \hat M(\gamma_{12}, \gamma_{14}) | \leq K({\rm Re}[\gamma_{12}], {\rm Re}[\gamma_{14}]) e^{- {\rm sup}_{\Theta_{CFT}} \left( {\rm arg}[u] {\rm Im}[\gamma_{12}] + {\rm arg}[v] {\rm Im}[\gamma_{14}]  \right)}, ~~~  (\gamma_{12}, \gamma_{14}) \in U_{CFT} + i {\mathbb R}^2 .
\eeq
Note that the Mellin amplitude $M(\gamma_{12}, \gamma_{14})$ is defined by
\beq
\hat M(\gamma_{12}, \gamma_{14}) = \left[ \Gamma(\gamma_{12}) \Gamma(\gamma_{14}) \Gamma(\Delta - \gamma_{12} - \gamma_{14}) \right]^2 M(\gamma_{12}, \gamma_{14}) .
\eeq
One can easily check that remarkably
\begin{eqnarray}
\label{eq:decaycorrect}
\left[ \Gamma(\gamma_{12}) \Gamma(\gamma_{14}) \Gamma(\Delta - \gamma_{12} - \gamma_{14}) \right]^2 &\sim& e^{- \pi \left( |{\rm Im}[\gamma_{12}]| +  |{\rm Im}[\gamma_{14}]| +  |{\rm Im}[\gamma_{12}+\gamma_{14}]|  \right)} \nn \\
&=& e^{- {\rm sup}_{\Theta_{CFT}} \left( {\rm arg}[u] {\rm Im}[\gamma_{12}] + {\rm arg}[v] {\rm Im}[\gamma_{14}]  \right)} ,
\end{eqnarray}
and therefore $M(\gamma_{12}, \gamma_{14})$ is polynomially bounded! Moreover, as we will show below its maximal power growth is controlled by the Regge limit.

\section{Analytic Properties Of Mellin Amplitudes} \label{sec: dif cont}

In section \ref{sec:MelGen} we analyzed the conditions for the existence of the Mellin amplitude of the correlator and put forward the subtractions necessary to define it. Next we would like to understand analytic properties of the Mellin amplitude $M(\gamma_{12}, \gamma_{14})$. This is the main purpose of this section. 

To attack this problem we find it useful to develop a different approach to define Mellin amplitudes, namely we split the integral over cross ratios in the definition of the Mellin transform into 3 regions mapped into each other by crossing. The integral of the correlator over a sub-region is manifestly convergent for certain values of $\gamma_{12}$ and $\gamma_{14}$. We then define the full Mellin amplitude by bringing the contributions from different pieces together, see \eqref{mellinfull} below. As a result the subtractions we postulated in the previous section arise very naturally. It is also clear that they can be absorbed into the definition of the contour in (\ref{eq:correlatorIntro}), up to non-essential subtleties that we discuss below. 

We then analyze analytic properties of the Mellin amplitude and argue that the only singularities of the Mellin amplitudes are the ones that correspond to the physical operators (\ref{OPEpoles}). This is discussed in more detail in appendix \ref{app: proof}.

\subsection{Auxillary $K$-functions}\label{sec: Kfunctions}

We split the integral in \eqref{mellindef} into three regions as shown in figure \ref{fig:uvregions}. 
We define
\beq \label{defK}
 K (\gamma_{12}, \gamma_{14}) \equiv K_{\I} (\gamma_{12}, \gamma_{14}) \equiv \int_0^{1} {d u d v \over u v} \ u^{\gamma_{12}} v^{\gamma_{14}} F(u,v) \ .
\eeq
Notice that $K_{\I}(\gamma_{12}, \gamma_{14}) = K_{\I}(\gamma_{14}, \gamma_{12})$ as the result of crossing \eqref{crossing}. 
The lightcone  behavior \eqref{lorentz} and the double lightcone bound \eqref{ineq useful 2} imply that the integral converges  for ${\rm Re}\, \gamma_{12} > \Delta$ and ${\rm Re} \, \gamma_{14} > \Delta$. Therefore, $K_{\I}(\gamma_{12}, \gamma_{14})$ is analytic in this region.

Similarly, we have
\beq \label{threeregionsothers}
 K_{\II} (\gamma_{12}, \gamma_{14}) \equiv \int_1^{\infty} {d u \over u } \int_0^{u}  {d v \over v} \ u^{\gamma_{12}} v^{\gamma_{14}} F(u,v) \ .
\eeq
Let us do a change of variables $u \to {1 \over  u}$ and $v \to {  v \over  u}$.
The measure stays invariant and using crossing $F({1 \over u}, {v \over u}) = u^{\Delta} F(u,v)$ we get  
\beq \label{relation}
 K_{\II}(\gamma_{12}, \gamma_{14}) = K_{\I} (\gamma_{13}, \gamma_{14}), \qquad\gamma_{13} = \Delta - \gamma_{12} - \gamma_{14} .
\eeq

Lastly, we get
\begin{align}
 \label{threeregionsothersB}
& K_{\III} (\gamma_{12}, \gamma_{14}) \equiv \int_1^{\infty}  {d v \over v} \int_0^{v} {d u \over u } \ u^{\gamma_{12}} v^{\gamma_{14}} F(u,v) \ , \\
&
 K_{\III}(\gamma_{12}, \gamma_{14}) =  K_{\I} (\gamma_{12}, \gamma_{13}), \qquad\gamma_{13} = \Delta - \gamma_{12} - \gamma_{14} ,
\end{align}
where in the last line we again made use of crossing symmetry.

Importantly, splitting the analytic function $F(u,v)$ into three non-analytic pieces \footnote{
$F(u,v) = F(u,v) \Theta(1-u)\Theta(1-v) +F(u,v) \Theta(u-1)\Theta(u-v) +F(u,v) \Theta(v-1)\Theta(v-u)$.
} leads to a dramatic effect on the convergence properties of the inverse Mellin transform. Instead of converging in the sectorial domain the inverse Mellin transforms above converge only for ${\rm arg}[u]={\rm arg}[v]=0$. Nevertheless we will find it useful to use the $K$-functions to describe subtractions and analytic properties of the full Mellin amplitude.

We conclude that in any CFT and for arbitrary correlation functions of scalar primaries, we have
\begin{align}\label{inverseMellin naive}
F(u,v)  &= \int_{{\rm Re}[\gamma_{12},\gamma_{14}] > \Delta} {d \gamma_{12} d \gamma_{14} \over (2 \pi i)^2}  K(\gamma_{12}, \gamma_{14}) u^{- \gamma_{12}} v^{- \gamma_{14}} \cr
&+ \int_{{\rm Re}[\gamma_{13},\gamma_{14}] > \Delta } {d \gamma_{12} d \gamma_{14} \over (2 \pi i)^2} K(\gamma_{13}, \gamma_{14}) u^{- \gamma_{12}} v^{- \gamma_{14}} \cr
&+ \int_{{\rm Re}[\gamma_{13},\gamma_{12}]> \Delta} {d \gamma_{12} d \gamma_{14} \over (2 \pi i)^2} K(\gamma_{13}, \gamma_{12}) u^{- \gamma_{12}} v^{- \gamma_{14}} .
\end{align}
where we denoted $ K = K_{\I}$ and $\gamma_{12} + \gamma_{13} + \gamma_{14} =\Delta$. 
The contours run parallel to the imaginary axis of $\gamma_{12}$ and $\gamma_{14}$ with real parts obeying the inequalities shown under the integral sign.

Next, we would like to bring the three integrals in (\ref{inverseMellin naive}) to the same contour. We discuss this procedure below and it will naturally lead to the subtractions that appeared in (\ref{eq:subtractionF}). For the moment we can rather formally define the Mellin amplitude as the sum of the three terms analytically continued to the whole complex plane,
\beq \label{mellinfull}
\hat M(\gamma_{12}, \gamma_{14}) = K (\gamma_{12}, \gamma_{14}) +  K (\gamma_{13}, \gamma_{14}) +  K (\gamma_{12}, \gamma_{13}) ,
\qquad\gamma_{13} = \Delta - \gamma_{12} - \gamma_{14} .
\eeq

\subsection{Analytic Structure of the $K$-function}

\begin{figure}[h]
  \centering
  \begin{tikzpicture}
	\begin{pgfonlayer}{nodelayer}
		\node [style=none] (0) at (0, -3) {};
		\node [style=none] (1) at (0, 3) {};
		\node [style=none] (2) at (-3, 0) {};
		\node [style=none] (3) at (3, 0) {};
		\node [style=textdot] (4) at (2.65, -0.3) {${\rm Re}[\gamma_{12}]$};
		\node [style=textdot] (5) at (0.6, 2.85) {${\rm Re}[\gamma_{14}]$};
		\node [style=none] (6) at (2, 1.5) {${\rm [a] }$};
		\node [style=none] (7) at (-1.75, 1.5) {${\rm [b]}$};
		\node [style=none] (8) at (2, -1.5) {${\rm [c]}$};
		\node [style=none] (9) at (-1.75, -1.5) {${\rm [d]}$};
		\node [style=point] (10) at (0, 0) {};
		\node [style=textdot] (11) at (0.5, -0.3) {$(\Delta, \Delta)$};
	\end{pgfonlayer}
	\begin{pgfonlayer}{edgelayer}
		\draw [style=arrow end] (2.center) to (3.center);
		\draw [style=arrow end] (0.center) to (1.center);
	\end{pgfonlayer}
\end{tikzpicture}
  \caption{We want to analytically continue $K(\gamma_{12}, \gamma_{14})$ into $\mathbb{C}^2$. We break $\mathbb{C}^2$ according to the four regions in the figure. For example, region [a] corresponds to ${\rm Re}[\gamma_{12}], {\rm Re}[\gamma_{14}]>\Delta$. In region [a], $K(\gamma_{12}, \gamma_{14})$ is completely analytic and is defined by the integral (\ref{defK}). In the other regions, it will be defined by analytic continuation.
  }
  \label{fig:regions gamma plane}
\end{figure}
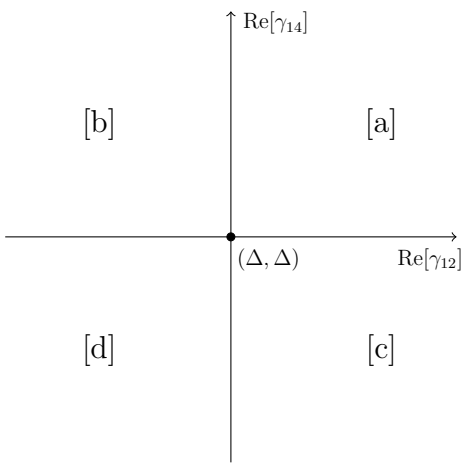

The function $K(\gamma_{12}, \gamma_{14})$  is analytic for ${\rm Re}[\gamma_{12}] >\Delta$ and ${\rm Re}[\gamma_{14}] >\Delta$. We would like to analytically continue this function to the rest of $\mathbb{C}^2$.
Different regions are shown in figure \ref{fig:regions gamma plane}.

In appendix \ref{app: sing var}, we explain how such analytic continuation is obtained for single-variable Mellin transforms. 
We shall see that we can use the same trick at fixed $\gamma_{14}$ to extend the domain in $\gamma_{12}$. The main trick is to \emph{add and subtract} the leading behavior at small $u$,
\begin{eqnarray}
K(\gamma_{12}, \gamma_{14})=\int_0^1\frac{du}{u} \int_0^1\frac{dv}{v} u^{\gamma_{12}} v^{\gamma_{14}} \left( \big(F(u, v)-   \sum_{\tau} u^{-\Delta+\frac{\tau}{2} } h_{\tau}(v)\big) + \sum_{\tau} 
 u^{-\Delta+\frac{\tau}{2}} h_{\tau}(v) \right),
\end{eqnarray} 
associated to the exchange of operators of twist $\tau$.
With this subtraction, we improved the convergence in $\gamma_{12}$ of the first term, without affecting the convergence in $\gamma_{14}$.
The second term just gives simples poles at 
\beq
\gamma_{12}=\Delta -\frac{\tau}{2}\,,\qquad \qquad
\mathcal{O}_{\tau} \in \mathcal{O}\times \mathcal{O}\,.
\eeq
These are just the usual OPE poles \eqref{OPEpoles} (notice that here we are not distinguishing between primaries and descendants). 
As we review below, CFT's have accumulation points in the twist spectrum.
For this reason one may need  infinite subtractions in order to analytically continue in $\gamma_{12}$ by a finite amount.
In appendix \ref{app: proof}, we explain in detail how to use OPE convergence to overcome this difficulty.
The conclusion is that  we can analytically continue $K(\gamma_{12}, \gamma_{14})$ into region [b] except for the OPE singularities at $\gamma_{12}=\Delta -\frac{\tau}{2}$.
 Similarly, we can  analytically continue $K(\gamma_{12}, \gamma_{14})$ into region [c] except for the OPE singularities at $\gamma_{14}=\Delta -\frac{\tau}{2}$.
 However, we cannot use the same strategy to analytically continue $K(\gamma_{12}, \gamma_{14})$ into region [d] because we do not have enough control over the double lightcone limit $u\sim v \to 0$.
In appendix (\ref{app: proof}), we give  strong evidence that $K(\gamma_{12}, \gamma_{14})$ can also  be extended to region [d], except for the same OPE singularities. 
Our arguments are based on Bochner's theorem (see appendix \ref{app: bochner}) and another conjectured theorem (see appendix \ref{app: theorem?}) for analytic functions of two complex variables. 
See also appendix \ref{app: examples K} for explicit formulas for $K(\gamma_{12}, \gamma_{14})$ in free theories and in the 2d Ising model.

The results of this section strongly support  the conjecture that the Mellin amplitude (\ref{mellinfull}) has singularities only at the OPE poles \eqref{OPEpoles}. We call this property {\it maximal Mellin analyticity} by analogy with the S-matrix. It would be interesting to understand the precise relation between the two.

\subsection{Twist spectrum}
\label{sec:twist}

The analytic structure of Mellin amplitudes (and $K$-functions) is controlled by the CFT twist spectrum. Let us review its basic properties  in $d=2$ and $d \ge 3$.

It is convenient to organize the CFT spectrum in terms of twist $\tau = \Delta - J$ and spin $J$.\footnote{We restrict our discussion to symmetric traceless operators.} Unitarity implies that $\tau \geq d-2$ for $J>0$, and $\tau \geq {d-2 \over 2
}$ for $J=0$. In other words, the twist spectrum of a unitary CFT is bounded from below. As shown in \cite{Caron-Huot:2017vep} operators organize themselves in the Regge trajectories $\tau(J)$, at least for $J>1$. It is interesting to ask what is the structure of the twist spectrum as $J \to \infty$. 

In $d\ge 3$ the twist spectrum exhibits additivity at large spin. Given operators with twists $\tau_1$ and $\tau_2$ (we can call them seed operators) there exists an infinite set of Regge trajectories with the property \cite{Fitzpatrick:2012yx,Komargodski:2012ek} 
\begin{eqnarray}\label{twist spectrum}
\lim_{J \to \infty} \tau_{1,2}^{(n)}(J) = \tau_1 + \tau_2 + 2 n .
\end{eqnarray}
The corrections to (\ref{twist spectrum}) at finite $J$ are given by powers of ${1 \over J}$. Existence of such Regge trajectories implies that the twist spectrum of an interacting CFT exhibits an intricate pattern of accumulation points. Indeed, as $J$ approaches infinity  (\ref{twist spectrum}) implies that there is an infinite number of operators with twist $| \tau - \tau_1 - \tau_2 | < \eps$, where $\eps$ is an arbitrary positive constant. These are the so-called double-twist Regge trajectories.

A double-twist operator $\tau_{1,2}^{(n)}(J_0)$ can itself serve as a seed that can be paired with another operator to produce triple-twist operators. In this case, (\ref{twist spectrum}) implies the existence of an infinite set of accumulation points
\begin{eqnarray}\label{tripletwist}
\lim_{J \to \infty} \tau_{\tau_{1,2}^{(n)}(J_0),\tau_3}^{(m)}(J) = \tau_{1,2}^{(n)}(J_0) + \tau_3 + 2 m .
\end{eqnarray}
Therefore, already at the second step we obtain an infinite number of accumulation points, for every $J_0$, in the space of twists (\ref{tripletwist}).
Moreover, there are also accumulation points of accumulation points at $\tau_1+\tau_2+\tau_3+2m$.

As we increase twist the number of available seed operators, as well as the number of the multi-twist Regge trajectories, quickly grows and therefore at high enough twist we expect the twist spectrum to become very thinly spaced. It is an open question if it really becomes dense in a finite interval.
One can get some intuition from the toy model 
\beq
\tau_{\rm toy}^{(k+1)}(J_1,\dots,J_{k}) = c -\sum_{i=1}^{k}  \frac{1}{J_i} \,, \qquad
\qquad J_i \in \mathbb{N}\,,
\eeq
for $(k+1)^{\rm th}$-twist operators.
In this model, the  twist spectrum contains many accumulation points but it is not dense in any interval of $\mathbb{R}$.\footnote{The only limit points of the set of twists 
$\tau_{\rm toy}^{(k+1)}$ is the set $\tau_{\rm toy}^{(k)}$ (see   \href{https://math.stackexchange.com/questions/142992/find-all-limit-points-of-m-left-frac1n-frac1m-frac1k-m-n-k}{math.stackexchange.com}).}
This suggest that the same is true for the twist spectrum of CFTs in $d\ge 3$ dimensions. Clearly this question has important implications for the analytic structure of Mellin amplitudes. Namely, if the twist spectrum becomes continuous then we expect branch cuts in the Mellin amplitudes.

This  issue has recently been made much more precise in the context of $d=2$ generic unitary (irrational, compact) CFTs. Using the Virasoro fusion kernel \cite{Ponsot:1999uf, Ponsot:2000mt, Teschner:2001rv}, it was argued in \cite{Collier:2018exn, Kusuki:2018wpa} that the twist spectrum becomes continuous for $\tau \ge \frac{c-1}{24}$. More precisely, for every $\tau > \frac{c-1}{24}$ there are infinitely many Regge trajectories that end in every interval $\delta \tau$.
Moreover, all these Regge trajectories appear in a single four-point correlation function. Similarly, in higher dimensions using the crossing symmetry of the quadruple discontinuity of the correlator one can elegantly show the necessity of the multi-twist operators in a given correlation function \cite{Caron-Huot:2017vep}, see also \cite{Fitzpatrick:2015qma}. %

There are situations where this complicated twist spectrum simplifies dramatically. As we review below one example is planar CFTs.\footnote{In planar gauge CFTs the twist of single-trace operators grows like $\log J$ therefore there are no accumulation points (for finite 't Hooft coupling). Double-trace operators do exhibit accumulation points, but they do not give rise to poles of the Mellin amplitude $M(\gamma_{i j})$ at the planar level.} In this case, the (single-trace) twist spectrum of the planar correlators becomes  simple. Accumulation points in the twist spectrum are also absent in  rational $d=2$ CFTs. In fact, the twist spectrum in this case is given by a finite set of non-trivial twists plus non-negative integers.
It is precisely in these contexts, when the twist spectrum simplifies, that Mellin amplitudes become particularly useful.

\subsection{Recovering the Straight Contour}\label{sec: contour}

We would like to bring the three integrals in (\ref{inverseMellin naive}) to the same straight contour. There are infinitely many ways to do this depending on the choice of the final contour. In the process of doing so we need to know the analytic structure of $K(\gamma_{12}, \gamma_{14})$ on $\mathbb{C}^2$. Above we argued that $K(\gamma_{12}, \gamma_{14})$ is an analytic function, with simple poles at 
\begin{eqnarray}
\label{polesposition}
\gamma_{12}= \Delta - \frac{\tau_i}{2}- n_1, ~~~\gamma_{14}= \Delta - \frac{\tau_j}{2}- n_2,
\end{eqnarray}
for each primary operator of twist $\tau_i$ or $\tau_j$ being exchanged in ${\cal O} \times {\cal O}$ and for each nonnegative integer $n_1$ and $n_2$.

At this point, we need to make a choice about the final contour for the Mellin representation. We consider two options: a straight contour and a deformed contour. %
We examine the second possibility in section \ref{sec: bent}.

Let us reunite the three integrals in (\ref{inverseMellin naive}) into a single integral with a straight contour. We will pick the straight contour at ${\rm Re}[\gamma_{12}]= {\rm Re}[\gamma_{14}] = \frac{\Delta}{3}$. This choice is very natural, since it is completely symmetric in $\gamma_{12}$, $\gamma_{13}$ and $\gamma_{14}$.

\begin{figure}[h]
  \centering
\begin{tikzpicture}
	\begin{pgfonlayer}{nodelayer}
		\node [style=none] (0) at (0, -3) {};
		\node [style=none] (1) at (0, 3) {};
		\node [style=none] (2) at (-3, 0) {};
		\node [style=none] (3) at (3, 0) {};
		\node [style=textdot] (4) at (2.65, -0.3) {${\rm Re}[\gamma_{12}]$};
		\node [style=textdot] (5) at (0.6, 2.85) {${\rm Re}[\gamma_{14}]$};
		\node [style=point] (10) at (0, 0) {};
		\node [style=textdot] (11) at (0.5, -0.3) {$(\Delta, \Delta)$};
		\node [style={lr_point}] (12) at (-1.5, -1.5) {};
		\node [style={lr_point}] (13) at (1.5, 1) {};
		\node [style=none] (14) at (-1.5, 1) {};
		\node [style=textdot] (15) at (-1.15, -1.8) {$({\Delta \over 3}, {\Delta\over 3})$};
		\node [style=none] (16) at (-.6, -3) {};
		\node [style=none] (17) at (-.6, 3) {};
		\node [style=none] (18) at (-3,-.6) {};
		\node [style=none] (19) at (3,-.6) {};
		\node [style=none] (20) at (-.3, 1) {};
		\node [style=none] (21) at (-1.5, -0.4) {};
	\end{pgfonlayer}
	\begin{pgfonlayer}{edgelayer}
		\draw [style=arrow end] (2.center) to (3.center);
		\draw [style=arrow end] (0.center) to (1.center);
		\draw [style=redline] (13) to (14.center);
		\draw [style=redline,style=arrow end] (13) to (20);
		\draw [style=redline] (14.center) to (12);
		\draw [style=redline,style=arrow end] (14.center) to (21);
		\draw [style=none] (16) to (17);
		\draw [style=none] (18) to (19);
	\end{pgfonlayer}
\end{tikzpicture}

  \caption{Picture of the contour manipulation that corresponds to formula (\ref{deform K general}). The contours run parallel to the imaginary axis of $\gamma_{12}$ and $\gamma_{14}$ and therefore correspond to a point in this figure. In red we display the change in the contour. We pick up poles along the way, which we denote by black lines. At the end, we arrive at the point $(\Re(\gamma_{12}), \Re(\gamma_{14}))=(\frac{\Delta}{3}, \frac{\Delta}{3})$.}
  \label{fig:contour K}
\end{figure}
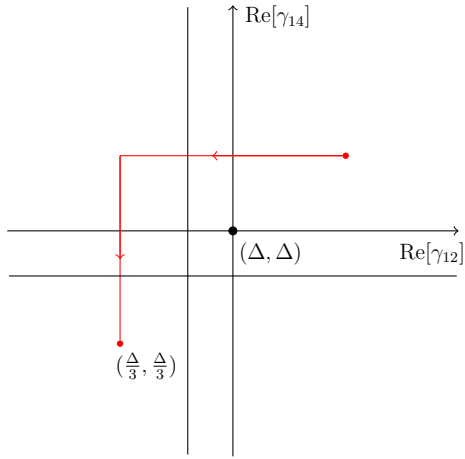

We need to deform the contours in the integrals (\ref{inverseMellin naive}). Let us see how this comes about. Consider the first integral. When we deform its contour we pick up poles. This will tell us the subtractions we need to make to the four point function, so as to have a Mellin representation with a straight contour. See figure \ref{fig:contour K}.

From figure \ref{fig:contour K}, we conclude that
\begin{eqnarray}\label{deform K general}
&&\int_{\Re(\gamma_{12})>\Delta} \frac{d \gamma_{12}}{2 \pi i} \int_{\Re(\gamma_{14})>\Delta} \frac{d \gamma_{14}}{2 \pi i} K(\gamma_{12}, \gamma_{14}) u^{-\gamma_{12}} v^{- \gamma_{14}} \\
&=&\int_{\Re(\gamma_{12})=\frac{\Delta}{3}} \frac{d \gamma_{12}}{2 \pi i} \int_{\Re(\gamma_{14})=\frac{\Delta}{3}} \frac{d \gamma_{14}}{2 \pi i} K(\gamma_{12}, \gamma_{14}) u^{-\gamma_{12}} v^{- \gamma_{14}} \nonumber \\
&+&\sum_{\tau < \frac{4 \Delta}{3}} \sum_{J} \sum_{n=0}^{ \[- \frac{\tau}{2} + \frac{2 \Delta}{3} \]} u^{- \Delta + \frac{\tau}{2}+n} \int_{\Re(\gamma_{14})>\Delta} \frac{d \gamma_{14}}{2 \pi i} \hat{K}_{\tau, J}(\gamma_{12}=\Delta- \frac{\tau}{2} -n , \gamma_{14})v^{- \gamma_{14}}   \nonumber \\
 &+& \sum_{\tau < \frac{4 \Delta}{3}} \sum_{J}  \sum_{m=0}^{ \[- \frac{\tau}{2} + \frac{2 \Delta}{3} \]} v^{- \Delta + \frac{\tau}{2}+m} \int_{\Re(\gamma_{12})=\frac{\Delta}{3}} \frac{d \gamma_{12}}{2 \pi i} \hat{K}_{\tau, J}(\gamma_{12}, \gamma_{14}=\Delta- \frac{\tau}{2}-m)u^{- \gamma_{12}},  \nonumber
\end{eqnarray}
where $\hat{K}_{\tau, J}(\gamma_{12}=\Delta- \frac{\tau}{2} -n , \gamma_{14})$ denotes the contribution from the operator $\mathcal{O}_{\tau,J}$ to the residue of $K$ at $\gamma_{12}=\Delta- \frac{\tau}{2} -n$.
 The symbol $[y]$ denotes the biggest integer that is smaller than $y$. 

Let us compute the residues of $K$. The four point function can be expanded as $F(u, v)= \sum_{\tau, J} C_{\tau, J}^2  \sum_{m=0}^{+\infty} u^{- \Delta + \frac{\tau}{2}+m} g^{(m)}_{\tau, J}(v)$, with the sum running over the primary operators $\O_{\tau, J}$ exchanged. Here $g_{\tau, J}^{(m)}(v)$ is what multiplies  $u^{\frac{\tau}{2}+m}$ in the small $u$ expansion of the conformal block. For example, $g_{\tau,J}^{(0)}(v)$ is the collinear block
\begin{eqnarray}
\label{collinearblock}
g_{\tau,J}^{(0)}(v) =(1-v)^J \ _2 F_1({\tau \over 2} + J, {\tau \over 2} + J,\tau + 2 J , 1 - v ) \ .
\end{eqnarray} 
Therefore, we can write
\begin{eqnarray}
\hat{K}_{\tau, J}(\gamma_{12}= \Delta-\frac{\tau}{2}-m, \gamma_{14})= C_{\tau, J}^2  \int_0^1 \frac{dv}{v} v^{\gamma_{14}} g_{\tau, J}^{(m)} (v).
\end{eqnarray}
This integral converges when $\Re(\gamma_{14})>0$. By inverting the Mellin transform, we can compute all the integrals in (\ref{deform K general}). We conclude that 
\begin{eqnarray}
\label{subtractionK}
&&\int_{\Re(\gamma_{12})>\Delta} \frac{d \gamma_{12}}{2 \pi i} \int_{\Re(\gamma_{14})>\Delta} \frac{d \gamma_{14}}{2 \pi i} K(\gamma_{12}, \gamma_{14}) u^{-\gamma_{12}} v^{- \gamma_{14}} \\
&=&\int_{\Re(\gamma_{12})=\frac{\Delta}{3}} \frac{d \gamma_{12}}{2 \pi i} \int_{\Re(\gamma_{14})=\frac{\Delta}{3}} \frac{d \gamma_{14}}{2 \pi i} K(\gamma_{12}, \gamma_{14}) u^{-\gamma_{12}} v^{- \gamma_{14}} \nonumber \\
&+&\sum_{\tau < \frac{4 \Delta}{3}} \sum_{J=0}^{J_{max}} \sum_{m=0}^{ \[- \frac{\tau}{2} + \frac{2 \Delta}{3} \]}  C_{\tau, J}^2  \left[  u^{- \Delta + \frac{\tau}{2}+m} \theta(1-v) g^{(m)}_{\tau, J}(v)+   v^{- \Delta + \frac{\tau}{2}+m} \theta(1-u) g^{(m)}_{\tau, J}(u) \right] \ , \nonumber
\end{eqnarray}
where we assumed that we only subtracted a finite number of operators.
In other words, the spin is bounded, $J\le J_{max}$. Situations where $J_{max} = \infty$ should be analyzed on the case by case basis. We will see such an example below when analyzing the free field theory and minimal models.

A similar exercise can be done to deform the other $K$ functions. We use the identity $u^{- \frac{\tau}{2}-m} g_{\tau, J}^{(m)}(\frac{1}{u})= g_{\tau, J}^{(m)}(u)$ (see \cite{Costa:2012cb}) valid for the exchange of operators of even spin. This way we get rid of the $\theta$ functions. We conclude that 
\begin{eqnarray}\label{subtractions finite}
F(u, v) &=& 1 + u^{-\Delta} + v^{-\Delta} + \int_{\Re(\gamma_{12})=\frac{\Delta}{3}} \frac{d \gamma_{12}}{2 \pi i} \int_{\Re(\gamma_{14})=\frac{\Delta}{3}} \frac{d \gamma_{14}}{2 \pi i} \hat M(\gamma_{12}, \gamma_{14}) u^{-\gamma_{12}} v^{- \gamma_{14}}  \\
&+& \sum_{0<\tau < \frac{4 \Delta}{3}} \sum_{J=0}^{J_{max}} \sum_{m=0}^{ \[ - \frac{\tau}{2} + \frac{2 \Delta}{3} \]} C_{\tau, J}^2  \left[ u^{- \Delta + \frac{\tau}{2}+m} g^{(m)}_{\tau, J}(v) + v^{- \Delta + \frac{\tau}{2}+m} g^{(m)}_{\tau, J}(u) + v^{- \frac{\tau}{2}-m} g^{(m)}_{\tau, J}\left(\frac{u}{v}\right) \right] \ , \nn
\end{eqnarray}
where the Mellin integral is to be done with a straight contour. We therefore recovered (\ref{eq:subtractionF}) and \eqref{eq:mellinsub}.

When we deformed the integration contour as described in figure \ref{fig:contour K} we assumed that moving the real part of the integration contour did not affect convergence of the integral \eqref{deform K general} at large imaginary values of the Mellin variables. In appendix \ref{Kasymp}, we discuss the asymptotic behavior of $K$-functions at large values of Mellin variables.
The main point is that $K$-functions decay as powers for large imaginary Mellin variables but their sum $\hat{M}$  decays exponentially.

\subsection{Subtractions with Unbounded Spin}

We do not have a general understanding of the case with an infinite number of subtractions with an unbounded spin. Here we simply present a couple of simple examples of this type: minimal models and free field theory correlators.

In the 2d Ising model, we know explicitly the correlator $\langle \sigma \sigma \sigma \sigma \rangle$, where $\Delta_{\sigma} = {1 \over 8}$. It is equal to $\frac{F(u, v)}{|x_1-x_3|^{\frac{1}{4}} |x_2-x_4|^{\frac{1}{4}} }$, where
\begin{eqnarray}\label{F in 2d Ising}
F^{2 {\rm dIsing}}(u, v)= \frac{\sqrt{\sqrt{u}+\sqrt{v}+1}}{\sqrt{2} \sqrt[8]{u v}}.
\end{eqnarray}
We also know a formula for the function $K(\gamma_{12}, \gamma_{14})$ for this correlator (see appendix \ref{app: K for 2d Ising}). So, we can implement the procedure outlined in this section to obtain a Mellin representation with straight contours. This is done in appendix (\ref{app: straight 2d Ising}). Note that in this case we do not need to use conformal blocks, since we know the function $K(\gamma_{12}, \gamma_{14})$. 

We conclude that if we define
\begin{eqnarray}\label{subtracted 2d Ising}
F^{2 {\rm dIsing}}_{sub} (u, v) \equiv F^{2 {\rm dIsing}}(u,v) - \frac{\sqrt{1+\sqrt{u}}+\sqrt{1+\sqrt{v}}+\sqrt{\sqrt{u}+\sqrt{v}}}{\sqrt{2}(u v)^{\frac{1}{8}}} + \frac{u^{- \frac{1}{8}}v^{- \frac{1}{8}}+ u^{- \frac{1}{8}} v^{\frac{1}{8}} + u^{\frac{1}{8}}v^{-\frac{1}{8}} }{\sqrt{2}} \nonumber,
\end{eqnarray}
then
\begin{eqnarray}\label{Mellin 2d Ising straight}
F_{sub}^{2 {\rm dIsing}}(u, v)= \int_{0 <\Re(\gamma_{12}) < {1 \over 8}} \frac{d \gamma_{12}}{2 \pi i} \int_{0<\Re(\gamma_{14})<{1 \over 8}} \frac{d{\gamma_{14}}}{2 \pi i} \hat M(\gamma_{12}, \gamma_{14}) u^{- \gamma_{12}} v^{- \gamma_{14}},
\end{eqnarray}
where the Mellin integral is evaluated with a straight contour and $\hat M(\gamma_{12}, \gamma_{14})$ is given by
\begin{eqnarray}\label{Mellin2dIsingMain}
\hat M(\gamma_{12}, \gamma_{14}) = - \sqrt{\frac{2}{\pi}} \Gamma \left(2 \gamma_{12}-\frac{1}{4}\right) \Gamma \left(2 \gamma_{14}-\frac{1}{4}\right) \Gamma (-2 \gamma_{12}-2 \gamma_{14}) \ .
\end{eqnarray}

Another example is the free scalar theory in which we consider the four-point function of $\O = \frac{1}{\sqrt{2} N} ( \vec{\phi}\,)^2$, where $\vec{\phi}$ has $N$ components.
In appendix \ref{app: free}, we show that  $\hat M(\gamma_{12}, \gamma_{14}) = 0$.
In this case,  (\ref{subtractionK}) takes the form
\begin{eqnarray}
&&\int_{\Re(\gamma_{12})>\Delta} \frac{d \gamma_{12}}{2 \pi i} \int_{\Re(\gamma_{14})>\Delta} \frac{d \gamma_{14}}{2 \pi i} K(\gamma_{12}, \gamma_{14}) u^{-\gamma_{12}} v^{- \gamma_{14}} \\
&=&\int_{\Re(\gamma_{12})=\frac{\Delta}{3}} \frac{d \gamma_{12}}{2 \pi i} \int_{\Re(\gamma_{14})=\frac{\Delta}{3}} \frac{d \gamma_{14}}{2 \pi i} K(\gamma_{12}, \gamma_{14}) u^{-\gamma_{12}} v^{- \gamma_{14}} +  u^{-\Delta} \theta(1-v) + v^{-\Delta} \theta(1-u)  \nonumber \\
&+& \frac{4}{N} \left( \theta(1-v) (u^{-\Delta/2} + u^{-\Delta/2} v^{-\Delta/2}  ) + \theta(1-u) u^{-\Delta/2} - \theta(u-1) u^{-\Delta/2} v^{-\Delta/2}  \right) \ , \nonumber
\end{eqnarray}
where $\Delta = d-2$.  
 Note the presence of the $\theta(u-1)$ term in the last line which is absent in (\ref{subtractionK}). This is related to the fact that in this case we have $J_{max} = \infty$. Therefore in this case the whole correlator comes from subtraction terms.

\subsection{Deformed Contour}\label{sec: bent}

Another option is to have a deformed contour. We will bring the three integrals in (\ref{inverseMellin naive}) into a single integral with a deformed contour (we explain this in detail in appendix \ref{app: sing var}, for the single-variable case). 
The integration contours in (\ref{inverseMellin naive}) can be deformed as long as we do not cross any OPE pole of the $K$-functions. 
In particular, if there is a deformed contour ${\cal C}$ that passes to the right of all OPE poles at $\gamma_{12}=\Delta-\frac{\tau_k}{2}$ in the $\gamma_{12}$ complex plane and similarly for $\gamma_{13}$ and $\gamma_{14}$, then we can bring the 3 integrals to the same contour.
Such a contour exists 
 unless three  poles  collide at 
\beq
\gamma_{12}=\Delta-\frac{\tau_k}{2}\,,\qquad
\gamma_{13}=\Delta-\frac{\tau_j}{2}\,,\qquad
\gamma_{14}=\Delta-\frac{\tau_i}{2}\,.
\eeq
When this happens, we say that the contour gets pinched. In order to deal with possible pinches, we employ a regularization procedure, that is encapsulated in the formula 
\begin{align}\label{Mellin pinched} 
F(u,v)  &= \lim_{\eps \to 0}\int_{{\cal C}} {\frac{d \gamma_{12}}{2 \pi i}  \frac{d \gamma_{14}}{2 \pi i} } \hat M^{\eps}(\gamma_{12}, \gamma_{14}) u^{- \gamma_{12}} v^{- \gamma_{14}} , \\
\hat M^{\eps}(\gamma_{12}, \gamma_{14}) &\equiv K (\gamma_{12} + \eps, \gamma_{14} + \eps) + K (\gamma_{13} + \eps, \gamma_{14} +\eps) + K (\gamma_{12} + \eps, \gamma_{13} + \eps).\nonumber
\end{align}
We introduce a regulator $\epsilon>0$ that allows us to separate poles that are pinched. Here we are assuming that the twist spectrum is discrete.  

\begin{figure}
  \centering
  \includegraphics[scale=1.3]{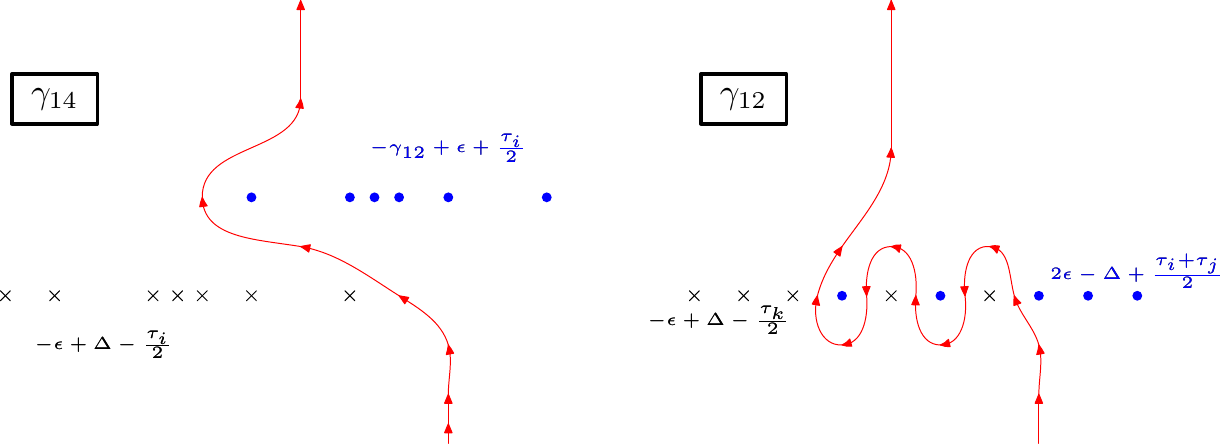} 
  \caption{Deformed integration contour $\mathcal{C}$.
  Firstly, we integrate over $\gamma_{14}$ as shown on the left keeping $\gamma_{12}$ fixed.
  Secondly,  we integrate over $\gamma_{12}$ as shown on the right.
  Pinching occurs if, as $\epsilon \to 0$, a pole marked with a black cross collides with a pole marked with a blue dot on the $\gamma_{12}$ complex plane.}
  \label{fig:deformedcontour}
\end{figure}

 The contour ${\cal C}$ can be described as follows. Firstly, we fix $\gamma_{12}$ and integrate over $\gamma_{14}$. The integrand as poles at 
\beq
\gamma_{14}=-\epsilon+\Delta-\frac{\tau_i}{2} \,,\qquad\qquad
\gamma_{14}=-\gamma_{12}+\epsilon+\frac{\tau_j}{2} \,,
\eeq
where $\tau_i,\tau_j$ are twists of exchanged operators (including descendants). %
The second set of poles originates from OPE poles in $\gamma_{13}$. 
The contour ${\cal C}$ splits the two set of poles to the left and to the right in the $\gamma_{14}$ complex plane as shown in figure \ref{fig:deformedcontour}. Notice that this is always possible for generic values of $\gamma_{12}$. For special values of $\gamma_{12}$ poles from the left set can collide with poles from the right set, pinching the integration contour and giving rise  to poles in $\gamma_{12}$.
Secondly, we consider the integral in $\gamma_{12}$. In this case, the integrand will have poles at
\beq \label{g12poles}
\gamma_{12}=-\epsilon+\Delta-\frac{\tau_k}{2} \,,\qquad\qquad
\gamma_{12}=2\epsilon-\Delta+\frac{\tau_i+\tau_j}{2} \,,
\eeq
where the second set of poles arises from pinching the contour integral over $\gamma_{14}$. Again, the contour ${\cal C}$ splits the two sets of poles two the left and to the right in the $\gamma_{12}$ complex plane as shown in figure \ref{fig:deformedcontour}.
Notice that this is always possible for arbitrarily small $\epsilon$ and a discrete twist spectrum. %
So, with $\epsilon\neq 0$, the integral in (\ref{Mellin pinched}) is always well defined. After evaluating the integral, we take the limit $\epsilon \to 0$. %

Unfortunately, equation (\ref{Mellin pinched}) is not very useful, since it involves a regularization procedure. Furthermore, it also involves the function $K(\gamma_{12}, \gamma_{14})$, which we expect to be more complicated than the Mellin amplitude $\hat M(\gamma_{12}, \gamma_{14})$. In what follows, we will consider a generic CFT\footnote{By a generic CFT, we have in mind an interacting and non-perturbative CFT$_d$ in $d>2$, like the 3d Ising model.} and evaluate the contribution from the pinches in (\ref{Mellin pinched}). We will then set $\epsilon=0$ and obtain a Mellin representation with a deformed contour, that  makes no reference to $K(\gamma_{12}, \gamma_{14})$.
  
\begin{figure}[h]
  \centering
  \includegraphics[scale=1.0]{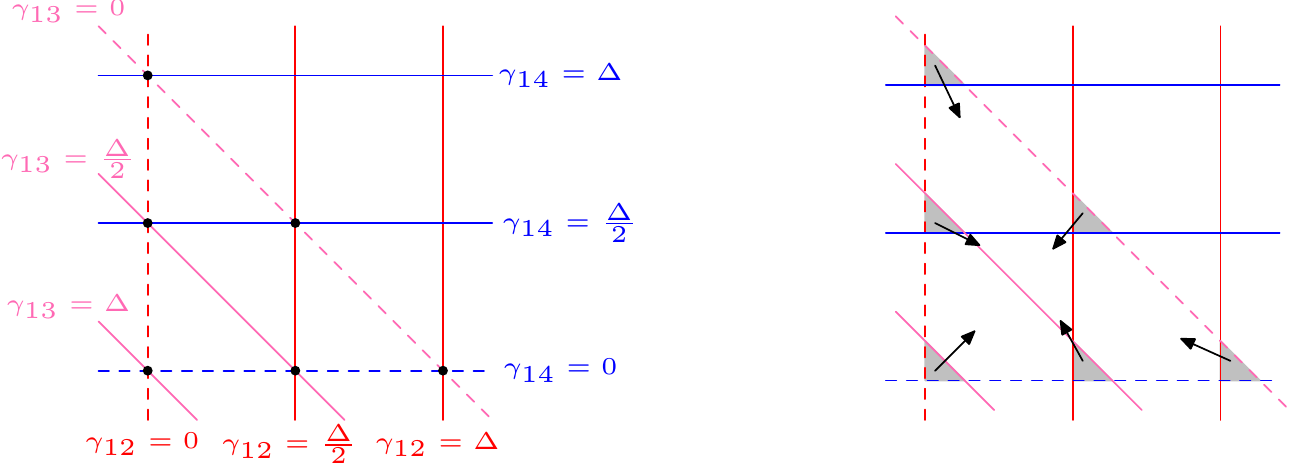} 
  \caption{Singularities in $\gamma_{12}$, $\gamma_{13}$ and $\gamma_{14}$ are represented by red, pink and blue lines respectively. We denote poles by continuous lines and accumulation points by dashed lines.  It is possible for three singularities to collide at the same point, thus causing a pinch. There are $6$ pinches that occur in generic CFTs. These are marked with black dots. Firstly, we have collisions between the identity pole and  accumulation points at $(\gamma_{12}, \gamma_{13},\gamma_{14}) =(\Delta,0,0)$ and permutations. Secondly, there can be collisions between the pole associated to the exchange of the external operator and accumulation points at  $(\gamma_{12}, \gamma_{13},\gamma_{14}) =(\frac{\Delta}{2},\frac{\Delta}{2},0)$ and permutations.
 On the right we consider the case in which we introduce the $\epsilon$ regulator. This removes the pinches  and the contour can go through the gray triangles. 
 To resolve the pinches we first move the contour out the gray triangles as indicated by the arrows and then send $\epsilon \to 0$.
  }
  \label{fig:trianglepicture}
\end{figure}

The first step is to understand when  there will be pinches in a generic CFT. From \eqref{g12poles}, we conclude that the deformed contour will be pinched as $\epsilon \to 0$, if the condition 
\begin{eqnarray}
\tau_i + \tau_j + \tau_k   = 4 \Delta
\end{eqnarray}
is satisfied.
There are $6$ pinches that occur in generic CFTs (see figure  \ref{fig:trianglepicture}). Firstly, we have collisions between the identity pole ($\tau=0$) and   accumulation points ($\tau=2\Delta$). These correspond to $(\gamma_{12}, \gamma_{13},\gamma_{14}) =(\Delta,0,0)$ and permutations. Secondly, there can be collisions between the pole associated to the exchange of the external operator ($\tau=\Delta$) and accumulation points. These correspond to 
  $(\gamma_{12}, \gamma_{13},\gamma_{14}) =(\frac{\Delta}{2},\frac{\Delta}{2},0)$ and permutations.

Suppose the contour goes through the rightmost shaded triangle  in figure \ref{fig:trianglepicture}. We deform the $\gamma_{12}$ contour to the left and we pick up the pole at $\gamma_{12}=\Delta-\epsilon$:
\begin{eqnarray}
\int_{{\rm Re}( \gamma_{12})>\Delta- \epsilon } \frac{d \gamma_{12}}{2 \pi i}  \int \frac{d \gamma_{14}}{2 \pi i}  \hat M^{\epsilon}(\gamma_{12}, \gamma_{14}) u^{- \gamma_{12}} v^{- \gamma_{14}} = u^{-\Delta + \epsilon} \int \frac{d \gamma_{14}}{2 \pi i} \hat{M}^{\epsilon}(\gamma_{12}=\Delta-\epsilon, \gamma_{14}) v^{- \gamma_{14}}\\
 +  \int_{{\rm Re}( \gamma_{12})<\Delta- \epsilon } \frac{d \gamma_{12}}{2 \pi i}   \int \frac{d \gamma_{14}}{2 \pi i}  \hat M(\gamma_{12}, \gamma_{14}) u^{- \gamma_{12}} v^{- \gamma_{14}}, \nonumber
\end{eqnarray}
where $\hat{M}^{\epsilon}(\gamma_{12}=\Delta-\epsilon, \gamma_{14})$ is the residue of the Mellin amplitude at $\gamma_{12}=\Delta-\epsilon$. In the second integral, we can drop the regularization. 
To evaluate $\hat{M}^\epsilon(\gamma_{12}=\Delta-\epsilon, \gamma_{14})$,  consider the contribution of the identity to $K(\gamma_{12}, \gamma_{14})$,
\begin{eqnarray}
\int_0^1 \frac{du}{u} \int_0^1 \frac{dv}{v} u^{\gamma_{12}} v^{\gamma_{14}} u^{-\Delta} = \frac{1}{\gamma_{14}(\gamma_{12}- \Delta)}. 
\end{eqnarray}
Similarly, $K(\gamma_{12}, \gamma_{13})$ also has a pole at $\gamma_{12}=\Delta$. We conclude that the regularised Mellin amplitude $\hat M^{\epsilon}(\gamma_{12}, \gamma_{14})$ has a pole at $\gamma_{12}=\Delta-\epsilon$ with residue given by 
\begin{eqnarray}
\hat{M}^\eps(\gamma_{12}, \gamma_{14}) \approx \frac{3 \epsilon}{(\gamma_{12}-\Delta+ \epsilon)(\gamma_{14}+ \epsilon)(- \gamma_{14}+2\epsilon)}\,,\qquad \qquad
\gamma_{12} \to \Delta- \epsilon \,.
\end{eqnarray}
Notice that this residue goes to $0$ in the limit $\epsilon \rightarrow 0$. We expected this from the fact that $F(u, v)$ cannot actually diverge due to a pinch.
Let us evaluate the finite contribution to the four point function given by the pinch:
\begin{eqnarray}
\lim_{\epsilon \rightarrow 0} u^{-\Delta + \epsilon} \int_{- i \infty}^{+ i \infty}  \frac{d\gamma_{14}}{2 \pi i} \frac{3 \epsilon}{(\gamma_{14}+ \epsilon)(- \gamma_{14}+2\epsilon)}
v^{- \gamma_{14}} = u^{- \Delta}. 
\end{eqnarray}
Notice that the integrand goes to $0$ when $\epsilon$ goes to 0, but at the same time the contour gets pinched between a pole from the left with a pole from the right. For this reason, the integration gives a finite result.

To summarize, we saw that the pinch at $\tau_i=2 \Delta$, $\tau_{j}=2\Delta$ and $\tau_k=0$ gives a finite contribution $u^{- \Delta}$ to the four point function. After taking $\epsilon \rightarrow 0$, the pole at $\gamma_{12}= \Delta$ disappears and the contour of the Mellin amplitude does not get pinched at all. By crossing symmetry, a similar discussion holds for permutations of the previous pinch condition. In the absence of other pinches, we conclude that
\begin{eqnarray}\label{conn}
F_{conn}(u, v) \equiv F(u, v)- (1+u^{- \Delta} + v^{- \Delta}) = \int \int_{{\cal C}'} \frac{d \gamma_{12}}{2 \pi i}  \frac{d \gamma_{14}}{2 \pi i} \hat M(\gamma_{12}, \gamma_{14}) u^{- \gamma_{12}} v^{- \gamma_{14}}.  
\end{eqnarray}
The Mellin integral computes the connected part of the four point function. The integral is to be taken with a deformed contour ${\cal C}'$ that differs from ${\cal C}$ as indicated by the arrows in figure \ref{fig:trianglepicture}.

Generically, $\langle \O \O \O \rangle \propto C_{\O \O \O} \neq 0$ and we also need to deal with the pinches at   $(\gamma_{12}, \gamma_{13},\gamma_{14}) =(\frac{\Delta}{2},\frac{\Delta}{2},0)$ and permutations.
In order to remove these pinches, we  consider the following function
\begin{eqnarray}\label{Fminsub}
\tilde F(u, v) \equiv F(u, v)- (1+u^{- \Delta} + v^{- \Delta}) - C_{\O \O \O}^2  \left[ u^{- {\Delta \over 2}} g^{(0)}_{\Delta, 0}(v) + v^{- {\Delta \over 2} } g^{(0)}_{\Delta, 0}(u) + v^{- \frac{\Delta}{2}} g^{(0)}_{\Delta, 0} \left(\frac{u}{v} \right) \right].  
\end{eqnarray}
The corresponding $K$-function $\tilde{K}(\gamma_{12},\gamma_{14})$ does not have poles at $\gamma_{12}=\Delta/2$ or $\gamma_{14}=\Delta/2$. This means that the contour ${\cal C}'$ can be shifted as indicated by the arrows in figure \ref{fig:trianglepicture}.
 Thus, we can write
\begin{eqnarray}\label{conn2}
\tilde F(u, v)  =  \int \int_{{\cal C'}} \frac{d \gamma_{12}}{2 \pi i}  \frac{d \gamma_{14}}{2 \pi i} \left( \hat M(\gamma_{12}, \gamma_{14}) + \delta \hat M(\gamma_{12}, \gamma_{14}) \right) u^{- \gamma_{12}} v^{- \gamma_{14}}
\end{eqnarray}
where $\delta \hat M$ is obtained from 
\begin{align}
\delta K(\gamma_{12}, \gamma_{14}) \equiv&
- \int_0^1 \frac{du}{u}  \int_0^1 \frac{dv}{v} u^{\gamma_{12}} v^{\gamma_{14}} \left[ 
1+u^{- \Delta} + v^{- \Delta} \right]\\
&- C_{\O \O \O}^2  \int_0^1 \frac{du}{u}  \int_0^1 \frac{dv}{v} u^{\gamma_{12}} v^{\gamma_{14}} \left[ u^{- {\Delta \over 2}} g^{(0)}_{\Delta, 0}(v) + v^{- {\Delta \over 2} } g^{(0)}_{\Delta, 0}(u) + v^{- \frac{\Delta}{2}} g^{(0)}_{\Delta, 0} \left(\frac{u}{v} \right) \right]\,.\nonumber
\end{align}
The first integral is elementary. The second integral converges when ${\rm Re}[\gamma_{12},\gamma_{14}] > {\Delta \over 2}$.
Using
\beq \label{int rep g}
g^{(0)}_{\Delta, 0}(v) = \int_{c - i \infty}^{c + i \infty} {d s \over 2 \pi i} v^{-s} {\Gamma(s)^2 \Gamma({\Delta \over 2}-s)^2 \Gamma(\Delta) \over \Gamma({\Delta \over 2})^4} \,,%
\eeq
with $0<c<\frac{\Delta}{2}$, we obtain
\begin{align}
&\delta K(\gamma_{12}, \gamma_{14}) =
-\frac{1}{\gamma_{12}\gamma_{14}}- 
\frac{1}{(\gamma_{12}-\Delta)\gamma_{14}}-
\frac{1}{\gamma_{12}(\gamma_{14}-\Delta)}
- C_{\O \O \O}^2 \int_{c- i \infty}^{c + i \infty} {d s \over 2 \pi i} 
\label{deltaK}\\
&
{\Gamma(s)^2 \Gamma({\Delta \over 2}-s)^2 \Gamma(\Delta) \over \Gamma({\Delta \over 2})^4}   
\left( {1 \over \gamma_{12} - {\Delta \over 2}} {1 \over \gamma_{14} - s} + {1 \over \gamma_{14} - {\Delta \over 2}} {1 \over \gamma_{12} - s} + {1 \over \gamma_{12} - s} {1 \over \gamma_{14} + s - {\Delta \over 2}}\right),  \nn 
\end{align}
with ${\rm Re}[\gamma_{12},\gamma_{14}] > {\Delta \over 2}$.
We can then explicitly compute \footnote{To see this write $s=\frac{\Delta}{4} +i\, x$ in \eqref{deltaK} and integrate over real $x$. 
Then   decrease the real part of  $\gamma_{12}$ and $\gamma_{14}$ from bigger than $\frac{\Delta}{2}$ to the neighbourhood of the crossing symmetric point $\gamma_{12}=\gamma_{14}= \frac{\Delta}{3}$. This can be done without any pole of the integrand crossing the $s$-integration contour.
Finally, sum the 3 $\delta K$'s and observe that the total integrand is an odd function of $x$.  }
\begin{align}
\delta \hat M(\gamma_{12}, \gamma_{14}) =
\delta K(\gamma_{12}, \gamma_{14}) +
\delta K(\gamma_{12}, \gamma_{13})+
\delta K(\gamma_{13}, \gamma_{14}) =0 \,.
\end{align}
We conclude that the subtractions do not affect the Mellin amplitude but only the integration contour. In particular, the subtractions in \eqref{Fminsub} did not remove the poles at $\gamma_{1i}=\frac{\Delta}{2}$ (for $i=2,3,4$) from the Mellin amplitude. This happens as follows. The function $\tilde{K}(\gamma_{12},\gamma_{14}) = K(\gamma_{12},\gamma_{14}) +\delta K (\gamma_{12},\gamma_{14})$ does not have poles at $\gamma_{12} =\frac{\Delta}{2}$ nor at 
$\gamma_{14} =\frac{\Delta}{2}$ like the original $K$-function $K(\gamma_{12},\gamma_{14})$. However, $\tilde{K}(\gamma_{12},\gamma_{14})$ has a pole at $\gamma_{13} =\frac{\Delta}{2}$ that was not present in $K(\gamma_{12},\gamma_{14})$. The same mechanism happens for the other subtractions in  (\ref{eq:subtractionF}). The exception being the exchange of the identity operator (or disconnected piece) that does not give rise to any poles in the Mellin amplitude.

In the end, we can simply write
\begin{eqnarray}
\label{eq:deformedfinal}
\tilde F(u, v)  =  \int \int_{{\cal C'}} \frac{d \gamma_{12}}{2 \pi i}  \frac{d \gamma_{14}}{2 \pi i} \hat M(\gamma_{12}, \gamma_{14})  u^{- \gamma_{12}} v^{- \gamma_{14}} \,.
\end{eqnarray}
Therefore most of the subtractions that we encountered in the straight contour formula (\ref{eq:subtractionF}) can be neatly absorbed into the deformation of the contour of integration. 
Moreover, the argument in this subsection is  valid even if  the straight contour formula requires an infinite number of subtractions with unbounded spin. 
In section \ref{sec:MM}, we show examples of such deformed contours for correlators  in 2d minimal models. 
If there are special relations among the scaling dimensions of the theory such that $\tau_i+\tau_j+\tau_k=4\Delta$ for some trio of operators, then there are extra pinches in the limit $\eps \to 0$ that must be analysed. This is relevant for perturbative CFTs.  
In appendix \ref{app:phi3}, we confirm that our general discussion works in the critical $\phi^3$ theory in $d=6+\epsilon$ spacetime dimensions to first order in $\epsilon$.

\section{Unitarity and Polynomial Boundedness}\label{sec: unitarity and regge}

In this section we analyze constraints on the Mellin amplitude coming from unitarity (or the OPE expansion) and boundedness of the correlator in the Regge limit \cite{Caron-Huot:2017vep,Maldacena:2015waa}. The OPE expansion dictates the form of the residues of the Mellin amplitude $M(\gamma_{12}, \gamma_{14})$ which are given by the Mack polynomials \cite{Mack:2009mi}, see appendix B in \cite{Costa:2012cb}. Bounds on the Regge limit restricts the rate of growth of the Mellin amplitude as one of the arguments of the Mellin amplitude becomes large.

\subsection{OPE expansion}
\label{sec:OPEexpansion}

The OPE expansion states that we can write the correlation function as a sum of conformal blocks
with positive coefficients
\begin{eqnarray}
\label{eq:OPE}
F(u,v) =u^{- \Delta} \sum_{\tau, J - {\rm even}} C_{\tau, J}^2 g_{\tau, J}(u,v)  .
\end{eqnarray}

As before it is convenient to write each conformal block as a sum in the powers of $u$ \cite{Dolan:2000ut}
\begin{eqnarray}
\label{eq:mfunctionsAg}
g_{\tau, J}(u,v) &=& u^{ \frac{\tau}{2}} \sum_{m=0}^{+\infty} u^{m} g^{(m)}_{\tau, J}(v) , \nonumber \\
g^{(0)}_{\tau, J}(v)  &=& (-1)^J (1-v)^J \ _2 F_1 ({\tau\over 2} + J, {\tau \over 2} + J, \tau + 2 J, 1 - v) .
\end{eqnarray}

From the definition of the Mellin transform (\ref{mellindef}) it is clear that powers of $u$ will lead to the poles in $\gamma_{12}$ dictated by the twists of the exchanged operators. In particular, a single primary operator with twist $\tau$ introduces an infinite series of poles at 
\beq
\label{eq:poles}
\gamma_{12}, \gamma_{14}, \gamma_{13} = \Delta - \frac{\tau}{2} - m, ~~~ m=0,1,2,... \ ,
\eeq
where $m$ is precisely the same $m$ that appears in (\ref{eq:mfunctionsAg}). Our next task is to fix the residue of the Mellin amplitude at a given OPE pole so that it  reproduces the OPE expansion (\ref{eq:OPE}).

To make contact with \cite{Costa:2012cb}, where this question was investigated in great detail we introduce {\it Mellin-Mandelstam variables}
\begin{align}\label{OPEconsequence}
t &= 2 \Delta - 2 \gamma_{12} , \cr
s &= 2 (\gamma_{12} + \gamma_{14} - \Delta) = - 2 \gamma_{13} .
\end{align}

In terms of these variables the residue of the pole takes the following form
\begin{align}\label{poles}
M(s,t) &\simeq {C_{\tau,J}^2 {\cal Q}_{J,m}^{\tau, d}(s)\over t - (\tau + 2 m)} + ... \ , ~~~ m = 0,1,2,... \ , \cr
{\cal Q}^{\tau, d}_{J,m}(s) &=- K(\tau, J, m) \ Q^{\tau, d}_{J,m}(s) ,
\end{align}
where $Q^{\Delta,\tau, d}_{J,m}(s) $ are {\it Mack polynomials} in $s$ of degree $J$.
$K(\Delta, J, m)$ is a non-negative kinematical pre-factor\footnote{Note a difference by a factor of $2^J$ compared to  \cite{Costa:2012cb} due to the normalization of conformal blocks that we adopt in this paper.}
\beq
K(\tau, J, m) = {2 \Gamma(\tau+2J) (\tau + J -1)_J \over 2^J \Gamma({\tau+2 J \over 2})^4} {1 \over m! (\tau + J  - {d \over 2} + 1)_m \Gamma \left(\Delta - {\tau \over 2} - m \right)^2} \ .
\eeq
The role of the $\left[ \Gamma(\gamma_{12}) \Gamma(\gamma_{13}) \Gamma(\gamma_{14}) \right]^2$ pre-factor in (\ref{eq:correlatorIntro}) is to correctly reproduce the collinear conformal blocks (see appendix A in \cite{Costa:2012cb}). In particular, double poles in $\Gamma(\gamma_{13})$ correctly capture the small $v$ expansion of $g^{(m)}_{\tau, J}(v)$. We find it quite remarkable that this product of Gamma-functions at the same time correctly encodes the analytic properties of the correlator through its behavior at infinity (\ref{eq:decaycorrect}), as well as the details of the OPE expansion.

Another feature worth mentioning is that $K(\tau, J, m)$ has double zeros at the position of double twist operators $\tau = 2 \Delta + 2 n$. This will play an important role in the consideration of dispersion relations in Mellin space below.

\subsection{Properties of Mack Polynomials}
\label{sec:MackPol}

Mack polynomials have many remarkable properties. Let us review some of them (we are largely following \cite{Costa:2012cb}). Similarly to Legendre polynomials Mack polynomials satisfy \cite{Dolan:2011dv,Costa:2012cb}
\beq
Q^{\tau, d}_{J,m}(s) = (-1)^J Q^{\tau, d}_{J,m}(-s-\tau -2m) \ . 
\eeq
For $m=0$ they are related to continuous Hahn polynomials
\beq \label{Macks m0}
Q^{\tau, d}_{J,0}(s) = {2^J \left( {\tau \over 2} \right)_J^2 \over (\tau + J - 1)_J} \ _3 F_2 (- J, J + \tau -1 , -{s \over 2} ; {\tau \over 2}, {\tau \over 2}; 1) .
\eeq
Higher $m$ polynomials can be computed recursively, see \cite{Costa:2012cb} for details.

In the large $s$ limit Mack polynomials behave as follows
\begin{align}\label{simplestlimit}
\lim_{s \to \infty} Q^{\Delta,\tau, d}_{J,m}(s) = s^J + O(s^{J-1}) .
\end{align}
This goes along well with the flat space scattering and $s$ being the usual Mandelstam variable. Indeed, given an exchange Witten diagram in AdS, its asymptotic behavior for large Mellin $s$ is controlled by the spin of the exchanged operator.

Another limit which is relevant to the flat space limit is $s, \tau , m \gg 1$ with $J$ fixed. In this case we get
\begin{align}\label{largetauregion}
Q^{\tau, d}_{J,m}(s) &\approx m^{ {J \over 2} } \left(m +\tau \right)^{ {J \over 2} } C^{({d-2 \over 2})}_{J} \left( {{\tau \over 2} +  m + s  \over  \sqrt{m (m + \tau)} } \right) + ... \ .
\end{align}
This asymptotic of Mack polynomials is relevant for recovering the flat space scattering amplitudes.

Mack polynomials have interesting positivity properties. We observed that for even $J$ and for general $s, \tau, m$ Mack polynomials are non-negative for 
\beq
\left| {\tau \over 2} + m + s \right| > \sqrt{m (m + \tau)} \ ,
\eeq
which again generalizes the familar property of Gegenbauer polynomials that emerge in the flat space limit (\ref{largetauregion}). We would like this to hold for any $m$ and any $\tau$ that satisfies the unitarity bound. Indeed, we observed that  
\beq
\label{eq:mackpos}
\pa_s^n Q^{\tau, d}_{J,m}(s) |_{s \geq 0} \geq 0 \ , ~~~ n \geq 0.
\eeq
This again parallels a famous property of Gegenbauer polynomials familiar from the S-matrix bootstrap considerations \cite{Martin:1965jj}. We observed it by studying many particular examples. It would be of course better to prove it rigorously. We will use this property in our considerations of Mellin space dispersion relations below.

It would be also interesting to understand more refined positivity properties of the Mack polynomials in the spirit of \cite{YuTin}. One difference compared to the flat space case is the existence of an extra quantum number $m$, in addition to ``mass'' $\tau + 2m$ and spin $J$. For example, fixing $m$ we observed that the Hankel matrices of the type \cite{YuTin} are indeed positive-definite. For the most generic case however (of general $\tau$'s and $m$'s) we did not find any positivity. We leave a more detailed exploration of these interesting properties for the future. It would be also interesting to understand if there is any relation between these positivity properties and the ones discussed in \cite{Arkani-Hamed:2018ign,Sen:2019lec,Huang:2019xzm}.

Another interesting limit is $J \gg 1$ with all other parameters being fixed\footnote{This limit is very different from the corresponding limit of Legendre polynomials that appear in the description of the flat space physics. It is responsible for intrinsically AdS effects, see appendix B in \cite{Sever:2017ylk}.}
\begin{align}\label{largespinresult}
{\cal Q}_{J,0}^{\tau, d}(s) =  {2^{2J+\tau} \over \sqrt{\pi} \Gamma^2(\Delta - {\tau \over 2})} \left( {J^{s+{1 \over 2}} \over \Gamma^2({s+\tau \over 2})} + (-1)^J {J^{-s-\tau+{1 \over 2}} \over \Gamma^2(-{s \over 2})} + O(J^{-1}) \right) \ .
\end{align}
We will use this asymptotic below in our considerations of the double twist operators in Mellin space.
\subsection{Boundedness at Infinity and the Regge limit}
\label{sec:reggebound}

Let us understand the behaviour of Mellin amplitudes at infinity. The relevant limit to consider is the Regge limit $s \to \infty$, $t$ - fixed. As explained in \cite{Costa:2012cb} this limit of the Mellin amplitude controls the Regge limit of the correlation function. Thanks to the OPE it is very easy to bound the Regge behaviour of the CFT correlation functions both nonperturbatively \cite{Caron-Huot:2017vep} and in the planar limit \cite{Maldacena:2015waa}. This leads to bounds on the Mellin amplitude $M(s,t)$ that we review in this section. 

\begin{figure}[h]
  \centering
  \includegraphics[width=0.5\textwidth]{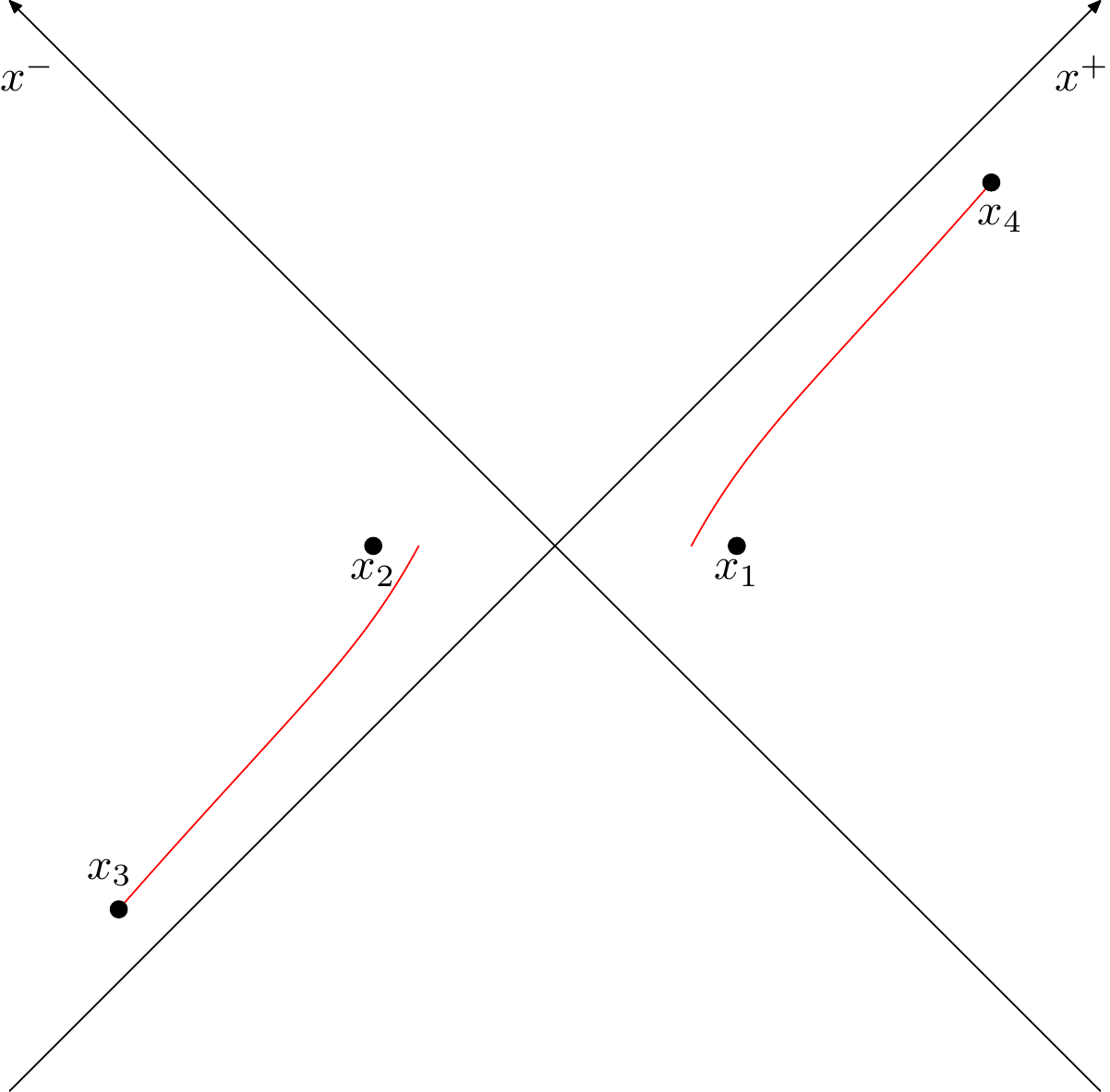}
      \caption{Kinematics (\ref{coordinates}). The Regge limit corresponds to taking $t' \to \infty$. In this limit $x_{34}^2 \to 0$.}
   \label{fig:MSS}   
\end{figure}

To describe the Regge limit consider a Lorentzian time-ordered four-point function 
\begin{align}\label{fourpoint}
F(t',\rho) = \la {\rm T} \left[ V(x_1) V(x_2) W(x_3) W(x_4) \right] \ra ,
\end{align}
where we restrict points to a Lorentzian plane and choose the following light-cone coordinates ($x^{\pm} \equiv t \pm x$)
\begin{align} \label{coordinates}
x_1^{\pm} = \pm 1, ~ x_2^{\pm} = \mp 1, ~ x_3^{\pm} = \mp e^{\rho \pm t'}, x_4^{\pm} = \pm e^{\rho \pm t'} \ .
\end{align}
see figure \ref{fig:MSS}.

In the Regge limit the cross ratios take the following values
\begin{eqnarray}\label{crossratios}
&&u = \sigma^2, ~~~ v \simeq 1 - 2 \sigma \cosh \rho , ~~~ \sigma = 4 e^{-t'}  .
\end{eqnarray}
As we increase $t'$,  $x_{14}^2 , x_{23}^2$ become time-like and $\sigma \to 0$. All other distances are space-like. The ordering of operators implies that we analytically continue $v \to v e^{2 \pi i}$ around $v=0$.

To discuss bounds on the Regge limit it is particularly convenient to set $\sigma = i \tilde \sigma = i 4 e^{-t}$. One can show that in this case $F(\tilde \sigma , \rho)$ is real and positive. Unitarity and the Euclidean OPE then imply that in any CFT \cite{Caron-Huot:2017vep}
\beq
\label{Reggebound}
\lim_{\tilde \sigma \to 0} {F(\tilde \sigma , \rho) \over F_{disc}(\tilde \sigma, \rho)} \leq 1 ,
\eeq
where $F_{disc} = \la {\rm T} \left[ V(x_1) V(x_2) \right] \ra  \la {\rm T} \left[W(x_3) W(x_4) \right] \ra  $. 

In the context of the Regge limit in large $N$ CFT, \cite{Maldacena:2015waa} considered $f(t, \rho) \equiv {F(\tilde \sigma , \rho) \over F_{disc}(\tilde \sigma, \rho)}$ and showed that the correlator obeys
\beq
\label{boundonchaos}
{| \partial_t f(t, \rho) | \over 1 - f} \leq 1 + O(e^{-2 (t-t_0)}) 
\eeq
for $t>t_0 = O(1)$.

Let us see how this comes about from the conformal Regge theory \cite{Costa:2012cb}. Assuming that the leading Regge behavior comes from a pole,\footnote{Having a more complicated singularity in the $J$-plane, say a cut, does not change a discussion since it only affects the sub-exponential terms.} we get the following behavior of the correlator in the Regge limit
\begin{eqnarray}\label{Reggelimit}
f(t, \rho) &= 1 - 2 \pi \int_{- \infty}^{\infty} d \nu \ \hat \alpha(\nu)  e^{[j(\nu) -1] t} \Omega_{i \nu} (\rho) + ...  ,
\end{eqnarray}
where $\hat \alpha(\nu)$ is related in a specific way to the product of the three-point couplings $c_{VV {\cal O}_{J}} c_{WW {\cal O}_{J}}$ with ${\cal O}_J$ being the operators of the leading Regge trajectory and $j(\nu) = j(- \nu)$. The integral over $\nu$ is then evaluated via a saddle point at $\nu = 0$. The location of the saddle at $\nu = 0$ follows from convexity properties of the Regge trajectories.\footnote{Convexity of $j_{full}(\nu)$ has been proven in \cite{Costa:2017twz}. Convexity of $j_{planar}(\nu)$ is simply assumed here.}  

In the language of the Regge trajectory $j(\nu)$ the bounds (\ref{Reggebound}) and (\ref{boundonchaos}) imply that
\begin{eqnarray}
\label{interceptbound}
j_{full}(0) \leq 1 \ , \cr
j_{planar}(0) \leq 2 \ ,
\end{eqnarray}
where $j_{full}(0)$ is the leading Regge trajectory in the finite $N$ CFT, whereas $j_{planar}(0)$ is the Regge trajectory of the single trace operators in the planar theory.  The bounds can be also generalized for non-zero complex $\nu$'s, see \cite{Kulaxizi:2017ixa,Costa:2017twz,Mezei:2019dfv}.

Let us now consider the Regge limit $s \to \infty$ of the reduced Mellin amplitude $M(s,t)$. For simplicity we consider identical operators $\Delta_V = \Delta_W = \Delta$. The relation between the correlator in the Regge limit and the  Mellin amplitude in the strip of holomorphy was worked out in \cite{Costa:2012cb} with the following result
\begin{eqnarray}
\label{eq:asymMellin}
f_{sub}(\tilde \sigma, \rho) = \int_{U_{CFT}} {d t \over 4 i} \tilde \sigma^t \Gamma\left(\Delta - {t \over 2} \right)^2  \int^{\infty} d x M({\rm Re}[s] + i x , t) \left( {x \over 2} \right)^{t-2} e^{ - x \tilde \sigma \cosh \rho} + ... \ , 
\end{eqnarray}
where the integration contour is the straight line along the imaginary axis with 
\begin{eqnarray}\label{eq:reggemellin}
U_{CFT}: ~~~~~ \tau_{sub} <\ {\rm Re}[t]&<&\tau_{sub}' ~,  ~ \nn \\
\tau_{sub} - 2 \Delta <\ {\rm Re}[s] &<& \tau_{sub}' - 2 \Delta ~, \nn \\
\tau_{sub}' - 2 \Delta<\ {\rm Re}[s]+{\rm Re}[t] &<& 2 \Delta - \tau_{sub} \ . 
\end{eqnarray}
We also suppressed terms that are subleading in the Regge limit.

Consider next the leading Regge pole contribution to the Mellin amplitude \cite{Costa:2012cb}
\begin{eqnarray}\label{Mellinregge}
M(s,t) &\simeq& \int_{- \infty}^{\infty} d \nu \ \beta(\nu) \omega_{\nu, j(\nu)}(t) {s^{j(\nu)} + (-s)^{j(\nu)} \over \sin \left( \pi j(\nu) \right)} + ... , 
\end{eqnarray}
where $\beta(\nu)$ is related in a known way to $\hat \alpha(\nu) $ in (\ref{Reggelimit}). 
Function $\omega_{\nu, J}(t)$ is given by
\begin{eqnarray}
\label{omega}
\omega_{\nu, J}(t) &=& {\Gamma({2 \Delta_V + J + i \nu - {d \over 2} \over 2}) \Gamma({2 \Delta_W + J + i \nu - {d \over 2} \over 2}) \Gamma({2 \Delta_V + J - i \nu - {d \over 2} \over 2}) \Gamma({2 \Delta_W + J - i \nu - {d \over 2} \over 2}) \over 8 \pi \Gamma( i \nu) \Gamma(- i \nu)}  \nn \\
&\times& { \Gamma({{d \over 2}+ i \nu - J - t \over 2 }) \Gamma({{d \over 2} - i \nu - J - t \over 2 }) \over \Gamma(\Delta_V - {t \over 2}) \Gamma(\Delta_W - {t \over 2})} .
\end{eqnarray}

As reviewed in the previous section as we vary $t$ the Mellin amplitude should exhibit poles at the positions of the physical operators. Let us review how they come about in (\ref{Mellinregge}) for the operators on the leading Regge trajectory. This expression has poles whenever $j(\nu) = 2 \mathbb{Z} $. Recall that $j(\nu) = j(-\nu)$ describes the leading Regge trajectory $\Delta(J)$ and is defined via
\beq
\nu^2 + \left(\Delta(j(\nu)) - {d \over 2} \right)^2 = 0. 
\eeq
Therefore, say for $j(\nu) = 2$ which corresponds to the stress tensor we have $\Delta(2) = d$ and $\nu = \pm i {d \over 2}$.

At the same time $\omega_{\nu, j(\nu)}(t)$ has poles at $t = 2 \mathbb{Z}_{\geq 0} + {d \over 2} - j(\nu) \pm i \nu$. These poles pinch the $\nu$-contour when $t$ crosses $\tau + 2m$, where $\tau$ is the twist of a physical operator with spin $j$. In this way (\ref{Mellinregge}) generates the expected poles in $t$.

Let us next plug (\ref{Mellinregge}) into (\ref{eq:asymMellin}). We get
\begin{eqnarray}
\label{eq:asymMellin}
f_{sub}(\tilde \sigma, \rho) &=&\pi \int_{- \infty}^{\infty} d \nu \beta(\nu)  {2^{j(\nu)} \over \sin {\pi j(\nu) \over 2}}  \tilde \sigma^{1-j(\nu)} \\
&&\int_{U_{CFT}} {d t \over 2 \pi i} \Gamma\left(\Delta_V - {t \over 2} \right) \Gamma\left(\Delta_W - {t \over 2} \right)  {\Gamma(j(\nu)+ t - 1) \over (2 \cosh \rho)^{j(\nu)+t -1}}  \omega_{\nu, J}(t) .
\end{eqnarray}

We next deform the $t$ contour to ${\rm Re}[t] =0$. As explained above in doing so the $\nu$-contour develops pinches at the position of the physical operators. These are precisely the operators with $0 < \tau \leq \tau_{sub}$ which cancel the subtractions that we made in defining $f_{sub}$.\footnote{Strictly speaking this only refers to the operators on the leading Regge trajectory in the analyticity in spin $J>J_0^{Regge}$ region.} After deforming the contour to ${\rm Re}[t]=0$ and doing the $t$ integral we arrive at (\ref{Reggelimit}) as shown in \cite{Costa:2012cb}. We should again contrast the Regge behavior of the Mellin amplitude which is controlled by the large $s$ behavior and the subtractions that originate from the poles in $t$.

After reviewing the relation to the coordinate space Regge limit, let us come back to the expression in Mellin space (\ref{Mellinregge}). As usual we assume that in the Regge limit the $\nu$ integral is dominated by the region $\nu = 0$,\footnote{This is not always true. For example, in the minimal models the trajectories are exactly linear and the integral is dominated by the closest pole in the upper half-plane.} 
and therefore we can use bounds on $j(0)$ (\ref{interceptbound}) to bound the growth of the Mellin amplitude.

An important question is for which values of ${\rm Re}[t]$ this argument holds? Above we made it for ${\rm Re}[t] = 0$. As we increase ${\rm Re}[t]$ the integral over $\nu$ can develop a pinch as we reviewed above and will not be dominated by $\nu = 0$ anymore. The relevant pinch corresponds to $J=2$ operator on the leading twist Regge trajectory, which for identical operators is the stress tensor, namely ${\rm Re}[t] = d-2$.

Taking into accounts bounds on $j(0)$, in that way we get the following conditions on the Mellin amplitude
\begin{eqnarray}\label{boundonmellin}
\lim_{|{\rm Im}[s]| \to \infty} | M_{full}(s,t) | &\leq& c |s| , ~~~ {\rm Re}[t]<d-2 , \\
\lim_{|{\rm Im}[s]| \to \infty} | M_{planar}(s,t) | &\leq& c |s|^2 , ~~~ {\rm Re}[t]<d-2, \nn 
\end{eqnarray}
where in writing the planar bound we implicitly allowed for slower than a power growing corrections. In terms of $\gamma_{i j}$ the bound corresponds to ${\rm Re}[\gamma_{12}]> \Delta - {d-2 \over 2}$ as we send ${\rm Im}[\gamma_{14}] \to \infty$.

\subsection{Extrapolation}
\label{sec:extrap}

The bounds in the previous section were derived only in the limit ${\rm Im}[s] \to + \infty$, or equivalently ${\rm arg}[s] =\pm {\pi \over 2}$, however we would like to relax this condition and apply the Regge bounds for any ${\rm arg}[s]$.\footnote{For ${\rm arg}[s]=0, \pm \pi$ we understand them in the tauberian \cite{Pappadopulo:2012jk,Mukhametzhanov:2018zja} or averaged sense. Indeed for these values of the arguments the Mellin amplitude has poles, however if we average the Mellin amplitude over $s$ then we assume that the Regge bound still hold.} As we change ${\rm arg}[s]$ away from $\pm {\pi \over 2}$ in principle some sort of Stokes phenomenon might occur. Here we assume that for physical correlators this does not happen and the Regge bound that we found along the imaginary axis holds everywhere in the complex $s$-plane.\footnote{This is consistent with all examples that we know, but of course it requires a proof.} At the same time as we observed above it is important to keep ${\rm Re}[t]$ in the region of holomorphy.

In this way we arrive at the following Regge bounds for the Mellin amplitude
\begin{eqnarray}
\label{eq:ReggeboundM}
\lim_{|\gamma_{12}| \to \infty} | M_{full}(\gamma_{12}, \gamma_{14}) | \leq c |\gamma_{12}|, ~~~  {\rm Re}[\gamma_{14}] > \Delta - {d-2 \over 2},  \nn \\
\lim_{|\gamma_{12}| \to \infty} | M_{planar}(\gamma_{12}, \gamma_{14}) | \leq c |\gamma_{12}|^{2}, ~~~   {\rm Re}[\gamma_{14}] > \Delta - {d-2 \over 2} \ .
\end{eqnarray}
Again we will assume that for ${\rm arg}[\gamma_{12}] = 0 , \pm \pi $ the Regge limit is still bounded in the averaged sense that the corresponding dispersion relations will converge. Imposing that the Regge bound holds at the crossing symmetric point leads to ${\Delta \over 3} > \Delta - {d-2 \over 2}$, which  gives us {\it the Regge lightness condition} 
\beq
\label{eq:ReggeLightness}
\Delta < {3 \over 4} (d-2) \ , 
\eeq
for the four-point function of identical operators. If we were to consider heavier operators we expect the Regge limit, for $\gamma_{14}$ around the crossing symmetric point, to be dominated by subtractions.
For simplicity below we will restrict our bootstrap analysis to the case $\Delta < {3 \over 4} (d-2)$.

\section{Polyakov Conditions}\label{sec: polyakov}

At this point the reader might be perplexed by the following two facts. On one hand, we have the crossing relation for the full correlator
\beq
\label{eq:OPEagain}
F(u,v) =u^{- \Delta} \sum_{\tau, J - {\rm even}} C_{\tau, J}^2 g_{\tau, J}(u,v) = v^{- \Delta} \sum_{\tau, J - {\rm even}} C_{\tau, J}^2 g_{\tau, J}(v,u)  =F(v,u) ,
\eeq
where the full correlator is reproduced by either the $s$-channel exchanges, or by the $t$-channel exchanges. On the other hand, we have the formula (\ref{eq:correlatorIntro}), where only the connected correlator is represented via a Mellin transform. Moreover, as we explained in the previous section, the Mellin amplitude has poles at the position of all operators (except the identity operator) designed in precisely such a way to reproduce the OPE expansion (\ref{eq:OPEagain}). Therefore, we seem to be running into a paradox: closing the, say $\gamma_{12}$, integration contour in (\ref{eq:correlatorIntro}) would produce the full correlator, instead of producing the connected correlator only! 

This confusion is related to the subtle nature of the twist accumulation points that we are bound to cross when trying to run into the paradox above. As we will see in the end everything is consistent, however the fact that the Mellin amplitude correctly reproduces the full correlator does lead to some subtle and nontrivial conditions on nonperturbative Mellin amplitudes, which we call Polyakov conditions \cite{Polyakov:1974gs}. Indeed, in spirit they are the same familiar conditions from the Mellin-Polyakov bootstrap program \cite{Gopakumar:2016wkt,Gopakumar:2016cpb,Gopakumar:2018xqi}. However, we will see that the nonperturbative nature of the Mellin amplitude makes them much more subtle.

\subsection{Reproducing the identity}

Let us try to run into the paradox described above. 
We  assume that the lowest twist $\tau_{gap}$ after the identity obeys ${4 \Delta \over 3} < \tau_{gap} < 2 \Delta$, so that the only subtractions in \eqref{eq:subtractionF} are the disconnected parts of the correlator.
Consider then the straight contour formula  
\begin{eqnarray}
F(u,v) &=& \left(1 + u^{- \Delta} + v^{- \Delta} \right) + \int_{\Re(\gamma_{12})=\Re(\gamma_{14})=\frac{\Delta}{3}} \frac{d \gamma_{12}}{2 \pi i} \frac{d \gamma_{14}}{2 \pi i} \hat M(\gamma_{12}, \gamma_{14}) u^{-\gamma_{12}} v^{- \gamma_{14}} ,  
\end{eqnarray}
Now keep the $\gamma_{14}$ contour fixed and deform the the $\gamma_{12}$ contour to the left picking up poles. Assuming that the resulting sum over residues converges and exchanging the sum and the $\gamma_{14}$ integration we get
\beq
F(u,v)  =  \left(1 + u^{- \Delta}  + v^{- \Delta} \right) + \sum_{\tau} \sum_{m=0}^\infty 
 u^{-\Delta+\frac{\tau}{2}+m} \int_{\Re(\gamma_{14})=\frac{\Delta}{3}} \frac{d \gamma_{14}}{2 \pi i} v^{- \gamma_{14}} {\rm Res}_{\gamma_{12}=\Delta-\frac{\tau}{2}-m} \hat M(\gamma_{12}, \gamma_{14})  ,  
\eeq
where $\tau$ are the twists of the primary operators and sum over $m$ is the sum over descendants.
As we explained in the previous section the residue is given in terms of OPE coefficients and Mack polynomials (\ref{poles})
\beq
{\rm Res}_{\gamma_{12}=\Delta-\frac{\tau_i}{2}-m} \hat M(\gamma_{12}, \gamma_{14}) =
- {1 \over 2} C_{\tau_i}^2 {\cal Q}_{J_i,m}^{\tau_i, d}(\gamma_{14})
\Gamma^2(\gamma_{14})\Gamma^2\left(\frac{\tau_i}{2} + m -\gamma_{14}\right)
\Gamma^2\left(\Delta- \frac{\tau_i}{2} - m \right) \ .
\eeq
A factor of $-{1 \over 2}$ in front comes from the fact that we take the residue in $\gamma_{12}$, see (\ref{poles}).

Notice that the contour   $\Re(\gamma_{14})=\frac{\Delta}{3}$ falls in between the series of poles produced by the Gamma functions (since ${\tau_{gap} \over 2} - {\rm Re}(\gamma_{14} ) +m > {\Delta \over 3} > 0$). Therefore the Mellin integral just reproduces the collinear blocks as required by the OPE. Thus, we find
\beq
F(u,v)  =   \sum_{\tau,J} 
 C_{\tau,J}^2  g_{\tau,J}(u, v) +\left(1  + v^{- \Delta} \right) 
 +
\sum_{n=0}^\infty
 u^{n} \int_{\Re(\gamma_{14})=\frac{\Delta}{3}} \frac{d \gamma_{14}}{2 \pi i} v^{- \gamma_{14}}  R_n(\gamma_{14}) ,  
\eeq
where
\beq
R_n(\gamma_{14})\equiv  {\rm Res}_{\gamma_{12}=-n} \hat M(\gamma_{12}, \gamma_{14})\,.
\eeq
Given that the first sum already gives the full correlator, the other two terms must cancel.
This gives the Polyakov conditions
\beq
\left(1  + v^{- \Delta} \right) +
  \int_{\Re(\gamma_{14})=\frac{\Delta}{3}} \frac{d \gamma_{14}}{2 \pi i} v^{- \gamma_{14}}  R_0(\gamma_{14})=0
  \label{impos}
\eeq
and $R_n=0$ for $n>0$. However, clearly \eqref{impos} is impossible to satisfy. Indeed, $1+v^{-\Delta}$ does not admit the usual Mellin representation as required by (\ref{impos}).

The resolution of this apparent paradox lies in the fact that our assumption about the convergence of the sum over $\gamma_{12}$ residues for ${\rm Re}[\gamma_{14}] = {\Delta \over 3}$ does not hold. To see this we note that the relevant divergence comes from the large $J$ fixed $\tau$ operators which are controlled by the light-cone bootstrap \cite{Fitzpatrick:2012yx,Komargodski:2012ek}. To leading order we can therefore simply use the mean field theory OPE data together with (\ref{largespinresult}) to get
\footnote{Since we are looking at poles accumulating at $\gamma_{12}=0$ when $J \to \infty$ we can use  $\gamma_{13} = \Delta - \gamma_{14}$. }
\begin{eqnarray}
\label{eq:sumoverspin}
&& 
- {1 \over 2} ( C^{GFF}_{\tau = 2 \Delta, J} )^2 {\cal Q}_{J,0}^{\tau, d}(\gamma_{14})
\Gamma^2(\gamma_{14})\Gamma^2\left(\frac{\tau}{2}-\gamma_{14}\right)
\Gamma^2\left(\Delta- \frac{\tau}{2} \right)\label{Mellin near acc}  \\
&&=  {4 \over \Gamma^2(\Delta)} {1 \over J} \left(J^{2 (\Delta - \gamma_{14}) } \Gamma^2(\gamma_{14}) + J^{2 \gamma_{14}} \Gamma^2(\Delta - \gamma_{14}) + ...  \right) \ ,
\qquad J\to \infty\,, \ \ \tau \to 2\Delta\,,
\nn
\end{eqnarray} 
where we omitted the terms that are suppressed at large spin $J$. 
Note that the sum over $J$ of the first term in the brackets in the second line of (\ref{Mellin near acc}) converges only for $\Re \,\gamma_{14} >\Delta $, while the second for $\Re\,\gamma_{14}<0$. Therefore, if we try to evaluate the Mellin integral by closing the $\gamma_{12}$ contour we run into a divergent sum for any value of $\gamma_{14}$. 

The resolution is that we should first deform (\ref{eq:sumoverspin}) into the region where the sum converges. Indeed, let us first deform the contour in the first term (\ref{eq:sumoverspin}) to $\Re\,\gamma_{14}>\Delta$ and in the second term to $\Re\, \gamma_{14}<0$. This makes the sum over $J$ convergent and thus we can exchange the order of the sum and the integral.  One way to do it is to split the Mellin amplitude back into the $K$-functions such that two powers of $J$ in (\ref{eq:sumoverspin}) appear in different $K$-functions. We then first deform the $\gamma_{14}$ contour before closing the $\gamma_{12}$ contour, essentially going back to (\ref{inverseMellin naive}). In doing so we encounter extra poles which cancel the disconnected piece and in this way the double counting is avoided.

A simpler way to see it is to note the following. To ensure the convergence of the integral we would like to deform the $\gamma_{14}$ for each of the two terms in (\ref{Mellin near acc}) to the region, where the sum over $J$ converges. Let us start with the second term. In this case we would like to deform the Mellin integral to the region $\Re\,\gamma_{14}<0$. It is easy to see that in doing so we encounter a pole. Indeed, $\sum_{J - {\rm even}} {1 \over J} J^{2 \gamma_{14}} \sim -{1 \over 4 \gamma_{14}}$. The residue of this pole produces $-1$. Similarly, in the first term when continuing to the region $\Re\,\gamma_{14}> \Delta$ we encounter the pole at $\gamma_{14} = \Delta$ with the residue being precisely $-v^{- \Delta}$. We see that by deforming the $\gamma_{14}$ region so that the sum over the $\gamma_{12}$ residues converges we precisely canceled the disconnected piece.

Let us quickly check that the deformed integral indeed correctly reproduces the expected light-cone singularity in the dual channel
\begin{eqnarray}
\label{eq:sumoverspinB}
\delta f(u, v) = \int_{{\rm Re}(\gamma_{14}) <0 } \sum_{J > J_0, J - {\rm even}}^{\infty}  {4 \over \Gamma^2(\Delta)} {1 \over J}   \frac{d \gamma_{14}}{2 \pi i} J^{2 \gamma_{14}} \Gamma^2(\Delta - \gamma_{14}) v^{- \gamma_{14}} \nn \\
+  \int_{{\rm Re}(\gamma_{14})> \Delta}  \sum_{J > J_0, J - {\rm even}}^{\infty}  {4 \over \Gamma^2(\Delta)} {1 \over J} \frac{d \gamma_{14}}{2 \pi i} J^{2 (\Delta - \gamma_{14}) } \Gamma^2(\gamma_{14}) v^{- \gamma_{14}} + ... \nonumber
\end{eqnarray}
Both integrals converge. They give 
\begin{eqnarray}
\label{eq:changeinf}
\delta f(u, v) = \sum_{J > J_0, J - {\rm even}}^{\infty} \frac{8 J^{-1+2\Delta}}{\Gamma^2(\Delta)} \Big( v^{- \Delta} K_0(\frac{J}{\sqrt{v}}) + K_0(2 J\sqrt{v}) \Big) + ... \ .
\end{eqnarray}

Since we are interested in the small $v$ asymptotic we can turn the sum into an integral. The fact that we sum over even $J$ produces an extra factor of ${1 \over 2}$ and we get
\begin{eqnarray}
\delta f(u, v) &=& 4 \int_0^\infty dJ \frac{J^{-1+2\Delta}}{\Gamma^2(\Delta)} \Big( v^{- \Delta} K_0(\frac{2 J}{\sqrt{v}}) + K_0(2 J\sqrt{v}) \Big) + ... \ \nonumber \\
&=&1+ v^{-\Delta} + ... 
\end{eqnarray}
as expected. 

Below we devise a toy model which demonstrates the issue discussed above in a simpler and more controlled setting.

\subsection{Toy Model}

We can illustrate the general ideas discussed above in a specific example. Consider the following function
\beq
f(u,v) = e^{- u} \sum_{J=1}^{\infty} u^{-\gamma(J)} e^{- J v} ,
\eeq
where $\lim_{J \to \infty} \gamma(J) = 0$. This mimicks the accumulation point in the $u \to 0$ OPE channel.
In the dual channel we have the following asymptotic
\beq
\label{eq:toymodelunit}
f(u,v) = {e^{-u} \over v} + ...\,,  \qquad\qquad v \to 0\ . 
\eeq

We can now compute the Mellin amplitude 
\beq
\hat M(\gamma_{12}, \gamma_{14}) \equiv \int_0^{\infty} {d u d v \over u v} u^{\gamma_{12}} v^{\gamma_{14}} f(u,v) = \Gamma(\gamma_{14}) \sum_{J=1}^{\infty} J^{- \gamma_{14} } \Gamma(\gamma_{12} - \gamma(J) ), 
\eeq
For ${\rm Re}[\gamma_{12}] > \gamma(J),\,\forall J\in \mathbb{N}$ and ${\rm Re}[\gamma_{14}]> 1$.
As expected the Mellin amplitude has a pole at $\gamma_{14} = 1$, namely $\hat M \sim {\Gamma(\gamma_{12}) \over \gamma_{14} - 1}$ which is of course consistent with  (\ref{eq:toymodelunit}).

We can also easily write down the expression for the analytic continuation of the Mellin amplitude to ${\rm Re}[\gamma_{14}]>0$
\beq
\label{eq:mellincontinued}
\hat M(\gamma_{12}, \gamma_{14}) =  \Gamma(\gamma_{12}) \Gamma(\gamma_{14}) \zeta(\gamma_{14}) 
+\Gamma(\gamma_{14}) \sum_{J=1}^{\infty} J^{- \gamma_{14} } \left( \Gamma(\gamma_{12} - \gamma(J) ) -  \Gamma(\gamma_{12} )  \right) ,
\eeq
where $\zeta(x)$ is the Riemann zeta function. Here we assumed that $\gamma(J) \to 0$ at large $J$ not slower than ${1 \over J}$.

We can write the inverse Mellin representation
\begin{eqnarray}
\label{eq:deformsub}
f(u,v) &=& \int_{{\rm Re}[\gamma_{12}] > \gamma(J)} {d \gamma_{12} \over 2 \pi i} \int_{{\rm Re}[\gamma_{14}] > 1} {d \gamma_{14} \over 2 \pi i} u^{- \gamma_{12}} v^{- \gamma_{14}} \hat M(\gamma_{12}, \gamma_{14}) \\
&=& {e^{-u} \over v} +  \int_{{\rm Re}[\gamma_{12}] > \gamma(J)} {d \gamma_{12} \over 2 \pi i} \int_{0 <{\rm Re}[\gamma_{14}] < 1} {d \gamma_{14} \over 2 \pi i} u^{- \gamma_{12}} v^{- \gamma_{14}} \hat M(\gamma_{12}, \gamma_{14}) , \nonumber
\end{eqnarray}
where in the second line we deformed the contour to extract the leading singularity in the dual channel (which is analogous to the disconnected piece of a CFT correlator).

Now let us try to evaluate the Mellin integral by closing the $\gamma_{12}$-contour. As above we can write the contribution of the physical operators as follows
\beq
\label{eq:toyressum}
\Gamma(\gamma_{14}) \sum_{J=1}^{\infty}  J^{- \gamma_{14}} u^{- \gamma(J)} .
\eeq
If we are to blindly exchange the sum and the $\gamma_{14}$ integral we would run into the double counting paradox as in the section above. The resolution of course is that the sum (\ref{eq:toyressum}) converges only for ${\rm Re}[\gamma_{14}] > 1$. Therefore, we can only close the $\gamma_{12}$ contour in the first line of (\ref{eq:deformsub}) and the double counting problem does not arise. 

We can also use (\ref{eq:mellincontinued}) to write for the residues as we deform the $\gamma_{12}$ contour to ${\rm Re}[\gamma_{12}]<0$
\beq
\label{eq:g12closing}
\Gamma(\gamma_{14}) \left( \zeta(\gamma_{14}) + \sum_{J=1}^{\infty}  J^{- \gamma_{14}} \left( u^{-  \gamma(J)} - 1 \right) \right) \ .
\eeq
Note that the expression above is formal in the sense that strictly speaking as we deform the contour we separately get $ \sum_{J=1}^{\infty}  J^{- \gamma_{14}} u^{-  \gamma(J)}$ and $- \sum_{J=1}^{\infty}  J^{- \gamma_{14}}$. However for $0 < {\rm Re}[\gamma_{14}] <1$ only the combined sum is well-defined. 

Plugging (\ref{eq:g12closing}) into (\ref{eq:deformsub}) and expanding it up to $O(u)$ we get
\begin{eqnarray}
f(u,v) &=& {1\over v} + \left( {1 \over e^v -1} - {1 \over v} \right) + \sum_{J=1}^{\infty}  \left( u^{-  \gamma(J)} - 1 \right) e^{- v J} + ... \\
&=& \sum_{J=1}^{\infty} u^{-  \gamma(J)}  e^{- v J} + ... ,
\end{eqnarray}
where ${1 \over e^v -1} - {1 \over v}$ is the Mellin transform of $\Gamma(\gamma_{14}) \zeta(\gamma_{14})$. Thus, we see that again
the double counting problem does not arise.

We can now ask what is the behavior of the Mellin amplitude close to the accumulation point. To this extent following the analogy to the light-cone bootstrap let us set $\gamma(J) = {\alpha \over J^{\beta}}$, $0 < \alpha <1$ and $\beta > 0$. For simplicity we can also consider an integral instead of the sum to get
\beq
\tilde M = \int_1^\infty d J \ J^{- \gamma_{14}} {1 \over \gamma_{12} - {\alpha \over J^{\beta}}} = { \ _2 F_1 (1, {\gamma_{14}-1 \over \beta}, {\gamma_{14} +\beta + 1 \over \beta}, {\alpha \over \gamma_{12}}) \over \gamma_{12} (\gamma_{14} - 1)} , ~~~ \quad {\rm Re}[\gamma_{14}]>1  \ .
\eeq

This has the following behavior close to the accumulation point (we can approach it from the regular direction which is ${\rm arg}[\gamma_{12}]  \neq 0$)
\beq
\label{eq:nonpertlim}
\lim_{|\gamma_{12}| \to 0, ~ {\rm arg}[\gamma_{12}]  \neq 0} \tilde M =  {\pi \over \beta \gamma_{12}} \left( - {\alpha \over \gamma_{12}} \right)^{{1 - \gamma_{14} \over \beta}} {1 \over \sin {\pi \over \beta} (\gamma_{14}-1) } + ... , ~~~ {\rm Re}[\gamma_{14}]>1 ,
\eeq
where we suppressed regular terms. Therefore the accumulation point behaves like a branch point with the asymptotic controlled by the ``large spin OPE data''. 
However, for $\Re[\gamma_{14}]>1$, it is not a branch point because there is no monodromy, \emph{i.e.} we can do a contour integral around it (going in between the poles for $\gamma_{12}>0$) and the result is just the convergent sum of the residues of the enclosed poles. 
Note also that in the region of convergence, namely ${\rm Re}[\gamma_{14}]>1$ we have
\beq
\label{eq:toycond}
\lim_{|\gamma_{12}| \to 0, ~ {\rm arg}[\gamma_{12}]  \neq 0} \gamma_{12} \tilde M = 0 ,~~~ {\rm Re}[\gamma_{14}]>1 .
\eeq
which states that we do not have a double trace operator at $\gamma_{12} = 0$. Below we will see that this is the relevant condition for the nonperturbative Mellin amplitudes.

This complicated behavior has to be contrasted with the perturbative behavior. Indeed, if we think of $\alpha = {1 \over c_T} \sim{1 \over N^2} \to 0$ we get order by order a very simple expansion  
\beq
\label{eq:largeNtoy}
\tilde M = {1 \over \gamma_{12} (\gamma_{14} - 1)} + {\alpha \over \gamma_{12}^2 (\gamma_{14}+\beta - 1)} +  {\alpha^2 \over \gamma_{12}^3 (\gamma_{14}+2 \beta - 1)}+ ... \ ,
\eeq
which is much simpler than the ``non-perturbative'' limit (\ref{eq:nonpertlim}). Notice also that the condition (\ref{eq:toycond}) is genuinely nonperturbative. If we try to plug the ``large $N$'' expansion (\ref{eq:largeNtoy}) in (\ref{eq:toycond}) we see that only the leading term produces a finite result ${1 \over \gamma_{14}-1}$, whereas the higher terms in $\alpha\sim {1 \over N^2}$ produce infinity.

\subsection{Nonperturbative Polyakov Conditions}

We are now ready to formulate Polyakov conditions for nonperturbative amplitudes. In light of the discussion above we consider the derivative of the correlator
\begin{eqnarray}
-u \pa_u F(u,v) &=& \Delta u^{- \Delta} + \int_{{\cal C}} \frac{d \gamma_{12}}{2 \pi i} \frac{d \gamma_{14}}{2 \pi i}  \gamma_{12} \hat M(\gamma_{12}, \gamma_{14}) u^{-\gamma_{12}} v^{- \gamma_{14}} . 
\end{eqnarray}
Note that since $- u \pa_u (1 + v^{- \Delta}) = 0$ if we are to close the $\gamma_{12}$ contour we will not run into the double counting problem described above. Correspondingly, close to the first twist accumulation point $\gamma_{12} = \Delta - {\tau_{[{\cal O}, {\cal O}]_{0,J}} \over 2} = - {\gamma_{[{ {\cal O}, {\cal O}]_{0,J}} } \over 2} \sim {1 \over J^{\tau_{gap}}}$ which improves the convergence of the sum over spins in (\ref{eq:sumoverspin}).\footnote{By $[{\cal O}, {\cal O}]_{n,J}$ we as usual denote a family of the double-twist operators that approach twist $2 \Delta+2n$ at infinite spin.} In particular, the sum over $J$ of both terms in (\ref{eq:sumoverspin}) multiplied by ${1 \over J^{\tau_{gap}}}$  converges for 
\begin{eqnarray}
\label{conditionPol}
{\tau_{gap} \over 2}> {\rm Re}[\gamma_{14}] > \Delta - {\tau_{gap} \over 2} .
\end{eqnarray}
Notice that our original assumption $\Delta < \frac{3}{4} \tau_{gap}$ guarantees that there are allowed values of $\gamma_{14}$ compatible with this condition.

To analyze the behavior of the Mellin amplitude close to the branch point we can use the toy model from the previous section. As in the toy model example above, we conclude that the presence of the double-twist trajectory makes $\gamma_{12} = 0$ look like a branch point, see (\ref{eq:nonpertlim}). Similarly, we conclude that 
\beq
\lim_{|\gamma_{12}| \to 0, ~ {\rm arg}[\gamma_{12}]  \neq 0} \gamma_{12} \left( \gamma_{12} \hat M(\gamma_{12}, \gamma_{14}) \right) = 0 ,~~~ \quad {\tau_{gap} \over 2}> {\rm Re}[\gamma_{14}] > \Delta - {\tau_{gap} \over 2}  .
\eeq
In other words, the higher spin tail produces a contribution which is softer than a pole. 

Let us now translate this condition to a statement about the Mellin amplitude $M(\gamma_{12}, \gamma_{14})$ itself. Recall that due to the pre-factor $\Gamma^2(\gamma_{12})\Gamma^2(\gamma_{13})\Gamma^2(\gamma_{14})$ that relates $\hat M$ to $M$ and which has a double pole at $\gamma_{12} = 0$ we can rewrite the condition above as follows
\beq
\label{eq:finalpolyakov}
M(\gamma_{12}=0, \gamma_{14})= 0 ,\qquad \qquad  {\tau_{gap} \over 2}> {\rm Re}[\gamma_{14}] > \Delta - {\tau_{gap} \over 2}  ,
\eeq
where we set $\gamma_{12} = 0$ by approaching the accumulation point from any direction with  ${\rm arg}[\gamma_{12}] \neq 0$. The condition (\ref{eq:finalpolyakov}) is the central result of this section. Note that the non-perturbative Polyakov condition is very subtle. In particular, we cannot argue that the Mellin amplitude has a double zero at $\gamma_{12} =0$ and similarly we cannot simply go to the accumulation points with twists $2 \Delta + 2n$. Finally, the condition is a genuinely nonperturbative (finite $N$) condition. We leave further exploration of these extra Polyakov conditions for the future.

\subsection{Scalar Ambiguity}
\label{sec:scalaramb}

Let us emphasize the importance of the condition (\ref{eq:finalpolyakov}) for the nonperturbative bootstrap. Indeed, it is very easy to construct a crossing-symmetric Regge bounded Mellin amplitude with correct analyticity properties by considering scalar exchanges in AdS.

Let us be more explicit. We take external fields to be identical scalars of dimensions $\Delta$ and the exchanged field of dimension $ \Delta_{exch}$. As was shown in \cite{Penedones:2010ue} the Mellin amplitude in this case takes the following form

\begin{eqnarray}\label{phicube}
M^{sc}(\gamma_{12}, \gamma_{14}) &=& M^{12-34}(\gamma_{12}, \gamma_{14}) + M^{13-24}(\gamma_{12}, \gamma_{14}) + M^{14-23}(\gamma_{12}, \gamma_{14}) \nn  , \\ 
M^{12-34}(\gamma_{12}, \gamma_{14}) &=& {g^2 \over 2} \sum_{m=0}^{\infty} {r_{m} \over \gamma_{12} - \Delta + {\Delta_{exch} \over 2} + m} \nn , \\
r_{m} &=& {\Gamma(\Delta + {\Delta_{exch} - d \over 2})^2 \over 2 \Gamma(2 \Delta - {d \over 2})} { (1 + {\Delta_{exch} - 2 \Delta \over 2})_m^2 \over m! \Gamma(\Delta_{exch} - {d \over 2} + 1 + m)} \geq 0 ,
\end{eqnarray}
where we explicitly wrote the contribution of an exchange in every channel. $M^{13-24}(\gamma_{12}, \gamma_{14})$ and $M^{14-23}(\gamma_{12}, \gamma_{14})$ are trivially obtained from $M^{12-34}(\gamma_{12}, \gamma_{14})$ by corresponding permutations. Note that as expected from unitarity all the residues $R_{m}$ are manifestly positive.

Consider now the large $\gamma_{12} \to \infty$ limit. We get
\beq
\lim_{\gamma_{12} \to \infty} M^{sc}(\gamma_{12}, \gamma_{14}) < A(\gamma_{14}) .
\eeq
In other words, $M^{sc}(\gamma_{12}, \gamma_{14})$ satisfies (\ref{eq:ReggeboundM}). 

It is however easy to see that $M^{sc}(\gamma_{12}, \gamma_{14})$ does not satisfy the nonperturbative Polyakov condition (\ref{eq:finalpolyakov}). Indeed, it is easy to check that each $m$ gives a positive contribution at $\gamma_{12}=0$. Therefore, when considering nonperturbative Mellin amplitudes we do not have an ambiguity of adding scalar exchanges in AdS as soon as we impose the Polyakov conditions. Note that similar ambiguity exists in the coordinate space as well. Given a crossing symmetric function that satisfies subtracted dispersion relations \cite{Carmi:2019cub} we can add to it the sum over exchanges above and it will still obey the same subtracted dispersion relations and obey crossing.

\section{Dispersion Relations in Mellin Space}\label{sec:application}

After establishing the basic properties of nonperturbative CFT Mellin amplitudes we finally would like to consider some applications. The strategy we adopt relies on all the properties that we established in the previous sections. We use analyticity and polynomial boundedness of Mellin amplitudes to write down subtracted dispersion relations. We then impose crossing to simplify them and we finally impose the nonperturbative Polyakov condition (\ref{eq:finalpolyakov}). The result of all this is a set of linear functionals that act on the OPE data and give zero. 

These functionals have some remarkable properties. They annihilate generalized free field theory. They are non-negative for heavy operators and have double zeros at the positions of the double twist operators $[{\cal O}, {\cal O}]_{n,J}$. We find that in this way they are particularly suitable for studies of large $N$ holographic CFTs.

We check the overall consistency of our construction by applying the functionals to the OPE data of the 3d Ising model. After this successful and nontrivial test we move on and apply them to some simple holographic theories. 


%

\subsection{Subtractions and Polyakov condition}

For simplicity in this section we consider a limited class of CFTs in $d>2$ for which the analysis is particularly simple. We assume that the theory admits a scalar primary operator with dimension $\Delta$ and the twist gap $\tau_{gap}$ such that $\Delta < {3 \over 4} \tau_{gap}$. As we argued above in this case we can write down the Mellin representation for the connected correlator with the straight contour
\begin{eqnarray}
F(u,v) &=&1 + u^{- \Delta} + v^{-\Delta} \nn \\
&+& \int_{{\cal C}} {d \gamma_{12} d \gamma_{14} \over (2 \pi i)^2} \ u^{- \gamma_{12}} v^{- \gamma_{14} }\Gamma(\gamma_{12})^2 \Gamma(\gamma_{14})^2 \Gamma(\Delta - \gamma_{12} - \gamma_{14})^2 M(\gamma_{12}, \gamma_{14}) \nn \\
{\cal C} &:& \Delta - {\tau_{gap} \over 2} < \Re(\gamma_{12}), \Re(\gamma_{14}) , 
~~~  \Re(\gamma_{12})+\Re(\gamma_{14}) < {\tau_{gap} \over 2} \ .
\end{eqnarray}
Furthermore, we shall assume that $\tau_{gap} =d-2$ corresponds to the stress tensor operator.

Let us consider dispersion relations for the Mellin amplitude $ M(\gamma_{12}, \gamma_{13}, \gamma_{14})$, where $\gamma_{12}+\gamma_{13}+\gamma_{14} = \Delta$, at fixed $\gamma_{13}$. According to (\ref{eq:ReggeboundM}), the Mellin amplitude is bounded by the linear growth $|\gamma_{12}|$ for fixed $\Re(\gamma_{13})>\Delta -\tau_{gap}/2$. In particular, this includes a neighbourhood of $\gamma_{13}=\frac{\Delta}{3}$.
Therefore we can write the fixed $\gamma_{13}$ subtracted dispersion relation as follows
\begin{eqnarray}
& &{M(\gamma_{12}, \gamma_{13}, \gamma_{14}) \over (\gamma_{12} - {\Delta \over 3}) (\gamma_{13} - {\Delta \over 3}) (\gamma_{14} - {\Delta  \over 3})} 
=\oint_{\gamma_{12}} \frac{d\gamma}{2\pi i} \frac{1}{\gamma-\gamma_{12}}
{M(\gamma, \gamma_{13}, \Delta-\gamma_{13}-\gamma) \over (\gamma - {\Delta \over 3}) (\gamma_{13} - {\Delta \over 3}) (\Delta-\gamma_{13}-\gamma - {\Delta  \over 3})} \nn\\
&&=- {M({\Delta  \over 3}, \gamma_{13}, {2\Delta \over 3} - \gamma_{13}) \over (\gamma_{13} - {\Delta \over 3})^2} \left( {1 \over \gamma_{12} - {\Delta \over 3}} + {1 \over \gamma_{14} - {\Delta \over 3}}  \right) \nn \\
&&-{1 \over 2} \sum_{\tau, J, m} {C_{\tau, J}^2 {\cal Q}_{J,m}^{\tau, d}(\gamma_{13}) \over (\Delta - {\tau \over 2} - m - {\Delta \over 3})(\gamma_{13}  - {\Delta \over 3})(\Delta - \gamma_{13} - (\Delta-{\tau \over 2} - m) - {\Delta \over 3})} \nn \\
&&\times \left( {1 \over \gamma_{12} - \Delta +{\tau \over 2} + m} + {1 \over \gamma_{14} - \Delta + {\tau \over 2} + m}  \right) \ ,
\end{eqnarray}
where the last expression was obtained by opening up the contour integral in the first line and picking up all the poles in the $\gamma$ complex plane.
In the last expression, we used the fact that $M(\gamma_{12}, \gamma_{13}, \gamma_{14})=M(\gamma_{14}, \gamma_{13}, \gamma_{12})$.
Crossing symmetry further implies that 
\beq
M \left({\Delta \over 3}, \gamma_{13}, {2\Delta \over 3} - \gamma_{13}\right) = G\left( \big(\gamma_{13} - {\Delta \over 3} \big)^2\right)\,.
\eeq
We can solve for derivatives of $G(x^2)$ at $x^2 = 0$ in terms of the OPE data. To do it we evaluate the formula above as $M ({\Delta \over 3} - x, {\Delta \over 3}, {\Delta \over 3} + x) = G(x^2)$, expand in $x$ and solve for derivatives of $G$. We should remark at this point that as usual doing subtractions in dispersion relations is a matter of choice. We will comment on other choices below.

We next consider the nonperturbative Polyakov condition (\ref{eq:finalpolyakov}). More precisely, we expand the condition   $M(0, \gamma_{13}, \Delta  - \gamma_{13}) = 0$  around $\gamma_{13} = {\Delta  \over 3}$. In this way, we get an infinite set of equations the simplest of which takes the following form
\begin{eqnarray}
\label{eq:magicfunctionals}
&&\sum_{\tau, J, m} C_{\tau, J}^2 \alpha_{\tau, J, m} = 0 ,  \\
&&\alpha_{\tau, J, m} = - {16 \Delta \over 3 (\tau - {2 \Delta \over 3} + 2m) (\tau - {4 \Delta \over 3} + 2m)} \left( {(\tau +2  m - \Delta) {\cal Q}_{J,m}^{\tau, d}({\Delta \over 3}) \over   (\tau - {2 \Delta \over 3} + 2m) (\tau - {4 \Delta \over 3} + 2m)} - {\Delta \over 3} {{\cal Q}_{J,m}^{\tau, d}({\Delta \over 3})' \over \tau +2  m - 2 \Delta } \right) \ , \nn \\
&&\alpha_{\hat 1} =\alpha_{0, 0, m}= 0 \ , \nn 
\end{eqnarray}
where in the last line we explicitly wrote that the identity operator does not contribute to (\ref{eq:magicfunctionals}).
In other words, we arrived at a particular set of linear functionals that act on the OPE data. They have very interesting properties that we describe below.

First, note that ${\cal Q}_{J,m}^{\tau, d}$ have double zeros at the position of double twist operators $[{\cal O}, {\cal O}]_{n,J}$. This directly translates to the fact that $\alpha_{\tau, J, m}$ have double zeros for $\tau = 2 \Delta + 2n$ with $n \geq 1$. For $n=0$ and $J \neq 0$, $\alpha_{2 \Delta, J,m}$ has a single zero due to the extra pole in the second term in (\ref{eq:magicfunctionals}) for $m=0$. Therefore, generalized free fields automatically satisfy the sum rule (\ref{eq:magicfunctionals}).

Second, we find that 
\begin{eqnarray}
\alpha_{\tau, J, m} \geq 0, ~~~ \qquad   \tau \geq 2 \Delta \ , \ \ \ J \in 2\mathbb{N}\ , \ \ \ m \ge 0\,,
\label{eq:magicproperty}
\end{eqnarray}
where the only zeros of $\alpha_{\tau, J, m}$ for $\tau \geq 2 \Delta$ are the ones at the positions of the double twist operators described above. We elaborate on evidence for this claim (that we do not prove) below in section \ref{sec:positivityofalpha}. Therefore we conclude that the functionals above are extremal (in the sense of \cite{ElShowk:2012hu, Mazac:2016qev}) for the following bootstrap problem: {\it 
Find the maximal value of $\tau_{0}$ for which there exist a unitary solution to the crossing equations with all twists $\tau \ge \tau_0$ for all spins $J$ (apart from the identity operator).
}

From the properties of the functionals described above it immediately follows that there are no nontrivial solutions to crossing equations that satisfy $\tau_0 > 2 \Delta$. For $\tau_0 = 2 \Delta$ the only solution to crossing with this property is GFF. Note that in CFTs presence of the stress tensor in the spectrum and unitarity bound for scalar operators immediately imply this, see e.g. \cite{Komargodski:2012ek}. However, the claim above applies as well for non-local CFTs, say AdS QFTs, which do not have a stress tensor.

We have to emphasize that we have not proven (\ref{eq:magicproperty}). But we did exhaustive tests to the best of our knowledge. These include $m=0$ and any spin and arbitrary $m$ for low spin. It would be helpful to prove (\ref{eq:magicproperty}) rigorously to put our results on a more solid ground.

Third, let us comment on the convergence of the sum (\ref{eq:magicfunctionals}). Using the standard results of the light-cone bootstrap we can explicitly check the convergence at large $J$. We can also check the convergence of the sum at large $\Delta$ and fixed $J$ using the results of \cite{Mukhametzhanov:2018zja}. Note that the contribution of the heavy operators to the sum rule (\ref{eq:magicfunctionals}) is suppressed like a power of $\tau$ independent of $\Delta$, see appendix \ref{sec:heavytails}.

Finally, let us mention that in the derivation above we can also find functionals that do not rely on the Polyakov conditions, solely from crossing symmetry. As an example we get
\begin{eqnarray}
\label{eq:crossingfunctional}
&&\sum_{\tau, J, m} C_{\tau, J}^2 \beta_{\tau, J, m} = 0 ,  \\
&&\beta_{\tau, J, m} = 6 {  {\cal Q}_{J,m}^{\tau, d}({\Delta \over 3})' \over (\tau - {4 \Delta \over 3} + 2m)^4 } + {  {\cal Q}_{J,m}^{\tau, d}({\Delta \over 3})'' \over (\tau - {4 \Delta \over 3} + 2m)^3 } \ , \nn \\
&& \beta_{\hat 1} =\beta_{0, 0, m} = 0 \ . \nn 
\end{eqnarray}
One can easily check, however, that this functional does not have the crucial positivity property (\ref{eq:magicproperty}). For that reason below we use (\ref{eq:magicfunctionals}). It would be interesting to explore further and systematically if one can derive interesting functionals with useful positivity properties based on crossing symmetry only. 

\subsubsection{Positivity of $\alpha_{\tau, J}$}
\label{sec:positivityofalpha}

Here we elaborate on our claim (\ref{eq:magicproperty}) above. We restrict our consideration only to the relevant case of $d \geq 3$. Let us emphasize that we do not prove (\ref{eq:magicproperty}) but present evidence for it to the best of our current knowledge. The reason being that computing high spin Mack polynomials up to arbitrary spin $J$ and checking positivity in the three-dimensional parameter space of $\Delta, \tau, m$ is a computationally difficult task. Therefore, we could only analyze (\ref{eq:magicproperty}) explicitly for low spins $0 \leq J \leq 40$, as well as for arbitrary spins in some limits, namely the flat space limit and for collinear Mack polynomials $m=0$.

Strictly speaking, what we care about is only the positivity properties of $\alpha_{\tau, J} = \sum_{m=0}^\infty \alpha_{\tau, J,m}$ and not positivity of each descendant $\alpha_{\tau, J,m}$ separately. In practice however we found it much easier to analyze $\alpha_{\tau, J,m}$ for fixed $m$. It would be very interesting to improve our analysis in this regard.

We start by analyzing (\ref{eq:magicproperty}) for low spins, namely $J = 0, ... , 26$. The simplest way to check positivity is to fix the external dimension to some particular value. Foreseeing our holographic consideration below we can fix $\Delta = {5 \over 8} (d-2)$. Then setting $d = 3 + \delta d$ and $\tau = 2 \Delta + \delta \tau$ we checked that all $\alpha_{\tau, J,m}$ are polynomials in $\delta d, \delta \tau, m \geq 0$ with positive coefficients.\footnote{In $d=4$ and $\Delta=\frac{5}{8}$ we checked positivity of the functional up to spin $40$.} Similarly, as is relevant for our case setting $\Delta = c_{\delta} (d-2)$ with ${1 \over 2} < c_{\delta} < {3 \over 4}$ we checked that the same property holds for $c_{\delta}=0.51, 0.52, ..., 0.74$, for spins $J=0, ..., 16$. Keeping $\Delta$ general and writing $\Delta = {d-2 \over 2} + \delta$ we observed that for $J=0,2,4$ the same manifest positivity holds (this time polynomial also includes powers of $\delta$). However, starting from $J=6$ the polynomial is not manifestly positive. Restricting to particular low values of $\Delta$ we have not observed any violations of positivity but the simple analytic argument that we presented above does not hold in this case. One simple analytic check in this more general case is to consider the $m \gg 1$ limit. We computed such a limit for the cases $J=0, ..., 12$ and found that the functional is positive for any $\delta d, \delta \tau \geq 0$ and $\Delta \geq \frac{d-2}{2}$.

Another test of (\ref{eq:magicproperty}) is the flat space limit. Indeed, we consider $m , \tau \gg 1$ with ${\tau \over m}$ fixed. In this case we can use (\ref{largetauregion}) to evaluate the functional. The result is that to leading order in the large $\tau, m$ it is proportional to $C^{({d-2 \over 2})}_{J} \left( {{\tau \over 2} +  m\over  \sqrt{m (m + \tau)} } \right) \geq 0$. 

Finally, we set $m=0$ and used collinear Mack polynomials to perform the large $J$ tests. At large $J$, the leading contribution to the functional is given by 
\begin{eqnarray}
\frac{\Delta ^2 J^{\frac{1}{2}-\frac{2 \Delta }{3}} \log (J) 2^{2 \Delta +\delta \tau +2 J+5}}{\sqrt{\pi } \delta \tau  (2 \Delta +3 \delta \tau ) (4 \Delta +3 \delta \tau ) \Gamma (-\frac{\delta \tau }{2})^2 \Gamma (\frac{2 \Delta }{3}+\frac{\delta \tau }{2})^2},
\end{eqnarray}
where we set $\tau=2\Delta+\delta \tau$. If $\delta \tau >0$, the functional is positive.

\subsection{3d Ising}
\label{sec:3dIsing}

One example of the situation above is given by the correlator $\langle \sigma \sigma \sigma \sigma \rangle$ in the 3d Ising model, which has scaling dimension $\Delta_{\sigma} \approx 0.518$ and where the twist gap is controlled by the stress tensor $\tau_{gap} = 1$, so that as required we have $\Delta_{\sigma} < {3 \over 4} $. We conclude that
\begin{eqnarray}
\label{3disingMellin}
{\cal C} &:& \Delta_{\sigma} - {1 \over 2} < \Re(\gamma_{12}), \Re(\gamma_{14}) , \nn \\
~~~ & & \Re(\gamma_{12})+\Re(\gamma_{14}) < {1 \over 2} \ .
\end{eqnarray}
and that the connected part of $\langle \sigma \sigma \sigma \sigma \rangle$ in the 3d Ising model admits the Mellin representation with a straight contour
\begin{eqnarray}
F^{3d}_{Ising}(u,v) &=&1 + u^{- \Delta_\sigma} + v^{-\Delta_\sigma} \nn \\
&+& \int_{{\cal C}} {d \gamma_{12} d \gamma_{14} \over (2 \pi i)^2} \ u^{- \gamma_{12}} v^{- \gamma_{14} }\Gamma(\gamma_{12})^2 \Gamma(\gamma_{14})^2 \Gamma(\Delta_\sigma - \gamma_{12} - \gamma_{14})^2 M^{3d} (\gamma_{12}, \gamma_{14}) \nn \\
{\cal C} &:& \Delta_{\sigma} - {1 \over 2} < \Re(\gamma_{12}), \Re(\gamma_{14}) , 
~~~  \Re(\gamma_{12})+\Re(\gamma_{14}) < {1 \over 2} \ .
\end{eqnarray}

Next we analyze the sum rules (\ref{eq:magicfunctionals}). For this particular value of $\Delta$, we find that $\alpha_{\tau, J=0, m}$ produce non-negative result for $\tau > \Delta_{\sigma}$.

\begin{figure}
  \centering
  \includegraphics[scale=0.6]{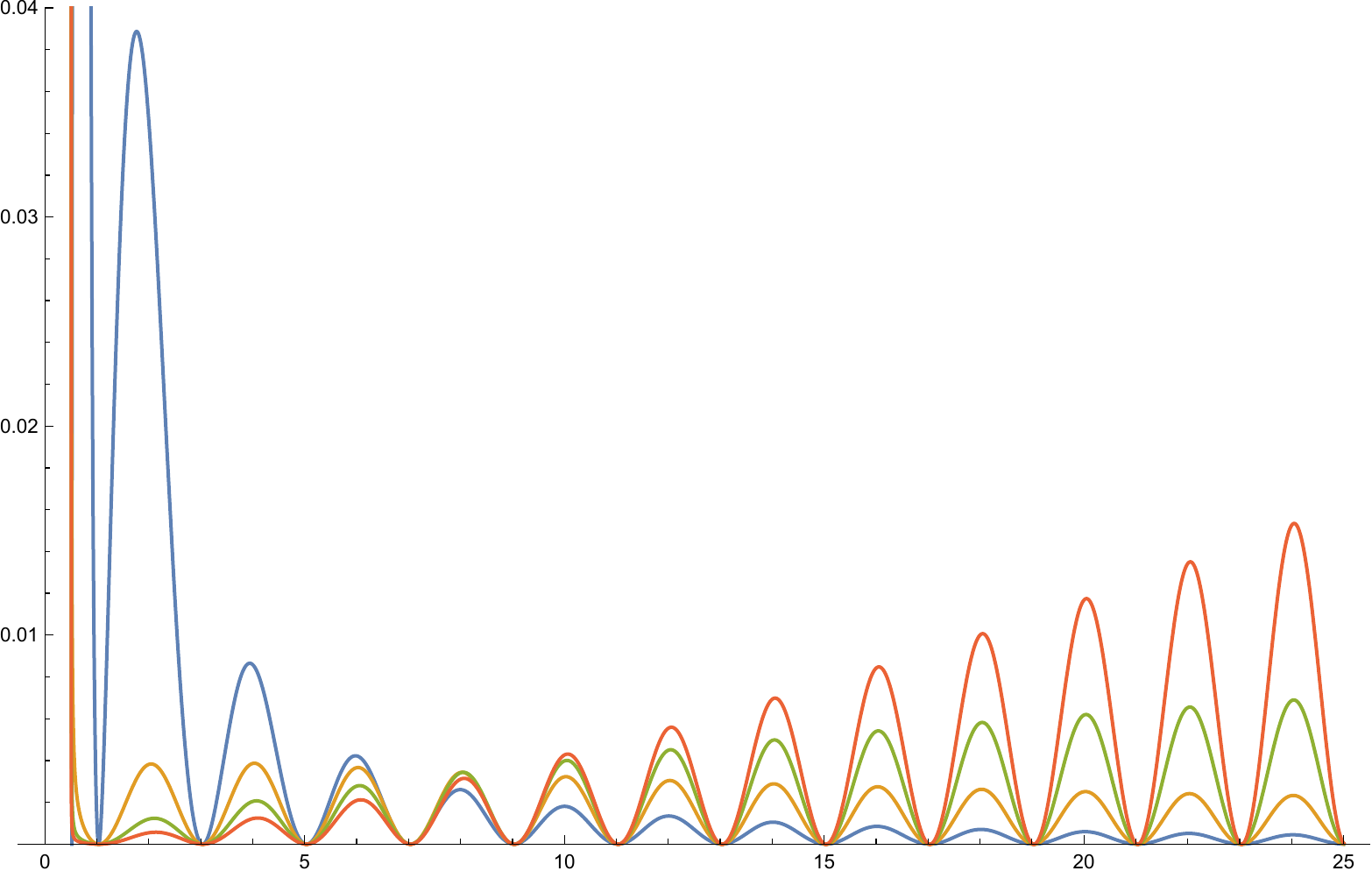} 
  \caption{Functionals $\alpha_{\tau, 0,m}$ as a function of twist $\tau$. They are non-negative with double zeros at the position of double trace operators $\tau_n = 2 \Delta_{\sigma}+2 n$. Different colors correspond to the contribution of descendants labeled by $m$. The external dimension is set to its numerical value in the 3d Ising model $\Delta_{\sigma} \simeq 0.518$.}
  \label{fig:leadtwist1}
\end{figure}

\begin{figure}
  \centering
  \includegraphics[scale=0.6]{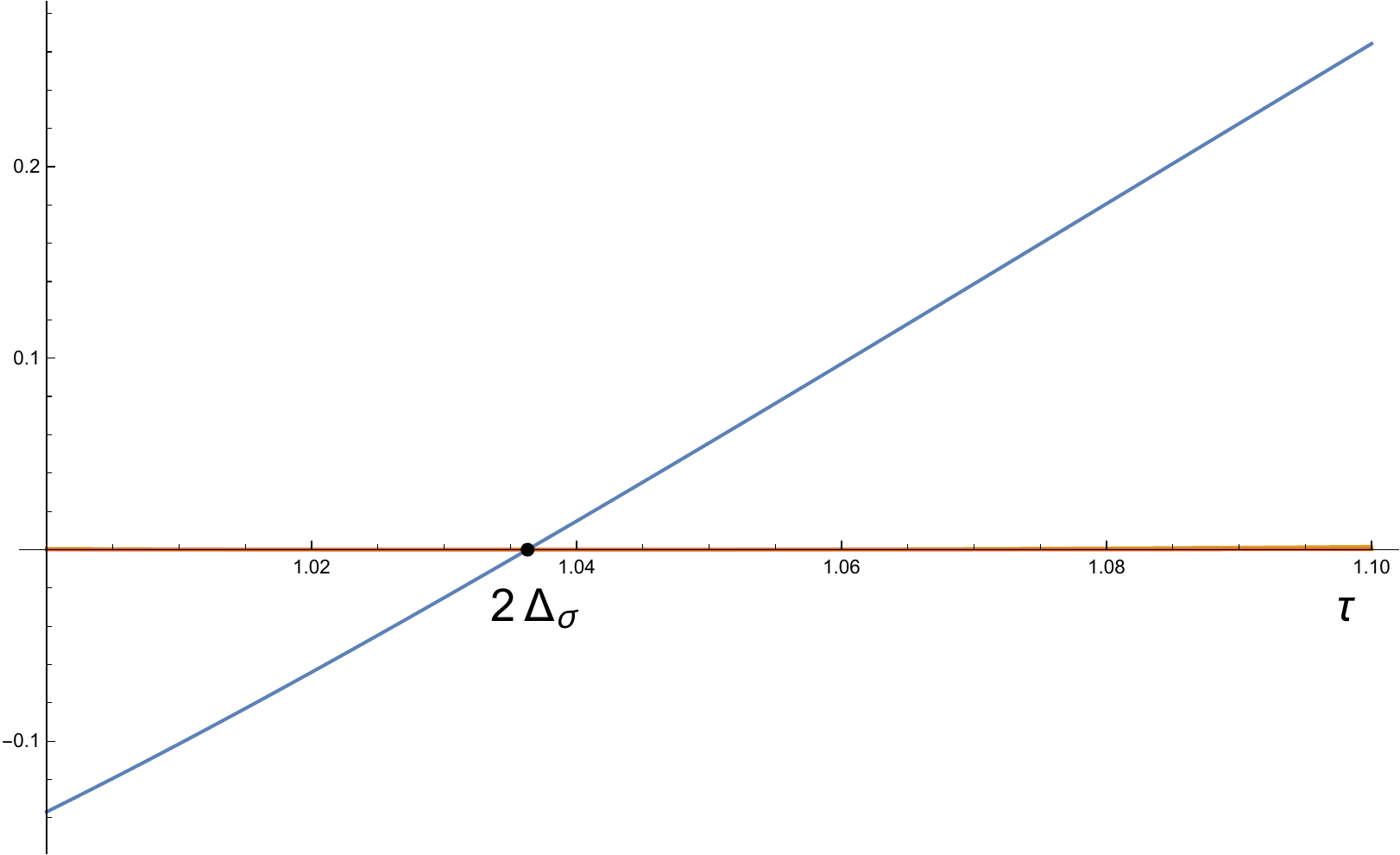} 
  \caption{When acting on operators with spin the functionals $\alpha_{\tau, J,m}$ are negative for operators with twist $\tau < 2 \Delta_\sigma$ and non-negative for $\tau > 2 \Delta_\sigma$ with double zeros at the position of double trace operators $\tau_n = 2 \Delta_{\sigma}+2 n$ with $n \geq 1$. Here we plot the result for $J=2$. Different colors correspond to the contribution of descendants labeled by $m$. The external dimension is set to its numerical value in the 3d Ising model $\Delta_{\sigma} \simeq 0.518$.}
  \label{fig:leadtwist1}
\end{figure}

\begin{figure}
  \centering
  \includegraphics[scale=0.8]{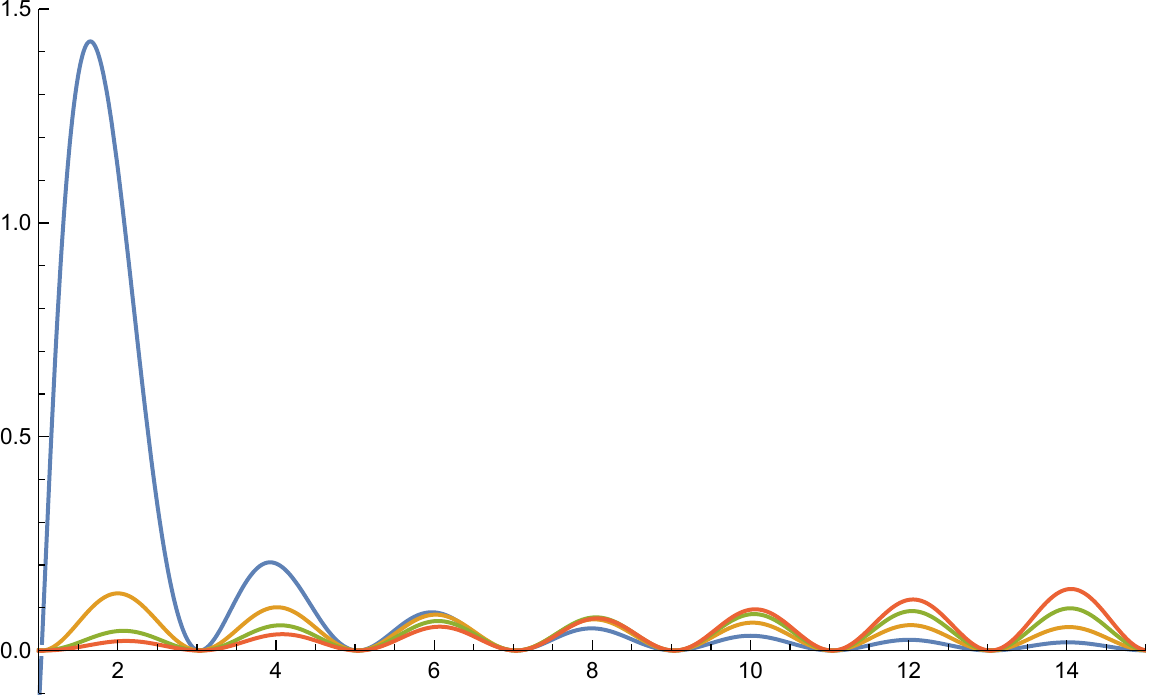} 
  \caption{Same as figure \ref{fig:leadtwist1} but with an extended range of twists $\tau$ plotted.}
  \label{fig:leadtwist2}
\end{figure}

In other words, we can rewrite the sum rule (\ref{eq:magicfunctionals}) as follows
\begin{eqnarray}
\label{eq:sumruleIs}
-\sum_{\tau< 2 \Delta_\sigma, J>0, m} C_{\tau, J}^2 \alpha_{\tau, J} &=& \sum_{\tau > 2 \Delta_\sigma, J>0, m} C_{\tau, J}^2 \alpha_{\tau, J}  + \sum_{\tau, J=0} C_{\tau, J}^2 \alpha_{\tau, J}  ,  \\
\alpha_{\tau, J} &=& \sum_{m=0}^{\infty} \alpha_{\tau, J, m} \ , \nn
\end{eqnarray}
where $m$ is a sum over descendants. We see that the leading twist Regge trajectory is mapped to the rest of the spectrum.

Using the results from \cite{Simmons-Duffin:2016wlq} we get the following numerical values for some terms in the relation above
\begin{eqnarray}
\label{eq:3dsum}
0.0924&=&0.028968_{T_{\mu \nu}}+0.012122_{J=4} + 0.029107_{6 \leq J \leq 30} + 0.0222_{J>30} \nn \\
&=& 0.084569_{\eps} + 0.0018_{[\sigma,\sigma]_{1}^{0 \leq J \leq 30} }+ 0.0016_{[\eps,\eps]_{0}^{4 \leq J \leq 30} } + 0.0014_{[\eps,\eps]_{0}^{J \geq 32} } +...
\end{eqnarray}
where we indicated explicitly contribution of which operators we took into account. In the first line we computed the contribution of $J>30$ currents using the light-cone bootstrap formulae from \cite{Simmons-Duffin:2016wlq}. Similarly, in the second line for the higher spin tail of $[\eps,\eps]_{0}$ we used the formulae from \cite{Kaviraj:2015xsa} and the contribution of descendants (terms with $m \geq 1$ in (\ref{eq:magicfunctionals})). All dropped operators in the second line of (\ref{eq:3dsum}) contribute positively. Note also that the contribution of the heavy operators is only suppressed by a power of $\Delta$. We consider therefore a $5 \%$ difference between the LHS and the RHS for the included operators to be reasonable. It would be great to check the sum rule above in the 3d Ising model with a greater precision by including more operators in the RHS of (\ref{eq:sumruleIs}). 

Similarly, we checked that the $\beta$ functionals (\ref{eq:crossingfunctional}) that do not receive contributions from the scalar operators lead to reasonable numbers. We also observed that the $\beta$ functional sum rules are more sensitive to higher spin operators.

\subsection{Bounds on holographic CFTs} \label{sec:boundsHolo}

Let us now apply (\ref{eq:magicfunctionals}) to holographic CFTs, namely a CFT with large central charge $c_T \gg 1$ \cite{Aharony:2016dwx, Alday:2017gde}. As the simplest example we can consider a free massive scalar in AdS coupled to another field dual to a single trace operator $\tilde {\cal \O}_{st}$ (for example, another scalar field or graviton). We restrict our consideration to external scalars which satisfy $\Delta < {3 \over 4} \tau_{ \tilde {\cal \O}_{st} }$. 

If we simply consider a free massive scalar in AdS the sum rule (\ref{eq:magicfunctionals}) is trivially satisfied. Indeed, as we emphasized above $\alpha_{2 \Delta + 2n , J, m} = 0$. However, as we weakly couple our free scalar field to another field it is not at all obvious that (\ref{eq:magicfunctionals}) is satisfied. As we emphasized several times above the sum rule (\ref{eq:magicfunctionals}) is essentially nonperturbative in $c_T$. For example, in deriving it we used the nonperturbative Regge bound as well as Polyakov conditions. Neither holds in perturbation theory in $c_T$. This is in a stark contrast with \cite{Heemskerk:2009pn} where perturbation theory in AdS was mapped to  solutions to crossing perturbative in $1/c_T$.

Due to the nonperturbative nature of (\ref{eq:magicfunctionals}) we cannot simply expand it in ${1 \over c_T}$. However, we can isolate some parts of it which can be safely computed using the low-energy physics from those sensitive to the details of the UV completion. To that extent we write the sum rule as follows
\begin{eqnarray}
\label{eq:holosumrule}
C_{\tilde {\cal O}_{st}}^2 \alpha_{\tau_{\tilde {\cal O}_{st}},J_{\tilde {\cal O}_{st}}} + \sum_{ J>0} C_{[{\cal O}, {\cal O}]_{0,J}}^2 \alpha_{\tau_{[{\cal O}, {\cal O}]_{0,J}},J} + {\rm rest}_{UV} = 0\ ,
\end{eqnarray}
where the details of the UV completion are in $ {\rm rest}_{UV}$ which is non-negative due to (\ref{eq:magicproperty}). 


We are, thus, left with computing the contribution due to the leading twist double traces. Note that due to a single zero of the functional at $\tau = 2 \Delta$, to leading order in $C_{\tilde {\cal O}_{st}}^2\sim {1 \over c_T}$ we get the result $\sim  (C_{[{\cal O}, {\cal O}]_{0,J}}^{GFF})^2 \gamma_{[{\cal O}, {\cal O}]_{0,J}} \sim {1 \over c_T}$ with $J>0$, where $\gamma_{[{\cal O}, {\cal O}]_{0,J}}$ is the anomalous dimension of double trace operators. 
\footnote{The notation $C_{\tilde {\cal O}_{st}}^2\sim {1 \over c_T}$ is only precise if the exchanged single-trace operator is the stress tensor.
In the other cases, we think of $1/c_T$ as the square of the small  cubic coupling in AdS.}
The anomalous dimensions
$\gamma_{[{\cal O}, {\cal O}]_{0,J}}$ are observables that can be reliably computed using the low-energy theory in AdS. 

Let us first consider an example when the external scalar is coupled to another scalar, namely $J_{\tilde {\cal O}_{st}}=0$.  The contribution from the scalar exchange to the double trace operators can be found for example in \cite{Albayrak:2019gnz}. In this paper the relevant OPE data was computed for all $J$  using the Lorentzian inversion formula instead of computing the relevant Witten diagrams. In this case we numerically observed that the contributions exactly cancel to leading order in ${1 \over c_T}$
\begin{eqnarray}
\label{eq:holoscalar}
C_{\tilde {\cal O}_{st}}^2 \alpha_{\Delta_{\tilde {\cal O}_{st}},0} + \sum_{J>0} C_{[{\cal O}, {\cal O}]_{0,J}}^2 \alpha_{\tau_{[{\cal O}, {\cal O}]_{0,J}},J} + O \left( {1 \over c_T^2} \right) = 0\ .
\end{eqnarray}
A reader might be confused by the identity above in light of our discussion of the AdS scalar exchange in section \ref{sec:scalaramb}. There we argued that a sum of tree-level scalar exchange diagrams in AdS does not satisfy the Polyakov condition. The identity above is closely related to the scalar exchange in AdS but is different from it. One way to understand the difference is that evaluating the action of the functional above to leading order in ${1 \over c_T}$ involves knowledge of the Mellin amplitude to the order ${1 \over c_T^2}$. More precisely, given a term ${1 \over c_T^2} {1 \over \gamma_{12} - \Delta + {\tau \over 2}}$ in the Mellin amplitude we get ${1 \over c_T}$ term in the functional when we set $\gamma_{12} = 0$ and $\tau = 2 \Delta + O({1 \over c_T})$. Therefore, sum of scalar exchanges considered in section \ref{sec:scalaramb} taken as a non-perturbative Mellin amplitude does not satisfy Polyakov condition. This is not surprising, since given a sum of scalar exchanges to order ${1 \over c_T}$ unitarity, which we use when deriving the functional, dictates that it has to be supplemented by the ${1 \over c_T^2}$ correction that emerges due to anomalous dimension of double trace operators \cite{Aharony:2016dwx}. When evaluating the functional to ${1 \over c_T}$ these two mix and produce $0$ in the formula above. We do not have an explanation why the sum of the two contributions above cancels.
We provide more details on this calculation in appendix \ref{sec:HoloCalc}. 

\begin{figure}[h]
  \centering
  \includegraphics[scale=0.3]{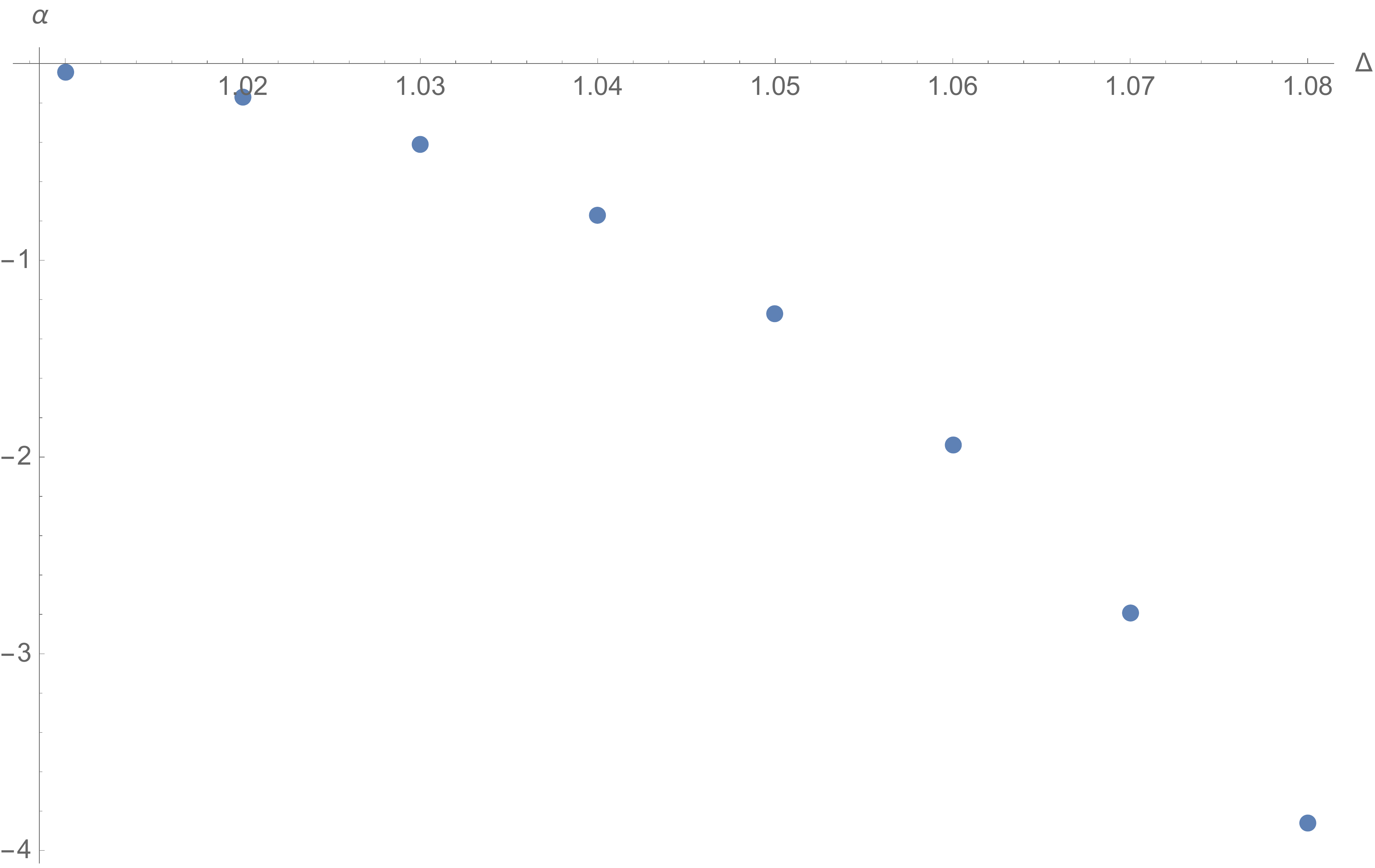} 
  \caption{We consider a scalar minimally coupled to gravity in $AdS_5$ or $CFT_4$. We imagine that the gravitational coupling is weak, or, equivalently, $c_T \gg 1$. The $\alpha$-functional (\ref{eq:magicfunctionals}) can be applied to $1 < \Delta < 1.5$. We plot the sum given by (\ref{eq:holostress}). We find that the sum is always negative within the region of applicability of the functional.}
  \label{fig:d4minscal}
\end{figure}

\begin{figure}[h]
  \centering
  \includegraphics[scale=0.4]{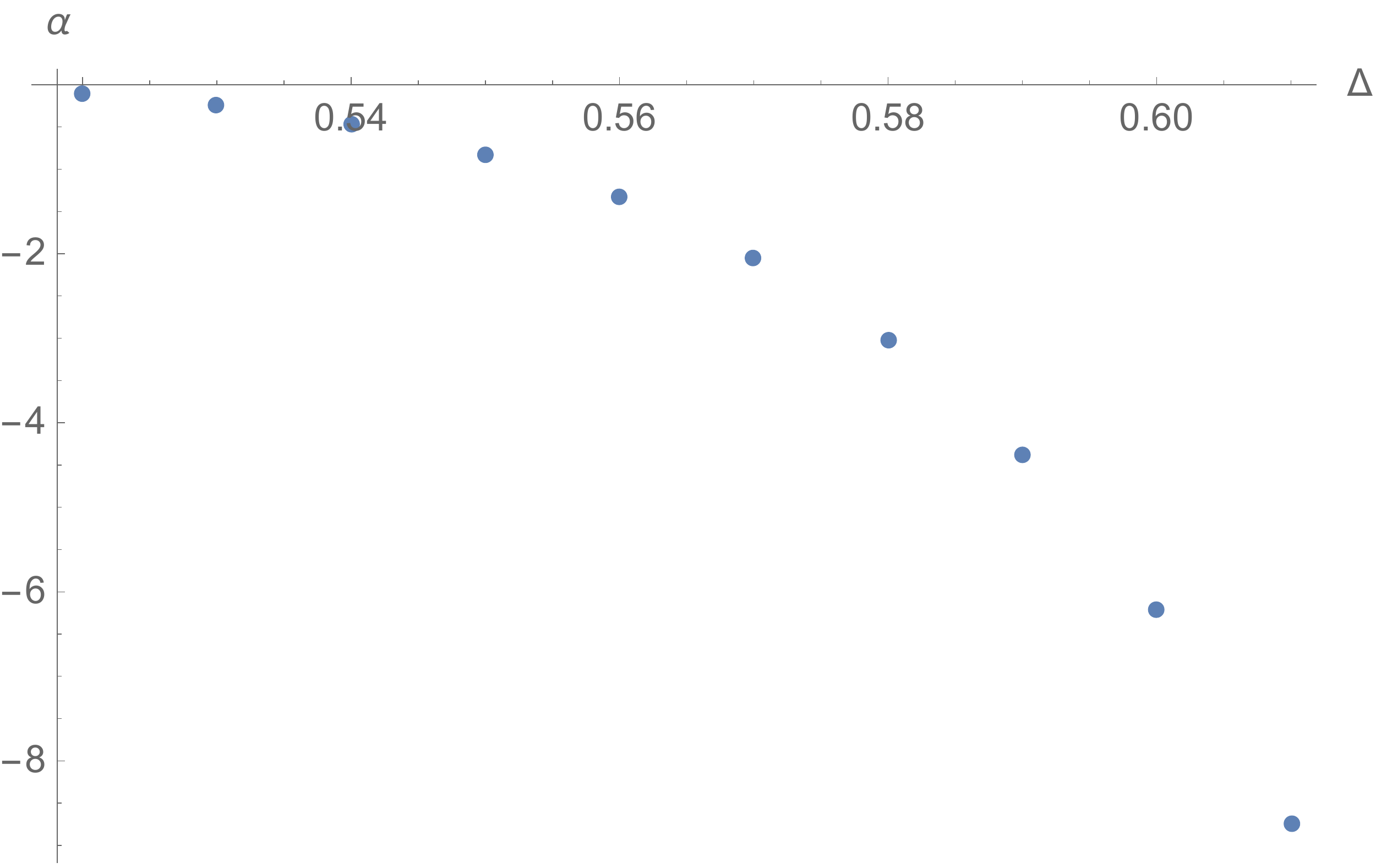} 
  \caption{We consider a scalar minimally coupled to gravity in $AdS_4$ or $CFT_3$. We imagine that the gravitational coupling is weak, or, equivalently, $c_T \gg 1$. The $\alpha$-functional (\ref{eq:magicfunctionals}) can be applied to ${1 \over 2}  < \Delta < {3 \over 4}$. We plot the sum given by (\ref{eq:holostress}).  We find that the sum is always negative within the region of applicability of the functional.}
  \label{fig:d3minscal}
\end{figure}

Next we consider a scalar minimally coupled to gravity. In this case we can use the results of \cite{Alday:2017gde,Costa:2014kfa} for the anomalous dimensions of double trace operators. One subtlety in this case is that there is a contribution at $J=0,2$ which is non-analytic in spin. In this case, (experimenting across $d$ and ${d-2 \over 2} < \Delta < {3 \over 2} {d-2 \over 2}$) we find that 
\begin{eqnarray}
\label{eq:holostress}
C_{T_{\mu \nu}}^2 \alpha_{d-2,2} + \sum_{ J>0} C_{[{\cal O}, {\cal O}]_{0,J}}^2 \alpha_{\tau_{[{\cal O}, {\cal O}]_{0,J}},J}= - \frac{a(d,\Delta)}{c_T} + O \left( {1 \over c_T^2} \right) \,,
\qquad a(d,\Delta)> 0\,.
\end{eqnarray}
The results for $d=4$ and $d=3$ are presented on fig. \ref{fig:d4minscal} and fig. \ref{fig:d3minscal} correspondingly. See appendix \ref{sec:HoloCalc} for more details on this. Therefore, we conclude that the rest of the sum in (\ref{eq:magicfunctionals}) must give
\beq
 {\rm rest}_{UV} = \sum_{J,\tau>2\Delta}C^2_{\tau,J} \alpha_{\tau,J}=  \frac{a(d,\Delta)}{c_T} + O \left( {1 \over c_T^2} \right)\,.
\eeq
It would be very interesting to understand what operators produce this contribution. Notice that any double-trace operator by itself contributes only $O \left( {1 \over c_T^2} \right)$. Thus, either we need some (heavy) single-trace operators or infinite sums of double-traces can enhance to $O \left( {1 \over c_T} \right)$.


%

%

\subsection{Bounds on classical AdS EFT}

Finally, let us consider a classical theory in AdS or equivalently a planar CFT correlator. In this case Mellin amplitude is simply a meromorphic function polynomially bounded at infinity according to (\ref{eq:ReggeboundM}). We can then consider the following simple dispersion integral
\beq
{1 \over n!} \pa_{\gamma_{12}}^{n} M(\gamma_{12}, \gamma_{13}) =  \oint {d \tilde \gamma_{12} \over 2 \pi i} {1 \over (\tilde \gamma_{12} - \gamma_{12})^{n+1}} M(\tilde \gamma_{12}, \gamma_{13}) ~ . 
\eeq
We can next deform the contour. Assuming we can drop arches at infinity (which is possible for large enough $n$) we get
\begin{eqnarray}
\label{eq:adsEFT}
{1 \over n!} \pa_{\gamma_{12}}^{n} M(\gamma_{12}, \gamma_{13}) &=&- \int d \tilde \gamma_{12} {\rm Disc}_{\tilde \gamma_{12}} M(\tilde \gamma_{12}, \gamma_{13}) \left( {1 \over (\tilde \gamma_{12} - \gamma_{12})^{n+1}} + {(-1)^{n} \over (\tilde \gamma_{12} - \gamma_{14})^{n+1}} \right) , \nn \\
{\rm Disc}_{ \gamma_{12}} M( \gamma_{12}, \gamma_{13}) &=& -{1 \over 2} \sum_{\tau, J, m} C_{\tau, J}^2 {\cal Q}_{J,m}^{\tau, d}(\gamma_{13}) \delta(\gamma_{12} - [\Delta - {\tau \over 2} - m ]) \ .
\end{eqnarray}
where we used crossing symmetry to combine the contribution of two channels. According to our definition ${\rm Disc}[ {1 \over \gamma_{12}}] = \delta(\gamma_{12})$. 
The positivity property of Mack polynomials  (\ref{eq:mackpos}) and \eqref{poles} implies 
\begin{eqnarray}
\label{eq:positivityDiscM}
{\rm Disc}_{ \gamma_{12}} M( \gamma_{12}, \gamma_{13})  \geq 0 , ~~~ \gamma_{13} \leq 0 \ . 
\end{eqnarray}
From this it immediately follows that
\beq
\label{eq:conditionAdSEFT}
\pa_{\gamma_{12}}^{2 n} M({\Delta - \gamma_{13} \over 2}, \gamma_{13}) \geq 0 , ~~~ \gamma_{13} < {\rm min}[0, \tau_{gap} - \Delta] \ ,
\eeq 
which is the AdS analog of the flat space result \cite{Adams:2006sv}. The condition $ \gamma_{13} < {\rm min}[0, \tau_{gap} - \Delta] $ guarantees that (\ref{eq:positivityDiscM}) holds, as well as  $- \left( {1 \over (\tilde \gamma_{12} - \gamma_{12})^{2n+1}} + {1 \over (\tilde \gamma_{12} - \gamma_{14})^{2n+1}} \right) \geq 0$ on the support of ${\rm Disc}_{ \gamma_{12}} M( \gamma_{12}, \gamma_{13})$ and for  $\gamma_{12} =\gamma_{14}= {\Delta - \gamma_{13} \over 2}$ which appears in (\ref{eq:conditionAdSEFT}). These constraints are mapped to bounds on the coefficients of higher derivative operators in AdS. Note also that in a theory with a large gap we naturally get a suppression by ${1 \over \Delta_{gap}}$ for each derivative in $\gamma_{12}$ due to the fact that the integral in (\ref{eq:adsEFT}) starts from $\Delta_{gap}$.

Similarly, we can derive a bound on the local growth of the Mellin amplitude in the upper half-plane  \cite{Caron-Huot:2017vep,Maldacena:2015waa}.
Let us fix  $\gamma_{13} \leq {\rm min}[0, \tau_{gap} - \Delta] $ and define 
\beq
M(z) \equiv M({\Delta - \gamma_{13} \over 2} + z, \gamma_{13}) 
\eeq
Notice that crossing implies that $M(z)=M(-z)$.
Then we can write a dispersion relation
\beq
{1 \over n!} \pa_{z}^{n} M(z) =  \oint {d \tilde z \over 2 \pi i} {1 \over (\tilde z - z)^{n+1}} M(\tilde z) ~ . 
\eeq
Using crossing we can write
\beq
{1 \over n!} \pa_{z}^{n} M(z) = (-1)^n {1 \over n!} \pa_{z}^{n} M(-z) =  \oint {d \tilde z \over 2 \pi i}  M(\tilde z)  \frac{1}{2}\left[ {1 \over (\tilde z - z)^{n+1}} + {(-1)^n \over (\tilde z + z)^{n+1}}\right] ~ . 
\eeq
In particular, for $n=1$ we find
\beq
 \pa_{z}  M(z) =  \oint {d \tilde z \over 2 \pi i}  M(\tilde z)    {2 z\tilde z \over (\tilde z^2 - z^2)^{2}}  ~ ,
\eeq
Therefore we can blow up the contour to $|\tilde z| \to \infty$ and drop the arcs at infinity as long as $M(\tilde z)/ |\tilde z|^2 \to 0$.
This leads to 
\begin{equation}
  \pa_{y}M({\Delta - \gamma_{13} \over 2} + i y, \gamma_{13}) = 4 y \int d \tilde \gamma_{12} {\rm Disc}_{\tilde \gamma_{12}} M(\tilde \gamma_{12}, \gamma_{13} ) {\tilde \gamma_{12} - {\Delta - \gamma_{13} \over 2} \over (y^2 + (\tilde \gamma_{12}-{\Delta - \gamma_{13} \over 2})^2)^2} 
\geq 0 .   
\end{equation}
In addition,
\beq
y \pa_y \log [  \pa_{y}M({\Delta - \gamma_{13} \over 2} + i y, \gamma_{13})] = 1-4  {\int d \tilde \gamma_{12} {\rm Disc}_{\tilde \gamma_{12}} M(\tilde \gamma_{12}, \gamma_{13}) {(\tilde \gamma_{12}-{\Delta - \gamma_{13} \over 2}) y^2 \over (y^2 + (\tilde \gamma_{12}-{\Delta - \gamma_{13} \over 2})^2)^3} \over \int d \tilde \gamma_{12} {\rm Disc}_{\tilde \gamma_{12}} M(\tilde \gamma_{12}, \gamma_{13} ) {(\tilde \gamma_{12}-{\Delta - \gamma_{13} \over 2}) \over (y^2 + (\tilde \gamma_{12}-{\Delta - \gamma_{13} \over 2})^2)^2}},
\eeq
where the last term can be thought of as an average $  \langle \frac{y^2 }{ y^2 + (\tilde \gamma_{12}-{\Delta - \gamma_{13} \over 2})^2   }\rangle$ with a positive measure over $\tilde \gamma_{12}$. Clearly this average must take values between 0 and 1. Therefore,
\beq
-3 \le y \pa_y \log  [ \pa_{y}M({\Delta - \gamma_{13} \over 2} + i y, \gamma_{13})]  \le 1\,, \label{boundlocalgrowth}
\eeq
which leads to the  bounds $-2\le \alpha \le 2$ for the asymptotic behavior $M\sim (\gamma_{12})^\alpha$.
%
In writing the formulas above we assumed that  $\gamma_{13} \leq {\rm min}[0, \tau_{gap} - \Delta] $ and that the $n=1$ sum rule converges, \emph{i.e.}  that $M(\tilde z)/ |\tilde z|^2 \to 0$  when $|z|\to \infty$.
Notice that this is not guaranteed by the Regge bounds \eqref{eq:ReggeboundM}.
We can implement subtracted dispersion relations whose convergence is guaranteed by \eqref{eq:ReggeboundM} but these lead to weaker bounds than \eqref{boundlocalgrowth}.

Note that a very similar reasoning was used in \cite{Alday:2016htq} to derive bounds on anomalous dimensions of double trace operators in the planar ${\cal N}=4$ SYM. Here we simply note that this reasoning can be generalized to arbitrary CFTs due to the underlying positivity properties of Mack polynomials (\ref{eq:mackpos}). It would be also interesting to use properties of Mack polynomials to derive the analogs of \cite{YuTin}.

\subsection{Applications to ${\cal N}=4$ SYM}

Mellin amplitudes of four-point functions of scalar half-BPS operators in $\mathcal{N}=4$ SYM \cite{Nirschl:2004pa, Dolan:2004mu, Heslop:2002hp, Dolan:2001tt, Eden:2000bk, Eden:2001ec, Beem:2013qxa, Beem:2016wfs}  were used in \cite{Belitsky:2013xxa,Belitsky:2013bja} to study energy-energy correlators.  In this section we will apply the machinery of sections \ref{sec:MelGen} and \ref{sec: dif cont} to the four-point functions of scalar half-BPS operators in $\mathcal{N}=4$ SYM. We argue that such correlators admit Mellin representation for any value of $N$ and of the t'Hooft coupling $a$, both in perturbation theory and for finite values of $N$ and $a$.

\subsubsection{Review of Mellin amplitudes in $\mathcal{N}=4$ at weak and strong coupling}

In this section we review Mellin amplitudes of the planar half-BPS correlator at one-loop at weak coupling and at strong coupling following \cite{Belitsky:2013xxa}. The precise four-point function under consideration is $\langle \tilde{\cO}(x_1) \tilde{\cO}(x_2)  \cO^{\dagger}(x_3) \cO(x_4)\rangle$, where
\begin{eqnarray}
\cO(x)= \frac{1}{\sqrt{c}} Y^I Y^J \tr\[ \Phi^I \Phi^J \] , ~~~ Y^I=(1, 0, 1, 0, i, i) \nonumber ,
\end{eqnarray}
and
\begin{eqnarray}
\tilde{\cO}(x_1) &=& \frac{1}{\sqrt{c}} 2 S'^{I J} \tr\[ \Phi^I \Phi^J \], ~~~ \tilde{\cO}(x_2) = \frac{1}{\sqrt{c}} 2 S^{I J} \tr\[ \Phi^I \Phi^J \]  , \nonumber  \\
S' &=& \diag{0, 0, 1, -1, 0, 0},~~~S = \diag{1, -1, 0, 0, 0, 0}  \nonumber  ,
\end{eqnarray}
where $c = {(N^2-1) \over (2 pi^4)}$.

The point of choosing such polarizations was to simplify the operator product expansions. In particular, scalar half-BPS operators transform under the $\textbf{20}'$ representation of the R-symmetry group and $\textbf{20}' \times \textbf{20}' = \textbf{1}+\textbf{15} +\textbf{20}'+\textbf{84}+\textbf{105}+\textbf{175}$. For this choice of polarizations, the OPE $\tilde{\cO} \times \tilde{\cO}$ only contains the $\textbf{105}$ representation. On the contrary, the other OPE's can contain any representation.

It was found in \cite{Belitsky:2013bja} that at weak coupling
\begin{eqnarray}\label{weak coupling N=4}
\langle\tilde{\cO}(x_1) \tilde{\cO}(x_2)  \cO^{\dagger}(x_3)  \cO(x_4) \rangle &=& \frac{N_c^2-1}{8 (2 \pi)^4} \Big( \frac{1}{x_{14}^4 x_{23}^4 }+ \frac{1}{x_{13}^4 x_{24}^4} \Big) + {F_{conn}(u,v) \over (x_{13}^2 x_{24}^2)^2} ,\nn \\ 
&=&\frac{5 (N_c^2 - 1)}{32 \pi^8} \frac{A_{\textbf{105}}(u, v)}{x_{12}^4 x_{34}^4} ,  \\
F_{conn}(u,v)= \frac{1}{2 (2 \pi)^4}{1 \over v} &+& \int_{{\cal C}_0} \frac{d \gamma_{12} d\gamma_{14}}{(2 \pi i)^2} \Gamma(\gamma_{12})^2 \Gamma(\gamma_{14})^2 \Gamma(2-\gamma_{12} - \gamma_{14})^2 M(\gamma_{12}, \gamma_{14}) u^{-\gamma_{12}} v^{-\gamma_{14}} \ , \nn
\end{eqnarray}
where $a=\frac{g_{YM}^2 N_c}{(4 \pi)^2}$ and $N_c$ denotes the number of colors. $A_{\textbf{105}}(u, v)$ signifies the fact that the only representation that appears in the OPE of $\tilde{\cO} \times \tilde{\cO}$ is $\textbf{105}$. The Mellin amplitude $M(\gamma_{12},\gamma_{14})$ satisfies crossing relations
\beq
M(\gamma_{12}, \gamma_{14}) = M(\gamma_{12}, \gamma_{13}) \ ,
\eeq
where $\gamma_{12} + \gamma_{13} + \gamma_{14} = 2$.

The leading order weak coupling result takes the form
\begin{eqnarray}\label{Mellin N=4 weak}
M(\gamma_{12}, \gamma_{14})=-\frac{a}{4(2 \pi)^4} {\gamma_{12}^2 \over (\gamma_{14}-1)^2(\gamma_{13}-1)^2} + O(a^2), %
\end{eqnarray}
Double poles at $\gamma_{13}, \gamma_{14} = 1$ encode anomalous dimension of twist $2$ operators.

At strong coupling, it was found in \cite{Belitsky:2013bja} that the correlation function has the same form as (\ref{weak coupling N=4}), but now we have 
\begin{eqnarray}\label{Mellin N=4 strong}
M(\gamma_{12}, \gamma_{14})=-{1 \over 2 (2 \pi)^4} {\gamma_{12}^2 (1+\gamma_{12}) \over (\gamma_{14}-1)(\gamma_{13} - 1)} \ .
\end{eqnarray}
Poles at $\gamma_{13}, \gamma_{14}=1$ encode exchanges of twist $2$ operators with spin $0,1,2$ that stay light at strong coupling.

Notice that both Mellin amplitudes (\ref{Mellin N=4 weak}) and (\ref{Mellin N=4 strong}) have been written with the same contour ${\cal C}_0$
\begin{eqnarray}\label{holomorphic strip}
{\cal C}_0: ~~~{\rm Re}(\gamma_{12})>-1, ~~~{\rm Re}(\gamma_{14})>1, ~~~{\rm Re}(\gamma_{12}+\gamma_{14})<1.
\end{eqnarray}

Let us now apply our general analysis to the strong coupling correlator. To write down the straight contour representation we should do subtractions of $\tau=2$ collinear blocks with $J=0,1,2$ in the $\gamma_{13}$ and $\gamma_{14}$ channel. In other words we consider  
\begin{eqnarray}
\label{eq:subtractionFN4}
&&F_{sub}(u,v) = F_{conn}(u,v) -\sum_{J=0}^{2} C_{2, J}^2  {1 \over v} \Big( g^{(0)}_{2, J}(u) + g^{(0)}_{2, J}(\frac{u}{v}) \Big) \ ,
\end{eqnarray}
where $(-1)^J$ in our definition (\ref{eq:mfunctionsAg}) of $ g^{(0)}_{2, J}(u)$ is important since for non-identical operators we have odd spin exchanges as well. We can then check that indeed the strong coupling result takes the form
\beq
\label{eq:n4res}
F_{sub}(u,v) = \int_{{\cal C}} \frac{d \gamma_{12} d\gamma_{14}}{(2 \pi i)^2} \Gamma(\gamma_{12})^2 \Gamma(\gamma_{14})^2 \Gamma(2-\gamma_{12} - \gamma_{14})^2 M(\gamma_{12}, \gamma_{14}) u^{-\gamma_{12}} v^{-\gamma_{14}} \ ,
\eeq
where the contour is given by (\ref{eq:Ucft})
\begin{eqnarray}\label{holomorphic strip}
{\cal C}: ~~~1>{\rm Re}(\gamma_{13}),{\rm Re}(\gamma_{14})>0, ~~~ 2>{\rm Re}(\gamma_{12})>0 \ .
\end{eqnarray}

Based on general arguments at finite coupling we expect to have the strip of holomorphy
\begin{eqnarray}\label{holomorphic strip}
{\cal C}: ~~~1>{\rm Re}(\gamma_{13}),{\rm Re}(\gamma_{14})>2 - {\tau_{sub}' \over 2}, ~~~ 2>{\rm Re}(\gamma_{12})>2 - {\tilde \tau_{sub}'  \over 2} \ ,
\end{eqnarray}
where $\tau_{sub}' $ and $\tilde \tau_{sub}'$ are the twist gaps above the twists of subtracted spins in the corresponding OPE channels. At strong coupling $\tau_{sub}' = \tilde \tau_{sub}' = 4$. At zero coupling $\tau_{sub}' = 2$ because higher spin currents become conserved. In the other channel we have at any coupling $\tilde \tau_{sub}' = 4$ as follows from superconformal symmetry. At finite coupling we have $\tau_{sub}' >2$ and therefore we have a finite strip to write down the Mellin representation.

The argument above was nonperturbative. What happens if we consider ${1 \over N}$ expansion of the correlator at finite 't Hooft coupling $a$? One might worry about the fact that the Regge limit of the correlator becomes more singular at every order in $\frac{1}{N}$ expansion, see \cite{Cornalba:2007zb,Meltzer:2019pyl}. However, it is always true at any order in $\frac{1}{N}$ that the perturbative Mellin amplitude $M(\gamma_{12}, \gamma_{14})$ is polynomially bounded at infinity. Similarly, by keeping the 't Hooft coupling $a$ fixed we keep the anomalous dimension of higher spin currents finite and therefore $\tau_{sub}' >2$ which is sufficient for the existence of the non-empty holomorphic strip. Therefore ${1 \over N}$ expansion of correlators for $a>0$ admit Mellin representation (\ref{eq:n4res}) as well. Note, however, if we are to do perturbation theory in $a$ or $g_{YM}$ then extra subtractions are needed due the presence of the higher spin conserved currents in the spectrum.

As emphasized by Mack \cite{Mack:2009mi}, Mellin amplitudes are very convenient to study analytic continuations of the correlator (assuming the Mellin integral converges which we will establish in some cases below). Indeed, the analytic continuation under the integral is very simple with the result for the Wightman function\footnote{We use mostly plus signature.}
\begin{eqnarray}\label{def Mellin general cont}
\langle \cO_1 (x_1) \dots \cO_n(x_n) \rangle = \int \[ d \gamma_{ij} \] M(\gamma_{ij})
\prod_{i<j}^n \frac{\Gamma(\gamma_{ij})}{(x_{i j}^2 + i \eps x^0_{i j})^{ \gamma_{ij}}}\, .
\end{eqnarray}

This was extensively used in perturbative computations of the energy-energy correlators \cite{Belitsky:2013xxa,Belitsky:2013bja,Belitsky:2013ofa}, as well as in the study of the Regge limit \cite{Costa:2012cb}, and the bulk point singularity \cite{Penedones:2010ue}. It would be interesting to see if Mellin space could shed some light on the OPE properties of the multi-point event shapes and the OPE of light-ray operators \cite{Kologlu:2019mfz}.

\section{Mellin Amplitudes in minimal models} 
\label{sec:MM}

In this section, we compute Mellin amplitudes in minimal models.\footnote{By minimal models, we mean the 2d CFT's with finite number of Virasoro primaries at $c<1$ and with diagonal partition function. } Minimal models were discovered in \cite{Belavin:1984vu} and formulas for correlation functions were found in \cite{Dotsenko:1984nm} and \cite{Dotsenko:1984ad}. These theories provide a context where we can explicitly compute Mellin amplitudes in interacting non-perturbative CFT's. We will see that in minimal models any correlator of scalar Virasoro primaries has a well defined Mellin amplitude. We will also check in several examples that Mellin amplitudes only have the singularities \eqref{OPEpoles} dictated by the  OPE.

This section is structured as follows. In section \ref{sec: coulomb}, we briefly review the Coulomb gas formalism of minimal models. The Coulomb gas technique is a way to determine correlation functions that is very suitable for computing the associated Mellin amplitudes. In section \ref{sec: phi12}, we compute the Mellin amplitude of $\langle \Phi_{1, 2} O \Phi_{1, 2} O \rangle$, where $\Phi_{1, 2}$ is a second-order degenerate Virasoro primary and $O$ is any Virasoro primary. This is a simple example of a Mellin amplitude that serves as warmup for section \ref{sec: mellin general 2d}, where we compute the Mellin amplitude of any correlation function of scalar Virasoro primaries. In section \ref{sec: phi13}, we consider the Mellin amplitude associated to $\langle \Phi_{1, 3} O \Phi_{1, 3} O \rangle$ and check that it is a meromorphic function with poles at the locations given by the OPE. This is a non-trivial check  that Mellin amplitudes only have the OPE poles, since the Mellin amplitude of $\langle \Phi_{1, 3} O \Phi_{1, 3} O \rangle$ is given by $5$ Mellin-Barnes integrals in our setup, so its pole structure is not apparent. In section \ref{sec: bulk point limit 2d}, we compute the bulk point limit of any correlator of Virasoro primaries in minimal models. It agrees with the expectations about the bulk point limit coming from \cite{Maldacena:2015iua}. Finally, in section \ref{sec: simple corr}, we exhibit examples of nonzero correlators with vanishing Mellin amplitude. This phenomenon might seem puzzling, but we explain it more generally in section \ref{sec: contour}. 

Upon completion of our work, we learned about the paper \cite{Furlan:1988qe}, whose results overlap with those of this section. To be precise, the idea of using Symanzik's formula in conjunction with the Coulomb gas formalism is already present in that paper. It also contains the computation of the Mellin amplitudes of $\langle \Phi_{1, 2} O \Phi_{1, 2} O \rangle$ and $\langle \Phi_{1, 3} O \Phi_{1, 3} O \rangle$ and an analysis of the respective pole structure (albeit with a different method). So, our sections \ref{sec: phi12} and \ref{sec: phi13} mostly reproduce results already contained in \cite{Furlan:1988qe}. By contrast, the other results we present in this section are new, to the best of our knowledge.\footnote{In \cite{Lowe:2016ucg} Mellin amplitudes in minimal models are also studied, but from a different point of view. In particular there it is proposed to define a Mellin amplitude as a transform of a chiral block, whereas we define Mellin amplitudes as transforms of the full correlation function.}

\subsection{The Coulomb gas formalism of minimal models}\label{sec: coulomb}

In this section, we review the Coulomb gas formalism of minimal models, following \cite{Dotsenko:2006}. Our goal is just to state the formulas we will need, in order to compute Mellin amplitudes. For a detailed review of this technique, see \cite{Dotsenko:2006,Dotsenko:1984nm,Dotsenko:1984ad,DiFrancesco:1997nk}.

The Coulomb gas formalism provides a representation of minimal models in terms of the theory of a deformed scalar field. The idea is to associate Virasoro primaries in minimal models with vertex operators in the deformed scalar field theory and thus compute correlation functions in minimal models using the correlation functions of vertex operators in the scalar theory.

To begin with, let us remind ourselves of basic facts about the theory of a massless two dimensional scalar field $\phi$. The action is
\begin{eqnarray}\label{action free}
S= \frac{1}{4 \pi} \int d^2 x  \partial  \phi \bar{\partial}  \phi
\end{eqnarray}
where we used complex coordinates $\partial=\partial_z$ and $\bar{\partial}=\partial_{\bar{z}}$.
The two point function is given by
\begin{eqnarray}
\langle \phi(x) \phi(x') \rangle =- 2 \log  |x-x'|^2\,.
\end{eqnarray}
Furthermore,
\begin{eqnarray}
T(z) \equiv T_{zz}(z)= - \frac{1}{4} : \partial  \phi \partial  \phi :,
\end{eqnarray}
from which it follows that $\langle T(z) T(z') \rangle= \frac{1/2}{(z-z')^4}$. 
 Thus, the theory has central charge $c=1$. For each real number $\alpha$ we can define a vertex operator $V_{\alpha}\equiv :e^{i \alpha \phi(x)}:$. From the OPE of $V_{\alpha}$ with $T(z)$, we conclude that $V_{\alpha}$ is a scalar Virasoro primary, of dimension $2 \alpha^2$. Correlation functions of vertex operators are given by
\begin{eqnarray}
\langle V_{\alpha_1}(x_1) ... V_{\alpha_n}(x_n) \rangle = \prod_{i<j} |x_i-x_j|^{4 \alpha_i \alpha_j},
\end{eqnarray}
if $\sum_{i=1}^n \alpha_i=0$, otherwise the correlation function vanishes.

Now imagine adding to the theory a background charge $-2\alpha_0$, where $\alpha_0$ is a real number that we can pick. 
More precisely, correlations functions in the deformed theory are defined by
\begin{align}\label{corr background free}
\langle V_{\alpha_1}(x_1) ...V_{\alpha_n}(x_n) \rangle_{-2 \alpha_0} &= \lim_{x_{n+1}\rightarrow \infty}|x_{n+1}|^{16 \alpha_0^2} \langle V_{\alpha_1}(x_1) ...V_{\alpha_n}(x_n) V_{- 2 \alpha_0}(x_{n+1}) \rangle \\
 &= \prod_{i<j} |x_i-x_j|^{4 \alpha_i \alpha_j} \nonumber,
\end{align}
if $\sum_{i=1}^n \alpha_i = 2 \alpha_0$, otherwise the correlation function vanishes. 
 $V_{\alpha}$ is still  a scalar Virasoro primary, but now with conformal dimension %
\begin{eqnarray}\label{conformal dimension vertex}
\Delta ({\alpha})= 2 \alpha^2 -4 \alpha_0 \alpha. 
\end{eqnarray}

This means that the stress tensor also changed. In fact, the stress energy tensor is
\begin{eqnarray}
T(z)= - \frac{1}{4} : \partial  \phi \partial  \phi : + i \alpha_0 \partial ^2 \phi.
\end{eqnarray}
As explained in \cite{Dotsenko:2006}, this follows from changing the boundary conditions we impose on $\phi$ in the derivation of Noether's theorem. We now have that $\langle T(z) T(z') \rangle= \frac{1-24 \alpha_0^2}{2(z-z')^4}$, so the deformed theory has central charge
\begin{eqnarray}\label{background charge}
c=1-24 \alpha_0^2.
\end{eqnarray} 

Formula (\ref{corr background free}) is still too simple to represent correlation functions in minimal models. So, besides introducing a background charge, we still need to modify the free field theory further. We notice that equation (\ref{conformal dimension vertex}) allows for the existence of vertex operators of dimension $2$, which we denote by $V_{+}$ and $V_{-}$. The corresponding $\alpha$'s obey
\begin{eqnarray}\label{def alpha+}
\alpha_+ + \alpha_-= 2\alpha_0\,, \qquad \qquad  \alpha_+ \alpha_- = -1 .
\end{eqnarray}
We modify action (\ref{action free}) by introducing interacting terms
\begin{eqnarray}\label{action deformed}
S= \frac{1}{4 \pi} \int d^2 x  \partial \phi \bar{\partial}  \phi - 
\int d^2 x \left(V_{-}(x) + V_{+}(x)\right)\,.
\end{eqnarray}
We will use theory (\ref{action deformed}) defined with a background charge $-2 \alpha_0$ to represent minimal models. In the theory (\ref{action deformed}), correlation functions of vertex operators are given by
\begin{eqnarray}\label{corr background interacting}
\langle V_{\alpha_1} (x_1)... V_{\alpha_n} (x_n) \rangle = \frac{1}{l! k!} \prod_{j=1}^l \int d^2 w_j \prod_{i=1}^k d^2 y_i \\
 \langle V_{\alpha_1}(x_1) ... V_{\alpha_n}(x_n) V_{-}(y_1)...V_{-}(y_l) V_+(w_1) ... V_{+}(w_k) \rangle_{-2 \alpha_0}, \nonumber
\end{eqnarray}
if we can find non-negative integers $l$ and $k$ such that $\sum_{i=1}^n \alpha_i=2 \alpha_0 - l \alpha_- - k \alpha_+$. Otherwise, the correlation function is equal to $0$. In the second line of (\ref{corr background interacting}) we can use (\ref{corr background free}).

Given a minimal model at central charge $c$, we associate to it a scalar field theory with background charge according to (\ref{background charge}) and to the Virasoro primaries in that theory, we associate vertex operators according to (\ref{conformal dimension vertex}), by matching the respective conformal dimensions. Since equation (\ref{conformal dimension vertex}) is quadratic in $\alpha$, to the same Virasoro primary of dimension $\Delta$, we can associate two vertex operators $V_{\alpha}$ and $V_{2\alpha_0-\alpha}$. In minimal models, both the central charge and the dimensions of operators are discretized. For the minimal model $\mathbb{M}(p, q)$ (with $p>q$), we have
\begin{eqnarray} \label{discretization central charge}
\alpha_+^2=\frac{q}{p}.
\end{eqnarray}
This fixes the central charge according to (\ref{background charge}) and (\ref{def alpha+}). Minimal models are composed of Virasoro primaries $\Phi_{m, n}$ that are degenerate, in the sense that they have (null) descendants that are themselves Virasoro primaries. The Virasoro primaries $\Phi_{m, n}$ have conformal dimension given by (\ref{conformal dimension vertex}), with
\begin{eqnarray}
\alpha_{m, n} = \frac{1-m}{2} \alpha_- + \frac{1-n}{2} \alpha_+, \label{discretization alpha}  
\end{eqnarray}
where $ 1\leq m \leq p-1$ and $ 1\leq n \leq q-1$. Correlation functions of Virasoro primaries in a minimal model  can then be computed using the vertex operator representation
\beq
\Phi_{m, n} (x)=  \frac{1}{N_{m,n}} V_{\alpha_{m,n}} (x) 
=  N_{m,n} V_{2\alpha_0-\alpha_{m,n}} (x)\,,
\eeq
where $N_{m,n}$ is a normalization constant given in equation (\ref{normalisation coulomb degenerate}) in appendix (\ref{app: norm Coulomb}).
Notice that the representations in terms of $V_{\alpha_{m,n}}$ and $
V_{2\alpha_0-\alpha_{m,n}}$ are equivalent but in a given correlation function one choice will typically lead to simpler computations.
For example, the two-point function is trivial to compute as follows
\beq
\langle \Phi_{m, n} (x) \Phi_{m, n} (y) \rangle = 
\langle  V_{\alpha_{m,n}} (x) V_{2\alpha_0-\alpha_{m,n}} (y) \rangle=
\frac{1}{|x-y|^{2\Delta(\alpha_{m,n})}}\,.
\eeq
On the other hand, the non-trivial computation (see \cite{Dotsenko:1984ad})
\beq
\langle \Phi_{m, n} (x) \Phi_{m, n} (y) \rangle = 
 \frac{1}{N_{m,n}^2} 
\langle  V_{\alpha_{m,n}} (x) V_{\alpha_{m,n}} (y) \rangle=
\frac{1}{|x-y|^{2\Delta(\alpha_{m,n})}}\,,
\eeq
determines the normalization constant $N_{m,n}$.

\subsection{Example: the $\langle \Phi_{1, 2} O \Phi_{1, 2} O\rangle $ correlator} \label{sec: phi12}

We consider the four point function $\langle \Phi_{1, 2} O \Phi_{1, 2} O \rangle$, where $\Phi_{1, 2}$ is a second order degenerate Virasoro primary, $O$ is an arbitrary Virasoro primary $\Phi_{m,n}$ and we consider such a correlator for general central charge.\footnote{For the Yang-Lee and the Ising model, the Mellin amplitude of $\langle \Phi_{1, 2} \Phi_{1, 2} \Phi_{1, 2} \Phi_{1, 2} \rangle$ was computed in \cite{Alday:2015ota}. 
} %
The point of considering such a correlator is that it has a simple Mellin amplitude. We will see that its pole structure is dictated by the operator product expansion.

We associate each Virasoro primary to vertex operators in the following way:
\begin{eqnarray}\label{correspondence simple vertex}
\Phi_{1, 2} \rightarrow \frac{1}{N_{1,2}}V_{\alpha_{1, 2}}, \qquad
O \rightarrow \frac{1}{N_{m,n}}V_{\alpha} ,\qquad
\Phi_{1, 2} \rightarrow \frac{1}{N_{1,2}}V_{\alpha_{1, 2}}, \qquad
O\rightarrow N_{m,n}V_{2\alpha_0 - \alpha},
\end{eqnarray}
where $\alpha=\alpha_{m,n}$.
With these choices, we only need to insert one positive screening charge to compute the correlator
\begin{eqnarray}\label{one screening}
\langle \Phi_{1, 2} O \Phi_{1, 2} O  \rangle =
 \frac{1}{N_{1,2}^2}
 \prod_{i<j}^4 |x_i-x_j|^{4 \alpha_i \alpha_j} \int d^2 w \prod_{k=1}^4 |x_i-w|^{4 \alpha_i \alpha_+}.
\end{eqnarray} 

We would like to compute the Mellin amplitude of the correlator (\ref{one screening}). In our work, the following formula due to Symanzik (see \cite{Symanzik:1972wj}) will prove very useful:
\begin{eqnarray} \label{Symanzik}
\frac{1 }{\pi^{\frac{d}{2}}} \int d^d u 
\prod_{i=1}^n\frac{\Gamma(y_i)}{ |x_i-u|^{2 y_i}} = \prod_{i<j} \int \lbrack d\gamma_{ij} \rbrack  \Gamma(\gamma_{ij}) |x_i-x_j|^{-2\gamma_{ij}},
\end{eqnarray}
where the Mellin integral is constrained to  $\sum_{j \neq i} \gamma_{ij} = y_i$ and the formula is only valid if $\sum_i y_i = d$\footnote{Formula (\ref{Symanzik}) was derived in \cite{Symanzik:1972wj} under the additional assumptions $0<Re(y_i)< \frac{d}{2}$. The condition $Re(y_i)< \frac{d}{2}$ ensures that the integral (\ref{Symanzik}) converges when $u \rightarrow y_i$. The condition $Re(y_i)>0$ enables us to use the formula
\begin{eqnarray}\label{Schwinger param}
\frac{\Gamma(y_i)}{|x_i-u|^{2 y_i}}= \int_0^{\infty} \frac{d \sigma_i}{\sigma_i}\sigma_i^{y_i}e^{- \sigma_i |x_i-u|^{2}}
\end{eqnarray}
which is used in deducing (\ref{Symanzik}). If $Re(y_i)>0$, then the rhs of (\ref{Schwinger param}) converges. The assumptions $0<Re(y_i)< \frac{d}{2}$ ensure that on the rhs of (\ref{Symanzik}) the contour of the Mellin variables $\gamma_{ij}$ can be straight and parallel to the imaginary axis.
In this paper, we will use formula (\ref{Symanzik}) in contexts where the assumptions $0<Re(y_i)< \frac{d}{2}$ are not met. In that case, we can think of the rhs of (\ref{Symanzik}) as an analytic continuation in $y_i$ of its lhs. Indeed, consider the integral on the rhs of (\ref{Symanzik}) and suppose we start shifting continuously the positions of the poles. At some point, it is no longer possible to use a straight contour and so the contour should bend in order to account for this. Notice that even with a bent contour the integral  the rhs is perfectly well defined, since it has the same asymptotics when $\gamma_{ij} \rightarrow i \infty$ as before. }.
Since $4 (\sum_{i=1}^4 \alpha_i) \alpha_+ = 4 \alpha_- \alpha_+ = -4 $, then we can apply (\ref{Symanzik}) in (\ref{one screening}). Further doing a change of variables, we get
\begin{eqnarray}\label{Mellin phi12 1}
\langle \Phi_{1,2}(x_1) O(x_2) \Phi_{1,2}(x_3)O(x_4) \rangle 
= C_0 \prod_{i<j} \int \lbrack d\gamma_{ij} \rbrack  \Gamma(\gamma_{ij}+2 \alpha_i \alpha_j) |x_i-x_j|^{-2\gamma_{ij}}.%
\end{eqnarray}
The Mellin constraints are $\sum_{j \neq i} \gamma_{ij}=2\alpha^2_i - 4 \alpha_i \alpha_0 =\Delta(\alpha_i)$, which indeed  is the dimension of the operator inserted at the position $x_i$. %
We associate an $\alpha_i$ to each operator according to (\ref{correspondence simple vertex}). 
$C_0$ is a constant given by
\begin{align}\label{C0 phi12}
C_0 =  \frac{\pi }{N_{1,2}^2\prod_{i=1}^n \Gamma(-2\alpha_i\alpha_+)} 
=\frac{\Gamma \left(2-2 \alpha_+ ^2\right)}{\Gamma^2 \left(1-\alpha_+ ^2\right) \Gamma \left(2 \alpha_+ ^2-1\right) \Gamma (-2 \alpha  \alpha_+ ) \Gamma \left(2-2 \alpha_+^2+2 \alpha  \alpha_+ \right)}.
\end{align}

From (\ref{Mellin phi12 1}) we can read off the Mellin amplitude 
\begin{align}\label{Mellin explicit phi12 1}
&\hat{M}(\gamma_{12}, \gamma_{14})=C_0 
\Gamma\Big(\gamma_{13}-\frac{2 \Delta(\alpha_{1, 2})-
\Delta(\alpha_{1, 1} ) }{2}   \Big)
\Gamma\Big(\gamma_{13}-\frac{2 \Delta(\alpha_{1, 2})-
\Delta(\alpha_{1, 3} ) }{2}   \Big)\\
 & \Gamma\Big(\gamma_{12}-\frac{\Delta(\alpha_{1, 2})+\Delta({\alpha})-\Delta(\alpha-\frac{1}{2} \alpha_+)}{2}\Big)
  \Gamma\Big(\gamma_{12}-\frac{\Delta(\alpha_{1, 2})+\Delta({\alpha})-\Delta(\alpha+\frac{1}{2} \alpha_+)}{2}\Big)
  \nonumber \\
    &\Gamma\Big(\gamma_{14}-\frac{\Delta(\alpha_{1, 2})+\Delta({\alpha})-\Delta(\alpha-\frac{1}{2} \alpha_+)}{2}\Big)
  \Gamma\Big(\gamma_{14}-\frac{\Delta(\alpha_{1, 2})+\Delta({\alpha})-\Delta(\alpha+\frac{1}{2} \alpha_+)}{2}\Big) , \nonumber
\end{align}
where $\gamma_{12}+\gamma_{13}+\gamma_{14}= \Delta(\alpha_{1,2})$. 
We conclude that the Mellin amplitude is a meromorphic function.
The position of its simple poles is dictated by the OPE as in
\eqref{OPEpoles}.
To see that, recall that
the OPE of an arbitrary Virasoro primary $\Phi_{m,n}$ with $\Phi_{1,2}$ only contains 2 Virasoro primaries, 
\begin{eqnarray}\label{OPE exp}
\Phi_{1, 2} \times \Phi_{m,n} = \Phi_{m,n+1}+\Phi_{m,n-1}\,.
\end{eqnarray}
Each $\Gamma$--function in \eqref{Mellin explicit phi12 1} encodes the poles associated to each Virasoro primary exchanged in a given channel.
In appendix \ref{app: blocks}, we compare the Mellin representation of this four-point function with its expansion in Virasoro conformal blocks.
In particular, we check that the overall normalization is correct by matching the contribution of the identity block.

The correlator $\langle \phi_{2,1} O \phi_{2,1} O \rangle$ can be similarly analysed. In the Coulomb gas formalism, we now insert a negative screening charge instead of a positive one. An expression analogous to (\ref{Mellin phi12 1}) holds for the correlator, with 
\begin{eqnarray}
C_0= \frac{\Gamma(2-\frac{2}{\alpha_+^2})}{\Gamma^2(1-\frac{1}{\alpha_+^2})\Gamma(-1+\frac{2}{\alpha_+^2})\Gamma(2-\frac{2}{\alpha_+^2}-2\frac{\alpha}{\alpha_+})\Gamma(\frac{2\alpha}{\alpha_+})}. 
\end{eqnarray}

The correlator $\langle \Phi_{1, 2} O \Phi_{1, 2} O \rangle$ obeys a BPZ differential equation, in virtue of the fact that $\Phi_{1,2}$ is a degenerate operator. This implies that the Mellin amplitude (\ref{Mellin explicit phi12 1}) obeys a recursion relation. We confirm this in appendix \ref{app: Mellin diff}. 

Consider now the more general correlator $\langle \Phi_{1, 2} O_2 O_3 O_4 \rangle$, where $O_2$, $O_3$ and $O_4$ are arbitrary Virasoro scalar primaries. Given expression (\ref{Mellin explicit phi12 1}), it is simple to guess a Mellin amplitude $M(\gamma_{12}, \gamma_{14})$ for this correlator: 
\begin{align}\label{Mellin gen Phi12 text}
&\hat{M} (\gamma_{12}, \gamma_{14})=
C_0 \times\\&
\times\prod_{j=2}^4\Gamma\Big(\gamma_{1j}-\frac{\Delta(\alpha_{1, 2})+\Delta({\alpha_j})-\Delta(\alpha_j-\frac{1}{2} \alpha_+)}{2}\Big)
\Gamma\Big(\gamma_{1j}-\frac{\Delta(\alpha_{1, 2})+\Delta({\alpha_j})-\Delta(\alpha_j+\frac{1}{2} \alpha_+)}{2}\Big)
\nonumber
\end{align}
where $\gamma_{13}+\gamma_{12}+\gamma_{14}= \Delta(\alpha_{1,2})$ and $\alpha_2$, $\alpha_3$ and $\alpha_4$ can be computed from the conformal dimensions of $O_2$, $O_3$ and $O_4$. We confirm this guess in appendix (\ref{app: Mellin diff}), where we check that the correlator $\langle \Phi_{1, 2} O_2 O_3 O_4 \rangle$ defined with the Mellin amplitude (\ref{Mellin gen Phi12 text}) obeys the appropriate BPZ differential equation. A formula for $C_0$ can be found in appendix (\ref{app: analytic continuation}), see formula (\ref{C0phi12}). 

\subsection{Mellin amplitude of any correlation function} \label{sec: mellin general 2d}

In this section, we write the Mellin amplitude of a general $n$ point correlator in minimal models. Our procedure to do so will be a generalisation of the computation in section \ref{sec: phi12}. The key ingredients are the Coulomb gas formalism and Symanzik formula (\ref{Symanzik}).

Consider then a general $n$ point correlator $\langle \Phi_1 (x_1) ... \Phi_n(x_n) \rangle$.   To each operator $\Phi_i$, we associate a vertex operator: $\Phi_i \rightarrow \frac{1}{N_i} V_{\alpha_i}$. 
We will assume that all operators are degenerate, i.e. each $\alpha_i$ is of the form \eqref{discretization alpha} as it is the case in minimal models. \footnote{Here, we have used the shorter notation $\Phi_i \equiv \Phi_{m_i,n_i}$ and $N_i \equiv N_{m_i,n_i}$.}
In the Coulomb gas formalism suppose we need to insert $p_1$ positive screening charges and $q_1$ negative screening charges to compute $\langle V_{\alpha_1}(x_1) ... V_{\alpha_n}(x_n) \rangle$. Then, this correlator has a Mellin representation (\ref{def Mellin general}), with Mellin amplitude given by
\begin{align}\label{Mellin minimal models}
\hat{M}(\gamma_{ij})&= C_0 \int \prod_ {r=1}^{z-1} 
\left[
\prod_{k_r<l_r}^{n+r} \[ d \xi_{k_r, l_r}^{r} \] \Gamma(\xi_{k_r, l_r}^r)
 \prod_{s_r=1}^{n+r-1} \frac{1}{\Gamma(-2 \alpha_{s_r} \alpha_{n+r} + \sum_{v=r}^{z-1} \xi_{s_r, n+r}^{v})}
 \right]
\\
&\times  \prod_{i<j}^n \Gamma(\gamma_{ij}- \sum_{r=1}^{z-1}\xi_{ij}^r +2\alpha_i \alpha_j), \nonumber
\end{align}
where  $\alpha_{n+1}= ... = \alpha_{n+p_1}=\alpha_+$ and $\alpha_{n+p_1+1}=...=\alpha_{n+p_1+q_1}=\alpha_-$ and $z=p_1+q_1$ is the total number of screening charges.\footnote{We assumed $p\geq 1$.} The measure is constrained by
\begin{eqnarray}
\sum_{j=1 \atop j\neq i}^{n+r} \xi_{ij}^r= -2 \alpha_i \alpha_{n+r+1} + \sum_{s=r+1}^{z-1} \xi_{i, n+r+1}^s\,,\qquad\qquad
i=1,2,\dots, n+r \label{measure general formula},
\end{eqnarray}
for $r=1$, ..., $z-1$. The constant $C_0$ can be determined from the normalisations,
\begin{eqnarray}\label{C0 general norm}
C_0= \frac{\pi^{z}}{\prod_{i=1}^{n } N_i \prod_{j=1}^{n+z-1}\Gamma( -2 \alpha_j \alpha_{n+z})}.
\end{eqnarray}

Formula (\ref{Mellin minimal models}) can be deduced by iteratively applying (\ref{Symanzik}). We show how to do this for the case in which we insert two screening charges in appendix \ref{app: 2 screening}.
The complexity of (\ref{Mellin minimal models}) grows very quickly with the total number $z$ of screening charges.\footnote{Namely, it involves
\beq
\sum_{r=1}^{z-1} \frac{(n+r )(n+r-3)}{2} =
(z-1)\frac{3n(n+z-3)+z(z-5)}{6}
\eeq
integrals of an integrand with
\beq
\sum_{r=1}^{z-1}\left[ \frac{(n+r )(n+r-1)}{2} + (n+r-1)\right] + \frac{n(n-1)}{2}=
\frac{z^3+3nz(n+z)-6n -7z+6}{6}
\eeq
$\Gamma$-functions.} Nevertheless, this explicit Mellin-Barnes representation  is useful to derive several general properties of the Mellin amplitude. %

Formula (\ref{Mellin minimal models}) is correct in every CFT for which the Coulomb gas technique of computing correlation functions applies. This is the case in diagonal minimal models (whether they are unitary or not). There are other CFT's  that can be viewed as limits of minimal models, like generalised minimal models and Liouville theory (see \cite{Ribault:2014hia} and \cite{Ribault:2015sxa}).\footnote{We thank Sylvain Ribault for discussions on this.} So, perhaps formula (\ref{Mellin minimal models}) can be useful in that context. See also \cite{Furlan:2004dq} and \cite{Furlan:2018jlv} for attempts to generalise the Coulomb gas formalism to higher dimensions. Identities of Dotsenko-Fateev integrals were used in \cite{Pasquetti:2019tix} and \cite{Pasquetti:2019uop} to derive properties of 3d supersymmetric theories, so maybe our formulas can be useful in that context.

\subsection{Analyticity of Mellin amplitudes
} \label{sec: phi13}

We conjecture that the Mellin amplitudes in 2D minimal models are meromorphic functions with only poles at the locations predicted by the OPE as in \eqref{OPEpoles}. Meromorphicity follows from the Mellin-Barnes representation  \eqref{Mellin minimal models}. However, we were not able to show in that the only singularities are the ones predicted by the OPE. In this section, we check this statement for a class of correlators in minimal models.

We consider the correlation function $\langle \Phi_{1, 3} O \Phi_{1, 3} O \rangle$, where $O$ is an arbitrary Virasoro primary. This correlation function can be computed in the Coulomb gas formalism by the prescription \begin{eqnarray}\label{correspondence simple vertex 2}
\Phi_{1, 3} \rightarrow \frac{1}{N_{1,3}}V_{\alpha_{1, 3}}, \qquad
O \rightarrow \frac{1}{N_{m,n}}V_{\alpha} ,\qquad
\Phi_{1, 3} \rightarrow \frac{1}{N_{1,3}}V_{\alpha_{1, 3}}, \qquad
O\rightarrow N_{m,n}V_{2\alpha_0 - \alpha},
\end{eqnarray}
with $\alpha=\alpha_{m,n}$ and inserting two positive screening charges.

Its Mellin amplitude can be obtained fom the general formulas (\ref{Mellin minimal models}) and (\ref{C0 general norm}). We can write the Mellin amplitude  in the form
\begin{align}\label{Mellin 5 int}
\hat{M}(\gamma_{12}, \gamma_{14})=C_0 \int \lbrack d\xi_{12} \rbrack  \int \lbrack d\xi_{15} \rbrack  \int \lbrack d\xi_{24} \rbrack  \int \lbrack d\xi_{34} \rbrack  \int \lbrack d\xi_{35}  \rbrack \Gamma (\xi_{12}) \Gamma (\xi_{15})  \\
 \Gamma (-2 \alpha \alpha_+ +\gamma_{12}-\xi_{12}) \Gamma (\xi_{24})  
 \Gamma (\xi_{15}-\xi_{24}-\xi_{34}+1) \Gamma \left(-2 \alpha_+^2+2 \alpha \alpha_+ +\gamma_{12}-\xi_{34}+2\right)\nonumber \\
\Gamma (\xi_{34}) \Gamma (-2 \alpha \alpha_+-\xi_{12}-\xi_{15}+\xi_{34}-1) 
\Gamma (-2 \alpha \alpha_++\gamma_{14}-\xi_{15}+\xi_{24}+\xi_{34}-1) \nonumber \\
\Gamma \left(-2 \alpha_+^2+2 \alpha \alpha_++\gamma_{14}+\xi_{12}+\xi_{24}-\xi_{35}+1\right) 
\Gamma \left(2 \alpha_+^2-\xi_{15}+\xi_{24}-\xi_{35}-1\right) \nonumber \\
\Gamma \left(\xi_{12}-\xi_{34}-\xi_{35}+2 \alpha \alpha_+-2 \alpha_+^2+1\right) 
\Gamma (\xi_{35}) \Gamma (-\xi_{12}-\xi_{24}+\xi_{35}+1) \nonumber \\
\frac{\Gamma \left(-\gamma_{12}-\gamma_{14}+\xi_{15}-\xi_{24}+\xi_{35}+4 \alpha_+^2-1\right)}{\Gamma \left(2 \alpha_+^2+\xi_{15}\right) \Gamma (-4 \alpha \alpha_+-\xi_{12}-\xi_{15}+\xi_{34}-1)}  \nonumber \\
\frac{\Gamma (-\gamma_{12}-\gamma_{14}-\xi_{24})}{\Gamma \left(2 \alpha_+^2+\xi_{35}\right) \Gamma \left(-4 \alpha_+^2+4 \alpha \alpha_++\xi_{12}-\xi_{34}-\xi_{35}+3\right)}. \nonumber
\end{align} 
The above formula is a complicated integral that the reader should not read carefully. Our point is just to consider an application of formula (\ref{Mellin minimal models}). In the rest of this section, we will check that (\ref{Mellin 5 int}) only gives rise to  singularities at the locations  \eqref{OPEpoles} predicted by the OPE.

In the OPE of $\Phi_{13}$ and $\Phi_{\alpha}$, there are three Virasoro primaries exchanged:
\begin{eqnarray}\label{OPE Phi13}
\Phi_{1, 3} \times \Phi_{\alpha} = \Phi_{\alpha + \alpha_+} + \Phi_{\alpha} + \Phi_{\alpha -\alpha_+}.
\end{eqnarray}
Each Virasoro family contains many global primaries, whose twists differ by even integers. So, we expect the Mellin amplitude to have $9$ sequences of poles: $3$ sequences of poles in $\gamma_{12}$, another $3$ in $\gamma_{14}$ and another $3$ in $\gamma_{13}$.

Let us now outline how we obtained the singularity structure of (\ref{Mellin 5 int}). We used the technique introduced in \cite{Yuan:2018qva} to compute the singularities of Mellin-Barnes integrals. The reader interested in understanding this technique can read appendix C of that paper. In the next sentences, we briefly describe the method. Suppose we have a multiple Mellin-Barnes integral and we ask what are its singularities. Given just one Mellin-Barnes integral, it diverges whenever its contour gets pinched by two poles of the integrand.  

When we have multiple integrals, we must take a more global perspective. Let us illustrate this with an example taken from \cite{Yuan:2018qva}. Consider the integral
\begin{eqnarray}
\int \frac{dx dy}{(2 \pi i)^2} \Gamma(a_1+x)\Gamma(a_2-x)\Gamma(b_1+x-y)\Gamma(b_2-x+y)\Gamma(c_1+y)\Gamma(c_2-y).
\end{eqnarray}
We might expect expect to obtain its singularities in the following manner. Suppose we first integrate in $x$ and later in $y$. For a single Melin-Barnes integral, we know how to predict its singularities. So, we might expect to determine the singularities of a multiple Mellin Barnes integral by applying the one dimensional technique in succession. It turns out that if we do that we end up with a lot of fake poles.

In this example, let us list the poles that we would obtain by iteratively applying the one dimensional technique. We can integrate first in $x$ and later in $y$, or vice versa. The set of singularities thus obtained is:
\begin{eqnarray}\label{list possible sings}
\Gamma_{ \{1, 2\} }(a_1+a_2), \Gamma_{ \{3, 4\} }(b_1+b_2), \Gamma_{ \{5, 6\} }(c_1+c_2), \Gamma_{ \{1, 4, 6\} }(a_1 + b_2 + c_2), \Gamma_{ \{2, 3, 5\} }(a_2+b_1+c_1) \\
\Gamma_{ \{1, 2, 3, 4 \} }(a_1+a_2+b_1+b_2), \Gamma_{ \{1, 2, 5, 6 \} }(a_1+a_2+c_1+c_2), \Gamma_{ \{3, 4, 5, 6\} }(b_1+b_2+c_1+c_2), \nonumber \\
 \Gamma_{ \{1, 2, 3, 4, 5, 6\} }(a_1+a_2+b_1+b_2+c_1+c_2).  \nonumber
\end{eqnarray}
Here, the use of the $\Gamma$ function is symbolic. We just mean that the integral is singular whenever the argument of the $\Gamma$ functions above is a nonpositive integer. We use a subscript to denote the position of the poles that gave rise to the singularity in the original integrand. For example, $\Gamma_{ \{1, 2\} }(a_1+a_2)$ came from the collision of $\Gamma(a_1+x)$ and $\Gamma(a_2-x)$.

Consider the $\Gamma$ functions in (\ref{list possible sings}). We will say that the poles coming from a certain $\Gamma$ function are \textit{composite} whenever the corresponding subscript contains as a subset the subscript of another $\Gamma$ function in (\ref{list possible sings}). The poles from $\Gamma_{ \{1, 2, 3, 4 \} }$, $\Gamma_{\{1, 2, 5, 6\} }$, $\Gamma_{ \{3, 4, 5, 6\} }$ and $\Gamma_{ \{1, 2, 3, 4, 5, 6\} }$ are composite. The rest of the poles are non composite.   

All composite poles are fake. This is a very important statement that allows to generate a fast algorithm to determine the real poles in a multiple Mellin-Barnes integral. \cite{Yuan:2018qva} contains some general arguments in favour of this statement, although it is not clear to the present authors that we can conclude from them that such a statement is rigorously proved. Sill, to the best of our knowledge the algorithm of \cite{Yuan:2018qva} works in every example.

With this technique, we predict the Mellin amplitude to have singularities whenever any of the following expressions is equal to $0$:
\begin{eqnarray}
\{ \gamma_{12}+2 \alpha \alpha_+-2 \alpha_+^2+2,\gamma_{13}-4 \alpha_+^2+2,\gamma_{14}+2 \alpha \alpha_+-2 \alpha_+^2+2, \nonumber \\
\gamma_{12}-2 \alpha_+^2+1,\gamma_{13}-2 \alpha_+^2+1,\gamma_{14}-2 \alpha_+^2+1,\nonumber \\
\gamma_{12}-2 \alpha \alpha_+,\gamma_{13}+2 \alpha_+^2,\gamma_{14}-2 \alpha \alpha_+ \},
\end{eqnarray} 
with $\gamma_{13}=-\gamma_{12}-\gamma_{14}+4 \alpha_+^2-2$. This precisely matches the expectations from (\ref{OPE Phi13}). In particular, the first three poles correspond to the exchange of $\Phi_{\alpha + \alpha_+}$, the next three to the exchange of $\Phi_{\alpha}$ and the final three poles to the exchange of $\Phi_{\alpha - \alpha_+}$.

\subsection{The bulk point limit}\label{sec: bulk point limit 2d}

Correlation functions in Lorentzian signature can have divergences due to the existence of a point that is null separated from the points where we insert the external operators (see \cite{Maldacena:2015iua}).

Such singularities can arise from pertubation theory in the boundary or from Landau diagrams in the bulk. In a fully nonperturbative CFT those singularities are expected to be absent. This was proven in $d=2$ in \cite{Maldacena:2015iua}.

We will show that the divergence in the bulk point limit is controlled by the limit of large Mellin variables of the Mellin amplitude. For simplicity we will consider the case of a four point function of equal primaries of conformal dimension $\Delta$. 

We have already shown in section (\ref{sec:MelGen}) that the Mellin amplitude is polynomially bounded at infinity. Consider the limit where $\gamma_{12}= \beta s$, $\gamma_{14}= (1-\beta) s$, where $\beta \in [ 0, 1 ]$ and we take $s$ very large.  Suppose that in that limit the Mellin amplitude $M(\gamma_{12}= \beta s, \gamma_{14}=(1-\beta)s) \sim s^b$. Suppose also that the four point function diverges in the bulk point limit like $\frac{1}{(z- \bar{z})^{2 a}}$ when $z \rightarrow \bar{z}$. In the next subsection we will demonstrate the formula
\begin{eqnarray}
a=-1+2\Delta+b .
\end{eqnarray}

Even if the correlation function is regular in the bulk point limit this still constrains the power with which the Mellin amplitude can grow. We carry out this analysis for the case of minimal models.

\subsubsection{Relating the behaviour of the Mellin amplitude for large Mellin variables and the bulk point limit}

\begin{figure}[h]
  \centering
  \includegraphics[scale=0.5]{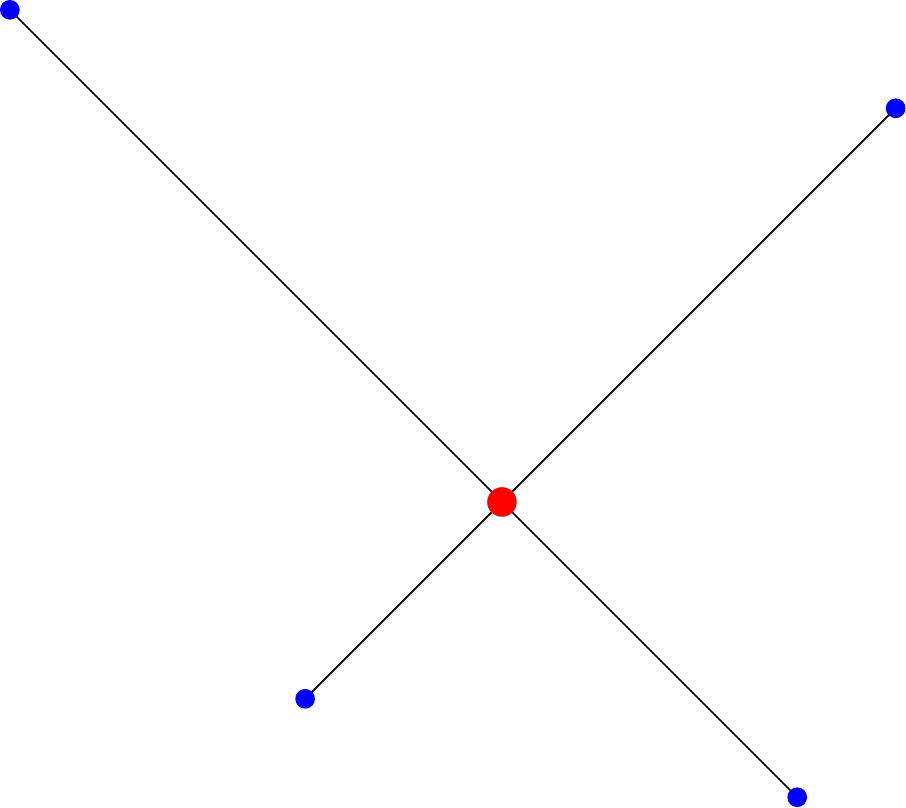} 
  \caption{Example of a boundary Landau diagram.}
  \label{fig:LandauD}
\end{figure}

Let us start by briefly reminding the reader about the bulk point limit with an example taken from \cite{Maldacena:2015iua}. Consider the following configuration of four point points in $d \geq 3$\footnote{The above configuration is an example of a Landau diagram. In $d=2$ there are no boundary Landau diagrams. However, there are bulk diagrams. It was shown in \cite{Maldacena:2015iua} that in $d=2$ for the full nonperturbative theory there are no singularities except for the lightcone singularities.}, see figure (\ref{fig:LandauD}):
\begin{eqnarray}
x_1= (-t, 0 , 1, \vec{0}), x_2= (t, 1 , 0, \vec{0}), x_3= (-t, 0 , -1, \vec{0}), x_4= (t, -1 , 0, \vec{0}).
\end{eqnarray}

We employ the $i \epsilon$ prescription $t = \tilde{t} -i \epsilon$, where $\tilde{t} \in [0, 1]$. When $\tilde{t}=0$ this configuration is Euclidean, but when $\tilde{t}=1$ the point $(0, 0, 0 , \vec{0})$ is null separated from the external points. This signals a divergence in the correlation function. 

The cross ratios $u$ and $v$ are given by
\begin{eqnarray}
u= \frac{(-(2 t)^2+2)^2}{16}, v= \frac{(-(2 t)^2+2)^2}{16}.
\end{eqnarray}
When we vary $\tilde{t}$ from $0$ to $1$ the cross ratios $u$ and $v$ circle around the origin in the complex plane.

Let us see how the Mellin representation of the correlation function changes as we vary $\tilde{t}$ from $0$ to $1$. We write
\begin{eqnarray}
\langle \O \O \O \O \rangle = \frac{f(u, v)}{x_{13}^{2 \Delta} x_{24}^{2 \Delta}},
\end{eqnarray}
where 
\begin{eqnarray}
f(u, v)= \int \frac{d \gamma_{12}}{2 \pi i} \int \frac{d \gamma_{14}}{2 \pi i}
M(\gamma_{12}, \gamma_{14}) \Gamma(\gamma_{12})^2 \Gamma(\gamma_{14})^2 \Gamma(\Delta-\gamma_{12}-\gamma_{14})^2 u^{- \gamma_{12}} v^{- \gamma_{14}} \\
\rightarrow \int \frac{d \gamma_{12}}{2 \pi i} \int \frac{d \gamma_{14}}{2 \pi i}
M(\gamma_{12}, \gamma_{14}) \Gamma(\gamma_{12})^2 \Gamma(\gamma_{14})^2 \Gamma(\Delta-\gamma_{12}-\gamma_{14})^2 u^{- \gamma_{12}} v^{- \gamma_{14}} e^{- 2 \pi i \gamma_{12}}  e^{- 2 \pi i \gamma_{14}}. \label{Mellin bulk point}
\end{eqnarray}
Notice that the factors $e^{- 2 \pi i \gamma_{12}}  e^{- 2 \pi i \gamma_{14}}$ grow exponentially in certain directions at infinity and cancel the exponential decay of the $\Gamma$ functions in the prefactor.

Let us suppose that $\langle \O \O \O \O \rangle$ diverges in the bulk point limit and let us see how this comes about from (\ref{Mellin bulk point}). Such a divergence can only come from the region where both $\gamma_{12}$ and $\gamma_{14}$ have a very big positive imaginary part. We basically follow section $(3.1)$ of \cite{Penedones:2010ue}.

Let us analyse the integral (\ref{Mellin bulk point}) in that region. Let us write $\gamma_{12}= \alpha_{12} + i m_1$ and $\gamma_{14}= \alpha_{14} + i m_2$, where $m_1, m_2 \gg 1$. $\alpha_{12}$ and $\alpha_{14}$ depend on the contour that we pick and we consider them arbitrary. Let us write furthermore $m_1= \beta s$, $m_2 = (1- \beta) s$, where $\beta \in [0, 1]$ and $s \gg 1$.  We can use Stirling's formula to approximate the $\Gamma$ functions in (\ref{Mellin bulk point}). Furthermore we assume that the Mellin amplitude goes like $g(\beta) s^b$ for large $s$, where $g(\beta)$ is a function that we do not know and $b$ is a power that will control the bulk point limit divergence. In this regime, the integral (\ref{Mellin bulk point}) can be written as
\begin{eqnarray} \label{Mellin bulk point 2}
\int_{s_0}^{\infty} \frac{ds}{s} \int_0^1 d\beta g(\beta) s^{-1 + 2 \Delta + b} (- 8 i) e^{- i \pi \Delta} \pi^3 u^{- \alpha_{12}- i s \beta} v^{-\alpha_{14} - i s(1- \beta)}\\
(1- \beta)^{-1 + 2 \alpha_{14} - 2 i s(-1 + \beta)} \beta^{-1 + 2 \alpha_{12} + 2 i s \beta},\nonumber
\end{eqnarray}
where we take $s_0$ very large.

The integrand goes like $\exp \Big(- i s \big( \beta \log(u) + (1- \beta) \log(v) - 2 (1-\beta) \log(1-\beta) - 2 \beta \log(\beta)   \big) \Big)$. The integral in $\beta$ is dominated by the saddle point $\beta_s = \frac{\sqrt{u}}{\sqrt{u}+ \sqrt{v}}$. At the saddle point the previous exponential becomes $e^{2 i s \log (\sqrt{u} + \sqrt{v})}$. The integral can only diverge when $\sqrt{u} + \sqrt{v}=1$. This happens precisely at the bulk point limit. Indeed, we have that $\log(\sqrt{u} + \sqrt{v}) \approx - \frac{(z - \bar{z})^2}{16 (1-\bar{z})\bar{z}}$ when $z \rightarrow \bar{z}$ and $0 \leq z \leq 1 , 0 \leq \bar{z} \leq 1$. So, expression (\ref{Mellin bulk point 2}) becomes
\begin{eqnarray}\label{Mellin bulk point 2}
\int_{s_0}^{\infty} \frac{ds}{s}  g(\beta_s)  s^{-1 + 2 \Delta + b} \frac{ 1}{ \sqrt{-i s} }  (- 4\sqrt{2} i) e^{- i \pi \Delta}  \pi^{\frac{7}{2}} u^{- \alpha_{12}} v^{-\alpha_{14} }\\
(1- \beta)^{-1 + 2 \alpha_{14}} \beta^{-1 + 2 \alpha_{12} } e^{i s \frac{(z - \bar{z})^2}{8 (1-\bar{z})\bar{z}} }   \nonumber
\end{eqnarray}
Equation (\ref{Mellin bulk point 2}) enables us to relate the rate of divergence of the correlator with the polynomial grow of the Mellin amplitude:
\begin{eqnarray}
M(s \gamma_{ij}) \sim s^b \implies f(u, v) \sim \frac{1}{(z-\bar{z})^{-3 + 4\Delta + 2 b } }
\end{eqnarray}

Let us consider the case in which the correlation function does not diverge in the bulk point limit. The previous analysis gives us the bound $b < \frac{3}{2} - 2 \Delta$. Furthermore it is reasonable to assume that also the derivatives with respect to $z$ of the correlation function should be analytic at the bulk point. Thus, let us assume that the correlation function is regular in $z- \bar{z}$. Then $b$ can only take the values $b= \frac{3}{2} - 2\Delta - \frac{n}{2}$ where $n$ is a positive integer. Indeed if it were not to take such values this would generate a divergence in some derivative of the correlation function.
A similar but more general analysis in general $d$ was recently performed in \cite{Dodelson:2019ddi} with similar conclusions (which are identical in $d=2$). It would be interesting to establish analyticity of the bulk point locus in higher $d$ rigorously. It would be also interesting to understand further implications of the bulk point analyticity for the high energy limit of the flat space scattering amplitudes.

\subsubsection{The bulk point limit in minimal models}

In minimal models we expect no divergence in the bulk point limit. As we saw in the last section this constrains the Mellin amplitude of a four point function of equal scalar primaries to behave like $M(\gamma_{12}= \beta s, \gamma_{14}=(1-\beta) s) \sim s^{1- 2 \Delta - \frac{n}{2}}$ where $n$ is a positive integer. In this subsection we will use our general formula (\ref{Mellin minimal models}) to confirm this prediction.   

The general formula (\ref{Mellin minimal models}) drastically simplifies  in the limit of large Mellin variables $\gamma_{ij} \rightarrow i  \infty$. Let us divide (\ref{Mellin minimal models}) by $\prod_{i<j}\Gamma(\gamma_{ij})$. We can use 
Stirling's approximation to get %
\begin{eqnarray}
\frac{\Gamma(\gamma_{ij}- \sum_{r=1}^{z-1}\xi_{ij}^r +2\alpha_i \alpha_j)}{\Gamma(\gamma_{ij})} =  \gamma_{ij}  ^{- \sum_{r=1}^{z-1}\xi_{ij}^r +2\alpha_i \alpha_j } \left[1+ O(\gamma_{ij}^{-1})\right].
\end{eqnarray}
If we consider the case where all Mellin variables are proportional to some parameter $s$ that is much bigger than any number  in our system, then the Mellin amplitude is proportional to
\begin{eqnarray}\label{power s 2d}
 s^{- \sum_{r=1}^{z-1} \sum_{i<j}^n \xi_{ij}^r +2 \sum_{i<j}^n \alpha_i \alpha_j}.
\end{eqnarray}

The sum in the exponent of (\ref{power s 2d}) simplifies drastically due to the measures (\ref{measure general formula}) and in fact does not depend on the integration variables $\xi_{ij}$. We do this sum in appendix (\ref{app: sums exponent}) and obtain that the Mellin amplitude grows with  
\begin{eqnarray}
s^{1- \frac{1}{2}\sum_{i} \Delta_i },
\end{eqnarray}
for a general $n$ point function of scalar Virasoro primaries in minimal models with conformal dimensions $\Delta_i$. For the case of a four point function of equal scalar primaries we find $s^{1- 2 \Delta}$ in agreement with what we predicted in the preceding subsection.

\subsection{Correlation functions with vanishing Mellin Amplitude}\label{sec: simple corr}

In this section, we provide examples of non zero correlation functions with vanishing Mellin amplitudes. This phenomenon might seem paradoxical, but is in fact entirely reasonable. An explanation of this can be found in section \ref{sec:MelGen}, where we explain how to construct Mellin amplitudes in general CFT's. It follows from that construction that there is no problem with having a Mellin amplitude that vanishes, even if the correlation function does not\footnote{\cite{Rastelli:2017udc} and \cite{Ponomarev:2017qab} already suggest this, but in section \ref{sec:MelGen} we provide a definition of the Mellin amplitude, which in our opinion settles this issue.}. In this section, we limit ourselves to giving a few simple examples of this.

One such example is the correlator $\langle \sigma \epsilon \sigma \epsilon \rangle$ in the 2d Ising model.
\begin{eqnarray}\label{sigma epsilon sigma epsilon}
\langle \sigma(x_1) \epsilon(x_2) \sigma(x_3) \epsilon(x_4) \rangle = \frac{1}{x_{13}^{\frac{1}{4}} x_{24}^2}  \frac{u+v-\frac{1}{2}}{2 \sqrt{u} \sqrt{v}}.
\end{eqnarray}
In the 2d Ising model, the operator $\sigma$ is a second order degenerate Virasoro primary $\Phi_{1, 2}$, whereas $\epsilon$ is $\Phi_{1, 3}$. The 2d Ising model is a minimal model at central charge $c=\frac{1}{2}$. We will consider the Mellin amplitude of $\langle \Phi_{1, 2} \Phi_{1, 3} \Phi_{1, 2} \Phi_{1, 3} \rangle$ at general central charge and then take the limit $c \rightarrow \frac{1}{2}$ to see that it vanishes.

From formulas (\ref{discretization alpha}), (\ref{Mellin explicit phi12  1}) and (\ref{C0 phi12}) it follows that 
\begin{align}\label{phi12 phi13}
\langle \Phi_{1, 2} \Phi_{1, 3} \Phi_{1, 2} \Phi_{1, 3} \rangle=& \frac{1}{|x_1-x_3|^{-2+3\alpha_+^2}}\frac{1}{|x_2-x_4|^{-4+8\alpha_+^2}} f(u, v), \\
f(u,v)=&\int\frac{d \gamma_{12}}{2 \pi i}  \frac{d \gamma_{14}}{2 \pi i}  M(\gamma_{12}, \gamma_{14})u^{- \gamma_{12}} v^{-\gamma_{14}} \label{f eps}, \\
M(\gamma_{12}, \gamma_{14})=& \frac{\Gamma \left(2-2 \alpha_+^2\right)}{\Gamma \left(2 \alpha_+^2\right) \Gamma \left(2-4 \alpha_+^2\right) \Gamma \left(1-\alpha_+^2\right)^2 \Gamma \left(2 \alpha_+^2-1\right)}\\
&\Gamma \left(-2 \alpha_+^2+\gamma_{12}+1\right) \Gamma \left(\alpha_+^2+\gamma_{12}\right) \Gamma \left(-2 \alpha_+^2+\gamma_{14}+1\right)\nonumber \\
 &\Gamma \left(\alpha_+^2+\gamma_{14}\right) \Gamma (-\gamma_{12}-\gamma_{14}) \Gamma \left(2 \alpha_+^2-\gamma_{12}-\gamma_{14}-1\right),\nonumber
\end{align}  
where we remind that $\alpha_+$ is related to the central charge according to $c=1-6(\alpha_+ - \frac{1}{\alpha_+})^2$. 

In the limit $\alpha_+^2 \rightarrow \frac{3}{4}$, expression (\ref{phi12 phi13}) should turn into (\ref{sigma epsilon sigma epsilon}). Notice that in that limit the Mellin amplitude $M(\gamma_{12}, \gamma_{14})$ goes to 0, thanks to the factor $\Gamma(2-4\alpha_+^2)$ in the denominator. But we will see that the contour gets pinched, which usually signals a divergence of the Mellin integral. But since the integrand is also going to $0$, the net result is a finite quantity, expression (\ref{sigma epsilon sigma epsilon}).

Let us see how this comes about explicitly. For the moment, we will consider $\alpha_+^2$ generic and later take the limit $\alpha_+^2 \rightarrow \frac{3}{4}$. For sufficiently small $u$, we can evaluate the integral  by deforming the $\gamma_{12}$ contour in (\ref{f eps}) to the left.  Consider the contribution coming from the poles originating from $\Gamma \left(-2 \alpha_+^2+\gamma_{12}+1\right)$. We conclude that $f(u, v)$ contains the following terms
\begin{eqnarray}\label{interm f}
\sum_{n=0}^{\infty}\frac{(-1)^n u^{1+n-2\alpha_+^2} \Gamma \left(2-2 \alpha_+^2\right) \Gamma \left(3 \alpha_+^2-n-1\right)}{\Gamma(n+1) \Gamma \left(2 \alpha_+^2\right) \Gamma \left(2-4 \alpha_+^2\right) \Gamma \left(1-\alpha_+^2\right)^2 \Gamma \left(2 \alpha_+^2-1\right)} \\
\int[d \gamma_{14}] \Gamma \left(-2 \alpha_+^2+\gamma_{14}+1\right) \Gamma \left(\alpha_+^2+\gamma_{14}\right) \Gamma (n-\gamma_{14}) \Gamma \left(-2 \alpha_+^2-\gamma_{14}+n+1\right) v^{-\gamma_{14}} \nonumber
\end{eqnarray}
Consider the product $\Gamma(-2 \alpha_+^2+\gamma_{14}+1)\Gamma(-2 \alpha_+^2 - \gamma_{14}+n+1)$. The factor $\Gamma(-2 \alpha_+^2+\gamma_{14}+1)$ gives rise a sequence of poles to the left, whereas $\Gamma(-2 \alpha_+^2 - \gamma_{14}+n+1)$ gives rise to a sequence of poles to the right. In the limit $\alpha_+^2 \rightarrow \frac{3}{4}$, the product becomes $\Gamma \left(-\gamma_{14}+n-\frac{1}{2}\right) \Gamma \left(\gamma_{14}-\frac{1}{2}\right)$. For $n=0$ and $n=1$, the two sequences collide. In that case, the contour in (\ref{interm f}) gets pinched, which compensates for the divergent $\Gamma \left(2-4 \alpha_+^2\right)$ appearing in the denominator.

In order to actually evaluate (\ref{interm f}) in the limit $\alpha_+^2 \rightarrow \frac{3}{4}$, we proceed in the following way. For general $\alpha_+^2$, we deform the contour to the left in $\gamma_{14}$, thus obtaining a double power series expansion in $u$ and $v$. The coefficients multiplying each term depend on $\alpha_+^2$. When we take the limit $\alpha_+^2 \rightarrow \frac{3}{4}$, all coefficients vanish but three of them, precisely matching (\ref{sigma epsilon sigma epsilon})\footnote{The occurrence of a vanishing Mellin amplitude for a non zero correlation function is not an isolated phenomenon. An infinite class of examples are correlators of type $\langle \Phi_{12} \Phi_{n1} \Phi_{12} \Phi_{n1}\rangle$.}.

\section{Discussion}

Let us briefly summarize our findings, review assumptions and discuss open problems and further directions.

\subsection{Summary of Results}

In this paper we discussed a four-point function of identical scalar primary operators in a CFT and clarified several aspects of  Mellin amplitudes. From our discussion the following general picture emerged:

\begin{itemize}
\item Connected correlators of light operators in a general CFT admit the  Mellin representation (\ref{eq:mellinsub}), (\ref{eq:Ucft}) with simple subtraction terms given by (\ref{eq:subtractionF}).
Two important properties went into establishing this claim. Analyticity and polynomial boundedness in a sectorial domain $\Theta_{CFT}$ (\ref{eq:thetaCFT}), (\ref{eq:polybound}). While analyticity can be easily established using unitarity and OPE, see appendix \ref{app: rhombus}, polynomial boundedness is more complicated. Indeed, as we explained in the main text, establishing the required polynomial boundedness amounts to proving new bounds on the double light-cone limit  $u,v \to 0$ of the correlator. We presented our best arguments for this claim in appendix \ref{sec:polboundapp}, but we do not have a proof. 

\item Analyticity in a sectorial domain $\Theta_{CFT}$ of the four-point function implies that the Mellin transform of the correlator decays exponentially fast in imaginary directions (\ref{eq:decaycorrect}). The standard $\Gamma$-functions pre-factor in (\ref{def Mellin general}) correctly captures analytic properties of the physical correlator (\ref{eq:decaycorrect}) making the   Mellin amplitude $M(\gamma_{12},\gamma_{14})$ polynomially bounded.

\item Mellin amplitude $M(\gamma_{12},\gamma_{14})$ has only simple pole singularities. The positions of the poles are given by the twist spectrum of the theory (\ref{eq:poles}). Due to the presence of the twist accumulation points in every CFT, the nonperturbative Mellin amplitude has an infinite number of pole accumulation points whose position is known, see section \ref{sec:twist}. 

This claim, which we can think of as {\it Maximal Mellin Analyticity} is a conjecture. Establishing it is closely related to the double light-cone limit issue mentioned above but goes beyond that.  In appendix \ref{app: proof}, we proved part of this analyticity and suggested a strategy to finish the proof.  We leave this exploration for the future.

\item The residues of the Mellin amplitude are fixed by the OPE and are given by known Mack polynomials times the square of the corresponding three-point functions, see section \ref{sec:OPEexpansion}. 

We describe some basic properties of Mack polynomials in section \ref{sec:MackPol}. Mack polynomials encode a great deal of information. Together with the $\Gamma$-functions pre-factor they reproduce the collinear block expansion upon doing an inverse Mellin transform. In the relevant limit (which is the flat space limit) they reduce to Gegenbauer polynomials (\ref{largetauregion}). They have positivity properties familiar in the context of flat space scattering amplitudes (\ref{eq:mackpos}). Finally, they encode purely AdS effects in the large spin limit (\ref{largespinresult}).

\item  Mellin amplitude $M(\gamma_{12}, \gamma_{14})$ is polynomially bounded in the Regge limit $|\gamma_{12}| \to \infty$, $\gamma_{14}$ - fixed, see section (\ref{sec:reggebound}).

The boundedness is the direct consequence of the boundedness of the CFT correlator in the coordinate space Regge limit \cite{Caron-Huot:2017vep}. An important subtlety is that strictly speaking the coordinate space Regge boundedness of the CFT correlator only bounds the Mellin amplitude in the Regge limit along the imaginary axis. We assume that there is no Stokes phenomenon and extrapolate this bound to any direction in the complex plane. This is an assumption and it would be great to make it rigorous. 

\item Non-perturbative Mellin amplitude $M(\gamma_{12}, \gamma_{14})$ satisfies Polyakov conditions, e.g. (\ref{eq:finalpolyakov}).

Note that the non-perturbative Mellin amplitude $M(\gamma_{12}, \gamma_{14})$ has an infinite number of accumulation points of poles at the positions of double twist operators \cite{Fitzpatrick:2012yx,Komargodski:2012ek}. This requires a certain care as we discuss in section \ref{sec: polyakov}. In particular approaching a twist accumulation point requires choosing a direction in the complex plane and the result depends on the value of the second Mellin variable. Nevertheless it is still possible to impose the condition that there are no isolated operators with twist $2 \Delta + 2 n$. In this paper we only explored the simplest condition of this type (\ref{eq:finalpolyakov}).

\item Crossing symmetry in Mellin space is given by (\ref{eq:crossingMellin}).

As usual by itself crossing symmetry is trivial to impose. However, then combined with all the other properties of the Mellin amplitude it becomes very nontrivial.

\item Mellin amplitude satisfy subtracted dispersion relations, see section {\ref{sec:application}}.

This simply follows from all the properties that we listed above and the standard Cauchy argument.\footnote{Note that this makes manifest the fact that CFT correlator can be re-constructed from its double discontinuity \cite{Turiaci:2018dht,Carmi:2019cub}. The double discontinuity of the correlator and the  simple discontinuity of the Mellin amplitude capture the same information.}
We then combined all the nonperturbative properties of the correlator listed above together with dispersion relations to arrive at a set of extremal functionals. An example of such a functional is given in (\ref{eq:magicfunctionals}). This functional has a few remarkable properties: a) it annihilates the identity operator; b) it is non-negative for $\tau > 2 \Delta$ and has double zeros at the positions of the double-twist operators, namely for $\tau = 2 \Delta + 2 n$ for $n \geq 1$. It has a single zero for $n=0$. These remarkable properties are well-suited to apply (\ref{eq:magicfunctionals}) to holographic CFTs. 

\item We tested our ideas on minimal models in section \ref{sec:MM}. 

\end{itemize}

\subsection{Future Directions}

There are many directions in which our work can be extended and possibly improved:

\begin{itemize}
\item Rigorously establish our assumptions.

This concerns our assumptions about the double light-cone limit (or polynomial boundedness in a sectorial domain), extrapolation of the Regge bound to the whole Mellin plane and the \emph{Maximal Mellin Analyticity} conjecture. This is necessary to put our functionals on a solid theoretical ground. In this paper we did our best to justify them but we could not prove them. It is however very re-assuring that (\ref{eq:magicfunctionals}) produces reasonable results when applied to the 3d Ising model, see section \ref{sec:3dIsing}, and we believe that these assumptions are true very generally. Proving these assumptions is required to put our conclusions on a firm theoretical basis.

\item Systematically study functionals obtained from subtracted dispersion relations.

In this work we only explored a very simple functional (\ref{eq:magicfunctionals}). Using the same techniques we can generate an infinite family of functionals. We leave systematic exploration of their properties and the corresponding bootstrap bounds for the future \cite{future}. One interesting question is to establish the precise relation between the present analysis and the recent works \cite{Mazac:2019shk,Carmi:2019cub}, see also \cite{Sleight:2019ive}. While some gross features look very similar the details seem to be different.

We can imagine considering other subtractions. For example, one natural choice is to start with ${M(\gamma_{12}, \gamma_{14}) \over \gamma_{12} \gamma_{13} \gamma_{14}}$. In this case, due to the Polyakov conditions, the amplitude itself does not contribute and we get directly the Mellin amplitude as a function of its discontinuity.

\item Extend our techniques to heavy external operators.

Throughout the paper we were very careful about tracing the conditions on the external dimensions and the spectrum of the CFT at hand for which our formulas are applicable. In particular, one serious limitation of our analysis is that it only applies to very light operators, namely operators that satisfy $\Delta < {3 \over 4} \tau_{gap}$. We believe it should be possible to extend the whole technology to higher $\Delta$'s provided suitable subtractions and/or contour deformations. \footnote{In section \ref{sec: bent}, we presented  an argument for the existence of a Mellin representation with a deformed contour, assuming analyticity.}
This is very important if nonperturbative Mellin amplitudes are to be thought as a   general CFT tool.

\item Extend our techniques to spinning and non-identical operators.

It is a well-known problem in the Mellin space literature: how to define Mellin amplitudes for spinning operators?\footnote{See e.g. \cite{Goncalves:2014rfa, Chen:2017xdz,Sleight:2018epi} for some work in this direction.} From the point of view of the present analysis Mellin space is simply a very efficient way to encode analyticity, the OPE expansion, and polynomial and Regge boundedness of the physical correlators. All of these carries through to spinning case as well. Therefore, we believe it should be possible to extend everything we say to the spinning case as well. It is however an open problem of what is a systematic and the most economic way of doing that. 

\item Higher-point correlation functions. 

The simplest picture of the higher-point correlation function that emerged from the present paper is that upon a simple subtraction of disconnected pieces it will admit Mellin representation (\ref{def Mellin general}). In particular, it would be interesting to establish the sectorial domain of analyticity for the higher-point CFT correlators and check that it is correctly captured by the $\Gamma$-functions pre-factor so that $M(\gamma_{ij})$ behaves polynomially in the Regge limit with only poles at the locations dictated by the OPE.

\item BCFT Mellin amplitudes.

One can also try to extend our work for BCFTs. Mellin amplitudes for BCFTs were considered in the past \cite{Rastelli:2017ecj,Goncalves:2018fwx} but the general analysis that we performed in this paper has not been done for BCFTs to the best of our knowledge.

\item Flat space limit.

Recently Mellin amplitudes were used to analyze some nonperturbative bounds on the flat space scattering amplitudes \cite{Dodelson:2019ddi,Haldar:2019prg}. It would be very interesting to understand if our analysis could shed any further light on this interesting question.

\item Cosmological Mellin amplitudes.

It would be very interesting to understand how analyticity and polynomial boundedness which we used to define the Mellin amplitude play out in the cosmological context, see e.g. \cite{Sleight:2019mgd,Sleight:2019hfp}. As we explained both constraints are essentially Lorentzian and it would be interesting to understand what are the analogous bounds for  cosmological correlators.

\item Connection to the Polyakov-Mellin bootstrap.

Needless to say the idea of doing bootstrap in Mellin space is not new \cite{Polyakov:1974gs,Gopakumar:2016wkt}. In this approach crossing symmetry is manifest but the OPE is not. The standard approach in this case, which was proven to be very useful in bootstrapping perturbative CFTs, relies on writing the CFT correlator as a sum of the so-called Polyakov blocks in three channels. While such a representation was recently rigorously established in 1d \cite{Mazac:2018qmi}, it is still an open problem to establish its validity in higher dimensions. Our approach is different in this regard. We do not assume the usual ansatz, and as a consequence crossing symmetry is not manifest. It would be interesting to better understand the relation between the present paper and the usual Mellin space analysis. 

\end{itemize}

Overall we find that the nonperturbative Mellin bootstrap is an exciting avenue which is worth further explorations. It has many appealing properties and allows to incorporate various physical constraints on the correlator rather easily. Analytic functionals that came out our analysis have truly remarkable properties and they are particularly fit for holographic theories. It would be interesting to explore this further. 



\section*{Acknowledgements}

We are grateful for discussions with Aninda Sinha, Dean Carmi, Simon Caron-Huot, Miguel Costa, Aditya Hebbar, Tobias Hansen, Raj Kunkolienkar, Dalimil Mazac, Miguel Paulos, Sylvain Ribault, Balt van Rees, Pedro Vieira, Xinan Zhou.

JP and JS are supported by the Simons Foundation grant 488649 (Simons Collaboration on the Nonperturbative Bootstrap) 
and by the Swiss National Science Foundation through the project
200021-169132 and through the National Centre of Competence in Research SwissMAP.

\appendix

\section{2D CFT calculations}

\subsection{Mellin amplitudes from BPZ differential equations} \label{app: Mellin diff}

Let us consider the four point function $\langle \Phi_{1, 2}(z_1, \bar{z}_1) O_2(z_2, \bar{z}_2) O_3(z_3, \bar{z}_3) O_4(z_4, \bar{z}_4) \rangle$, where $O_2$, $O_3$ and $O_4$ are arbitrary scalar Virasoro primaries and $\Phi_{1, 2}$ is a degenerate Virasoro primary. $\langle \Phi_{1, 2}(z_1, \bar{z}_1) O_2(z_2, \bar{z}_2) O_3(z_3, \bar{z}_3) O_4(z_4, \bar{z}_4) \rangle$ is annihilated when acted upon by the differential operator $\mathcal{L}_{sing}$:
\begin{eqnarray}\label{L sing}
\mathcal{L}_{sing} = \sum_{i=2}^4 \frac{1}{z_1-z_i}\partial_{z_i} + \frac{h_i}{(z_1-z_i)^2} -\frac{3}{2(2 h_1+1)}\partial^2_{z_1}, 
\end{eqnarray}
where we used the usual notation in 2d: $\Delta_i= h_i + \bar{h}_i, l_i=h_i - \bar{h}_i$ and in (\ref{L sing}) we assume $\Phi_{1, 2}$ to be at position $1$.

The application of $\mathcal{L}_{sing}$ to $\Phi_{1, 2}$ increases $h$ and leaves $\bar{h}$ fixed. Thus, it will be useful for us to apply both $\mathcal{L}_{sing}$ and $\mathcal{\bar{L}}_{sing}$ to $\langle \Phi_{1, 2}(z_1, \bar{z}_1) O_2(z_2, \bar{z}_2) O_3(z_3, \bar{z}_3) O_4(z_4, \bar{z}_4) \rangle$, so as to get a null scalar Virasoro primary, which we denote by $\xi_{1, 2}$. In equations,
\begin{eqnarray}\label{corr sing}
\mathcal{\bar{L}}_{sing} \mathcal{L}_{sing} \langle \Phi_{1, 2} O_2 O_3 O_4 \rangle = \langle \xi_{1, 2} O_2 O_3 O_4 \rangle =0.
\end{eqnarray}

In (\ref{corr sing}), we use the Mellin representation:
\begin{eqnarray}
 \langle \Phi_{1, 2} O_2 O_3 O_4 \rangle = \int \Big( \prod_{i<j}\lbrack d\gamma_{ij}\rbrack (X^2_{ij})^{-\gamma_{ij}} \Big) M_{\Phi_{(1, 2)}}( \{ \gamma_{ij} \}).
\end{eqnarray}
We see that the differential operators only act on $(X^2_{ij})^{-\gamma_{ij}}$, which can be factorized in holomorphic and antiholomorphic parts. Afterwards, we use $u$ and $v$ variables and do a change of variables in order to get a recursion relation for the Mellin amplitude:
\begin{eqnarray} \label{Mellin recursion}
\sum_{p=0}^2 \sum_{q=0}^2 c_{p, q}(\gamma_{12}, \gamma_{14}) M_{\Phi_{1, 2}}(\gamma_{12}+p, \gamma_{14} +q) =0. 
\end{eqnarray}
Expressions for the coefficients $c_{p, q}$ are not very important, but let us register one such expression for concreteness:
\begin{eqnarray}
c_{1, 1}= \frac{1}{9} \Big(  (2+4 h_1)h_2 + (-1+4 h_1 -3 \gamma_{12})\gamma_{12} \Big)\Big(  (2+4 h_1)h_4 + (-1+4 h_1 -3 \gamma_{14})\gamma_{14} \Big).
\end{eqnarray}

Equation (\ref{Mellin recursion}) is solved by
\begin{align}\label{Mellin gen Phi12}
M_{\Phi_{(1, 2)}}(\gamma_{12}, \gamma_{14})=C_0 \Gamma(-\frac{\Delta(\alpha_{12})+\Delta({\alpha_2})-\Delta(\alpha_2-\frac{1}{2} \alpha_+)}{2}+\gamma_{12})\nonumber \\
\Gamma\Big(-\frac{\Delta(\alpha_{12})+\Delta({\alpha_2})-
\Delta(\alpha_2+\frac{1}{2}\alpha_+) }{2} +\gamma_{12} \Big)  \Gamma(-\frac{\Delta(\alpha_{12})+\Delta({\alpha_3})-\Delta(\alpha_3-\frac{1}{2} \alpha_+)}{2}+\gamma_{13})\nonumber \\
\Gamma(-\frac{\Delta(\alpha_{12})+\Delta({\alpha_3})-\Delta(\alpha_3+\frac{1}{2} \alpha_+)}{2}+\gamma_{13})  \Gamma(-\frac{\Delta(\alpha_{12})+\Delta({\alpha_4})-\Delta(\alpha_4-\frac{1}{2} \alpha_+)}{2}+\gamma_{14}) \nonumber \\
\Gamma(-\frac{\Delta(\alpha_{12})+\Delta({\alpha_4})-\Delta(\alpha_4+\frac{1}{2} \alpha_+)}{2}+\gamma_{14}), 
\end{align}
where $\gamma_{13}=\Delta(\alpha_{12})- \gamma_{12}-\gamma_{14}$ and $\Delta({\alpha})$ is given by (\ref{conformal dimension vertex}). $C_0$ is not fixed by equation (\ref{Mellin recursion}). We compute it in \ref{app: analytic continuation} (see formula (\ref{C0phi12})). Equation (\ref{Mellin gen Phi12}) is a simple generalisation of (\ref{Mellin explicit phi12 1}). We just wrote a product of six $\Gamma$ functions, with poles prescribed by the OPE $\Phi_{1, 2} \times \Phi_{\alpha_i}=\Phi_{\alpha_i-\frac{1}{2} \alpha_+}+\Phi_{\alpha_i+\frac{1}{2} \alpha_+}$.

\subsection{Comparison with conformal block expansion} \label{app: blocks}

We start by establishing some notation. Consider $4$ scalar Virasoro primaries. We write their four point function in the Mellin representation as
\begin{eqnarray}
 \langle O_1 O_2 O_3 O_4 \rangle = |x_1-x_3|^{-2 \Delta_1}  |x_2-x_3|^{\Delta_1-\Delta_2-\Delta_3+\Delta_4}  |x_2-x_4|^{- \Delta_1- \Delta_2+ \Delta_3-\Delta_4} \nonumber \\ 
|x_3-x_4|^{\Delta_1 + \Delta_2 -\Delta_3 - \Delta_4}   \int_{C_1} \frac{d\gamma_{12}}{2 \pi i} \int_{C_2} \frac{d\gamma_{14}}{2 \pi i} M(\gamma_{12}, \gamma_{14} ) u^{-\gamma_{12}} v^{-\gamma_{14}},
\end{eqnarray}
where $M(\gamma_{12}, \gamma_{14})$ is the Mellin amplitude. Let us consider the usual kinematics:
\begin{eqnarray}\label{defG}
G_{34}^{21}(z, \bar{z})= \lim_{z_1 \rightarrow \infty , \bar{z_1}\rightarrow \infty} z_1^{2 h_1} \bar{z}_1^{2 h_1} \langle \O_1(z_1, \bar{z}_1), \O_2(1, 1), \O_3 (z, \bar{z}), \O_4(0, 0) \rangle,
\end{eqnarray}
where we use the 2d notation $h_i= \frac{\Delta_i}{2}$. So,
\begin{eqnarray}\label{G with M}
G_{34}^{21}(z, \bar{z})= v^{h_1-h_2-h_3+h_4} u^{h_1 +h_2 -h_3 -h_4} \int_{C_1} \frac{d\gamma_{12}}{2 \pi i} \int_{C_2} \frac{d\gamma_{14}}{2 \pi i} M(\gamma_{12}, \gamma_{14} ) u^{-\gamma_{12}} v^{-\gamma_{14}}.
\end{eqnarray}

We write the four point function as a sum over Virasoro blocks in the s-channel:
\begin{eqnarray}\label{Gdecomp}
G_{34}^{21}(z, \bar{z}) = \sum_{p} C_{34}^p C_{12}^{p} \mathcal{F}_{34}^{21}(p | z) \bar{\mathcal{F}}_{34}^{21}(p | \bar{z}),
\end{eqnarray}
where  $C_{12}^{p}$ denotes the OPE coefficient for a Virasoro primary exchanged in $O_1 \times O_2$.
$\mathcal{F}_{34}^{21}(p | z)$ is a kinematical function that can be expressed as a power series.
\begin{eqnarray}
\mathcal{F}_{34}^{21}(p | z) = z^{h_p - h_3 -h_4} \sum_{k=0}^{\infty} \mathcal{F}_k z^k,
\end{eqnarray}
where $h_p$ is half the conformal dimension of the exchanged primary. An analogous expansion exists for $\bar{\mathcal{F}}_{34}^{21}(p | \bar{z})$. Expression for the first three coeficients are
\begin{eqnarray}
\mathcal{F}_0=1, \\
\mathcal{F}_1 = \frac{(h_p + h_3 - h_4)(h_p +h_2 - h_1)}{2 h_p}, \\
\mathcal{F}_2 = \frac{A}{B}+\frac{C}{B}, \label{c term}
\end{eqnarray}
where 
\begin{eqnarray}
A=(h_p+h_2-h_1)(h_p+h_2-h_1+1)\\
\times \Big((h_p+h_3-h_4)(h_p+h_3-h_4+1)(4 h_p + \frac{c}{2})-6 h_p( h_p+2h_3 -h_4)\Big), \nonumber\\
B=4 h_p (2 h_p +1) (4 h_p + \frac{c}{2}) - 36 h^2_p,\\
C=(h_p+2h_2-h_1)\Big( 4 h_p (2h_p+1)(h_p+2h_3-h_4)\\
-6h_p(h_p+h_3-h_4)(h_p+h_3-h_4+1)\Big).\nonumber
\end{eqnarray}

Let us now specialise to the case $\langle \Phi_{1, 2} \Phi_{1, 2} O O \rangle$, where $O$ is an arbitrary Virasoro primary. Its Mellin amplitude is given by
\begin{eqnarray}\label{Mellin explicit phi12 2}
M(\gamma_{12}, \gamma_{14})= C_0 \Gamma(\gamma_{13}-\alpha \alpha_+)\Gamma( \gamma_{13}+\alpha  \alpha_+ - \alpha_+^2 +1)\Gamma(\gamma_{12}+\frac{\alpha_+^2}{2}) \\
\Gamma(\gamma_{12}+1-\frac{3\alpha_+^2}{2}) \Gamma(\gamma_{14}-\alpha \alpha_+)\Gamma( \gamma_{14}+\alpha  \alpha_+ - \alpha_+^2 +1),\nonumber
\end{eqnarray}
where $\gamma_{13}= \frac{3 \alpha_+^2}{2}-1-\gamma_{12}-\gamma_{14}$. We obtained this expression from considering (\ref{Mellin explicit phi12 1}) and doing $\gamma_{12} \leftrightarrow \gamma_{13}$.

In order to match the Mellin representation with the conformal block expansion, we need to expand the rhs of (\ref{G with M}) for small $z$ and $\bar{z}$. In particular, we want to match with the contribution of the identity
\begin{eqnarray}
(z \bar{z})^{-h_3 -h_4} \subset G_{34}^{21}(z, \bar{z}).
\end{eqnarray}

For the case of $\langle \Phi_{1, 2} \Phi_{1, 
2} O O \rangle$, the rhs of (\ref{G with M}) is 
\begin{eqnarray}\label{aux G}
u^{\frac{3}{2}\alpha_+^2-1-2 \alpha^2+4\alpha_0 \alpha }\int_{C_1} \frac{d\gamma_{12}}{2 \pi i} \int_{C_2} \frac{d\gamma_{14}}{2 \pi i}  C_0 \Gamma(\gamma_{13}-\alpha \alpha_+) \Gamma( \gamma_{13}+\alpha  \alpha_+ - \alpha_+^2 +1) \\
\Gamma(\gamma_{12}+\frac{\alpha_+^2}{2}) \Gamma(\gamma_{12}+1-\frac{3\alpha_+^2}{2})\Gamma(\gamma_{14}-\alpha \alpha_+)
\Gamma( \gamma_{14}+\alpha  \alpha_+ - \alpha_+^2 +1) u^{- \gamma_{12}}v^{-\gamma_{14}},\nonumber
\end{eqnarray}  
We want to reproduce the term $(z \bar{z})^{-h_3 -h_4}= u^{-2 \alpha^2+4\alpha_0 \alpha}$. In order to do so, let us consider the limit $u \rightarrow 0$ and $v \rightarrow 1$ in (\ref{aux G}). We take the residue of the integrand at $\gamma_{12}= \frac{3}{2}\alpha_+^2-1$ and set $v=1$ to compute the remaining integral:
\begin{eqnarray}
u^{-2 \alpha^2+4 \alpha_0 \alpha} \Gamma(-1+2\alpha_+^2)C_0 \int [\frac{d \gamma_{14}}{2 \pi i}] \Gamma(\gamma_{14}- \alpha \alpha_+)\Gamma(\gamma_{14}+ \alpha \alpha_+ - \alpha_+^2+1)\\
\Gamma(- \gamma_{14}-\alpha \alpha_+) \Gamma(- \gamma_{14}+\alpha \alpha_+ - \alpha_+^2 +1)  \nonumber \\
= u^{-2 \alpha^2+4 \alpha_0 \alpha}C_0 \Gamma(-1+2\alpha_+^2)\Gamma(-2 \alpha \alpha_+) \Gamma^2(1-\alpha_+^2)\frac{\Gamma(2-2\alpha_+^2+2 \alpha \alpha_+)}{\Gamma(2-2\alpha_+^2)}. \nonumber
\end{eqnarray}
Equating this to $u^{-2 \alpha^2+4 \alpha_0 \alpha}$ fixes the value of $C_0$ according to (\ref{C0 phi12}). 

\subsection{Normalisation of $\langle \Phi_{1, 2} O_2 O_3 O_4 \rangle$}\label{app: analytic continuation}

In \ref{app: Mellin diff} we computed the Mellin amplitude of $\langle \Phi_{1, 2} O_2 O_3 O_4 \rangle$, up to a constant $C_0$ (see \ref{Mellin gen Phi12}). In this appendix, we determine the value of $C_0$, which is in formula (\ref{C0phi12}). Conformal Virasoro primaries are normalised so as to have a two point function "equal" to $1$.

This problem was already analysed in \cite{Zamolodchikov:2005fy}. In that paper, following a technique explained in \cite{Teschner:1995yf}, an analytic continuation of three point functions for general central charge from the ones in minimal models is proposed. Here, we just transcribe that result into an expression for $C_0$.

We direct the reader interested in understanding the details of this analytic continuation to \cite{Zamolodchikov:2005fy}. In what follows, we just define some conventions and then write formula (\ref{C0phi12}) for $C_0$.

In this appendix, we change notation, so as to match the one in \cite{Zamolodchikov:2005fy}. Conformal dimensions are given by
\begin{eqnarray}
h_{\alpha} = \alpha( \alpha - q),
\end{eqnarray}
where
\begin{eqnarray}
q= \frac{1}{\beta} - \beta.
\end{eqnarray}
Note that $h_{\alpha}$ is invariant under $\alpha \rightarrow q - \alpha$. The central charge is related to $\beta$ by
\begin{eqnarray}
c= 1-6 q^2.
\end{eqnarray}
In the notation used in the rest of the paper, $\beta=-\alpha_+$.

Our formula for $C_0$ will depend on a $\Upsilon$ function. Its properties can be found in \cite{Zamolodchikov:2005fy}. We just briefly remind some basic facts. A representation for the $\Upsilon$ function is
\begin{eqnarray}\label{Upsilon rep}
\log \Upsilon(x) = \int_0 ^{\infty} \frac{dt}{t} \big( (\frac{Q}{2}-x)^2 e^{-t} - \frac{\sinh^2((\frac{Q}{2}-x)t)}{\sinh (\frac{\beta t}{2}) \sinh (\frac{ t}{2 \beta})}  \big).
\end{eqnarray}
This representation is valid for $0 < Re(x) < Q$. We note that $Q=\beta + \beta^{-1}$. For values of $x$ outside this representation, we need to use the shift relations
\begin{eqnarray}\label{relations1 Upsilon}
\Upsilon(x+\beta)=\gamma(\beta x) \beta^{1-2 \beta x} \Upsilon(x), \\
\Upsilon(x+\frac{1}{\beta})=\gamma(\frac{x}{\beta} ) \beta^{2 \frac{x}{\beta} -1} \Upsilon(x). \label{relations2 Upsilon}
\end{eqnarray}
Also note the identities
\begin{eqnarray}\label{Upsilon identities}
\Upsilon (x) = \Upsilon(Q-x),\\
\Upsilon(\frac{Q}{2}) = 1.
\end{eqnarray}

Now, the formula for $C_0$. 
\begin{eqnarray}\label{C0phi12}
C_0 = \prod_{k=2}^4 \frac{\tilde{\Upsilon}(\frac{\beta}{2}+\bar{\alpha}-2\alpha_k) }{\sqrt{\Upsilon(\beta+2\alpha_k) \Upsilon(-\frac{1}{\beta}+2\beta+2\alpha_k)  } } \tilde{\Upsilon}(-\frac{1}{\beta}+\frac{3}{2}\beta+\bar{\alpha}) \\
\frac{\gamma \left(\frac{1}{\beta ^2}-1\right) \gamma \left(\beta ^2\right)}{\sqrt{\gamma \left(\frac{1}{\beta ^2}-2\right)} \sqrt{\gamma \left(2 \beta ^2\right)} \Upsilon (\beta )}\nonumber,
\end{eqnarray}
where $\bar{\alpha}=\sum_{k=2}^4 \alpha_k$ and
\begin{eqnarray}
\gamma(x) \equiv \frac{\Gamma(x)}{\Gamma(1-x)} , \\
\tilde{\Upsilon}(x)=\frac{\Upsilon(x)}{\Gamma(1-\beta x)} \beta^{-\beta x + \frac{1}{4 \beta ^2}+\frac{3}{4}}.
\end{eqnarray}
(\ref{C0phi12}) is symmetric in the transformation $\alpha_i \rightarrow q- \alpha_i$, for each $i=2, 3, 4$, where $q=\frac{1}{\beta}-\beta$.

\subsection{Normalisations in the Coulomb gas formalism}\label{app: norm Coulomb}

Normalisations in the Coulomb gas formalism are computed in section $9$ of \cite{Dotsenko:2006}. The result is
\begin{eqnarray}\label{normalisation coulomb degenerate}
N^2_{m, n}= \frac{(\alpha_+ ^2-1)^2 \pi ^{m+n} \gamma (1-\frac{1}{\alpha_+ ^2})^m \alpha_+ ^{4 m-4 n-2} \gamma (1-\alpha_+ ^2)^n \gamma (\frac{m}{\alpha_+ ^2}-n)(-1 )^{4m -4n -2 +1}}{\left(\pi ^2 \alpha_+ ^2\right) \gamma \left(m-\alpha_+ ^2 n\right)}, 
\end{eqnarray}
where $\gamma(x)\equiv \frac{\Gamma(x)}{\Gamma(1-x)}$.

A simple check can be done on formula (\ref{normalisation coulomb degenerate}). Consider expression (\ref{C0phi12}) for $C_0$ in the correlator $\langle \Phi_{1, 2} \Phi_{\alpha_2} \Phi_{\alpha_3} \Phi_{\alpha_4} \rangle$. Consider now the case in which $\alpha_2+\alpha_3+\alpha_4= 2\alpha_0 - \alpha_{12} - \alpha_+$ and eliminate the variable $\alpha_{4}$. This corresponds to inserting one positive screening charge in the Coulomb gas model. In that case, we can simplify the expression for $C_0$ using the $\Upsilon$ identities (\ref{relations1 Upsilon}) and (\ref{Upsilon identities}), in such a way that the expression for $C_0$ depends only on $\Gamma$ functions: 
\begin{eqnarray}\label{C0 simp 1 screening}
C_0= \frac{1}{\Gamma(-2 \alpha_2 \alpha_+) \sqrt{\gamma(\frac{1}{\alpha_+^2}+\frac{2 \alpha_2}{\alpha_+})} \sqrt{\gamma(-1-2\alpha_2 \alpha_+ + \alpha_+^2)}} \\
\frac{1}{\Gamma(- 2 \alpha_3 \alpha_+) \sqrt{\gamma(\frac{1}{\alpha_+^2}+\frac{2 \alpha_3}{\alpha_+})} \sqrt{\gamma(-1-2\alpha_3 \alpha_+ + \alpha_+^2)}} \nonumber \\
\frac{1}{\Gamma(2-\alpha_+^2 +2 \alpha_2 \alpha_+ +2 \alpha_3 \alpha_+) \sqrt{\gamma(1-\frac{1}{\alpha_+^2}-\frac{2 \alpha_2}{\alpha_+}-\frac{2 \alpha_3}{\alpha_+})} \sqrt{\gamma(1+2\alpha_+ \alpha_2 +2 \alpha_+ \alpha_3)} } \nonumber \\
\frac{ \gamma \left(\frac{1}{\alpha_+ ^2}-1\right) \gamma \left(\alpha_+^2\right)}{\sqrt{\gamma \left(\frac{1}{\alpha_+ ^2}-2\right)} \sqrt{\gamma \left(2 \alpha_+ ^2\right)} \Gamma \left(1-\alpha_+^2\right)}. \nonumber
\end{eqnarray}
For that same case, it follows from (\ref{C0 general norm}) that 
\begin{eqnarray}\label{C0 normalization 1 screening}
C_0= \frac{\pi}{\prod_{i=1}^n \Gamma(-2 \alpha_i \alpha_+) N({\alpha_i}) }.
\end{eqnarray}

Let us now compare expressions (\ref{C0 normalization 1 screening}) and (\ref{C0 simp 1 screening}). The value of the normalisation of $N({\alpha_{1, 2}})$ is given in (\ref{normalisation coulomb degenerate}). We also use the shift symmetry $N(2\alpha_0 -\frac{ \alpha_{+}}{2} - \alpha_2 - \alpha_3)=\frac{1}{N(\alpha_2 + \alpha_3 + \frac{1}{2} \alpha_+)}$. We conclude that
\begin{eqnarray}\label{eq normalization coulomb gas}
\frac{ N(\alpha_2) N(\alpha_3) }{N( \alpha_2 + \alpha_3 + \frac{1}{2} \alpha_+)} = 
 \sqrt{\gamma(\frac{1}{\alpha_+^2}+\frac{2 \alpha_2}{\alpha_+})} \sqrt{\gamma(\frac{1}{\alpha_+^2}+\frac{2 \alpha_3}{\alpha_+})} \sqrt{\gamma(1-\frac{1}{\alpha_+^2}-\frac{2 \alpha_2}{\alpha_+}-\frac{2 \alpha_3}{\alpha_+})}  \\
\sqrt{\gamma(-1-2\alpha_2 \alpha_+ + \alpha_+^2)} \sqrt{\gamma(-1-2\alpha_3 \alpha_+ + \alpha_+^2)} \sqrt{\gamma(1+ 2\alpha_+ \alpha_2 +2 \alpha_+ \alpha_3)}   \nonumber \\
\frac{ \sqrt{\gamma \left(\frac{1}{\alpha_+^2}-2\right)} \sqrt{\gamma \left(2 \alpha_+ ^2\right)} } 
{\gamma(\frac{1}{\alpha_+^2}-1) \gamma(\alpha_+^2) \sqrt{\gamma(-1+2\alpha_+^2)} }  \sqrt{\pi}. \nonumber
\end{eqnarray}
We can now put formula (\ref{normalisation coulomb degenerate}) into the left hand side of (\ref{eq normalization coulomb gas}) and see if we obtain the right hand side. We checked this for several cases and indeed it is so.

Formula (\ref{normalisation coulomb degenerate}) is correct when $m$ and $n$ are both positive. We also need to consider normalisations for cases in which $n$ and $m$ are both negative. In that case, formula (\ref{normalisation coulomb degenerate}) does not apply. However, when $n$ and $m$ are both negative we can still find the correct normalisations. Since we normalise two point functions such that $\langle  O O \rangle = 1$, then it follows that $N({\alpha_{-m, -n}})=\frac{1}{N({\alpha_{m, n}})}$. 

\subsection{Two screening charges}
\label{app: 2 screening}

In this section, we consider a general $n$ point correlator $\langle \Phi_{\alpha_1} ... \Phi_{\alpha_n} \rangle$ for which we need to insert two positive screening charges in the Coulomb gas formalism and compute its Mellin amplitude.

As usual in the Coulomb gas formalism, we associate to each operator $\Phi_{\alpha_i}$ a vertex operator: $V_{\alpha_i}= N_{\alpha_i} \Phi_{\alpha_i}$, where $N_{\alpha_i}$ is a normalisation that for degenerate operators is equal to (\ref{normalisation coulomb degenerate}). Thus,
\begin{eqnarray} \label{2 screening 1}
\langle \Phi_{\alpha_1} ... \Phi_{\alpha_n} \rangle = \prod_{i<j}^n |x_i - x_j|^{4 \alpha_i \alpha_j} \int d^2 x_{n+1} \int d^2 x_{n+2} \\
\prod_{i=1}^n \frac{|x_i-x_{n+1}|^{4 \alpha_i \alpha_+}|x_i-x_{n+2}|^{4 \alpha_i \alpha_+} }{N_{\alpha_i}} |x_{n+1}-x_{n+2}|^{4 \alpha_+^2} \nonumber
\end{eqnarray} 

We use formula (\ref{Symanzik}) to transform the integral in $x_{n+2}$ into a Mellin integral. Notice that we can apply that formula since the integral in $x_{n+2}$ is conformal. Indeed, $4 \sum_{i=1}^n \alpha_i \alpha_+ + 4 \alpha_{+}^2 = 4 (- \frac{1}{\alpha_+} - \alpha_+) \alpha_+ + 4 \alpha_+^2=-4$. So, (\ref{2 screening 1}) is equal to 
\begin{eqnarray}
\frac{\pi^\frac{d}{2}}{\Gamma(-2 \alpha_+^2)} \prod_{i=1}^n \frac{1}{\Gamma(-2 \alpha_i \alpha_+) N_{\alpha_i}} \int d^2 x_{n+1} \prod_{j=1}^n |x_j - x_{n+1}|^{4 \alpha_j \alpha_+} \\
\prod_{i_1<j_1}^{n+1} \[ d\xi_{i_1, j_1}^1 \] \Gamma(\xi_{i_1, j_1}^1) |x_{i_1}-x_{j_1}|^{-2 \xi_{i_1, j_1}^1}, \nonumber
\end{eqnarray}
where we used the measure $\sum_{j_1 \neq i_1} \xi_{i_1, j_1}^1 = -2 \alpha_{i_1}\alpha_+$. 

The integral in $x_{n+1}$ can also be done by use of \eqref{Symanzik}, since $4 \sum_{i=1}^n \alpha_i \alpha_+ -2 \sum_{i=1}^n \xi_{i, n+1}^1 = -4 - 4 \alpha_+^2 +4 \alpha_+^2=-4$. Further doing a shift of integration variables, we obtain
\begin{eqnarray}\label{Mellin 2 screening}
\langle \Phi_{\alpha_1} ... \Phi_{\alpha_n} \rangle = \frac{\pi}{\Gamma(-2 \alpha_+^2)} \frac{1}{\prod_{k=1}^n \Gamma(-2 \alpha_k \alpha_+) N(\alpha_k)} \prod_{i<j}^n  \int \[ d\gamma_{ij} \] \prod_{i_1<j_1}^{n+1} \int \[ d \xi_{i_1, j_1}^1 \] \\
 \Gamma(\xi_{i_1, j_1}^1) \frac{1}{\prod_{k_1=1}^n \Gamma(-2 \alpha_{k_1} \alpha_+ + \xi_{k_1, n+1}^1) }  \Gamma(\gamma_{ij}+2\alpha_i \alpha_j- \xi_{ij}^1) |x_i- x_j|^{-2 \gamma_{ij}}, \nonumber \\
\end{eqnarray}
where the measure is $\sum_{j \neq i} \gamma_{ij}= \Delta_{\alpha_i}$ and $\sum_{j_1 \neq i_1}^{n+1} \xi_{i_1 j_1}^1=-2\alpha_{i_1}\alpha_+$. Formula (\ref{Mellin 2 screening}) is a particular case of (\ref{Mellin minimal models}). 

\subsection{Sums in exponent of the Mellin amplitude}\label{app: sums exponent}

In this subsection we work out the sums in formula (\ref{power s 2d}), using the measure (\ref{measure general formula}). 

The first term can be rewritten as
\begin{eqnarray}\label{sums exponent 1}
-\sum_{r=1}^{z-1} \sum_{i=1}^n \sum_{j=i+1}^n \xi_{ij}^r =- \frac{1}{2} \sum_{r=1}^{z-1} \sum_{i=1}^n \sum_{j \neq i}^n \xi_{ij}^r.
\end{eqnarray}
Notice that $\xi_{i j}$ with $i>j$ does not exist as an integration variable, so when we write the expression above we mean that $\xi_{i j}= \xi_{j i}$ when $i>j$. It is useful to write things as (\ref{sums exponent 1}) in order to use the measure (\ref{measure general formula}). Using the measure (\ref{measure general formula}) we get that
\begin{eqnarray}\label{sums exponent 2}
\sum_{j \neq i}^n \xi_{ij}^r = - \sum_{j=n+1}^{n+r} \xi_{ij}^r - 2 \alpha_i \alpha_{n+r+1} + \sum_{s=r+1}^{z-1} \xi_{i, n+r+1}^s.
\end{eqnarray} 
Let us plug (\ref{sums exponent 2}) into (\ref{sums exponent 1}). We have that 
\begin{eqnarray}
 \sum_{r=1}^{z-1} \sum_{i=1}^n \alpha_i \alpha_{n+r+1} =  (2 \alpha_0 - p_1 \alpha_+ - q_1 \alpha_-)\big( (p_1-1)\alpha_+ + q_1 \alpha_- \big)
\end{eqnarray} 
and furthermore
\begin{eqnarray}
-\frac{1}{2} \sum_{r=1}^{z-1} \sum_{i=1}^n \Big( - \sum_{j=n+1}^{n+r} \xi_{ij}^r +   \sum_{s=r+1}^{z-1} \xi_{i, n+r+1}^s \Big) =  \frac{1}{2}\sum_{r=1}^{z-1}\sum_{i=1}^n \xi_{i, n+1}^r\\
=  \frac{1}{2}\sum_{r=1}^{z-1} \Big( -  \cancel{\sum_{j=n+2}^{n+r} \xi_{n+1, j}^r }   -2 \alpha_{n+1} \alpha_{n+r+1} + \cancel{\sum_{s=r+1}^{z-1} \xi_{n+1, n+r+1}^s} \Big) \nonumber \\
= - \alpha_{+} \big( (p_1-1)\alpha_+ + q_1 \alpha_- \big),
 \end{eqnarray} 
where we used $\alpha_{n+1}=\alpha_+$. We conclude that (\ref{sums exponent 1}) is equal to
\begin{eqnarray}
- \big( (p_1-1)\alpha_+ + q_1 \alpha_- \big) \big( p_1 \alpha_+ + (q_1-1) \alpha_- \big).\label{first term}
\end{eqnarray}

Now let us work out the second sum in (\ref{power s 2d}). 
\begin{eqnarray}
\sum_{i=1}^n \sum_{j \neq i}^n \alpha_i \alpha_j  = \sum_{i=1}^n (2 \alpha_0 \alpha_i- \alpha_i^2 ) + \sum_{i=1}^n \alpha_i ( - p_1 \alpha_+ - q_1 \alpha_-) \nonumber \\
=  -\frac{1}{2}\sum_{i=1}^n \Delta(\alpha_i) - (p_1 \alpha_+ + q_1 \alpha_-)\big( (1-p_1)\alpha_+ + (1-q_1)\alpha_-  \big)\label{second term}.
\end{eqnarray}

(\ref{first term}) $+$ (\ref{second term}) is equal to 
\begin{eqnarray}
1 - \frac{1}{2} \sum_i \Delta(\alpha_i),
\end{eqnarray}
like we wanted to show.

\section{Single-variable Mellin transform}\label{app: sing var}

The standard definition of the Mellin transform $\varphi(s)$ of a function $f(z)$ is given by
\beq
\varphi(s)=\int_0^\infty dz\,f(z) \,z^{s-1} \,,\qquad\qquad
f(z)=\int_{c-i\infty}^{c+i\infty} \frac{ds}{2\pi i} \varphi(s) \,z^{-s}\,.
\label{eq:1varMellindef}
\eeq
Notice that this definition of $\varphi(s)$ only makes sense if the first integral converges for at least some values of $s$. 
Assuming that $f(z)$ does not have any (non-integrable) singularity for $z>0$, the convergence region  is determined by the asymptotic behavior \footnote{We focus on power-like asymptotic behavior because that is the most relevant in the case of Mellin amplitudes. However, the discussion easily generalizes for more general asymptotics. For example, exponential decay as $z\to \infty$ leads to convergence for $ \Re\, s  >a_1$ and logarithmic  behavior  like $z^{-a}(\log z)^n$ does not change the convergence region but leads to higher order poles of $\varphi(s)$.}
\begin{align}
 z\to 0 \,:\qquad  &f(z)=A_1 z^{-a_1} +A_2 z^{-a_2} + \dots\,,\qquad  a_1>a_2>\dots 
 \label{asym0}\\
z\to \infty\,:\qquad   &f(z)=B_1 z^{-b_1} +B_2 z^{-b_2} + \dots\,,\qquad  b_1<b_2<\dots
 \label{asyminfty}
\end{align}
Clearly, the first integral converges in the strip $a_1< \Re\, s <b_1$. In this case, for the contour of the second integral we can pick any $c$ such that $a_1< c <b_1$.

\begin{figure}[h]
  \centering
  \includegraphics[scale=1.0]{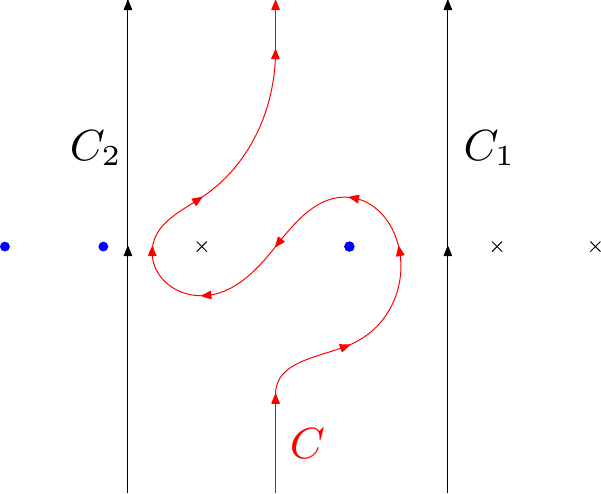} 
  \caption{Singularities in the complex plane of $s$. Blue balls represent the poles of $\psi(s)$, black crosses represent the poles of $\tilde{\psi}(s)$. Notice that we can gather the two straight contours $C_1$ and $C_2$ into a bent contour $C$, that separates poles to the left from poles to the right.}
  \label{fig:bentcontour1var}
\end{figure}

What if $b_1<a_1$ and therefore the first integral in \eqref{eq:1varMellindef} never converges? We shall now argue that even in this case the Mellin transform can still be defined by allowing a bent contour in the second integral in \eqref{eq:1varMellindef}.
The idea is very simple. We just split the first integral in two parts
\beq
\psi(s)=\int_0^1 dz\,f(z) \,z^{s-1} \,,\qquad 
\tilde{\psi}(s)=\int_1^\infty dz\,f(z) \,z^{s-1}\,.
\label{psiintegrals}
\eeq
The asymptotics \eqref{asym0} imply that $\psi(s)$ is defined and analytic for $  \Re\, s  >a_1$ and the asymptotics \eqref{asyminfty} imply that $\tilde{\psi}(s)$ is defined and analytic for $  \Re\, s  <b_1$. Therefore, we can write
\beq
f(z)=\int_{c_1 -i\infty}^{c_1 +i\infty} \frac{ds}{2\pi i} \psi(s) \,z^{-s}
+\int_{c_2 -i\infty}^{c_2+i\infty} \frac{ds}{2\pi i} \tilde{\psi}(s) \,z^{-s}
\eeq
with $c_1>a_1$ and $c_2<b_1$.
The next step is to deform the contours of these two integrals to the same bent contour $C$ without crossing any singularity of the respective integrands. This is depicted in figure \ref{fig:bentcontour1var}. 
If this is possible then we can write
\beq
f(z)=\int_{C} \frac{ds}{2\pi i} \varphi(s) \,z^{-s}\,,\qquad\qquad
\varphi(s) = \psi(s)+ \tilde{\psi}(s) \,.
\eeq

Bending the contours requires analytic continuation of $\psi$ and $\tilde{\psi}$ beyond the region of convergence of the integrals \eqref{psiintegrals}.
This is easily done by adding and subtracting the asymptotic behaviour of $f(z)$. For example, 
\beq
\psi(s)=\int_0^1 dz \,z^{s-1} \left[f(z) - A_1 z^{-a_1} + A_1 z^{-a_1} \right]=
\int_0^1 dz \,z^{s-1} \left[f(z) - A_1 z^{-a_1}   \right] + \frac{A_1}{s-a_1}
\eeq
where the last integral converges in the larger region $\Re\, s > a_2$.
By adding and subtracting more terms in the asymptotic expansion of $f(z)$ we can further analytically continue $\psi$ and $\tilde{\psi}$. Furthermore, we conclude that the asymptotic behaviour \eqref{asym0} and \eqref{asyminfty} gives rise to simple poles of the Mellin transform $\varphi(s)$ at $s=a_i$ and $s=b_i$ as shown in figure \ref{fig:bentcontour1var}. Thus, it is possible to bend the contours without crossing any singularity if and only if none of the poles of $\psi$ (at $s=a_i$) coincides with a pole of $\tilde{\psi}$ (at $s=b_i$). If there is a pole coincidence ($a_i=b_j$ for some $i$ and $j$)   
one can introduce a small parameter $\epsilon$ and define
\beq
f(z)=\lim_{\epsilon \to 0} \int_{C} \frac{ds}{2\pi i} \varphi_\epsilon(s) \,z^{-s}\,,\qquad\qquad
\varphi_\epsilon(s) = \psi(s+\epsilon)+ \tilde{\psi}(s-\epsilon)\,.
\eeq
Notice that the limit does not commute with the integral when the contour is pinched by two  poles that collide in the limit $\epsilon \to 0$. In fact, in this case, we obtain
\beq \label{f pinches}
f(z)=  \int_{C'} \frac{ds}{2\pi i} \varphi (s) \,z^{-s}+
\sum_i  A_i z^{-a_i}
\eeq
where the sum runs over the set of colliding poles.
The contour $C'$ passes to the left of all poles of $\tilde{\psi}$ and to the right of all poles of $\psi$, except those that are common.

Consider the simple example $f(z)=z^{-r}$. Then
\beq
\psi(s)=\frac{1}{s-r}\,,\qquad
\tilde{\psi}(s)=\frac{1}{r-s}\,,\qquad
\varphi(s)=0\,,\qquad
\varphi_\epsilon(s)=\frac{2\epsilon}{\epsilon^2-(s-r)^2}\,,
\eeq
and indeed
\begin{align}
f(z)&=\lim_{\epsilon \to 0} \int_{r-i\infty}^{r+i\infty}\frac{ds}{2\pi i} \frac{2\epsilon}{\epsilon^2-(s-r)^2} \,z^{-s}
=z^{-r}\lim_{\epsilon \to 0} \int_{-\infty}^{\infty}\frac{dy}{2\pi } 
\frac{2\epsilon}{\epsilon^2+y^2} \,z^{-i y}\\
&=z^{-r} \int_{-\infty}^{\infty} dy \delta(y)
  \,z^{-i y}
  =z^{-r}\,. \nonumber
\end{align}

\section{Theorems in complex analysis}

\subsection{Bochner's theorem}\label{app: bochner}

For completeness we reproduce the statement and proof of Bochner's theorem from the book \cite{Garrett}. Suppose $\Omega_0$ is a non-empty, connected, open set in $\mathbb{R}^n$, with $n>1$, not necessarily convex. Let $f$ be a holomorphic function of $n$ complex variables, defined on $\Omega_0 + i \mathbb{R}^n$. Suppose that $f$ does not grow too much at infinity, i. e. that for $x \in \Omega_0$ there is an $N \in \mathbb{N}$ such that
\begin{eqnarray}\label{condition growth}
|f(x+iy)| \ll e^{|y|^{2N}},
\end{eqnarray}
where $|y|^2=y_1^2 + ... + y_n^2$. Then, $f$ can be holomorphically entended to $\Omega + i \mathbb{R}^n$, where $\Omega$ is the convex hull of $\Omega_0$. 

The proof consists of an application of Cauchy's residue theorem in $1$ dimension. Let $x$ and $\xi$ be two points in $\Omega_0$. Consider the complex plane parametrized by $j(s) \equiv x + s (\xi- x)$ with $s \in \mathbb{C}$. 
We can define a rectangle $R$ with sides $s=it$ and $s=1-it$ for $- t_0 \leq t \leq t_0$ and $t_0$ is a positive number that we will eventually take to $\infty$. The top and bottom of the rectangle are given by
$s=\pm i t_0 +u$ with $0<u<1$.
Notice that the rectangle $R$ passes through $x=j(0)$ and $\xi=j(1)$. 
Now, we consider a point
$j(\zeta)= x + \zeta (\xi- x)$ inside this rectangle. In other words,  $\zeta$ is a complex number with $0<{\rm Re}\, \zeta <1$. Then we can write
\begin{align}\label{eq:Bochner}
W\left( j(\zeta) \right) f\left( j(\zeta) \right) =&  \lim_{t_0 \to \infty} \int_R \frac{ds}{2\pi i}\frac{W\left( j(s) \right) f\left( j(s) \right)}{
\zeta-s}\\
=&\frac{1}{2 \pi  } \int_{- \infty}^{+ \infty}dt \frac{
W\left( j(it )\right) f\left( j(it )  \right)}{ \zeta-it} 
+  \frac{1}{2 \pi } \int_{- \infty}^{+\infty} dt \frac{ W\left( j(1-it )\right) f\left( j(1-it )  \right)}{\zeta-1+it}, \nonumber
\end{align}
where $W$ is an auxiliary analytic function that ensures that the top and the bottom of the rectangle do not contribute in the limit of large $t_0$. For example, we can take
\beq
W(x) = e^{x^{2Q}}
\eeq
for $Q$ odd and $Q>N$.  In fact, the condition (\ref{condition growth}) could be weakened by choosing another $W$.
The first line of \eqref{eq:Bochner} is valid if the rectangle is contained in $\Omega_0 + i \mathbb{R}^n$. Remarkably, the second line of \eqref{eq:Bochner} is valid more generally. It is sufficient that both $x,\xi \in \Omega_0$ because the limit $t_0 \to \infty$ removed the top and the bottom of the rectangle. In this way we extended the domain of analyticity to the convex hull of $\Omega_0$.

\subsection{Conjectured Theorem}
\label{app: theorem?}
We would like to establish the  theorem below. Unfortunately, our rudimentary mathematical tools are not sufficient to formulate a rigorous proof.

{\bf Theorem:}
Consider a function $f(x,y)$ analytic for $y \in D \subset \mathbb{C} $ and $x \in \mathbb{C}/S$, where $D$ is a connected region independent of $x$ and $S \subset \mathbb{R}$ is a countable set of real numbers independent of $y$. $S$ may contain accumulation points but it is not dense in any interval of $\mathbb{R}$ and $f$ is single valued as a function of $x$ (no branch points). Then, the maximal region of analyticity of $f$ is  $x \in \mathbb{C}/S$ and $y \in D_{\rm max}$, with $D_{\rm max}$  independent of $x$. In particular, analytic continuation in $y$ cannot generate new singularities in $x$.

{\bf Proof:} Our proof is essentially by lack of imagination. Nevertheless, let us try to fake a rigorous argument. Suppose that the theorem fails. This means that when we analytically continue $y$ outside $D$ we find a singularity %
at $x=x_s(y)\notin S$.
By assumption this singularity must hide or disappear when $y$ approaches $D$ from the outside. Firstly consider the hiding scenario. In this case, we must have $x_s(y)=x_{0}$ for all $y \in \partial D$, with $x_{0}\in S$. Notice that as $y$ moves along the boundary of $D$, $x_s(y)$ must be continuous and therefore cannot jump between different points in $S$. But then by analytic continuation we conclude that $x_s(y)=x_{0}\in S$ for all $y \in \mathbb{C}$ and we find a contradiction.
Secondly, consider the disappearing scenario. Focus on the leading singular behaviour of $f$ at $x=x_s(y)$. For concreteness, we can imagine that $f(x,y)\approx R(y) (x-x_s(y))^{\alpha(y)}$ but the argument applies to other types of singularities.
In order for the singularity to disappear when $y$ approaches $D$ from the outside,  we must  have $R(y)=0$ or $\alpha(y)\in \mathbb{N}_0$ for $y \in \partial D$.
By continuity, we must have either $R(y)=0$ for all $y \in \partial D$ or $\alpha(y)=n\in \mathbb{N}_0$ for all $y \in \partial D$. Analytic continuation then implies that either $R(y)=0$ for all $y \in \mathbb{C}$ or $\alpha(y)=n\in \mathbb{N}_0$ for all $y \in \mathbb{C}$ and we find a contradiction.

{\bf Counter-example:} Let $D$ to be the region $|y|>2$, $S$ to be the segment $[0,1]$ along the real axis of the $x$ complex plane and
\beq
f(x,y) = \int_0^1dz \frac{1}{(x-z)(y-z)} = \frac{1}{x-y}\log\frac{y(1-x)}{x(1-y)}\,.
\eeq
This function is analytic for $y\in D$ and $x\in \mathbb{C}/S$. However, if we analytically continue in $y$ across the branch cut from $y=0$ to $y=1$, we will see a pole at $x=y$ emerge from the cut at $[0,1]$ in the $x$ complex plane.
Of course this is not a true counter-example because we violated the theorem condition that $S$ is not dense in any interval of $\mathbb{R}$.

\section{Analytic continuation of $K(\gamma_{12}, \gamma_{14})$} \label{app: proof}
We would like to show that 
\begin{eqnarray}\label{K for argument}
K(\gamma_{12}, \gamma_{14}) \equiv \int_0^{1} \frac{du}{u}\int_0^{1} \frac{dv}{v} u^{\gamma_{12}} v^{\gamma_{14}} f(u, v)\,,
\end{eqnarray}
can be defined for all $\gamma_{12}$, $\gamma_{14}$ in the complex plane, except at the OPE singularities: $\gamma_{12}, \gamma_{14}=\Delta-\frac{\tau}{2}-m$, where $\tau$ is the twist of an exchanged primary and $m$ is a nonnegative integer.
The integral above is well-defined for $\Re(\gamma_{12}) > \Delta$ and $\Re(\gamma_{14}) > \Delta$. Our task is to analytically continue (\ref{K for argument}) beyond this region.
Firstly, we will show that this can be done for regions [b] and [c] of figure \ref{fig:regions gamma plane}. Secondly, we will discuss the case of region [d] where we do not have a rigorous proof.
Finally, we discuss the asymptotic behaviour of $K$-functions.

\subsection{Regions [b] and [c]}

It is convenient to use the following expansion of the four point function
\begin{eqnarray}
f(u, v)= \sum_{\tau,l} a_{\tau,l}  u^{-\Delta +\frac{\tau}{2}} (z^l + \bar{z}^l) \label{expansion f z zb}
\end{eqnarray}
that holds almost everywhere in the integration region $(u,v) \in [0, 1] \times [0,1] $.  We sum over exchanged operators (both primaries and descendants) of twist $\tau$ and spin $l$ (see \cite{Hartman:2015lfa}) and the variables $z,\bar{z}$ are defined by the usual relations $u= z\bar{z}$ and $v=(1-z)(1-\bar{z})$.
The only points in that square where the expansion does not work are at $v=0$ and $u=1$. The coefficients $a_{\tau,l}$ are positive. When it converges, the series (\ref{expansion f z zb}) converges absolutely in each point.

Equation (\ref{expansion f z zb}) will be an essential ingredient in our argument for analyticity of $K$ functions and consequently of Mellin amplitudes. So, we can say that analyticity of Mellin amplitudes follows from the fact that CFT correlation functions enjoy an operator product expansion, whose coefficients have a definite sign for unitary theories. We expect this to be also true for non-identical operators though.

Let us suppose that $\Re(\gamma_{14}) > \Delta$ and let us attempt to analytically continue in $\gamma_{12}$. The first step consists in dividing the integration region into two regions.
\begin{eqnarray}\label{division to u=1/2}
K(\gamma_{12}, \gamma_{14}) \equiv \int_0^{\frac{1}{2}} \frac{du}{u}\int_0^{1} \frac{dv}{v} u^{\gamma_{12}} v^{\gamma_{14}} f(u, v) \\
+ \int_{\frac{1}{2}}^{1} \frac{du}{u}\int_0^{1} \frac{dv}{v} u^{\gamma_{12}} v^{\gamma_{14}} f(u, v) \nonumber.
\end{eqnarray}
The second integral is completely analytic in $\gamma_{12}$. From now on, we will refer ourselves only to the first integral. The usefulness of using these regions will be clear in a moment.

In order to analytically continue beyond the region $\Re(\gamma_{12})  > \Delta$ we add and subtract the twist contributions up to to some twist $\tau_{max}$:
\begin{align}
\int_0^{\frac{1}{2}} \frac{du}{u}\int_0^{1} \frac{dv}{v} u^{\gamma_{12}} v^{\gamma_{14}} f(u, v) &=
\int_0^{\frac{1}{2}} \frac{du}{u}\int_0^{1} \frac{dv}{v} u^{\gamma_{12}} v^{\gamma_{14}} f_{sub}(u, v) \label{fsub K}   \\ 
&+ \int_0^{\frac{1}{2}} \frac{du}{u}\int_0^{1} \frac{dv}{v} u^{\gamma_{12}} v^{\gamma_{14}} \sum_l\sum_{\tau<\tau_{max}} a_{\tau,l}  \,u^{-\Delta +\frac{\tau}{2}} (z^l + \bar{z}^l) \label{fsub K 2}   \,,
\end{align}
where
\beq 
f_{sub}(u, v) =f(u,v) -\sum_l \sum_{\tau<\tau_{max}} a_{\tau,l} \, u^{-\Delta +\frac{\tau}{2}} (z^l + \bar{z}^l)\,. 
\eeq

Let us show that the first term in the rhs of (\ref{fsub K}) is analytic in $\Re(\gamma_{12})> \Delta - \frac{\tau_{max}}{2}$ and $\Re(\gamma_{14})> \Delta$. In order to do this, we need to bound $f_{sub}(u, v)$ in the lightcone limits $u \rightarrow 0$, $v\rightarrow 0$ and the double lightcone limit $u, v \rightarrow 0$. 

Start by noticing that $f_{sub}(u, v) \sim u^{- \Delta + \frac{\tau_{max}}{2}}$ in the lightcone limit $u \rightarrow 0$, due to the subtractions that we made. In the lightcone limit $v \rightarrow 0$, the function $f_{sub}(u, v)$ cannot be more singular than $f(u, v)$. This is because in that limit $f(u, v)$ is a sum of positive terms (see (\ref{expansion f z zb})) and to get $f_{sub}(u, v)$ we just subtracted some of these terms. Thus, in the limit $v \rightarrow 0$, $f_{sub}(u, v)$ cannot be more singular than $v^{-\Delta}$.

Finally, we need to bound $f_{sub}(u, v)$ in the double lightcone limit. $f_{sub}(u, v)$ has the following series expansion
\begin{eqnarray}
f_{sub}(u, v)= \sum_l \sum_{\tau > \tau_{max}} a_{\tau,l}  u^{-\Delta +\frac{\tau}{2}} (z^l + \bar{z}^l). \label{expansion fsub z zb}
\end{eqnarray}
We proceed like in section (\ref{sec: convergence}). Let us switch to $z, \bar{z}$ coordinates. We find that
\begin{eqnarray}
(z_1 \bar{z}_1)^{\Delta - \frac{\tau_{max}}{2}} f_{sub}(z_1 , \bar{z}_1) < (z_2 \bar{z}_1)^{\Delta - \frac{\tau_{max}}{2}} f_{sub}(z_2 , \bar{z}_1),
\end{eqnarray}
if $0< z_1 < z_2$. Now let us suppose $z_1 \sim 0$. This corresponds to $u \rightarrow 0$ on the lhs. Furthermore, let us take the limit $\bar{z}\rightarrow 1$. This correponds to the double lightcone on the lhs and to the lightcone limit $v \rightarrow 0$ on the rhs. We conclude that 
\begin{eqnarray}
f_{sub}(u, v) \sim u^{-\Delta + \frac{\tau_{max}}{2}}v^{-\Delta},
\end{eqnarray}
where by $\sim$ above we mean that the lhs is not more singular than the rhs. Thus, the rhs of (\ref{fsub K}) is analytic in $\Re(\gamma_{12})> \Delta - \frac{\tau_{max}}{2}$ and $\Re(\gamma_{14})> \Delta $.

Let us consider now the second term in (\ref{fsub K 2}). It is clear that this term is well-defined for $\Re(\gamma_{12})$, $\Re(\gamma_{14})>\Delta$. We will show that we can commute the sum with the integral. Afterwards, we will analytically continue into the region $\Re(\gamma_{12})> \Delta- \frac{\tau_{max}}{2}$, $\Re(\gamma_{14})>\Delta$, except at the points where $\gamma_{12}=\Delta - \frac{\tau}{2}$, for any twist $\tau$ exchanged. Those are the OPE singularities.

We make use of the Fubini-Tonelli theorem, which says that commuting the sum with the integral is allowed in case of absolute convergence
\begin{eqnarray}
\int_0^{\frac{1}{2}} \frac{du}{u}\int_0^{1} \frac{dv}{v} u^{\gamma_{12}} v^{\gamma_{14}} 
\sum_l\sum_{\tau<\tau_{max}} a_{\tau,l}  \,u^{-\Delta +\frac{\tau}{2}} |z^l + \bar{z}^l| < \infty \ .
\label{inttaumax}
\end{eqnarray}
We divide the integral into two parts, the Lorentzian region $v \leq \ (1-\sqrt{u})^2$, and the Euclidean region $ \ (1-\sqrt{u})^2 \leq v \leq 1$ (see figure \ref{fig:u v plane}). 

In the Lorentzian region, $z$ and $\bar{z}$ are real and positive. So, the modulus in  (\ref{inttaumax}) does nothing. Thus, we can commute the sum with the integral over there. Consider now the Euclidean region. In the Euclidean region, $z$ and $\bar{z}$ are the complex conjugate of each other. Note that $|z^l + \bar{z}^l|_{(u, v)} \leq (z^l + \bar{z}^l)_{(u, (1-\sqrt{u})^2)}$. This is because $(u, v)$ and $(u, (1-\sqrt{u})^2)$ have the same value of $|z|$, but in the second case $z$ and $\bar{z}$ are positive real numbers. In the Euclidean region
\begin{align}
&\int_0^{\frac{1}{2}} \frac{du}{u} \int_{(1-\sqrt{u})^2}^{1} \frac{dv}{v} u^{\gamma_{12}} v^{\gamma_{14}} \sum_{\tau,l} a_{\tau,l}\,  u^{-\Delta +\frac{\tau}{2}} |z^l + \bar{z}^l| \leq 
\int_0^{\frac{1}{2}} \frac{du}{u} \int_{(1-\sqrt{u})^2}^{1} \frac{dv}{v} u^{\gamma_{12}} v^{\gamma_{14}} f(u, (1-\sqrt{u})^2) \nonumber \\&
= \int_0^{\frac{1}{2}} \frac{du}{u} u^{\gamma_{12}}f(u, (1-\sqrt{u})^2) \frac{1}{\gamma_{14}}(1- (1-\sqrt{u})^{2 \gamma_{14}})\label{eucld bound k}. 
\end{align}
If $\Re(\gamma_{12})>\Delta$, then this integral is well defined. Note that it was important that the integral in (\ref{eucld bound k}) did not go up to $u=1$. This was why we made the separation (\ref{division to u=1/2}). We conclude that 
\begin{eqnarray}
\int_0^{\frac{1}{2}} \frac{du}{u}\int_0^{1} \frac{dv}{v} u^{\gamma_{12}} v^{\gamma_{14}} 
\sum_l\sum_{\tau<\tau_{max}} a_{\tau,l}  \,u^{-\Delta +\frac{\tau}{2}} ( z^l + \bar{z}^l ) \\
= \sum_l\sum_{\tau<\tau_{max}} a_{\tau,l} \kappa_l (\gamma_{12}-\Delta+\frac{\tau}{2}, \gamma_{14}), \nonumber
\end{eqnarray}
where
\begin{eqnarray}
\kappa_l (\gamma_{12}, \gamma_{14})= \int_0^{\frac{1}{2}} \frac{du}{u}\int_0^{1} \frac{dv}{v} u^{\gamma_{12}}  v^{\gamma_{14}} (z^l + \bar{z}^l). 
\label{kappadef}
\end{eqnarray}

We conclude that when $\Re(\gamma_{12}),$  $\Re(\gamma_{14}) > \Delta$, then $K(\gamma_{12}, \gamma_{14})$ can be written as
\begin{align} \label{K series rep2}
K(\gamma_{12}, \gamma_{14})&= \int_\frac{1}{2}^{1} \frac{du}{u}\int_0^{1} \frac{dv}{v} u^{\gamma_{12}} v^{\gamma_{14}} f(u, v)+
\int_0^\frac{1}{2} \frac{du}{u}\int_0^{1} \frac{dv}{v} u^{\gamma_{12}} v^{\gamma_{14}} f_{sub}(u, v) \\
&+\sum_l  \sum_{\tau<\tau_{max}} a_{\tau,l}\,
\kappa_{l} \left(\gamma_{12}-\Delta+\frac{\tau}{2}, \gamma_{14}\right) \,.
\nonumber
\end{align}
Let us consider analytic continuation of (\ref{K series rep2}) in $\gamma_{12}$, keeping $\Re(\gamma_{14}) > \Delta$ fixed.
The first two terms in (\ref{K series rep2}) are analytic for $\Re(\gamma_{12}) >\Delta -\frac{\tau_{max}}{2}$. We will now argue that the analytic continuation of the last sum in (\ref{K series rep2})
is analytic for all $\gamma_{12}$ in the complex plane as long as 
\beq
\Re(\gamma_{12}) > \Delta-\frac{\tau_{max}}{2} \,,\qquad \qquad
\left|\gamma_{12} -\Delta +\frac{\tau}{2}+m\right| >\epsilon\,, %
\label{awayfromsing}
\eeq
where $\epsilon>0$ is a small regulator to avoid the OPE poles,  $\tau$ is the twist of any exchanged operator and $m$ is a non-negative integer.
Firstly, we will discuss the analytic continuation of each term $\kappa_{l} \left(\gamma_{12}-\Delta+\frac{\tau}{2}, \gamma_{14}\right)$. Secondly, we will discuss the convergence of the (infinite) sum in
(\ref{K series rep2}). 

Notice that when $l$ is a positive integer, then $z^l + \bar{z}^l$ is a polynomial of degree $l$ in $u$ and $v$. In fact, one can write
\begin{eqnarray}
\kappa_{l}(\gamma_{12}, \gamma_{14}) = 2^{-\gamma_{12}} \sum_{m=0}^l\frac{r_{l,m}(\gamma_{14})}{\gamma_{12}+m}\,, \label{kappaexplicit}
\end{eqnarray}
where $r_{l,m}(\gamma_{14})$ is a rational function.
This shows that the analytic continuation of each term $\kappa_{l} \left(\gamma_{12}-\Delta+\frac{\tau}{2}, \gamma_{14}\right) $ 
generates only OPE singularities at $\gamma_{12}=-\frac{\tau}{2}+\Delta-m$, where $\tau$ is the twist of an exchanged primary operator and $m$ is a nonnegative integer.

Now we would like to show that the sum over twists and spin in (\ref{K series rep2}) converges for any $\gamma_{12}$ as long as \eqref{awayfromsing}
is satisfied. %
We start by writing an upper bound
\begin{align}
\left|\sum_l  \sum_{\tau<\tau_{max}} a_{\tau,l}\,
\kappa_{l} \left(\gamma_{12}-\Delta+\frac{\tau}{2}, \gamma_{14}\right) \right|
&\le \sum_l  \sum_{\tau<\tau_{max}} a_{\tau,l}\,
\left|\kappa_{l} \left(\gamma_{12}-\Delta+\frac{\tau}{2}, \gamma_{14}\right)  \right|\\
&\le \sum_l  \sum_{\tau<\tau_{max}} a_{\tau,l}\,
\max_{{\rm Re}\, x >(\tau-\tau_{max})/2 \atop | x+m| > \epsilon} \left|\kappa_{l} \left( x , \gamma_{14}\right)  \right|\nonumber \\
&\le \sum_l  A_{l}\,
\max_{{\rm Re}\, x > -\tau_{max} /2 \atop | x+m| > \epsilon} \left|\kappa_{l} \left( x , \gamma_{14}\right)  \right|
\label{Alsum}
\end{align}
where $A_l = \sum_{\tau<\tau_{max}} a_{\tau,l}$.
It is clear from \eqref{kappaexplicit} that the maximum is finite for every value of the spin $l$.
Therefore, convergence of the sum follows from the large $l$ behaviour of the summand. To understand this it is convenient to make a small detour into the Lorentzian region.

Consider
the convergent sum (for $\Re(\gamma_{12}),\Re(\gamma_{14})  > \Delta$)
\begin{eqnarray}
W\equiv \sum_l\sum_{\tau<\tau_{max}} a_{\tau,l}\, \kappa_{l}^{\rm Lor}\left(\gamma_{12}-\Delta+\frac{\tau}{2}, \gamma_{14}\right)  < 
\int_0^{\frac{1}{2}} \frac{du}{u}\int_0^{(1-\sqrt{u})^2} \frac{dv}{v} u^{\gamma_{12}} v^{\gamma_{14}} f(u,v) \ ,
\end{eqnarray}
where
\begin{eqnarray}  
\kappa_{l}^{\rm Lor}(\gamma_{12}, \gamma_{14}) = \int_0^{\frac{1}{2}} \frac{du}{u}\int_0^{(1-\sqrt{u})^2} \frac{dv}{v} u^{\gamma_{12}} v^{\gamma_{14}}   (z^l + \bar{z}^l).
\end{eqnarray}
Since all terms are positive and $u\le \frac{1}{2}$ in the integration region, we have a lower bound
\beq
W>%
\sum_l A_l\, \kappa_{l}^{\rm Lor}(\gamma_{12}-\Delta+\frac{\tau_{max}}{2}, \gamma_{14})\,.
\eeq
Convergence of this sum implies a bound on the asymptotic growth of $A_l$ at large spin. 
Let us compute the large spin behavior of $\kappa_{l}^{\rm Lor}$.
When $l\rightarrow \infty$, we find
\begin{align}
\kappa_{l}^{\rm Lor}(\gamma_{12}, \gamma_{14})&= \int_0^{\frac{1}{2}} \frac{du}{u}\int_0^{(1-\sqrt{u})^2} \frac{dv}{v} u^{\gamma_{12}} v^{\gamma_{14}}  (z^l + \bar{z}^l)\nonumber \\ &
= \int_0^1 dz \int_0^{\min \{ 1, \frac{1}{2z} \} } d\bar{z} |z-\bar{z}|
\big( z^{\gamma_{12}-1 +l} (1-z)^{\gamma_{14}-1} \bar{z}^{\gamma_{12}-1 }(1-\bar{z})^{\gamma_{14}-1} \big)  \nonumber \\ & 
\approx \frac{1}{ l^{\gamma_{14}}} \Gamma (\gamma_{14}) B_{\frac{1}{2}}\left(\gamma_{12} ,\gamma_{14}+1\right) \label{limit l big},
\end{align}
where we used the incomplete $\beta$-function. Therefore,
\beq
\sum_l A_l\,\frac{1}{ l^{\gamma_{14}}} < \infty \,, \qquad\qquad
\Re(\gamma_{14})  > \Delta \,. \label{convAl}
\eeq

It turns out that this is sufficient to prove convergence of \eqref{Alsum} for $\Re(\gamma_{14})  > \Delta$. The reason is that 
\beq
\kappa_{l} \left( \gamma_{12}, \gamma_{14}\right) \approx
\kappa_{l}^{\rm Lor} \left(\gamma_{12}, \gamma_{14}\right) \,,\qquad
l\to \infty\,,
\eeq
up to exponential corrections of order $2^{-l/2}$ coming from the Euclidean region.
When $l \rightarrow \infty$, the integral in \eqref{kappadef}
is dominated by the region near $v=0$, since $z$ (or $\bar{z}$ depending on our choice) achieves its maximum value $1$ there.

In this way we established analyticity in region [b] up to OPE poles. The same analyticity in region [c] follows from  crossing symmetry $K(\gamma_{12} , \gamma_{14})=K(\gamma_{14} , \gamma_{12})$.

\subsection{Region [d]}

Analyticity of $K(\gamma_{12},\gamma_{14})$ in region [d] (up to OPE poles) follows from discreteness of the twist spectrum and the conjectured theorem in \ref{app: theorem?}. The argument goes as follows. Let $\gamma_{12}$ play the role of $x$, $\gamma_{14}$ play the role of $y$, $S$ be the set of points $\Delta-\frac{\tau}{2}$ for all  twists $\tau$ of operators appearing in the $\O \times \O$ OPE and $D$ be the domain ${\rm Re}( \gamma_{14}) > \Delta$. Then, the conditions of the theorem apply and the analytic continuation in $\gamma_{14}$ cannot generate new singularities in $\gamma_{12}$. Moreover, the maximal analyticity domain in $\gamma_{14}$ does not depend on $\gamma_{12}$. But for ${\rm Re}( \gamma_{12})>\Delta$ we know that $K$ is analytic for $\gamma_{14} \in \mathbb{C}/S$. Therefore, $K(\gamma_{12},\gamma_{14})$ is analytic for all
$\gamma_{12} \in \mathbb{C}/S$ and $\gamma_{14} \in \mathbb{C}/S$.

\begin{figure}[h]
  \centering
  \begin{tikzpicture}
	\begin{pgfonlayer}{nodelayer}
		\node [style=none] (0) at (0, -3) {};
		\node [style=none] (1) at (0, 6) {};
		\node [style=none] (2) at (-3, 0) {};
		\node [style=none] (3) at (6, 0) {};
		\node [style=textdot] (4) at (5.5, -0.3) {${\rm Re}[\gamma_{12}]$};
		\node [style=textdot] (5) at (0.6, 5.75) {${\rm Re}[\gamma_{14}]$};
		\node [style=none] (6) at (5.6, 5.6) {${\rm [a] }$};
		\node [style=none] (7) at (-2.6, 5.6) {${\rm [b]}$};
		\node [style=none] (8) at (5.6, -2.6) {${\rm [c]}$};
		\node [style=none] (9) at (-2.6, -2.6) {${\rm [d]}$};
		\node [style=point] (10) at (1, 1) {};
		\node [style=none] (11) at (3, -3) {};
		\node [style=none] (12) at (3, 6) {};
		\node [style=none] (13) at (-3, 3) {};
		\node [style=none] (14) at (6, 3) {};
		\node [style=none] (15) at (-1.5, -3) {};
		\node [style=none] (16) at (-1.5, 6) {};
		\node [style=none] (17) at (-3, -1.5) {};
		\node [style=none] (18) at (6, -1.5) {};
		\node [style=textdot]  at (-0.35, -0.2) {$(0,0)$};
		\node [style=textdot]  at (3.5, 3.2) {$(\Delta, \Delta)$};
		\node [style=textdot]  at (1.5, 1.2) {$\left(\frac{\Delta}{3}, \frac{\Delta}{3}\right)$};
		\node [style=textdot]  at (-2.1, 2.7) {$\left(-\frac{\Delta}{2}, \Delta \right)$};
		\node [style=textdot]  at (2.4, -1.8) {$\left( \Delta,-\frac{\Delta}{2} \right)$};
		\node [style=point] (19) at (-1, 5.5) {};
		\node [style=point] (20) at (4.5, -1) {};
		\node [style=point] (21) at (1.6, 2.42727) {};
		\node [style=textdot] at  (-1, 5.75)   {$\vec{\gamma}(0)$};
		\node [style=textdot] at  (4.9, -1) {$\vec{\gamma}(1)$};
		\node [style=textdot] at  (2, 2.5)  {$\vec{\gamma}(\zeta)$};
	\end{pgfonlayer}
	\begin{pgfonlayer}{edgelayer}
	\draw [style=red] (19.center) to (20.center);
		\draw [style=arrow end] (2.center) to (3.center);
		\draw [style=arrow end] (0.center) to (1.center);
		\draw [style=blue,thick] (13.center) to (14.center);
		\draw [style=blue,thick] (11.center) to (12.center);
		\draw [style=dashed] (17.center) to (18.center);
		\draw [style=dashed] (15.center) to (16.center);
		\draw[fill=blue!50, nearly transparent]  (-1.5,3) -- (3,-1.5) -- (3,3)  -- cycle;
		\draw[fill=blue!50, nearly transparent]  (-3,3) -- (6,3) -- (6,6) -- (-3,6) -- cycle;
		\draw[fill=blue!50, nearly transparent]  (3,-3) -- (3,3) -- (6,3) -- (6,-3) -- cycle;

	\end{pgfonlayer}
\end{tikzpicture}
  \caption{ Construction that leads to equations \eqref{BochnerforKwithpoles} and \eqref{BochnerforKwithpoles2}. We represent the case $\Delta =\tau_{lightest}$ and therefore $\tau_*=2\Delta$. We can establish analyticity in the shaded domain without crossing the accumulation point of accumulation points of triple-twist operators (marked with dashed lines).
  }
  \label{fig:KBochner}
\end{figure}
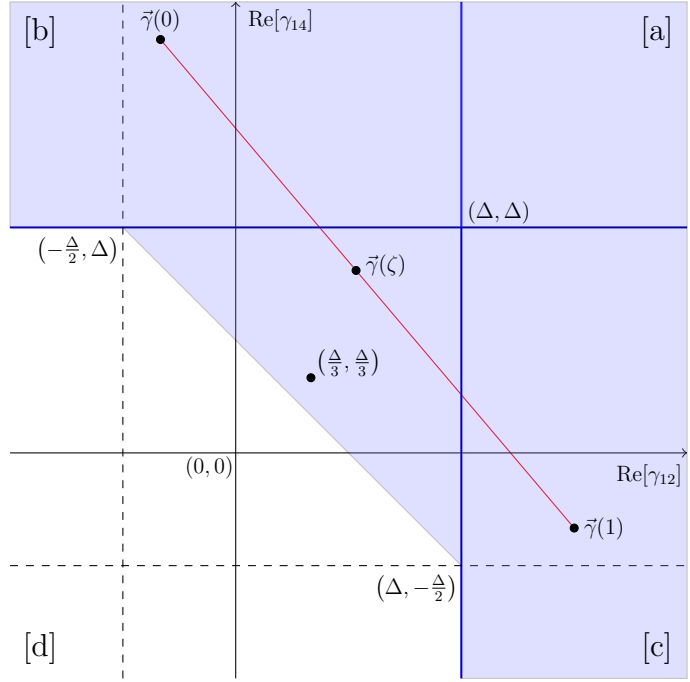

Unfortunately, neither the discreteness of the twist spectrum nor the conjectured theorem in \ref{app: theorem?} are  established. Therefore, in this section, we will present a more explicit argument that allows us to prove analyticity in a small corner of region [d] (see figure \ref{fig:KBochner}).

Let us use the trick of Bochner's theorem as reviewed in \ref{app: bochner}.  We introduce a complex plane parametrized by $s$ embedded in $\mathbb{C}^2$ as 
\beq
\vec{\gamma}(s) =
\vec{\gamma}(0)+ s (\vec{\gamma}(1) -\vec{\gamma}(0)) \,,\qquad
\vec{\gamma}(0) = (\gamma_{12}^{(0)}, \gamma_{14}^{(0)})\,, \qquad
\vec{\gamma}(1) = (\gamma_{12}^{(1)}, \gamma_{14}^{(1)})
\eeq
First we choose $\gamma_{12}^{(1)}>\gamma_{12}^{(0)}>\Delta$ and 
$\gamma_{14}^{(0)}>\gamma_{14}^{(1)}>\Delta$ so that both $\vec{\gamma}(0)$ and $\vec{\gamma}(1)$ are in region [a].
This allows us to write the following representation for the $K$-function,
\beq
KW\left(\vec{\gamma}(\zeta)\right) = \int \frac{dt}{2\pi} \frac{KW\left(\vec{\gamma}(it)\right)}{\zeta-it} +
\int \frac{dt}{2\pi} \frac{KW\left(\vec{\gamma}(1-it)\right)}{\zeta-1+it}
\label{BochnerforK}
\eeq
where $0<{\rm Re} \zeta <1$ and  $KW$ denotes the product of the $K$-function by an holomorphic function $W$  that decays fast along the imaginary direction as explained in \ref{app: bochner}.  

Notice that the function $K\left(\vec{\gamma}(s)\right)$ has poles at
\beq
s=s_\tau^{(12)}\equiv \frac{\Delta-\tau/2 - \gamma_{12}^{(0)}}{\gamma_{12}^{(1)}-\gamma_{12}^{(0)}}\,,
\eeq
and 
\beq
s=s_\tau^{(14)}\equiv \frac{\Delta-\tau/2 - \gamma_{14}^{(0)}}{\gamma_{14}^{(1)}-\gamma_{14}^{(0)}}=1+
\frac{\Delta-\tau/2 - \gamma_{14}^{(1)}}{\gamma_{14}^{(1)}-\gamma_{14}^{(0)}}\,.
\eeq
For our choice of $\vec{\gamma}(0)$ and $\vec{\gamma}(1)$, the poles obey $s_\tau^{(12)}<0$ and $s_\tau^{(14)}>1$ as required by the conditions of Bochner's theorem.

Now let us move $\vec{\gamma}(0)$ into region [b] and $\vec{\gamma}(1)$ into region [c] as depicted in figure \ref{fig:KBochner}. In other words we decrease
$\gamma_{12}^{(0)}$ and $\gamma_{14}^{(1)}$ below $\Delta$. Under this deformation, there are poles that  cross the contours along $s=it$ and $s=1-it$ for $t \in \mathbb{R}$. This will change equation \eqref{BochnerforK} into 
\begin{align}
KW\left(\vec{\gamma}(\zeta)\right) &= \int \frac{dt}{2\pi} \frac{KW\left(\vec{\gamma}(it)\right)}{\zeta-it} +
\int \frac{dt}{2\pi} \frac{KW\left(\vec{\gamma}(1-it)\right)}{\zeta-1+it}
\label{BochnerforKwithpoles}\\
&+ \sum_{s_\tau^{(12)}>0} \frac{{\rm Res}_{s=s_\tau^{(12)}}KW\left(\vec{\gamma}(s)\right)}{\zeta-s_\tau^{(12)}} 
+ \sum_{s_\tau^{(14)}<1} \frac{{\rm Res}_{s=s_\tau^{(14)}}KW\left(\vec{\gamma}(s)\right)}{\zeta-s_\tau^{(14)}} \nonumber
\end{align}
Notice that if we denote $\vec{\gamma}(\zeta) =(\gamma_{12},\gamma_{14})$ then the last equation can be written as
\begin{align}
KW\left(\gamma_{12},\gamma_{14}\right) &= \int \frac{dt}{2\pi} \frac{KW\left(\vec{\gamma}(it)\right)}{\zeta-it} +
\int \frac{dt}{2\pi} \frac{KW\left(\vec{\gamma}(1-it)\right)}{\zeta-1+it}
\nonumber \\
&+ 
 \sum_{\tau< 2(\Delta-\gamma_{12}^{(0)}) } \frac{{\rm Res}_{\gamma_{12}=\Delta-\tau/2}KW\left(\gamma_{12},\gamma_{14}(s_\tau^{(12)})\right)}{\gamma_{12}-\Delta+\tau/2} 
 \label{BochnerforKwithpoles2} \\
&+ 
 \sum_{\tau< 2(\Delta-\gamma_{14}^{(1)})} \frac{{\rm Res}_{\gamma_{14}=\Delta-\tau/2}KW\left(\gamma_{12}(s_\tau^{(14)}),\gamma_{14}\right)}{\gamma_{14}-\Delta+\tau/2}
 \nonumber  
\end{align}
Equations \eqref{BochnerforKwithpoles} or \eqref{BochnerforKwithpoles2} imply analyticity of the $K$-function at $\vec{\gamma}(\zeta) =(\gamma_{12},\gamma_{14})$ as long as: {\bf i.} we place 
$\vec{\gamma}(0)$ and $\vec{\gamma}(1)$ in an analytic domain inside region [b] and region [c], respectively; {\bf ii.} the sums converge.
Condition {\bf i.} is easy to satisfy if the twist spectrum is not continuous. Although this is an open question for high twist in CFT$_d>2$, it is clear that at least until the first accumulation point of accumulation points (triple-twist operators) the spectrum is discrete.
Condition {\bf ii.} is more non-trivial. Let us consider several cases of increasing difficulty:
\begin{itemize}
\item The sums converge because they contain finite number of terms. This is the case if we do not cross any accumulation point, \emph{i.e.} for $\Delta-\tau_*/2< \gamma_{12}^{(0)}, \gamma_{14}^{(1)}<\Delta$. 

\item We cross only one twist accumulation point $\tau_*$. For example, take
$ \gamma_{12}^{(0)}<\Delta-\tau_*/2$.
In this case, the infinite sum over twists accumulating at $\tau_*$ only converges if the intersection of $\gamma_{12}=\Delta-\tau_*/2$ with the straight line through $\vec{\gamma}(0)$ and $\vec{\gamma}(1)$ has $\gamma_{14}=\gamma_{14}(s_{\tau_*}^{(12)})>\Delta$ because of \eqref{convAl}.
Fortunately, this last condition can be relaxed by choosing a function $W$ that vanishes like $\left( \gamma_{12}-\Delta+\tau_*/2\right)^p $ at $\gamma_{12}=\Delta-\tau_*/2$ for some integer $p>0$. This makes the residues in \eqref{BochnerforKwithpoles2} smaller as we approach the accumulation point and therefore extends the convergence of the sum to $\gamma_{14}(s_{\tau_*}^{(12)})>\Delta - p\, \tau_{\rm gap}$.\footnote{Here we used the large spin behaviour 
$\tau_*-\tau(l)\sim l^{-\tau_{gap}}$ from the lightcone bootstrap.}
 
\item We cross only a finite number of double-twist accumulation points 
$\tau_*+2n$ in both region [b] and region [c].  This case can be treated similarly to the previous one. It is enough to  choose a function $W$ that vanishes sufficiently fast at $\gamma_{12}=\Delta-\tau_*/2-n$  and  $\gamma_{14}=\Delta-\tau_*/2-n$ for a finite set of integers $n$. Notice that if $\Delta = \tau_{lightest}$ then $\tau_*=2\Delta$ and this allows us to prove analyticity in the corner of region [d] with $\gamma_{13}<\frac{\Delta}{2}$ (see figure \ref{fig:KBochner}). This region contains the crossing symmetric point 
$\gamma_{12}=\gamma_{13}=\gamma_{14}=\frac{\Delta}{3}$.

\item We cross an infinite number of twist accumulation points. For example, we cross the triple-twist accumulation point of accumulation points at $\gamma_{12} = \Delta- \frac{1}{2} \tau_{triple}$. Conservatively, the sums converge as long as the intersection $\gamma_{14}(s_{3\tau_*/2}^{(12)})>\Delta$. This is not sufficient to extend the region of analyticity beyond the previous case.

 \end{itemize}

\subsection{Asymptotic behavior of $K$ functions and of the Mellin amplitude}
\label{Kasymp}

$K$ functions decay polynomially at infinity. This can be seen starting from their definition:
\begin{eqnarray}\label{K asymptotic 1}
K(\gamma_{12}, \gamma_{14})= \int_0^1 \frac{du}{u} \int_0^1 \frac{dv}{v} F(u, v) u^{\gamma_{12}} v^{\gamma_{14}}.
\end{eqnarray}
When $\gamma_{12}, \gamma_{14} \rightarrow i \infty$ this integral is dominated by $u=v=1$ and so we obtain $K(\gamma_{12}, \gamma_{14}) \sim \frac{1}{\gamma_{12} \gamma_{14}}$. Subleading corrections to this behaviour can be computed by expanding the correlation function close to the crossing symmetric point $u=v=1$.

By contrast, the Mellin amplitude 
\begin{eqnarray}\label{mellin asymptotic k}
M(\gamma_{12}, \gamma_{14})= K(\gamma_{12}, \gamma_{14}) + K(\gamma_{13}, \gamma_{14}) + K(\gamma_{12}, \gamma_{13})
\end{eqnarray}
decays exponentially at infinity. We proved this for the cases in which the theorem of section (\ref{sec:MelGen}) applies. This also happens in every example. 

Since the Mellin amplitude can be written as a sum of functions that decay polynomially, it is not obvious how come it can decay exponentially from the point of view of $K$ functions. Let us see that crossing symmetry implies that it does not decay polynomially. We check this in some simple examples but not in full generality in the sense that we will see next. Indeed consider expression (\ref{K asymptotic 1}) and expand
\begin{eqnarray}
F(u,v) = \sum_{n,m=0}^N a_{n,m} (1-u)^n (1-v)^m
\end{eqnarray}
where $N$ is some positive integer. Crossing relates different $a_{n,m}$ to each other. If we plug the function $K(\gamma_{12}, \gamma_{14})$ thus obtained into (\ref{mellin asymptotic k}), we seem to obtain that $M(\gamma_{12}, \gamma_{14})$ decays polynomially. However notice that the coefficients $a_{n, m}$ are not all arbitrary and they are constrained by crossing symmetry\footnote{A similar idea was pursued in \cite{Li:2017ukc}.}. For this reason many cancellations occur and one obtains that $M(\gamma_{12}, \gamma_{14}) \sim \frac{1}{\gamma_{12}^N \gamma_{14}^N}$. We checked this up to $N=10$ and we believe that it holds for arbitrary $N$.

\section{Examples}
\subsection{Examples of $K$-functions}\label{app: examples K}%

\subsubsection{Free fields}\label{app: free}

Consider a free scalar field $\phi$ of conformal dimension $\Delta$. Then, for $\langle \phi \phi \phi \phi \rangle$ we have
\begin{eqnarray}
F(u, v)=1+u^{-\Delta}+v^{-\Delta}, \\
K(\gamma_{12}, \gamma_{14})= \frac{1}{\gamma_{12} \gamma_{14}}+\frac{1}{\gamma_{12}(\gamma_{14}-\Delta)}+\frac{1}{(\gamma_{12}-\Delta)\gamma_{14}}.
\end{eqnarray}
The corresponding Mellin amplitude is $0$.

For the case $O=\frac{1}{\sqrt{2}N}\sum_{i=1}^N \phi^i \phi^i$ in free scalar theory, we have
\begin{eqnarray}
F(u,v)=1+u^{-\Delta}+v^{-\Delta}+\frac{4}{N} ( u^{- \frac{\Delta}{2}} + v^{- \frac{\Delta}{2}} + u^{- \frac{\Delta}{2}} v^{- \frac{\Delta}{2}}),\\
K(\gamma_{12}, \gamma_{14})= \frac{1}{\gamma_{12} \gamma_{14}}+\frac{1}{\gamma_{12}(\gamma_{14}-\Delta)}+\frac{1}{(\gamma_{12}-\Delta)\gamma_{14}}\\
+\frac{4}{N} \left(\frac{1}{(\gamma_{12}-\frac{\Delta}{2})(\gamma_{14}-\frac{\Delta}{2})}+\frac{1}{\gamma_{12}(\gamma_{14}-\frac{\Delta}{2})}+\frac{1}{(\gamma_{12}-\frac{\Delta}{2})\gamma_{14}}\right).
\end{eqnarray}
The corresponding Mellin amplitude is $0$.

\subsubsection{The correlator $\langle \sigma \sigma \sigma \sigma \rangle$ in 2D Ising} \label{app: K for 2d Ising}

In the 2d Ising model, 
\begin{eqnarray}
\langle \sigma \sigma \sigma \sigma \rangle= \frac{1}{x_{13}^{\frac{1}{4}} x_{24}^{\frac{1}{4}}} F(u, v) 
= \frac{1}{x_{13}^{\frac{1}{4}} x_{24}^{\frac{1}{4}}} \int \frac{d\gamma_{12}}{2 \pi i} \int \frac{d\gamma_{14}}{2 \pi i} \hat{M}(\gamma_{12}, \gamma_{14}) u^{-\gamma_{12}} v^{-\gamma_{14}} ,
\end{eqnarray} 
with
\begin{eqnarray}
F(u, v)= \frac{\sqrt{\sqrt{u}+\sqrt{v}+1}}{\sqrt{2} \sqrt[8]{u v}} \label{f 4sigma},\\
\hat{M}(\gamma_{12}, \gamma_{14}) = - \sqrt{\frac{2}{\pi}} \Gamma \left(2 \gamma_{12}-\frac{1}{4}\right) \Gamma \left(2 \gamma_{14}-\frac{1}{4}\right) \Gamma (-2 \gamma_{12}-2 \gamma_{14})\label{Mellin 2d Ising}.
\end{eqnarray}

We will compute $K(\gamma_{12}, \gamma_{14})$ in two ways. The first way is the following. Consider 
\begin{eqnarray}\label{def Q}
Q(\gamma_{12}, \gamma_{14})= \int_0^1 \frac{dv}{v} \int_0^v \frac{du}{u} u^{\gamma_{12}} v^{\gamma_{14}} F(u, v).
\end{eqnarray}
Then, $F(u,v)=F(v,u)$ implies
\begin{eqnarray}\label{K from Q}
K(\gamma_{12}, \gamma_{14})= Q(\gamma_{12}, \gamma_{14})+ Q(\gamma_{14}, \gamma_{12}).
\end{eqnarray}
Our goal is to compute $Q(\gamma_{12}, \gamma_{14})$ by expanding $F(u, v)$ in a power series expansion around $u=0$.

For a generic CFT, we would proceed in the following manner. We write \footnote{Presumably, expansion (\ref{def collinear}) converges on the square $(u, v) \in [0, 1] \times [0, 1]$, except for $v=0$. But we could not prove it.}
\begin{eqnarray}\label{def collinear}
F(u, v)= \sum_k C_{OO O_k}^2 u^{\frac{\tau_k}{2}-\Delta} \sum_{m=0}^{\infty} u^m g_m(v),
\end{eqnarray}
where $\Delta$ is the conformal dimension of the external scalar, $\tau_k$ is the twist of an exchanged primary,  $C_{OO O_k}^2$ is an OPE coefficient and finally $g_m(v)$ is a collinear block. Suppose we put equation (\ref{def collinear}) into (\ref{def Q}). Notice that the integral (\ref{def Q}) does not involve $v=0$, which is where the expansion (\ref{def collinear}) should fail. We find
\begin{eqnarray}\label{collinear k}
K(\gamma_{12}, \gamma_{14})= \sum_k C_{OO O_k}^2 \sum_{m=0}^{\infty} \big( \frac{1}{\gamma_{12}+\frac{\tau_k}{2}-\Delta+m} \\ + \frac{1}{\gamma_{14}+\frac{\tau_k}{2}-\Delta+m} \big) f_m^{\tau_k}(\gamma_{13})\nonumber,
\end{eqnarray}
where 
\begin{eqnarray}\label{int gm}
f^{\tau_k}_m(x)= \int_0^1 \frac{dy}{y} y^{-x+\frac{\tau_k}{2}+m} g_m(y)
\end{eqnarray}
 is a kinematical function. In practice, it was difficult to compute it for general $m$. We register the result for $m=0$:
 \begin{eqnarray}
g_0(v) = (v-1)^J \, _2F_1\left(\frac{1}{2} (2 J+\tau ),\frac{1}{2} (2 J+\tau );2 J+\tau ;1-v\right),
\end{eqnarray} 
\begin{eqnarray}
f^{\tau_k}_0(\gamma_{13})=(-1)^J 2^{2 J+\tau -1} \Gamma (J+1) \Gamma \left(J+\frac{\tau }{2}\right) \Gamma \left(J+\frac{\tau }{2}+\frac{1}{2}\right) \Gamma \left(\frac{\tau }{2}-\gamma_{13}\right) \\
\, _3F_2\left(J+1,J+\frac{\tau }{2},J+\frac{\tau }{2};J+\frac{\tau }{2}-\gamma_{13}+1,2 J+\tau ;1\right)\frac{1}{\sqrt{\pi } \Gamma (2 J+\tau ) \Gamma \left(J-\gamma_{13}+\frac{\tau }{2}+1\right)}. \nonumber
\end{eqnarray}

In the case of the 2d Ising model, $F(u, v)$ is simple enough to admit a power series expansion in $u$. From (\ref{f 4sigma}), we find
\begin{eqnarray}
F(u, v)= \sum_{n=0}^{\infty} \frac{(-1)^{n+1} \Gamma(n -\frac{1}{2})   \left(\sqrt{v}+1\right)^{\frac{1}{2}-n} u^{n/2}}{2 \sqrt{2 \pi} \Gamma(n+1) (u v)^{\frac{1}{8}} }
 \end{eqnarray}
where the expansion converges in $|\sqrt{u}|<|1+\sqrt{v}|$. 

We can now do the integrals to find a series expression for $Q(\gamma_{12}, \gamma_{14})$ and $K(\gamma_{12}, \gamma_{14})$. We find
\begin{eqnarray}\label{K for Ising}
K(\gamma_{12}, \gamma_{14})=\sum_{n=0}^{\infty} \left(\frac{1}{2 \gamma_{12}+n-\frac{1}{4}}+\frac{1}{2\gamma_{14}+n-\frac{1}{4}}\right) \\ \frac{2^{\frac{5}{2}-n} (-1)^{n+1} \Gamma(- \frac{1}{2}+n) \, _2F_1\left(1,\frac{5}{4}-2 \gamma_{13};n-2 \gamma_{13}+\frac{3}{4};-1\right)}{ \sqrt{2 \pi} \left(-2 \gamma_{13}+n-\frac{1}{4}\right) \Gamma(1+n)}, \nonumber
\end{eqnarray}
where $\gamma_{13}=\frac{1}{8}-\gamma_{12}-\gamma_{14}$. This series converges exponentially fast\footnote{Note that (\ref{K for Ising}) has no poles in $\gamma_{13}$, but $Q$ would have poles in $\gamma_{13}$. Each term term of the series (\ref{K for Ising}) does have poles in $\gamma_{13}$, but the residues of successive terms cancel.}. We checked for several values of $\gamma_{12}$ and $\gamma_{14}$ that using (\ref{K for Ising}), then $K(\gamma_{12}, \gamma_{14})+K(\gamma_{12}, \frac{1}{8}-\gamma_{12}-\gamma_{14})+K(\frac{1}{8}-\gamma_{12}-\gamma_{14}, \gamma_{14})$ numerically matches (\ref{Mellin 2d Ising}).

Let us outline another way to compute $K(\gamma_{12}, \gamma_{14})$. We first do the $u$ integral. Then, we attempt to do the $v$ integral. It is more complicated, since the integrand is also more complicated. So, we series expand the integrand around $v=0$. We spare the reader the details. We find
\begin{eqnarray} \label{K for Ising 2}
K(\gamma_{12}, \gamma_{14})=\sum_{n=0}^{\infty} g_n f_n(\gamma_{12}) f_n(\gamma_{14}),
\end{eqnarray}
where
\begin{eqnarray} \label{K for Ising 3}
g_n=- \frac{\Gamma \left(n-\frac{1}{2}\right)}{2 ^{2n-\frac{3}{2}}\sqrt{\pi } \Gamma (n+1)} , \\
f_n(\gamma_{12})= \, _2\tilde{F}_1\left(1,n-\frac{1}{2};2 \gamma_{12}+n+\frac{3}{4};\frac{1}{2}\right) \Gamma \left(2 \gamma_{12} +n-\frac{1}{4}\right) \\
\equiv\, _2F_1\left(1,n-\frac{1}{2};2 \gamma_{12} +n+\frac{3}{4};\frac{1}{2}\right) \frac{4}{8 \gamma_{12} +4 n-1}. \nonumber
\end{eqnarray}
This expansion also converges exponentially fast and with it we can obtain the correct value of $M(\gamma_{12}, \gamma_{14})$.

\subsection{Mellin representation with a straight contour for $ \langle \sigma \sigma \sigma \sigma \rangle $ in the 2d Ising model} \label{app: straight 2d Ising}

We provide a Mellin amplitude with a straight contour for $\langle \sigma \sigma \sigma \sigma \rangle$ in the 2d Ising model. The four point function is given by $\frac{1}{x_{13}^{\frac{1}{4}} x_{24}^{\frac{1}{4}}} F(u, v)$, with $F(u, v)$ given by (\ref{f 4sigma}). For this correlator, we have explicit expressions for the function $K(\gamma_{12}, \gamma_{14})$, see (\ref{K for Ising 2}) and (\ref{K for Ising 3}).

According to the discussion in section (\ref{sec: dif cont}), we have that
\begin{eqnarray} \label{F 3 integrals}
F(u, v)= \int_{Re(\gamma_{12})=\frac{1}{8}+0^+} \frac{d \gamma_{12}}{2 \pi i}  \int_{Re(\gamma_{14})=\frac{1}{8}+0^+} \frac{d \gamma_{14}}{2 \pi i}  K(\gamma_{12}, \gamma_{14}) u^{-\gamma_{12}} v^{- \gamma_{14}} \\
+ \int_{Re(\gamma_{12})=\frac{1}{8}+0^+}\frac{d \gamma_{12}}{2 \pi i}  \int_{Re(\gamma_{13})=\frac{1}{8}+0^+} \lbrack d \gamma_{13} \rbrack K(\gamma_{12}, \gamma_{13}) u^{-\gamma_{12}} v^{- \gamma_{14}} \nonumber \\
 + \int_{Re(\gamma_{13})=\frac{1}{8}+0^+} \lbrack d \gamma_{13} \rbrack \int_{Re(\gamma_{14})=\frac{1}{8}+0^+} \frac{d \gamma_{14}}{2 \pi i}  K(\gamma_{13}, \gamma_{14}) u^{-\gamma_{12}} v^{- \gamma_{14}}. \nonumber
\end{eqnarray}  
Each of these three integrals should be done with a straight contour. Consider the first integral. We can deform its contour until we reach the crossing symmetric point $Re(\gamma_{12})=\frac{1}{24}$, $Re(\gamma_{14})=\frac{1}{24}$. In the process, we pick up some poles, which will give us the appropriate subtractions to perform to the correlator. We do the same procedure for all three integrals on the rhs of (\ref{F 3 integrals}). At the end, we reunite the three integrals into a single integral with a straight contour at the crossing symmetric point $Re(\gamma_{12})=\frac{1}{24}$, $Re(\gamma_{14})=\frac{1}{24}$.

Let us work out this procedure for the first integral on the rhs of (\ref{F 3 integrals}).
\begin{eqnarray}
\int_{Re(\gamma_{12})=\frac{1}{8}+0^+} \frac{d \gamma_{12}}{2 \pi i} \int_{Re(\gamma_{14})=\frac{1}{8}+0^+} \frac{d \gamma_{14}}{2 \pi i} K(\gamma_{12}, \gamma_{14}) u^{-\gamma_{12}} v^{- \gamma_{14}} \\
= \int_{Re(\gamma_{12})=\frac{1}{24}} \frac{d \gamma_{12}}{2 \pi i} \int_{Re(\gamma_{14})=\frac{1}{24}} \frac{d \gamma_{14}}{2 \pi i} K(\gamma_{12}, \gamma_{14}) u^{-\gamma_{12}} v^{- \gamma_{14}} \nonumber \\
+ u^{- \frac{1}{8}} \int_{Re(\gamma_{14})=\frac{1}{8}+0^+} \frac{d \gamma_{14}}{2 \pi i} \hat{K}(\gamma_{12}=\frac{1}{8}, \gamma_{14}) v^{- \gamma_{14}} \nonumber \\
+ v^{- \frac{1}{8}} \int_{Re(\gamma_{12})=\frac{1}{24}} \frac{d \gamma_{12}}{2 \pi i} \hat{K}(\gamma_{12}, \gamma_{14}=\frac{1}{8}) u^{- \gamma_{12}}, \nonumber 
\end{eqnarray}
where we used hats to denote the residues. 

Let us now evaluate the integrals.
\begin{eqnarray}
u^{- \frac{1}{8}} \int_{Re(\gamma_{14})=\frac{1}{6}} \frac{d \gamma_{14}}{2 \pi i} \hat{K}(\gamma_{12}=\frac{1}{8}, \gamma_{14}) v^{- \gamma_{14}} \nonumber \\
= u^{- \frac{1}{8}} \int_{Re(\gamma_{14})=\frac{1}{6}} \frac{d \gamma_{14}}{2 \pi i} \frac{_2 F_1 (1, - \frac{1}{2}, 2 \gamma_{14}+\frac{3}{4}, \frac{1}{2})}{\gamma_{14}-\frac{1}{8}} v^{- \gamma_{14}}  = \theta(1-v) \frac{\sqrt{1+\sqrt{v}}}{\sqrt{2}u^{\frac{1}{8}} v^{\frac{1}{8}}} \nonumber
\end{eqnarray}
and 
\begin{eqnarray}
v^{- \frac{1}{8}} \int_{Re(\gamma_{12})=\frac{1}{24}} \frac{d \gamma_{12}}{2 \pi i} \hat{K}(\gamma_{12}, \gamma_{14}=\frac{1}{8}) u^{- \gamma_{12}} \\
=\theta(1-u) \frac{\sqrt{1+\sqrt{u}}}{\sqrt{2}u^{\frac{1}{8}} v^{\frac{1}{8}}} - \frac{1}{\sqrt{2}} \frac{1}{u^{\frac{1}{8}} v^{\frac{1}{8}} }. \nonumber
\end{eqnarray}

We proceed similarly concerning the other two integrals in (\ref{F 3 integrals}). We conclude that if we define
\begin{eqnarray}
F_{sub} (u, v)= F(u,v) - \frac{\sqrt{1+\sqrt{u}}+\sqrt{1+\sqrt{v}}+\sqrt{\sqrt{u}+\sqrt{v}}}{\sqrt{2}(u v)^{\frac{1}{8}}} + \frac{u^{- \frac{1}{8}}v^{- \frac{1}{8}}+ u^{- \frac{1}{8}} v^{\frac{1}{8}} + u^{\frac{1}{8}}v^{-\frac{1}{8}} }{\sqrt{2}} \nonumber,
\end{eqnarray}
then
\begin{eqnarray}\label{Mellin 2d Ising straight 2}
F_{sub}(u, v)= \int_{Re(\gamma_{12})=\frac{1}{24}} \frac{d \gamma_{12}}{2 \pi i} \int_{Re(\gamma_{14})=\frac{1}{24}} \frac{d{\gamma_{14}}}{2 \pi i} M(\gamma_{12}, \gamma_{14}) u^{- \gamma_{12}} v^{- \gamma_{14}},
\end{eqnarray}
where the Mellin integral is evaluated with a straight contour at $Re(\gamma_{12})=Re(\gamma_{14})=\frac{1}{24}$ and $M(\gamma_{12}, \gamma_{14})$ is given by (\ref{Mellin 2d Ising}). We checked equation (\ref{Mellin 2d Ising straight 2}) for several values of $u$ and $v$ by performing the Mellin integral numerically. $F_{sub}(u, v)$ is crossing symmetric and is softer than $F(u, v)$ in the lightcone limit as well as in the double lightcone limit.

\subsection{$\phi^3$ in $6+\epsilon$ dimensions}\label{app:phi3}

In this section we check some statements in section \ref{sec: dif cont} for the example of $\phi^3$ theory in $6 + \epsilon$ dimensions at the critical point.

\begin{figure}[h]
  \centering
  \includegraphics[scale=1.0]{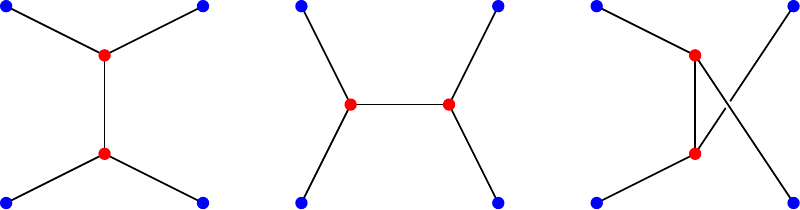} 
  \caption{The connected piece of $\langle \phi \phi \phi \phi \rangle$ at tree level. To first order in $\epsilon$, the scalar $\phi$ has dimension $\Delta=2 + \frac{5}{9} \epsilon$.}
  \label{fig:diagramsphi3}
\end{figure}

Consider the contribution to $\langle \phi \phi \phi \phi \rangle$ given by the three diagrams in figure \ref{fig:diagramsphi3}.

The main results of section \ref{sec: dif cont} are equations (\ref{subtractions finite}) and (\ref{eq:deformedfinal}). Our goal in this section is to show that such equations are correct for the four point function in figure \ref{fig:diagramsphi3}.

$\phi^3$ in $6 + \epsilon$ dimensions was studied in \cite{Goncalves:2018nlv} using the skeleton expansion. It was found in \cite{Goncalves:2018nlv} that the first diagram in figure \ref{fig:diagramsphi3} is equal to 
\begin{eqnarray}\label{diagram Vasco}
C_{\phi \phi \phi}^2 \frac{ \Gamma(\Delta) \Gamma(\frac{d-\Delta}{2})^2 }{\pi^{\frac{d}{2}} \Gamma(\frac{\Delta}{2})^2 \Gamma(\frac{d-2\Delta}{2})  } \frac{1}{ x_{12}^{3\Delta-d  } x_{34}^{\Delta} } \int \frac{ d^d x_5 }{x_{12}^{d-\Delta} x_{25}^{d-\Delta} x_{35}^{\Delta} x_{45}^{\Delta} }\\
=C_{\phi \phi \phi}^2 \frac{\Gamma(\Delta)}{\Gamma(\frac{d-2\Delta}{2}) \Gamma(\frac{\Delta}{2})^4  } \frac{1}{x_{12}^{2\Delta} x_{34}^{2\Delta}} u^{\frac{d-\Delta}{2}} \bar{D}_{\frac{d-\Delta}{2}, \frac{d-\Delta}{2}, \frac{\Delta}{2}, \frac{\Delta}{2}} (u, v), \nonumber
\end{eqnarray}
where $C_{\phi \phi \phi}$ is an OPE coefficient.

\begin{figure}
  \centering
  \includegraphics[scale=1.0]{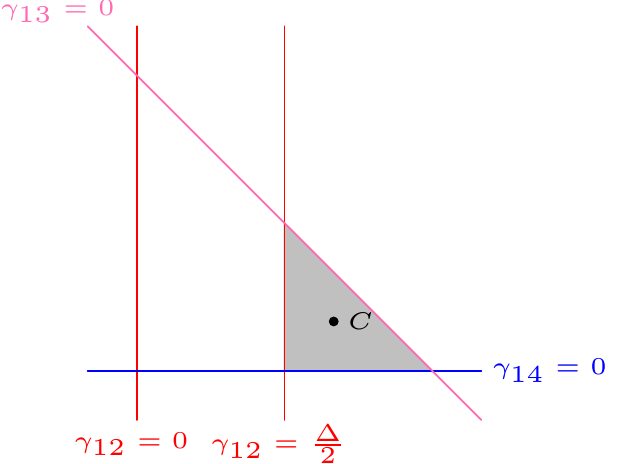} 
  \caption{According to expression (\ref{Mellin diagram 2}), the contour must be placed to the right of $\Re (\gamma_{12})=1$, above $\Re (\gamma_{14})=0$ and to the bottom of $Re (\gamma_{13})=0$. We are thus led to the shaded triangle in this figure.}
  \label{fig:phi3firstriangle}
\end{figure}

We can obtain a Mellin representation for (\ref{diagram Vasco}) using Symanzik's trick (\ref{Symanzik}). Expression (\ref{diagram Vasco}) is equal to 
\begin{eqnarray}\label{Mellin for diagram phi3}
C_{\phi \phi \phi}^2 \frac{\Gamma(\Delta)}{\Gamma(\frac{\Delta}{2})^4 \Gamma(\frac{d-2\Delta}{2})} \frac{1}{x_{13}^{2\Delta}x_{24}^{2\Delta} } \int_{\Re (\gamma_{12}) = \frac{2}{3} \Delta } \frac{d \gamma_{12}}{2 \pi i} \int_{ \Re (\gamma_{14}) = \frac{1}{6} \Delta } \frac{d\gamma_{14}}{2 \pi i}  M_{diag}(\gamma_{12}, \gamma_{14}) u^{-\gamma_{12}} v^{-\gamma_{14}},
\end{eqnarray}
where
\begin{eqnarray}
M_{diag}(\gamma_{12}, \gamma_{14} )= \Gamma(\gamma_{12}- \frac{\Delta}{2})\Gamma(\frac{d}{2} - \frac{3}{2}\Delta + \gamma_{12} )\Gamma(\gamma_{13})^2 \Gamma(\gamma_{14})^2.
\end{eqnarray}
If we set $\epsilon=0$, then 
\begin{eqnarray}\label{Mellin diagram 2}
M_{diag}(\gamma_{12}, \gamma_{14} )= \Gamma(\gamma_{12}-1)\Gamma(\gamma_{12}) \Gamma^2(\gamma_{13})\Gamma(\gamma_{14})^2.
\end{eqnarray}
So, in expression (\ref{Mellin for diagram phi3}) the contour is straight and can be placed anywhere in the shaded triangle in figure \ref{fig:phi3firstriangle}.

In order to make contact with formula (\ref{subtractions finite}) we would like to displace the contour in (\ref{Mellin for diagram phi3}) to ($\Re(\gamma_{12})= \frac{\Delta}{3}, \Re(\gamma_{14})= \frac{\Delta}{3} $). In order to do this we need to pick up the pole at $\gamma_{12}= \frac{\Delta}{2}$.  Expression (\ref{Mellin for diagram phi3}) is equal to
\begin{eqnarray}\label{mellin diagram subtraction}
C_{\phi \phi \phi}^2 \frac{\Gamma(\Delta)}{\Gamma(\frac{\Delta}{2})^4 \Gamma(\frac{d-2\Delta}{2})} \frac{1}{x_{13}^{2\Delta}x_{24}^{2\Delta} } \int_{Re (\gamma_{12}) = \frac{\Delta}{3} } \frac{d \gamma_{12}}{2 \pi i} \int_{ Re (\gamma_{14}) = \frac{\Delta}{3} } \frac{d\gamma_{14}}{2 \pi i}  M_{diag}(\gamma_{12}, \gamma_{14}) u^{-\gamma_{12}} v^{-\gamma_{14}} \nonumber \\
+ C_{\phi \phi \phi}^2 \frac{u^{-\frac{\Delta}{2}}}{x_{13}^{2 \Delta} x_{24}^{2 \Delta}  } {}_{2} F_1 (\frac{\Delta}{2}, \frac{\Delta}{2} ; \Delta, 1-v).
\end{eqnarray}
We see that the subtraction is precisely the collinear block, like we expected. 

\begin{figure}
  \centering
  \includegraphics[scale=1.0]{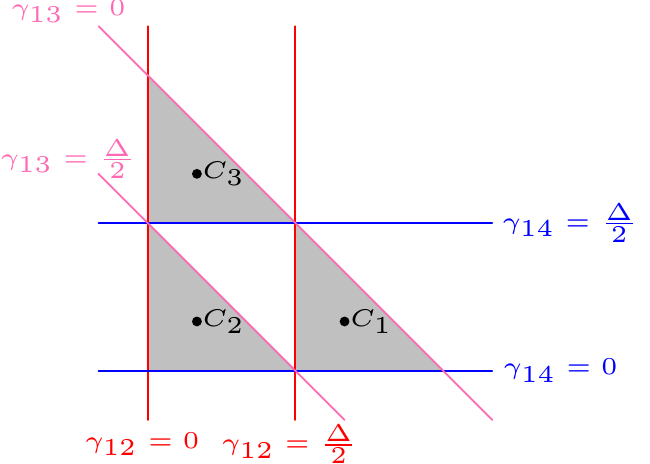} 
  \caption{The shaded triangles represent the regions where we can put the integration contours for each of the integrals in (\ref{Mellin for diagram phi3}). If we want to gather all three contours into a single deformed contour, then we run into the problem of having pinches. For example, consider the point in the picture where $\Re(\gamma_{14})=0$ and $\Re(\gamma_{14})=\frac{\Delta}{2}$. In order to have a deformed contour, the contour must pass to the right of $\Re(\gamma_{12})=\frac{\Delta}{2}$, above $\Re(\gamma_{14})=0$ and below $\Re (\gamma_{13})=\frac{\Delta}{2}$. This is impossible without introducing some regularization of the integrals. }
  \label{fig:phi3secondtriangle}
\end{figure}

The full expression for the connected piece of $\langle \phi \phi \phi \phi \rangle$ at tree level is
\begin{eqnarray}\label{phi3 3 integrals}
\langle \phi \phi \phi \phi \rangle = C_{\phi \phi \phi}^2 \frac{\Gamma(\Delta)}{\Gamma(\frac{\Delta}{2})^4 \Gamma(\frac{d-2\Delta}{2})} \frac{1}{x_{13}^{2\Delta}x_{24}^{2\Delta} } \Big( \int \int_{C_1} \frac{ d\gamma_{12} d\gamma_{14}}{(2 \pi i)^2}  M_{diag}(\gamma_{12}, \gamma_{14}) u^{-\gamma_{12}} v^{-\gamma_{14}} \\
+\int \int_{C_2} \frac{ d\gamma_{12} d\gamma_{14}}{(2 \pi i)^2}  M_{diag}(\gamma_{13}, \gamma_{14}) u^{-\gamma_{12}} v^{-\gamma_{14}} \nonumber \\
+ \int \int_{C_3} \frac{ d\gamma_{12} d\gamma_{14}}{(2 \pi i)^2}  M_{diag}(\gamma_{14}, \gamma_{13}) u^{-\gamma_{12}} v^{-\gamma_{14}} \Big) \nonumber.
\end{eqnarray}
Each of the 3 Mellin integrals has a different integration contour as in figure \ref{fig:phi3secondtriangle}..

In order to gather all three integrals in (\ref{phi3 3 integrals}) into a single integral there are two equivalent ways of proceeding. One way is to introduce an $\epsilon$ regularization in order to write a deformed contour, pick up some poles and then set $\epsilon=0$. Another way is to use (\ref{mellin diagram subtraction}) (and its equivalent for the other) diagrams. Our final formula is
\begin{eqnarray}\label{mellin phi3 tree}
\langle \phi \phi \phi \phi \rangle - \frac{C_{\phi \phi \phi}^2 }{x_{13}^{2 \Delta} x_{24}^{2 \Delta} } \Big( u^{-\frac{\Delta}{2}} {}_{2} F_1 (\frac{\Delta}{2}, \frac{\Delta}{2} ; \Delta, 1-v) \\ + v^{-\frac{\Delta}{2}} {}_{2} F_1 (\frac{\Delta}{2}, \frac{\Delta}{2} ; \Delta, 1-u)
+ v^{-\frac{\Delta}{2}} {}_{2} F_1 (\frac{\Delta}{2}, \frac{\Delta}{2} ; \Delta, 1-\frac{u}{v})  \Big) \nonumber \\ = \frac{C_{\phi \phi \phi}^2  \Gamma(\Delta)}{\Gamma(\frac{\Delta}{2})^4 \Gamma(\frac{d-2\Delta}{2})} \frac{1}{x_{13}^{2\Delta}x_{24}^{2\Delta} }  \int_{\Re(\gamma_{12})=\frac{\Delta}{3}} \frac{d \gamma_{12}}{2 \pi i}  \int_{\Re(\gamma_{14})=\frac{\Delta}{3}} \frac{d \gamma_{14}}{2 \pi i}    M(\gamma_{12}, \gamma_{14}) u^{-\gamma_{12}} v^{-\gamma_{14}} , 
\end{eqnarray}
for the connected piece of  $\langle \phi \phi \phi \phi \rangle$ at tree level, where
\begin{eqnarray}
M(\gamma_{12}, \gamma_{14}) = \Gamma^2(\gamma_{12})\Gamma^2(\gamma_{13})\Gamma^2(\gamma_{14}) \\
\times \Big( \frac{\Gamma(-\frac{\Delta}{2} + \gamma_{12}) \Gamma(\frac{d}{2} -\frac{3\Delta}{2} + \gamma_{12})  }{\Gamma^2(\gamma_{12})} + (\gamma_{12} \leftrightarrow \gamma_{13}) + (\gamma_{12} \leftrightarrow \gamma_{14}) \Big) \nonumber
\end{eqnarray}
Formula (\ref{mellin phi3 tree}) agrees with the equations (\ref{subtractions finite}) and (\ref{eq:deformedfinal}) in the main text.

\section{Analyticity in a Sectorial Domain $\Theta_{CFT}$}\label{app: rhombus}

In this appendix we establish the claim made in the main text about the region of analyticity of the correlator. The idea is to use the convergent OPE to bound the analytically continued correlator and its derivatives. We start by stating some preliminaries. Afterwards we give a proof that the correlation function is analytic inside the rhombus, see figure \ref{fig:thetaCFT}. Finally, we comment on the case of $\langle \sigma \sigma \sigma \sigma \rangle$ in the 2d Ising model to illustrate our claims.

\subsection{Preliminaries}

It is convenient to introduce the standard $(z,\bar z)$ coordinates for the cross ratios
\begin{eqnarray}
&&u = z \bar z  , \nn \\
&&v = (1-z)(1-\bar z) \ .
\end{eqnarray}
Let us briefly discuss the relationship between the two coordinates. We first consider the principal Euclidean sheet which corresponds to $u,v \geq 0$. It is convenient to split this in two regions, see figure \ref{fig:u v plane}.

In the gray region $(z,\bar z)$ coordinates are complex conjugate 
\beq
{\rm Grey~region} :~~~\bar z = z^* , ~~~ {\rm Im}[z] \neq 0 .
\eeq
In addition to this, in the colored regions we have for $z, \bar z \in {\mathbb R}$
\begin{eqnarray}\label{z zbar Lorentzian}
{\rm Red~region} &:&~~~z ,\bar z  \leq 0 , \nn \\
{\rm Blue~region} &:&~~~0 \leq z , \bar z \leq 1 , \nn \\
{\rm Pink~region} &:&~~~1 \leq z , \bar z < \infty .
\end{eqnarray}

In going from $(z, \bar z)$ to $(u,v)$ there is a square root ambiguity and we have to choose a branch of the continuation
\begin{eqnarray}\label{zzbar definition}
z &= {1 \over 4} \left( \sqrt{(1+ \sqrt{u})^2 -v} + \sqrt{(1- \sqrt{u})^2 -v}  \right)^2 , \nn \\
\bar z &= {1 \over 4} \left( \sqrt{(1+ \sqrt{u})^2 -v} - \sqrt{(1- \sqrt{u})^2 -v}  \right)^2 \ .
\end{eqnarray}
The branch point is located at $z  = \bar z$. 

It is also useful to recall $(\rho, \bar \rho)$ variables \cite{Pappadopulo:2012jk}
\beq
\rho(z) = { z \over (1+ \sqrt{1-z})^2} ,  
\eeq
which map the $[1,\infty)$ cut $z$-plane into a unit disc. Using the $\rho$ variable we can expand the correlator as follows \cite{Pappadopulo:2012jk,Hartman:2015lfa}
\beq
\label{rhoOPE}
F(z, \bar z) = \sum_{h, \bar h} b_{h, \bar h} \rho(z)^h \bar \rho(\bar z)^{\bar h}, ~~~  b_{h, \bar h} \geq 0.
\eeq
This expansion converges for $|\rho|, |\bar \rho|<1$ and makes the analytic structure in the $z$-plane manifest. The correlator has branch points at $z, \bar z = 0$ which correspond to crossing the light-cone. Moreover, analytic continuation around the origin simply introduces phases in the expansion (\ref{rhoOPE}). Similarly, we can use unitarity to bound the analytically continued correlator by its value on the principal Euclidean sheet
\beq
\label{boundOPE}
{\rm OPE \ bound}:~~~| F(e^{i \alpha}  z, e^{i \beta} \bar z) | \leq F(z^* ,\bar z^*) ,
\eeq
where $| \rho \left( e^{i \alpha}  z \right) | \equiv \rho(z^*)$ and $| \bar \rho \left(e^{i \beta}  \bar z \right) | \equiv \rho(\bar z^*)$. Analogous statements apply for analytic continuation around $z=1$ (if we use the same argument in the crossed channel). We, however, do not have a corresponding argument in $d>2$ for analytic continuation simultaneously around $z=0$ and $z=1$. In $d=2$ the analytic structure of the correlation function is the same on every sheet as was shown in \cite{Maldacena:2015iua}, thanks to the convergence properties of the so-called $q$-expansion. In $d>2$ we expect to have extra singularities, a full classification of which is not known. One simple example discussed in \cite{Gary:2009ae,Maldacena:2015iua} is the $z = \bar z$ singularity, where continuation on the second sheet in one of the variables is implicitly assumed. The $z = \bar z$ singularity corresponds to a very simple Landau graph where external points lie on a light-cone emanating from a point, two in the past and two in the future. It is also a singularity of an individual conformal block in $d>2$, see \cite{Maldacena:2015iua} for more details.

\subsection{Analyticity from OPE}

For ${\rm arg}(u) = {\rm arg}(v) = 0$ the s-channel OPE converges for all $u>0$ and $v>0$, except for the region in which $v \leq (1-\sqrt{u})^2$ and $u \geq 1$, this is the pink region on figure \ref{fig:u v plane}.  Let us consider $u, v >0$ inside the region where the s-channel OPE converges and consider the analytic continuation $u \rightarrow |u| e^{i \alpha}$, $v \rightarrow |v| e^{i \beta}$, where we are interested in $0 < |u|, |v| < \infty$.  We ask for which values of $\alpha$ and $\beta$ does the s-channel OPE still converges.

The s-channel OPE will cease to converge whenever we have either $z$ or $\bar{z}$ bigger or equal to $1$ and real. Without loss of generality, suppose that $\bar{z}\geq 1$ and real. Then we have the following relation
\begin{eqnarray}\label{u v boundary convergence 2}
|v| e^{i \beta} = (1- \bar{z})(1-\frac{|u| e^{i \alpha}}{\bar{z}}).
\end{eqnarray}
Taking the real and imaginary part of this equation, we find
\begin{eqnarray}\label{u v boundary convergence}
(|u|, |v|) = \big( - \bar{z} \frac{\sin(\beta)}{\sin(\alpha-\beta)},  (1- \bar{z}) \frac{\sin(\alpha)}{\sin(\alpha-\beta)} \big).
\end{eqnarray} 

\begin{figure}[h]
  \centering
  \includegraphics[width=0.5\textwidth]{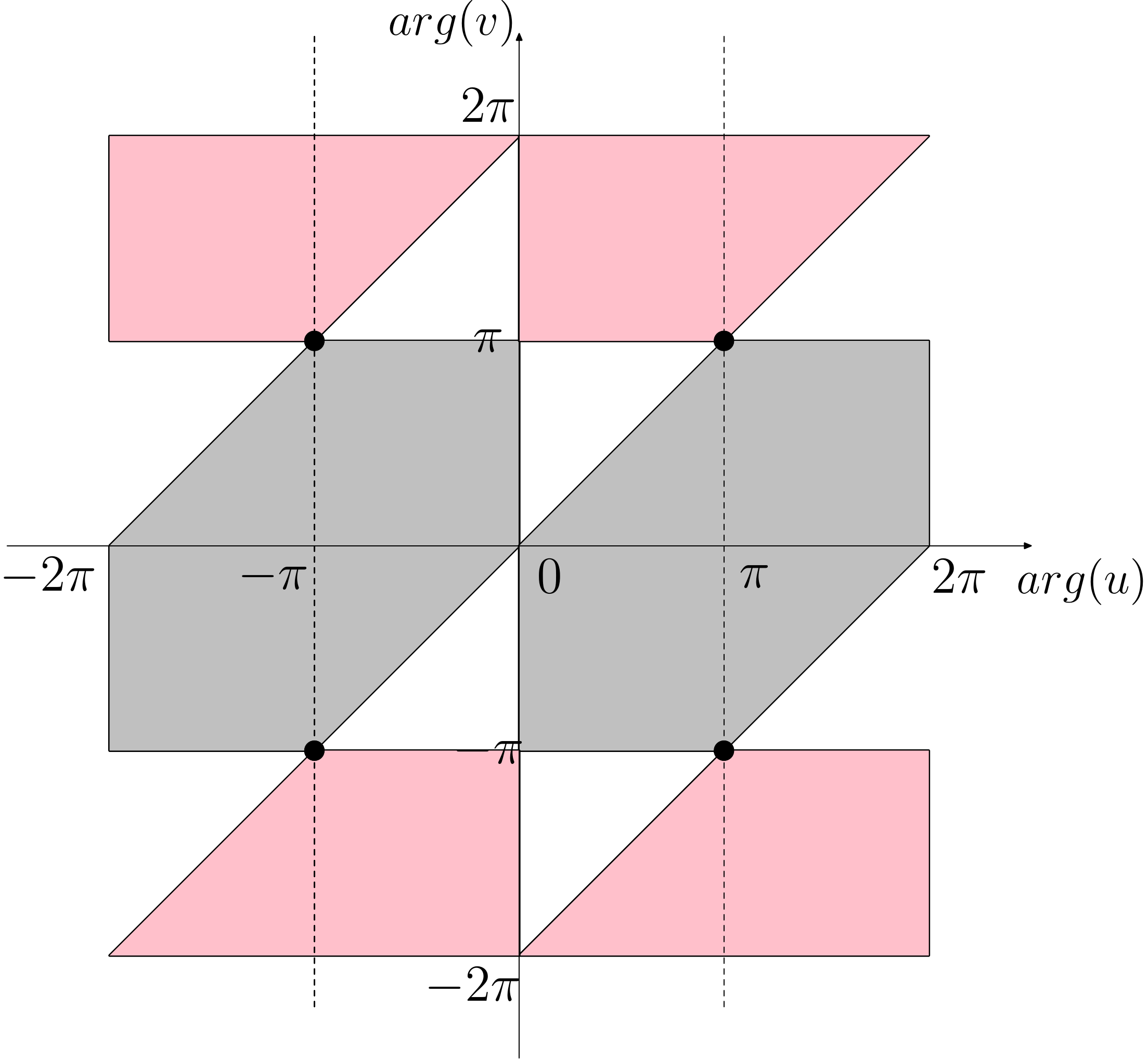}   
  \caption{Conditions (\ref{alpha beta}) are verified in the pink and grey region. They are not verified in the white region. The s-channel OPE does not converge at special points that connect the grey and the pink region at $(- \pi, \pi)$, $(- \pi, -\pi)$, $( \pi, -\pi)$, $( \pi, \pi)$. This reflects the fact that when continuing from the grey to the pink region we necessarily cross the $[1, \infty]$ cut in the $z$ or $\bar{z}$ plane. Therefore, the $s$-channel OPE cannot be used in the pink region.}
  \label{fig:conditionsalpha}
\end{figure}

We should read equation (\ref{u v boundary convergence}) in the following manner. Given $(\alpha, \beta)$ it tells us for which values of $|u|$ and of $|v|$ does the s-channel OPE cease to converge. As explained above, $|u|$ and $|v|$ must be positive. So, if equation (\ref{u v boundary convergence}) implies that $|u|$ and $|v|$ are negative, then the s-channel OPE converges for such values of $\alpha$ and $\beta$ (as long as to get to such values the s-channel OPE converged along the way of the analytic continuation). We thus obtain the following conditions for convergence of the s-channel OPE
\begin{eqnarray}\label{alpha beta}
\big( \sin(\alpha) \sin(\beta) \big)  <0 \lor \Big( \sin(\alpha-\beta)>0 \land \sin(\alpha)>0 \land \sin(\beta)>0 \Big)	 \\
\lor \Big( \sin(\alpha-\beta)<0 \land \sin(\alpha)<0 \land \sin(\beta)<0 \Big), \nonumber
\end{eqnarray}
see figure \ref{fig:conditionsalpha}. At $\alpha = \beta + n \pi$, with $n \in \mathbb{N}$, condition (\ref{u v boundary convergence 2}) does not hold, unless at possible special points where $\alpha= m \pi$, where $m \in \mathbb{N}$.

Our argument shows that the s-channel OPE converges in the grey region of figure (\ref{fig:conditionsalpha}) for all $0 < |u|, |v| < \infty$. Indeed, in this region both $|\rho|, |\bar \rho|<1$ and since the OPE converges exponentially fast \cite{Pappadopulo:2012jk}, both $F(u,v)$ and any of its $\pa_u$, $\pa_v$ derivatives are finite. This establishes analyticity of $F(u,v)$ inside the grey region. The situation is slightly different on a boundary of the grey region. Consider for example ${\rm arg}(u) = {\rm arg}(v) = 0$. As explained above in this case the $s$-channel OPE converges only for $|v| \leq (1-\sqrt{|u|})^2$ and $|u| \geq 1$.

Next we combine the argument above with crossing. Applying crossing symmetry to the grey region of figure (\ref{fig:conditionsalpha}) we find the correlation function is analytic in the whole sectorial domain given by the rhombus of figure \ref{fig:thetaCFT}. The rhombus in figure \ref{fig:thetaCFT} is the minimal crossing symmetric region that contains the grey regions in figure \ref{fig:conditionsalpha}.

One comment is in order regarding the special point ${\rm arg}(u) = {\rm arg}(v) = 0$ which is the common boundary point of analyticity of all three OPE channels. In this case as we mentioned above each of the channels converges only in some subspace of the sectorial domain $0 < |u|, |v| < \infty$. However, the union of them covers it fully and thus we have established the desired analyticity in $\Theta_{CFT}$.

\subsection{2d Ising}

For $\langle \sigma \sigma \sigma \sigma \rangle $ in the 2d Ising model we have that $F(u, v)= \frac{\sqrt{\sqrt{u}+\sqrt{v}+1}}{\sqrt{2} \sqrt[8]{u v}}$. Suppose that we start analytically continuing $u \rightarrow |u| e^{i \alpha}$, $v \rightarrow |v| e^{i \beta}$. We reach the boundary of the region of analyticity when  $1+ |u|^{1/2} e^{i  \frac{\alpha}{2}} + |v|^{1/2} e^{i \frac{\beta}{2}}=0$. Given $\alpha$ and $\beta$ the boundary of the region of analyticity is achieved when we start at $u, v >0$ such that 
\begin{eqnarray}
(\sqrt{u}, \sqrt{v})= (\frac{\sin (\frac{\beta}{2})}{ \sin ( \frac{\alpha-\beta}{2})  }, - \frac{\sin (\frac{\alpha}{2})}{\sin(\frac{\alpha-\beta}{2})}).
\end{eqnarray}
We are only interested in the situations where the RHS is positive. It is not possible for the RHS to be positive inside the rhombus of figure \ref{fig:thetaCFT}. So we conclude that $\langle \sigma \sigma \sigma \sigma \rangle $ is analytic inside the rhombus. 

Let us also comment on the branch point above in relation to the results of  \cite{Maldacena:2015iua}. It was argued in  \cite{Maldacena:2015iua} that 2d CFT correlators has only branch point singularities at $z, \bar z = 0,1, \infty$ on every sheet. The branch point above on the other hand is at $z = \bar z$. By switching to the $(z, \bar z)$ one can indeed check that the correlator is fully analytic at this point. The branch point originates from going between the $(z,\bar z)$ and $(u,v)$ variables, see (\ref{zzbar definition}), in full agreement with the results of  \cite{Maldacena:2015iua}.

\section{Polynomial Boundedness in a Sectorial Domain $\Theta_{CFT}$}\label{sec:polboundapp}

In this section we present some arguments in favor of the bound on the double light-cone limit that we used in the main text. Recall that the double light-cone limit is defined as $u,v \to 0$ with ${u \over v}$ fixed (or some more general path of approaching the origin in the $(u,v)$ plane). This limit is not controlled by the OPE and therefore an extra analysis is required.\footnote{See \cite{Alday:2010zy} for discussion of this limit in the planar gauge theories and \cite{Alday:2015ota} for the corresponding limit in the vector model.}

\subsection{Subtractions and a Bound on the Double Light-Cone Limit} 
\label{sec: better bound}

Here we use the asymptotic light-cone expansion on the second sheet to derive a better bound on the double light-cone limit
of the double discontinuity of the correlator. 

Consider the full correlation function
\begin{eqnarray}
\label{unitaritysum}
F(u, v) = u^{- \Delta} \sum_{h, \bar{h}}a_{h, \bar{h}} z^h \bar{z}^{\bar{h}}.
\end{eqnarray}
We will be interested in the double discontinuity of the connected correlator
\beq
{\rm dDisc}_u[F](u,v) \equiv F(u,v) - {1 \over 2} \left( F(u e^{2 \pi i} , v ) + F(u e^{-2 \pi i} , v) \right) .
\eeq

Recall that the connected correlator is equal to $F_{conn}(u,v) = F(u,v) - (1 + u^{- \Delta} + v^{- \Delta})$. Therefore we get
\begin{eqnarray}
\label{eq:ddiscsubtraction}
{\rm dDisc}_u[F_{conn}](z,\bar z) =2 u^{- \Delta} \sum_{h, \bar h \geq {\tau_{gap} \over 2}} \sin^2 \pi (h - \Delta) a_{h, \bar h} z^{h} \bar z^{\bar h} ,
\end{eqnarray}
where we used that ${\rm dDisc}_u[1 + v^{- \Delta}]= 0$ and wrote explicitly that sum goes only over operators above the vacuum. Using unitarity we then get
\beq
\label{ineqdDisc}
{{\rm dDisc}_u[F_{conn}](z_1,\bar z_1)  \over (z_1 \bar z_1)^{ {\tau_{gap} \over 2} - \Delta}} \leq {{\rm dDisc}_u[F_{conn}](z_2,\bar z_1)  \over (z_2 \bar z_1)^{{\tau_{gap} \over 2} - \Delta}} , ~~~ 0< z_1 < z_2 < 1 \ .
\eeq

We know take the limit $\bar z_1 \to 1$ and use in the RHS asymptotic light-cone expansion on the second sheet. As reviewed in detail in \cite{Kologlu:2019bco} for a general four-point correlator it is an assumption. For the case of identical scalars at hand, however, it was argued for in \cite{Hartman:2015lfa}. We can then take double discontinuity block by block. Since a double discontinuity of each individual block is zero the leading effect will come from the first twist accumulation point at some 
\beq
\tau_{gap} \leq \tau_* \leq 2 \Delta .
\eeq 

Using this fact we can write 
\beq
{{\rm dDisc}_u[F_{conn}](z_1,\bar z_1)  \over (z_1 \bar z_1)^{{\tau_{gap} \over 2} - \Delta}} \leq c { \left[ (1 - z_2) (1- \bar z_1) \right]^{ {\tau_{*} \over 2}- \Delta } \over (z_2 \bar z_1)^{{\tau_{gap} \over 2}- \Delta}} .
\eeq

In terms of $(u,v)$ the bound becomes
\beq
\label{doubledisccondition}
{\rm dDisc}_u[F_{conn}](u, v) \leq {c \over u^{\Delta - {\tau_{gap} \over 2}}  v^{\Delta - {\tau_{*} \over 2}}} ,  ~~~ 0< u,v <  c_0 .
\eeq
Of course, by applying crossing we can write an analogous bound for ${\rm dDisc}_v[F_{conn}](u, v) \geq  {c \over v^{\Delta - {\tau_{gap} \over 2}}  u^{\Delta - {\tau_{*} \over 2}}}$.

Note that if we try to remove ${\rm dDisc}$ the argument above fails. Indeed, in this case we cannot write (\ref{ineqdDisc}) because taking the connected part is equivalent to introducing terms with negative coefficients in the sum (\ref{unitaritysum}). These are due to double twist operators present in the OPE decomposition of the disconnected piece. Without non-negativity of the expansion (\ref{eq:ddiscsubtraction}) we cannot write (\ref{ineqdDisc}). Instead we can use similar arguments to derive the bound for the full correlator
\beq
\label{fulldoublelightcone}
F(u, v) \leq {c \over (u v)^{\Delta}} .
\eeq

Given the bound on ${\rm dDisc}_u[F_{conn}]$ we would like to derive a bound on $F_{conn}$ itself. A natural guess which is consistent with (\ref{doubledisccondition})
\beq
\label{eq:bound}
\left| F_{conn}(u,v) \right| \leq {c \over (u v)^{\Delta - {\tau_{gap} \over 2}}} , ~~~ 0< u,v <  c_0 . 
\eeq
Two examples where this bound is saturated are minimal models in $d=2$ and free field theories in $d>2$. In both of these cases $\tau_{gap} = \tau_*$. 

This, however, does not immediately follow from (\ref{doubledisccondition}). As a way to violate (\ref{eq:bound}), while having (\ref{doubledisccondition}) being satisfied, we can imagine that the spectrum contains special operators with twist $2 \Delta + 2 m$ which will contribute to $F(u,v)$ but will not contribute to the double discontinuity. We will see a nontrivial example of such a function below. In an interacting CFT we, however, do not expect such a problem to occur and therefore we believe that (\ref{eq:bound}) is a correct bound.

To avoid this problem and to generalize the argument above we consider subtractions. Assuming the number of low lying twist operators is finite we can improve the argument by considering additional subtractions. Indeed, let us consider 
\begin{eqnarray}
\label{eq:subtractionFb}
&&F_{sub}(u,v) = F(u,v) - (1 + u^{- \Delta} + v^{- \Delta})  \\
&-& \sum_{\tau_{gap} \leq \tau \leq \tau_{sub}} \sum_{J=0}^{J_{max}} \sum_{m=0}^{ \[ \frac{\tau_{sub} - \tau}{2}\]} C_{\tau, J}^2  \Big( u^{- \Delta + \frac{\tau}{2}+m} g^{(m)}_{\tau, J}(v) + v^{- \Delta + \frac{\tau}{2}+m} g^{(m)}_{\tau, J}(u) + v^{- \frac{\tau}{2}-m} g^{(m)}_{\tau, J}(\frac{u}{v}) \Big) \nn \ ,
\end{eqnarray}
where the role of the first subtraction term in the brackets is to make the sum (\ref{eq:ddiscsubtraction}) to start from $\sum_{h, \bar h \geq {\tau_{sub}' \over 2}} $. Going through the same argument (and the same assumptions) we conclude that
\beq
\label{eq:subddiscbound}
\left| {\rm dDisc}_u F_{sub}(u,v) \right| , \left| {\rm dDisc}_v F_{sub}(u,v) \right| \leq {c \over (u v)^{\Delta - {\tau_{sub}' \over 2}}} , ~~~ 0< u,v <  c_0 . 
\eeq
Note that since the number of subtractions is finite, we have necessarily $\tau_{sub}  < \tau_* < 2 \Delta$. In writing (\ref{eq:subddiscbound}) we used the fact that the second and the third subtraction terms in the brackets in (\ref{eq:subtractionF}) trivially have double discontinuity which satisfies (\ref{eq:subddiscbound}).

An interesting example of the function with zero double discontinuity in the $s$ and $t$-channel but singular in the double light-cone limit is provided by the $u$-channel subtractions above. Consider for example $v^{- \frac{\tau}{2}} g^{(0)}_{\tau, J}(\frac{u}{v})$ which corresponds to a collinear conformal block exchanged in the $u$-channel. It is easy to check that ${\rm dDics}_u \Big( v^{- \frac{\tau}{2}} g^{(0)}_{\tau, J}(\frac{u}{v}) \Big) = {\rm dDics}_v \Big( v^{- \frac{\tau}{2}} g^{(0)}_{\tau, J}(\frac{u}{v}) \Big) = 0$,\footnote{This is just to say the usual thing that the $u$-channel exchange decomposes into purely double trace operators in the $s$- and $t$-channel.} whereas its double light-cone limit is given by
\beq
\label{eq:subdLC}
\lim_{u, v \to 0, {u \over v} ~ - ~ {\rm fixed}}v^{- \frac{\tau}{2}} g^{(0)}_{\tau, J}(\frac{u}{v}) \sim v^{- \frac{\tau}{2}} .
\eeq
This is to be contrasted with a much more regular behavior of the same function in the light-cone limit
\begin{eqnarray}
&&\lim_{u \to 0, v ~ - ~ {\rm fixed}}v^{- \frac{\tau}{2}} g^{(0)}_{\tau, J}(\frac{u}{v}) \sim \log u , \nn \\
&&\lim_{v \to 0, u ~ - ~ {\rm fixed}}v^{- \frac{\tau}{2}} g^{(0)}_{\tau, J}(\frac{u}{v}) \sim \log v . 
\end{eqnarray}

In the argument above we expect that in an interacting CFT only subtraction terms provide examples of such functions. Therefore we conclude that (\ref{eq:subddiscbound}) implies 
\beq
\label{eq:subbound}
\left| F_{sub}(u,v) \right| \leq {c \over u^{\gamma_{12}} v^{\gamma_{14}}} , ~~~ 0< u,v <  c_0 , 
\eeq
where $\gamma_{12}, \gamma_{14} > \Delta - {\tau_{sub}' \over 2}$ and $\gamma_{12} + \gamma_{14} > \tau_{sub}$ due to  subtractions (\ref{eq:subdLC}). In a sense we would like to say that by improving the light-cone limits in the $u$- and $v$- channels we have also improved the corresponding double light-cone limit in the $s$- and $t$-channels but we make it worse due to the subtractions in the $u$-channel. 

Let us comment on another piece of intuition behind (\ref{eq:subbound}). As we take the double light-cone limit what could happen is that we have an infinite number of operators with approximately fixed twist given by $2 \Delta + 2n$ operators with arbitrarily large spin that develop an enhanced singularity in the double light-cone limit. However, from the usual light-cone bootstrap picture we expect those operators to control the double light-cone limit in the dual channel. Therefore, by improving the light-cone limit in both channels via subtractions we expect that we also improve the double light-cone limit at least for a few light isolated operators as above.

To go from the statement about double discontinuity to the correlator itself rigorously we can use recently derived CFT dispersion relations \cite{Carmi:2019cub}. We can consider
\beq
{\cal G}(z, \bar z) = (z \bar z)^\Delta F_{sub}(z, \bar z)
\eeq
and observe that $G(z, \bar z)$ satisfy all the conditions to apply subtracted dispersion relations to it. The subtracted dispersion relation expresses $F_{sub}(z, \bar z)$ via an integral of its double discontinuities. We can then use (\ref{eq:subddiscbound}) and try to derive   (\ref{eq:subbound}) via dispersion relations. We comment on this idea and related difficulties in appendix G.  To summarize, at the moment we were not able to rigorously prove (\ref{eq:subbound}) and leave it as an assumption hoping to improve on that in the future.

\subsection{Dangerous Limits in the Sectorial Domain}

In the subsections above we considered different small $0< u,v< c_0$ limits with $u$ and $v$ being real. For the purpose of deriving the Mellin amplitude we would like however to generalize this argument for analytically continued $u$ and $v$. As usual we would like to use the OPE to bound the correlator. However due to subtractions we do not have the OPE expansion representation of the correlator with positive coefficients. Therefore we cannot simply bound the analytically continued correlator by its value on the principal sheet using the Cauchy-Schwarz argument. Nevertheless we believe that our polynomial bounds on the double light-cone limit still apply in the region of analyticity of the correlator. In some sense this is a generalization of the idea that we can use the light-cone OPE on the second sheet as an asymptotic expansion.

To sum up, we would like now to say that with the region of analyticity $F_{sub}(u,v)$ satisfies the same bound as above
\beq
| F_{sub}(u,v) | \leq C(\gamma_{12}, \gamma_{14}) {1 \over |u|^{\gamma_{12}}} {1 \over |v|^{\gamma_{14}}}  , ~~~ (u,v) \in \Theta_{CFT}, ~~~ (\gamma_{12} , \gamma_{14}) \in U_{CFT} \ ,
\eeq

We already discussed the origin of this bound for small $u$ and $v$  in the previous section and we can use crossing (\ref{crossing}) to derive the bound at large $u$ and $v$.

\subsection{Dispersion Relations}

In this appendix we explain the relation between the double light-cone limit of the double discontinuity and the corresponding double light-cone limit of the correlation function using recently derived dispersion relation \cite{Carmi:2019cub}. The punchline is that we will not be able to derive the desired bound due to the fact dispersion relations involve the integral of double discontinuity over the Regge limit which we do not have enough control of.

Let us first review briefly dispersion relations without subtractions.The result of \cite{Carmi:2019cub} states that given a suitable complex function of two variables $G(z, \bar z)$ we can write it via its double discontinuity
\begin{align}
\label{eq:dispnosub}
G(z , \bar z ) &= \int_0^{1} d w d \bar w K(z, \bar z, w , \bar w) {\rm dDisc}[G(w, \bar w)] \nonumber \\
&+  \int_0^{1} d w d \bar w K({z \over z-1}, {\bar z \over \bar z - 1}, w , \bar w) {\rm dDisc}[G({w \over w-1}, {\bar w \over \bar w - 1} )] .
\end{align}
The explicit form of the kernels can be found in \cite{Carmi:2019cub}.

We will be interested in the double light-cone limit of $G(z, \bar z)$. We would like to argue that it is controlled by the corresponding double light-cone limit of the dDisc. 

As a warm-up let us consider a simple example of the generalized free field
\beq
G(z, \bar z) = (z \bar z)^{p_1} ([1-z] [1-\bar z])^{p_2}
\eeq
with $ {\rm dDisc}[G(z, \bar z)] =2 \sin^2 (\pi p_2) (z \bar z)^{p_1} ([1-z] [1-\bar z])^{p_2}$. As written above the dispersion relation converges for $p_1 > {1 \over 2}$ and $- {3 \over 4} < p_2 < {3 \over 4} - p_1$.

We are interested in the double light-cone limit $z \to 0$, $\bar z \to 1$. As we will see in a second it comes solely from the corresponding limit $w \to 0$, $\bar w \to 1$ of ${\rm dDisc}[G(w, \bar w)]$.

To demonstrate this recall that the structure of the kernel is 
\begin{align}
K(z, \bar z, w , \bar w) = K_{B}(z, \bar z, \rho_w, \bar \rho_w) \theta(\rho_z \bar \rho_z \bar \rho_w - \rho_w) + K_C {d \rho_w \over d w} \delta(\rho_w - \rho_z \bar \rho_z \bar \rho_w). 
\end{align}

Let us first discuss the bulk kernel. We write it in $\rho$ coordinates together with the measure
\begin{align}
\label{eq:integrationmeasure}
\int_0^1 d \rho_w d \bar \rho_w K_{B}(z, \bar z, \rho_w, \bar \rho_w) \theta(\rho_z \bar \rho_z \bar \rho_w - \rho_w)  . 
\end{align}
We then consider the following change of variables. We first rescale $\rho_w \to (\rho_z \bar \rho_z \bar \rho_w) \rho_w$ to remove the $\theta$-function. We then consider the following change of variables
\begin{align}
z &\to \eps z, ~~~\bar z \to 1- \eps \hat z, \nonumber \\
\rho_w &\to 1 - 2 \sqrt{\eps} \sqrt{\hat z} r_w, ~~~\bar \rho_w \to 1 - 2 \sqrt{\eps} \sqrt{\hat z}  \bar r_w . 
\end{align}
This change of variables essentially zooms in to the $w, \bar w \sim {1 \over 1 - \bar z}$ region. We then take $\eps \to 0$. The integration measure (\ref{eq:integrationmeasure}) becomes $\int_0^{\infty} d r_w d \bar r_w K_B(r_w, \bar r_w)$
\begin{align}
&K_B(r_w, \bar r_w) =- {1 \over \pi}   {\bar r_w \over (2 + r_w)^{3/2} (r_w +2 \bar r_w)^{3/2}} \ _2 F_1 \Big({1 \over 2}, {3 \over 2}, 2, {r_w (2 + r_w + \bar r_w) \over (2 + r_w)^{3/2} (r_w +2 \bar r_w)^{3/2}} \Big) .
\end{align}
Note that all the dependence on $(z, \bar z)$ disappeared from the kernel. Similarly for the contact kernel in the same limit we get $\int_0^{\infty} d \bar r_w K_C(\bar r_w)$
\begin{align}
K_C(\bar r_w) = {1 \over \pi} {\sqrt{\bar r_w} \over 1 + \bar r_w}.
\end{align}

Doing the corresponding change of variables on ${\rm dDisc}[G(w, \bar w)]$ above we get
 \begin{align}
{\rm dDisc}[G(w, \bar w)]|_{\rho_w \to (\rho_z \bar \rho_z \bar \rho_w) \rho_w} =\eps^{p_1 + p_2} 2 \sin^2 (\pi p_2)  z^{p_1} \hat z^{p_2} \bar r_w^{2 p_2} .
\end{align}

We see that the double light-cone limit of the correlation function is manifestly reproduced. We still however need to check that the coefficient also matches. We get the following condition
 \begin{align}
2 \sin^2 (\pi p_2) \left(\int_0^{\infty} d r_w d \bar r_w K_B(r_w, \bar r_w) \bar r_w^{2 p_2} + \int_0^{\infty} d \bar r_w K_C(\bar r_w)  \bar r_w^{2 p_2}   \right)= 1 .
\end{align}
Convergence of the integral requires $p_2 > - {3 \over 4}$. This becomes
 \begin{align}
\int_0^{\infty} d r_w d \bar r_w K_B(r_w, \bar r_w) \bar r_w^{2 p_2} = {1 \over 2 \sin^2 (\pi p_2) \cos 2 \pi p_2} .
\end{align}
It is easy to check that this is indeed a true identity. 

Let us now consider a more general situation where we start with a four-point function with a bounded double discontinuity. We can bound the four-point function as follows
 \begin{align}
 \label{eq:boundcorrelDer}
| G(z, \bar z) | &< \int_0^1 d w d \bar w | K(z, \bar z , w , \bar w) | |{\rm dDisc}[G(w, \bar w)]| \\
&+ \int_0^1 d w d \bar w | K({z \over z-1}, {\bar z \over \bar z -1 } , w , \bar w) | |{\rm dDisc}[G({w \over w-1}, {\bar w \over \bar w - 1})]|
\end{align}

Let us start with the first term.The integral over $(w, \bar w)$ involves several dangerous regions. Light-cone limit, Regge limit $(w, \bar w \to 0)$, as well as the double light-cone limit discussed above. In writing (\ref{eq:dispnosub}) it has been already assumed that $G(w, \bar w)$ decays faster then $(w \bar w)^{1/2}$ in the Regge limit. We would like to assume that
\begin{align}
\label{eq:bounddiscToy}
| {\rm dDisc}[G(w, \bar w)] | \leq c_0 (w \bar w)^{p_1} ([1-w] [1-\bar w])^{p_2}, ~~~ , 0 < w, \bar w < 1 ,
\end{align}
and as mentioned above it is assumed that $ p_1 > {1 \over 2}$ and we are interested in situations where $p_2 < 0$.

In this way we can bound the first term as follows
 \begin{align}
 \int_0^1 d w d \bar w | K(z, \bar z , w , \bar w) | |{\rm dDisc}[G(w, \bar w)]| \leq c_0 \int_0^1 d w d \bar w | K(z, \bar z , w , \bar w)| (w \bar w)^{p_1} ([1-w] [1-\bar w])^{p_2} .
 \end{align}
 Therefore we simply need to bound the integral in the RHS. Since $K$ involves $K_B$ and $K_C$ which have different sign in the region of support we should bound them separately. In this way we shoulds make sure that there are no singular terms in the double light-cone limit that cancel between $K_B$ and $K_C$. Therefore it is enough to bound $K_C$. We can split the integral into two pieces
 \begin{align}
 \int_0^{1-\eps_0} d \bar w K_C(z, \bar z, \bar w) +  \int_{1-\eps_0}^{1} d \bar w K_C (z , \bar z , \bar w) ,
 \end{align}
 where $\eps_0$ is some fixed small number. The double light-cone limit of the second term  is controlled by the rescaled kernel that we analyzed above. In the first integral we can simply take 
the $z \to 0$ and $\bar z \to 1$ limit to get the asymptotic $z^{p_1} (1 - \bar z)^{- {1 \over 4}}$ behavior which is subleading for $p_{1}$ and $p_{2}$ within the range of validity of the unsubtracted dispersion relations that we discuss here. The conclusion of this discussion is that given the bound on the double discontinuity (\ref{eq:bounddiscToy}) we bounded
 \begin{align}
  \int_0^1 d w d \bar w | K(z, \bar z , w , \bar w) | |{\rm dDisc}[G(w, \bar w)]| \leq \tilde c_0 (z \bar z)^{p_1} ([1-z] [1-\bar z])^{p_2}, ~~~ , 0 < z, \bar z < 1 \ .
 \end{align}
 
 Next we would like to bound the $u$-channel contribution. For simplicity as in the main body of the paper we imagine that we are dealing with the crossing symmetric function so that $G(z, \bar z) = G({z \over z-1}, {\bar z \over \bar z -1})$ and the same for ${\rm dDisc}$. We need to therefore only consider the behavior of the $u$-channel kernel $K({z \over z-1}, {\bar z \over \bar z -1 } , w , \bar w)$ in the double light-cone limit $z \to 0$, $\bar z \to 1$. We observed that the effect of $\bar z \to  {\bar z \over \bar z -1 }$ is to make the kernel more regular in the double light-cone limit and to remove the enhanced scaling limit considered above. Therefore for crossing symmetric functions that satisfy the unsubtracted dispersion relations and satisfy (\ref{eq:bounddiscToy}) we conclude that
 \begin{align}
| G(z, \bar z) | \leq \tilde c_0 (z \bar z)^{p_1} ([1-z] [1-\bar z])^{p_2}, ~~~ , 0 < z, \bar z < 1 \ .
 \end{align}
In other words we are simply saying that the double light-cone limit of the double discontinuity controls the double light-cone limit of the correlator. 

To make the argument relevant we have to address two issues
\begin{itemize}
\item subtractions ;
\item the keyhole contour .
\end{itemize}

Doing subtractions are required to relax the condition $p_1 > {1 \over 2}$ above which is not satisfied by physical correlators which can behave as $p_1 = 0$ in the Regge limit. As in  \cite{Carmi:2019cub} to solve this problem we can simply consider ${z \bar z \over (1-z) (1-\bar z)} G(z, \bar z)$ correlator instead. This function is bounded as $z \bar z$ in the Regge limit and therefore the integral above will converge.

The second issue of the keyhole contour is related to relaxing the condition on $p_2$ which is required to ensure the convergence of the integral close to $\bar w = 1$. It also correctly captures the distributional terms localized at $\bar w =1$. In the example above we can could have defined the integrals in the vicinity of $\bar w = 1$ by analytically continuing in $p_2$ to get the correct result. We expect this to be true in general but it would be nice to show this more explicitly.  

Now we can easily understand the problem with the idea of using the dispersion relations to prove the bound on the double light-cone limit in the main body of the paper is the following. By doing more and more subtractions we will improve the light-cone limit of the correlator and the behavior of the double discontinuity in the scaling regime analyzed in this section. Dispersion relations however involve also an integral away from the scaling limit discussed above which in particular involves the Regge limit. By doing subtractions that we discussed in the main text we improve the light-cone and the double light-cone limits of the double discontinuity. We however do not claim to improve the Regge limit. Therefore in deriving the bounds above (\ref{eq:boundcorrelDer}) we will always have a universal contribution from the Regge limit or more generally the region away from the scaling regime discussed in this section. This universal contribution estimate does not improve as we make further subtractions and therefore we cannot use dispersion relations in a simple and direct way to prove that improved bound on the double discontinuity in the double light-cone limit implies a better behavior of the correlator itself.

\section{Graviton Exchange in AdS}\label{sec:gravexchange}

In section (\ref{sec: contour}) we predicted the subtractions needed to have a Mellin representation with a straight contour, see formula (\ref{subtractions finite}). In this section we consider a simple example with nontivial subtractions and check that (\ref{subtractions finite}) is obeyed. 

We consider a four-point correlator of four identical scalars with dimension $\Delta = 4$. The connected correlator $\frac{F_{conn}(u,v)}{x_{13}^8 x_{24}^8}$ is given by a sum of three graviton exchange diagrams. 

Based on the OPE (\ref{eq:poles}), we expect the Mellin amplitude to have poles at $\gamma_{12}, \gamma_{13}, \gamma_{14} = 3-n$, where $n$ is a non-negative integer. Since the crossing-symmetric point is at $(\gamma_{12}=\frac{4}{3}, \gamma_{14}=\frac{4}{3})$ formula (\ref{subtractions finite}) predicts that
\begin{eqnarray}\label{subtractions graviton}
&&F_{sub}(u,v) = F_{conn}(u, v) - C_{OO T_{\mu \nu}}^2 \sum_{m=0}^1 \Big( u^{m-3} g^{(m)}_{\tau=4, J=2}(v) + v^{m-3} g^{(m)}_{\tau=4, J=2}(u) + v^{-m-1} g^{(m)}_{\tau=4, J=2} (\frac{u}{v})  \Big)  \ , \nn \\
&&g^{(0)}_{\tau=4, J=2}(v) =\frac{30 (-3 v^2+(v^2+4 v+1) \log v+3)}{(v-1)^3} \ , \nn \\
&&g^{(1)}_{\tau=4, J=2}(v) =\frac{30 ((v^3+11 v^2+11 v+1) \log v -4 (v^3+3 v^2-3 v-1))}{(v-1)^5} \ ,
\end{eqnarray}
has a Mellin representation with a straight contour at $\gamma_{12}=\frac{4}{3}, \gamma_{14}=\frac{4}{3}$.  

The relevant Mellin amplitude was computed in \cite{Penedones:2010ue}, see formula (8). We then have
\begin{eqnarray}\label{graviton three integrals}   
F_{conn}(u, v)= \int \int_{\mathcal{C}_1} \frac{d \gamma_{12} d\gamma_{14}}{(2 \pi i)^2} \Gamma(\gamma_{12})^2 \Gamma(\gamma_{13})^2 \Gamma(\gamma_{14})^2 u^{- \gamma_{12}} v^{- \gamma_{14}}m(\gamma_{12}, \gamma_{14})\\
+\int \int_{\mathcal{C}_2} \frac{d \gamma_{12} d\gamma_{14}}{(2 \pi i)^2} \Gamma(\gamma_{12})^2 \Gamma(\gamma_{13})^2 \Gamma(\gamma_{14})^2 u^{- \gamma_{12}} v^{- \gamma_{14}}m(\gamma_{13}, \gamma_{14}) \nonumber \\
+\int \int_{\mathcal{C}_3} \frac{d \gamma_{12} d\gamma_{14}}{(2 \pi i)^2} \Gamma(\gamma_{12})^2 \Gamma(\gamma_{13})^2 \Gamma(\gamma_{14})^2 u^{- \gamma_{12}} v^{- \gamma_{14}}m(\gamma_{14}, \gamma_{12}) \nonumber
\end{eqnarray}
where
\begin{eqnarray}
m(\gamma_{12}, \gamma_{14})= \frac{1}{2} \frac{(-\gamma_{12}-\gamma_{14}+4)^2-2 \gamma_{14} (-\gamma_{12}-\gamma_{14}+4)+\gamma_{14}^2-1}{\gamma_{12}-1}  \\
+\frac{4 (-2 (-\gamma_{12}-\gamma_{14}+4)-\gamma_{12}+4)^2}{\gamma_{12}-2} \nonumber \\
+\frac{3 (-2 (-\gamma_{12}-\gamma_{14}+4)-\gamma_{12}+4)^2+1}{-3+ \gamma_{12}}-\frac{15 \nonumber (\gamma_{12}-4)}{2}-\frac{55}{2}.
\end{eqnarray}
The contours $\mathcal{C}_1$, $\mathcal{C}_2$ and $\mathcal{C}_3$ are all straight and different from each other. We can take
\begin{eqnarray}
\mathcal{C}_1: 3<\Re(\gamma_{12})<4 ,&& 0<\Re(\gamma_{14})<4-\Re(\gamma_{12})-\Re(\gamma_{14}) \\
\mathcal{C}_2: 3<\Re(\gamma_{14})<4 ,&& 0<\Re(\gamma_{12})<4-\Re(\gamma_{12})-\Re(\gamma_{14}) \nonumber \\
\mathcal{C}_3: 3<\Re(\gamma_{13})<4 ,&& 0<\Re(\gamma_{14})<\Re(\gamma_{12}). \nonumber 
\end{eqnarray}

Bringing different contours together we get
\begin{eqnarray}
\label{eq:resultMellinExch}
F_{sub}(u,v) = \int \int_{\Re(\gamma_{12}), \Re(\gamma_{14})=\frac{4}{3}} \frac{d \gamma_{12} d\gamma_{14}}{(2 \pi i)^2} \Big(m(\gamma_{12}, \gamma_{14})  + m(\gamma_{13}, \gamma_{14}) + m(\gamma_{14}, \gamma_{12}) \Big) \ .
\end{eqnarray}
To match (\ref{eq:resultMellinExch}) with (\ref{subtractions graviton}) we used $C_{OO T_{\mu \nu}}^2= \frac{8}{15}$.

\section{Heavy Tails in Dispersion Relations}\label{sec:heavytails}

In the main text we discussed several functionals. Here we would like to comment on the convergence of the corresponding OPE sums at large $\Delta$. The relevant formulae for the asymptotic of the OPE coefficients can be found in \cite{Mukhametzhanov:2018zja}. 

For example, let us fix $J=0$. In this case we have
\begin{eqnarray}
\alpha_{\tau,0} = \sum_{m=0}^\infty  {1 \over m!} {2  \Gamma(\tau) \over \Gamma({\tau \over 2})^4 \Gamma(\Delta - m - {\tau \over 2})^2 (\tau - {d \over 2} +1)_m } {\tau - \Delta + 2 m \over (\tau - {4 \Delta \over 3} + 2m)^2 (\tau - {2 \Delta \over 3} + 2m)^2} .
\end{eqnarray}

We are interested in the asymptotic of this sum when $\tau \to \infty$. 

At large $\tau$ the density of primaries multiplied by their three-point couplings asymptotes to \cite{Mukhametzhanov:2018zja} (strictly speaking this asymptotic is only true on average)
\beq
\lim_{\tau \to \infty} \rho_J^{primary} C_J^2 \sim 4^{- \tau} \tau^{4 \Delta - {3 d \over 2}} .
\eeq

Combining this with $\alpha_{\tau, 0}$ we find that
\beq
\lim_{\tau \to \infty} \rho_0^{primary} C_0^2 \alpha_{\tau,0} \sim {1 \over \tau^{d+2}} \ ,
\eeq
and therefore the sum over the heavy scalar tail converges.  Note that the power does not depend on the dimension of the external operator.
Repeating the exercise for $J=2$ we get the same power law behavior. We expect that the same holds for any finite $J$ but we do not prove it here.

\section{Holographic Calculations} \label{sec:HoloCalc}

In this appendix we spell out the terms that enter into (\ref{eq:holoscalar}) and (\ref{eq:holostress}). The functionals $\alpha_{\tau, J, m}$ were written explicitly in (\ref{eq:magicfunctionals}), up to the expressions for the Mack polynomials. In the cases of interest,
\begin{align}
{\cal Q}^{\tau, d}_{J=0,m}(\gamma_{13}) = -\frac{2 \Gamma (\tau )}{\Gamma (m+1)
   \Gamma \left(\frac{\tau}{2}\right)^4 \left(-\frac{d}{2}+\tau +1\right)_m \Gamma \left(-m+\Delta -\frac{\tau}{2}\right)^2}, \\
{\cal Q}^{\tau, d}_{J=2,m}(\gamma_{13})= - \Big(  4 \gamma_{13}^2 -2 \gamma_{13} (2 m+\tau) \\ 
+\frac{m^2 \left(d^2 (\tau +1)-d\left(\tau ^2+5 \tau+4\right)+(\tau +2)^2\right)}{d(\tau +1) (d-\tau -3)}+\frac{\tau^2 (\tau +2)}{4 (\tau +1)} \Big) \nonumber \\
\times \frac{(\tau +1) (\tau +2) \Gamma (\tau+4)}{2 \Gamma (m+1) \Gamma \left(\frac{\tau +4}{2}\right)^4 \left(-\frac{d}{2}+\tau +3\right)_m \Gamma \left(-m+\Delta -\frac{\tau}{2}\right)^2}. \nonumber
\end{align}

Regarding the leading Regge trajectory, we can use 
\beq
\alpha_{\tau_{[{\cal O}, {\cal O}]_{0,J}},J} = \frac{d \alpha_{\tau = 2\Delta,J}}{d\tau} \gamma_{[{\cal O}, {\cal O}]_{0,J}} + O(1/c_T^2).
\eeq
Furthermore, only the $m=0$ term contributes to $\frac{d \alpha_{\tau = 2\Delta,J}}{d\tau}$. For $m=0$ a general expression for Mack polynomials is given by (\ref{Macks m0}). In this manner one obtains
\begin{align}
\frac{d \alpha_{\tau = 2\Delta,J}}{d\tau}= -\frac{\Gamma (2 (J+\Delta )) _3 F_2^{(\{0,1,0\},\{0,0\},0)} ( -J, \frac{\Delta }{3},2 \Delta +J-1 ; \Delta ,\Delta ;1 )}{\Gamma (\Delta )^2 \Gamma (J+\Delta )^2},
\end{align}
where the superscript in the hypergeometric function means a derivative with respect to the appropriate entry.

 As to the anomalous dimensions, for the exchange of a scalar we used expressions $(2.35)$, $(2.36)$ and $(2.37)$ in \cite{Albayrak:2019gnz}.\footnote{See also \cite{Cardona:2018qrt, Cardona:2018dov} for a similar discussion for a generic exchange of a spin $J$ primary.} For the stress tensor exchange in $d=4$, \cite{Alday:2017gde}
 \begin{align}
\gamma_{[{\cal O}, {\cal O}]_{0,J}} = - C_{\mathcal{O} \mathcal{O} T}^2 \frac{60 (\Delta -1)^2}{(J+1) (2
   \Delta +J-2)}, ~~~ J>2,  \\
   \gamma_{[{\cal O}, {\cal O}]_{0,J=2}} = - C_{\mathcal{O} \mathcal{O} T}^2 \frac{10 \left(-4 \Delta ^4+9 \Delta
   ^2+7 \Delta -12\right)}{\Delta  (2
   \Delta +1) (2 \Delta +3)},
 \end{align} 
 where $C_{\mathcal{O} \mathcal{O} T}^2$ is the OPE coefficient between the two external scalars and the stress tensor.\footnote{This OPE coefficient is fixed by Ward identities, but actually it is not necessary to know it explicitly in order to determine the sign of the functional, since it enters both in the direct exchange and in the anomalous dimensions of the double twist operators.} In general $d$ we use the results of \cite{Costa:2014kfa}\footnote{Formula (\ref{gamma 3D}) differs from $(172)$ in \cite{Costa:2014kfa} by a factor of $\frac{1}{2}$. This comes from the fact that we consider the exchange of identical scalar operators.}
 \begin{align}\label{gamma 3D}
\gamma_{[{\cal O}, {\cal O}]_{0,J}} = - \int_{-i \infty + c_1}^{+ i \infty + c_1} \frac{dt}{4 \pi i} M(s=0, t)  \Gamma(\frac{t}{2})^2 \Gamma(\frac{-t}{2} + \Delta)^2 \\
 \times _3 F_2 (-J, J+ 2 \Delta -1, \frac{t}{2}; \Delta, \Delta; 1), ~~ 0 < c_1 < 2 \Delta. \nonumber
\end{align}
To get the complete result graviton exchange diagrams in three channels should be added. These can be easily obtained by applying crossing to the result $(164-166)$ in \cite{Costa:2014kfa}. Two of the exchange diagrams produce results that are identical and analytic in spin. The third one only contributes to the anomalous dimension $\gamma_{[{\cal O}, {\cal O}]_{0,J}}$ for $J=0,2$.
  
For a generic exchange of a single trace operator of twist $\tau$ and spin $J$, our sum rule is valid for $\frac{d-2}{2} < \Delta < \frac{3 \tau}{4}$. It is interesting to study the behaviour of the sum rule when we take $\Delta \rightarrow \frac{3 \tau}{4}$. There are two terms that diverge like $\frac{1}{(\Delta-\frac{3 \tau}{4} )^2}$. One term comes from the direct exchange of the single trace operator. The other comes from the tail of the leading Regge trajectory. The sum of the two terms is equal to
\begin{align}\label{double pole}
\frac{1}{(\Delta- \frac{3}{4}\tau)^2} 
\frac{9 \Gamma (2 J+\tau )}{2 \Gamma \left(\frac{\tau
   }{4}\right)^2 \Gamma \left(\frac{\tau }{2}\right)^2
   \Gamma \left(J+\frac{\tau }{2}\right)^2}
   \left[
    \,   _3F_2\left(-J,\frac{\tau  }{4},J+\tau -1;\frac{\tau   }{2},\frac{\tau }{2};1\right)
   -1
   \right]
\end{align}
When $J=0$, (\ref{double pole}) vanishes. This agrees with the fact that a scalar exchange in AdS does not contribute to the sum rule. When $J>0$ we numerically find that (\ref{double pole}) is negative. Furthermore we checked this analytically for spins $J=2, 4$, ... $50$ and any positive $\tau$. Note that this implies that when $\Delta \rightarrow  \frac{3 \tau}{4}$ there is a UV contribution to the sum rule which is divergent and positive. It would be very interesting to understand the origin of this.


\bibliography{Mellin}
\bibliographystyle{utphys}

\end{document}